\documentclass[onecolumn]{aa}
\usepackage{txfonts}
\usepackage{natbib}
\bibpunct{(}{)}{;}{a}{}{,}

\usepackage[english]{babel}
\usepackage[dvips]{graphicx}

\begin{document}

\title{
  Residual noise covariance for {\sc Planck} low-resolution data analysis
}
\author{
  R.~Keskitalo\inst{1,2}
  \and
  M.A.J.~Ashdown\inst{3,4}
  \and
  P.~Cabella\inst{5,6}
  \and
  T.~Kisner\inst{7}
  \and
  T.~Poutanen\inst{1,2,8}
  \and
  R.~Stompor\inst{9}
  \and
  J.G.~Bartlett\inst{9}
  \and
  J.~Borrill\inst{7,10}
  \and
  C.~Cantalupo\inst{7}
  \and
  G.~de~Gasperis\inst{6}
  \and
  A.~de~Rosa\inst{11}
  \and
  G.~de~Troia\inst{6}
  \and
  H.K.~Eriksen\inst{12,13}
  \and
  F.~Finelli\inst{11,14,15}
  \and
  K.M.~G\'orski\inst{16,17}
  \and
  A.~Gruppuso\inst{11,14}
  \and
  E.~Hivon\inst{18}
  \and
  A.~Jaffe\inst{19}
  \and
  E.~Keih\"anen\inst{1}
  \and
  H.~Kurki-Suonio\inst{1,2}
  \and
  C.R.~Lawrence\inst{16}
  \and
  P.~Natoli\inst{6,20}
  \and
  F.~Paci\inst{21}
  \and
  G.~Polenta\inst{5,22,23}
  \and
  G.~Rocha\inst{16}
}
\institute{
  University of Helsinki, Department of Physics,
  P. O. Box 64, FIN-00014 Helsinki, Finland.
  \and
  Helsinki Institute of Physics,
  P. O. Box 64, FIN-00014 Helsinki, Finland.
  \and
  Astrophysics Group, Cavendish Laboratory,
  J J Thomson Avenue, Cambridge CB3 0HE, United Kingdom.
  \and
  Institute of Astronomy, Madingley Road,
  Cambridge CB3 0HA, United Kingdom.
  \and
  Dipartimento di Fisica, Universit\'a di Roma ``La Sapienza'',
  p.le A. Moro 2, I-00185 Roma, Italy
  \and
  Dipartimento di Fisica, Universit\'a di Roma ``Tor Vergata'',
  via della Ricerca Scientifica 1, I-00133 Roma, Italy.
  \and
  Computational Cosmology Center,
  Lawrence Berkeley National Laboratory, Berkeley CA 94720, U.~S.~A.
  \and
  Mets\"ahovi Radio Observatory, Helsinki University of Technology,
  Mets\"ahovintie 114, 02540 Kylm\"al\"a, Finland
  \and
  Laboratoire Astroparticule \& Cosmologie (APC) - UMR 7164, CNRS,
  Universit\'e Paris Diderot
  10, rue A. Domon et L. Duquet, 75205 Paris Cedex 13, France 
  \and
  Space Sciences Laboratory,
  University of California Berkeley,
  Berkeley CA 94720, U.~S.~A.
  \and
  INAF-IASF Bologna, Istituto di Astrofisica Spaziale e Fisica Cosmica
  di Bologna
  Istituto Nazionale di Astrofisica, via Gobetti 101, I-40129 Bologna, Italy 
  \and
  Institute of Theoretical Astrophysics,
  University of Oslo, P.O.\ Box 1029 Blindern, N-0315 Oslo, Norway
  \and
  Centre of Mathematics for Applications, University of
  Oslo, P.O.\ Box 1053 Blindern, N-0316 Oslo, Norway
  \and
  INFN, Sezione di Bologna,
  Via Irnerio 46, I-40126 Bologna, Italy
  \and
  INAF-OAB, Osservatorio Astronomico di Bologna
  Istituto Nazionale di Astrofisica, via Ranzani 1, I-40127 Bologna, Italy
  \and
  Jet Propulsion Laboratory, California Institute of Technology,
  4800 Oak Grove Drive, Pasadena CA 91109, U.~S.~A
  \and
  Warsaw University Observatory,
  Aleje Ujazdowskie 4, 00478 Warszawa, Poland
  \and
  Institut d'Astrophysique de Paris,
  98 bis Boulevard Arago, 75014 Paris, France
  \and
  Department of Physics, Blackett Laboratory, Imperial College London,
  South Kensington campus, London, SW7 2AZ, United Kingdom.
  \and
  INFN, Sezione di ``Tor Vergata'',
  Via della Ricerca Scientifica 1, I-00133 Roma, Italy
  \and
  Instituto de Fisica de Cantabria,
  CSIC-Universidad de Cantabria, Avda. Los Castros s/n, 39005 Santander, Spain
  \and
  ASI Science Data Center, c/o ESRIN,
  via G. Galilei snc, I-00044 Frascati, Italy
  \and
  INAF-Osservatorio Astronomico di Roma,
  via di Frascati 33, I-00040 Monte Porzio Catone, Italy
}

\date{Received date / Accepted date}

\abstract{
  Cosmic microwave background (CMB) data analysis
}{
  Develop and validate tools to estimate residual noise covariance
  in {{\sc Planck}} frequency maps. Quantify signal error effects
  and compare different techniques to produce low-resolution maps.
}{
  We derive analytical estimates of covariance of the residual
  noise contained in low-resolution maps produced using a number of
  map-making approaches. We test these analytical predictions using
  Monte Carlo simulations and their impact on angular power
  spectrum estimation. We use simulations to quantify the level of 
  signal errors incurred in different resolution downgrading schemes
  considered in this work.
}{
  We find an excellent agreement between the optimal residual noise
  covariance matrices and Monte Carlo noise maps. For destriping 
  map-makers, the extent of agreement is dictated by the knee frequency
  of the correlated noise component and the chosen baseline offset length.
  The significance of signal striping is shown to be insignificant when
  properly dealt with. In map resolution downgrading, we find that
  a carefully selected window function is required to reduce aliasing to
  the sub-percent level at multipoles, $\ell>2N_\mathrm{side}$,
  where $N_\mathrm{side}$ is the HEALPix resolution parameter.
  We show that sufficient characterization of the residual noise
  is unavoidable if one is to draw reliable contraints on large scale
  anisotropy.
}{
  We have described how to compute the low-resolution maps, with a
  controlled sky signal level, and a reliable estimate of covariance
  of the residual noise. We have also presented a method to smooth the
  residual noise covariance matrices to describe the noise correlations
  in smoothed, bandwidth limited maps.
}

\keywords{
  cosmic microwave background --- Cosmology:observations ---
  Methods:data analysis --- Methods:numerical
}

\maketitle

\section{Introduction}

Over the last two decades observations of the cosmic microwave background
(CMB) have led the way towards the high precision cosmology of today ---
a process best emphasized recently by the high quality data set delivered
by the {\sc WMAP} satellite.
The next major and nearly imminent step in a continuing exploitation of the 
CMB observable will be data analysis of data sets anticipated from another
satellite mission, {\sc Planck}. {\sc Planck} will observe
the entire sky in multiple frequency channels, promising to improve over the
recent {\sc WMAP} constraints on many fronts. In particular {\sc Planck},
as a satellite, will provide us with a unique access to the largest
angular scales, in which the total intensity has proven controversial
and difficult for theoretical interpretation and is still poorly measured
and exploited in the polarization. {\sc Planck} will be the only CMB satellite
deployed in the next decade. It is therefore particularly important that the
constraints at large angular scale derived from the anticipated {\sc Planck}
data are not only robust but also efficiently exploit the information
contained in them. This will be certainly necessary if {\sc Planck} is to
set strong constraints on the CMB B-mode power spectrum \citep{Efstathiou2009} 
--- one of the most attractive potential science targets of the mission.
 
The analysis of constraints on the largest angular scales requires robust
statistical estimators accounting for a proper description of the
statistical properties of the sought-after sky signal, instrumental noise
and other residuals due to the instrument, astrophysical signals and/or
data processing. This paper focuses on two of those ingredients ---
instrumental noise and so-called pixel noise, the latter due to residual
sky power on sub-pixel scales. In the standard data analysis pipeline
the measured time ordered data are first projected onto the sky, an 
operation called map-making, producing map-like estimates of the sky signal,
which are subsequently analyzed, e.g., in order to derive constraints upon the power spectrum. The map-making process is usually understood to be a linear
operation on the input, measured data and therefore a statistical uncertainty
of the produced sky maps can be straightforwardly obtained given known
characteristics of the time-domain data. Those also usually involve
assumptions about piece-wise stationarity of the instrumental noise, assumed
to conform with Gaussian statistics. In the realm of the {\sc Planck}
analysis such a straightforward route is not however plausible. This is
because of the high resolution of the {\sc Planck} instruments, the full
sky coverage and length of the mission combined with the high sampling rate
of the sky signals. That leads to the data set which is large in terms
of the numbers of both the sky pixels contained in the maps and the directly
registered measurements. The map-making procedures developed in the context
of {\sc Planck} \citep{Poutanen:2005yg, Ashdown:2007ta, Ashdown:2006ey,
  Ashdown:2008ah} have been demonstrated to be capable of dealing with 
the expected volumes of the data, producing high-quality maps; nevertheless
the calculation or even just storage of their respective noise covariance
matrices at their full resolution is beyond the limits of even the largest
currently available supercomputing facilities. This is because, unlike
maps --- sizes of which scale linearly with a number of pixels, $N_\mathrm{pix},$
--- the noise covariance matrices scale as a square of it and their
inversion involves ${\cal O}(N_\mathrm{pix}^3)$ floating point operations.
We emphasize that due to a combination of its scanning strategy and
noise-like contributions correlated over long periods of time, we expect
that non-negligible large scale noise correlations will be present in the
maps derived from the {\sc Planck} data and will be particularly important in
the analysis of the polarized signals given their lower amplitudes.

In this paper we develop tools necessary for the statistically sound analysis
of constraints on large angular scales. These include robust approaches to 
producing low-resolution maps and techniques for estimating pixel-pixel
correlations due to their residual noise.  The low-resolution
maps are expected to compress  nearly all the information relevant to the large
angular scale in fewer pixels and are therefore more readily manageable.  
Given our Gaussian noise assumption the full statistical description of the map
uncertainty is given by its pixel-pixel noise covariance matrix (NCM).
This is defined as
\begin{equation} \label{eq:ncvm}
  \vec N
  =\left\langle
    \left(\hat{\vec s}-\vec s\right)
    \left(\hat{\vec s}-\vec s\right)^\mathrm T
  \right\rangle,
  \quad\textrm{where}\quad
  \left\langle\left(\hat{\vec s}-\vec s\right)\right\rangle = 0,
\end{equation}
and $\vec s$ is the $3N_\mathrm{pix}$ input sky map of Stokes I, Q, and U
parameters and $\hat{\vec s}$ is our estimate of $\vec s$. In the absence
of signal errors, the difference, $\left(\hat{\vec s}-\vec s\right)$,
contains only instrument noise. We note that $\vec N$ is a symmetric and
usually dense matrix, which in general will depend on the map-making method
that produced the estimate. In the following we will consider a number of
numerical approaches to calculate such a matrix for each of the studied
low-resolution maps and then test their consistency with the actual noise
in the derived maps.

The full noise covariance matrices have been commonly computed and used in
the analysis of the small-scale balloon-borne,
\citep[e.g.][]{Hanany2000,deBernardis2000} and ground-based experiments,
\citep[e.g.,][]{Kuo2004}. The {\sc COBE}-DMR team also used them to derive
low-$\ell$ constraints, \citep[ e.g.,][]{Gorski1996,Wright1996}. In all
those cases, however, no resolution downgrading has been required, unlike
with {\sc Planck}, as the calculations for those experiments could be done
at a full resolution. To this date, only the {\sc WMAP} team has encountered
a similar problem. The instrument noise model employed by them is in fact
similar to the one used for {\sc Planck}. It is
parametrized, however, in the time domain rather than in the frequency domain
\citep{Jarosik:2003fe,Jarosik:2006ib}. Calculation of the {\sc WMAP} NCM is
formulated in exactly the same manner as for our optimal map-making
method and the {\sc WMAP} likelihood code\footnote{\tt
  http://lambda.gsfc.nasa.gov/} ships with an NCM very similar to what
we present here, although without the II, IQ and IU covariance
blocks. The simplification is motivated by the high S/N ratio of
the low temperature multipoles and weak coupling between temperature
and polarization pixel noise.

Our analysis is made unique by the differences in the experiment design: 
{\sc WMAP} pseudo-correlation receivers are differencing assemblies (DA)
with two mirrors, whereas {\sc Planck} will use a single mirror design
(HFI, the high frequency instrument) or has a reference load in place
of the second mirror (LFI, the low frequency instrument)
\citep{Planck:2006uk}. Between these, the pixel-pixel
correlations are different. In principle the balanced load systems
of {\sc COBE} and {\sc WMAP} should bring less correlated noise than the
unbalanced {\sc Planck} LFI. On the other hand, differencing experiments
generate pixel-pixel noise correlations even from white noise,
whereas in {\sc Planck} they originate from the correlated noise alone. 
In addition we also study here the so-called destriping algorithms, which have
been proposed as a {\sc Planck}-specific map-making approach 
\citep{Delabrouille1998, Maino2002, Keihanen2004a, Keihanen:2004yj}.

Residual noise covariance for {\sc Planck}-like scanning and instrument
noise has been studied before, either via some simplified toy-models
\citep{StomporWhite2004} or in more realistic circumstances
\citep{Efstathiou:2004eu,Efstathiou:2006wt}. Those studies approached the
problem in a semi-analytic way and thus needed to employ some simplifications,
which we avoid in our work. They were also based solely on the
destriping approximation, assuming that noise can be accurately modeled by
relatively long (one hour) baseline offsets and white noise, and did not
consider any other approaches. In this work we extend
those analyses into cases where modeling the noise correlations requires
shorter baselines and compare those with optimal solutions.

\section{Algebraic background}

\subsection{Maps and their covariances} \label{sec:maps}

To formulate map-making as a maximum likelihood problem we start with a
model of the timeline:
\begin{equation}
  \label{eq:tod}
   \vec d = \vec A\vec s + \vec B\vec x + \vec n,
\end{equation}
where the underlying microwave sky signal, $\vec s$, is to be estimated.
Here $\vec A$ is the pointing matrix, which encodes how the sky
is scanned and $\vec n$ is a Gaussian, zero mean noise vector. 
$\vec x$ denotes some extra instrumental effects, usually taken
hereafter to be constant baseline offsets, which we will use to model the
correlated part of the instrumental noise and $\vec B$ --- a `pointing'
matrix for $\vec x$ describing how it is added to the time domain data.
Convolution with an instrumental beam, assumed here to be axially symmetric, 
is already included in $\vec  s$.
 
The signal part of the uncalibrated data vector, $\vec d$, is the detector
response to the sky emission observed in the direction of pixel $p$. For a
total power detectors, e.g., the ones on {\sc Planck}, it is a linear 
combination of polarized and unpolarized contributions:
\begin{equation}
  \label{eq:det_response}
  d_t = \mathcal K\left\{
    (1+\epsilon)s_{p\mathrm I} +
    (1-\epsilon)\left(
      s_{p\mathrm Q}\cos (2\chi_t)
      + s_{p\mathrm U}\sin (2\chi_t)
    \right)
  \right\}
  + n_t,
\end{equation}
where it is implied that sample $t$ is measured in pixel $p$ and $\chi_t$
is a detector polarization angle with respect to the polarization basis and
we have dropped the baseline term for simplicity. Eq.~\ref{eq:det_response}
includes an overall calibration factor $\mathcal K$, and a cross polar
leakage factor $\epsilon$,  however, in what follows, we only consider the
case of perfect calibration, setting ${\mathcal K} = 1$ with no loss of generality, and no leakage,
$\epsilon = 0$.

To simplify future considerations we introduce a generalized pointing
matrix, $\vec A'$, and a generalized map, $\vec s'$. They are defined as
\begin{equation}
  \vec A' \equiv \left[ \vec A, \vec B\right],
  \quad\mathrm{and}\quad
  \vec s' \equiv \left[ 
    \begin{array}{c}
      \vec s\\
      \vec x
    \end{array}
  \right].
\end{equation}
Using those we can rewrite our data model in a more common form,
\begin{equation}
  \label{eq:todprime}
  \vec d = \vec A'\vec s' + \vec n.
\end{equation}
The detector noise has a time-domain covariance matrix $\mathcal
N=\langle \vec n \vec n^\mathrm T \rangle$ and the probability for
the observed timeline, $\vec d$, becomes the Gaussian probability of
a noise realization $\vec n = \vec d - \vec A' \vec s'$:
\begin{equation}
  \label{eq:map_prob}
  P(\vec d) = P(\vec n) = 
  \left[(2\pi)^{\mathrm{dim}\,\vec n}\det\mathcal N\right]^{-1/2}
  \exp \left(
    -\frac{1}{2}{\vec n}^\mathrm T\mathcal N^{-1}\vec n
  \right) P\left(\vec x\right),
\end{equation}
where the last factor is a prior constraint on the noise offsets, $\vec x$,
which hereafter we will take to be a Gaussian with a zero mean and 
some correlation matrix, ${\mathcal P}$, i.e.,
\begin{equation}
  P\left(\vec x\right) \propto \exp\left(
    -{1\over2}\vec x^\mathrm T\mathcal P^{-1}\vec x
  \right)
\end{equation}
By maximizing this likelihood with respect to the sky signal and baselines
contained in  $\vec s'$, we find an expression for a maximum likelihood
estimate which reads,
\begin{equation}
  \label{eq:min_var_map}
  \hat{\vec s}' = 
  \left( 
    \mathcal R^{-1} + {\vec A'}^\mathrm T\mathcal N^{-1}\vec A'
  \right)^{-1}{\vec A'}^\mathrm T\mathcal N^{-1}\vec d,
\end{equation}
where $\mathcal R^{-1}$ is defined as,
\begin{equation}
  \mathcal R^{-1} \equiv
  \left[
    \begin{array}{c c}
      \vec 0 & \vec 0\\
      \vec 0 & \mathcal P^{-1}
    \end{array}
  \right]
\end{equation}
The first part of the vector $\hat{\vec s}'$ is an estimate of the actual
sky signal, $\hat{\vec s}$, while all the rest is an estimate of the baseline
offsets, $\hat{\vec x}$. The map $\vec A'^{\mathrm T}\mathcal N^{-1}\vec d$ is
called the noise-weighted map. In case of flat prior for $\vec x$, expressions
identical to Eq.~(\ref{eq:min_var_map}) can be also derived from minimum
variance or generalized least square considerations and we will refer to
$\hat{\vec s}$ as either a minimum variance or optimal map in the future.
We note that we have ignored here any pixelization effects that cause
differences between $\vec d$ and $\vec A'\vec s'$ even for a noise free
experiment. This is usually true in the limit of the pixel size significantly
smaller than the beam resolution of the instrument. If this condition is not
fulfilled, the pixelization effects may be important and special methods
may be needed to minimize them. We discuss specific proposals in
Sect.~\ref{downgrading}. In the absence of such effects, the difference
between a map estimate, $\hat{\vec s}'$, and the input map, $\vec s'$, is
called \emph{residual pixel domain noise}.

Let us now consider first the prefactor matrix in Eq.~(\ref{eq:min_var_map}), 
\begin{equation}
  \vec M'\equiv
  \left({\mathcal R}^{-1}
  + {\vec A'}^\mathrm T\mathcal N^{-1}\vec A'\right)^{-1},
\end{equation}
a weight matrix combining both the baseline prior and the noise variance
weights. It acts on the generalized noise-weighted map, producing estimates
of the pixels and baselines.

Given that our generalized map is made of two parts: the actual sky signal
and the baseline offsets, the matrix $\vec M'$ has four blocks: two diagonal
blocks, denoted $\vec M$ and $\vec M_\mathrm x$, and two off-diagonal blocks,
each of which is a transposed version of the other and one of which is
referred here to as  $\vec M_\mathrm o$. Using inversion by partition we can
write an explicit expression for each of these blocks. For example, for the
sky-sky diagonal blocks we get,
\begin{eqnarray}
  \vec M & = &
  \left[\vec A^\mathrm T\mathcal N^{-1}\vec A 
    -\left(\vec A^\mathrm T\mathcal N^{-1}\vec B\right)
    \left(\mathcal P^{-1}+\vec B^\mathrm T\mathcal N^{-1}\vec B\right)^{-1}
    \left(\vec A^\mathrm T\mathcal N^{-1}\vec B\right)^\mathrm T
  \right]^{-1}
  \label{eq:MmatSS0}
  \\
  & = & 
  \left(\vec A^\mathrm T\mathcal N^{-1}\vec A\right)^{-1}
  +\left(\vec A^\mathrm T\mathcal N^{-1}\vec A\right)^{-1}
  \left(\vec A^\mathrm T\mathcal N^{-1}\vec B\right)
  \vec M_\mathrm x
  \left(\vec A^\mathrm T\mathcal N^{-1}\vec B\right)^\mathrm T
  \left(\vec A^\mathrm T\mathcal N^{-1}\vec A\right)^{-1},
  \label{eq:MmatSS}
\end{eqnarray}
while for the offset-offset part,
\begin{equation}
  \vec M_\mathrm x = 
  \left[
      \mathcal P^{-1}
      + \vec B^\mathrm T\mathcal N^{-1}\vec B
      - \left(\vec B^\mathrm T\mathcal N^{-1}\vec A\right)
      \left(\vec A^\mathrm T\mathcal N^{-1}\vec A\right)^{-1}
      \left(\vec B^\mathrm T\mathcal N^{-1}\vec A\right)^\mathrm T
  \right]^{-1}.
  \label{eq:MmatXX}
\end{equation}
With help of these equations we can now write explicit separate expressions
for the estimated sky signal and offsets. The former is given by,
\begin{equation}
  \hat{\vec s} = 
  \left(
    \vec M\vec A^\mathrm T + \vec M_\mathrm oB^\mathrm T
  \right)\mathcal N^{-1}\vec d
\label{eq:destripeMap1}
\end{equation}
while the latter,
\begin{equation}
  \hat{\vec x} =
  \left(
    \vec M_\mathrm o^\mathrm T \vec A^\mathrm T + \vec M_\mathrm xB^\mathrm T
  \right)\mathcal N^{-1}\vec d.
\end{equation}
We can also combine these two equations to derive an alternative expression
for the sky signal estimate, which makes a direct use of the offsets
assumed to be estimated earlier,
\begin{equation}
  \hat{\vec s}
  = \left(\vec A^\mathrm T\mathcal N^{-1}\vec A\right)^{-1}
  \left(\vec A^\mathrm T\mathcal N^{-1}\vec d
    - \vec A^\mathrm T\mathcal N^{-1}\vec B\hat{\vec x}
  \right)
\label{eq:destripeMap2}
\end{equation}

If the assumed data model, Eq.~(\ref{eq:tod}), and the time domain 
noise and baseline covariances are all correct, then the covariance of the
residual pixel domain noise is
\begin{equation}
  \label{eq:residual_covariance}
  \vec N' = 
  \langle (\hat{\vec s}'-\vec s')(\hat{\vec s}'-\vec s')^\mathrm T \rangle
  = \vec M'.
\end{equation}
In particular, Eq.~(\ref{eq:ncvm}), the pixel-pixel residual noise covariance
matrix, $\vec N$, is
equal to $\vec M$ and given by Eq.~(\ref{eq:MmatSS}).

We note that a sufficiently high quality estimate of the inverse time domain
correlations, $\mathcal N^{-1}$, is required in order to calculate both the
minimum-variance map and its noise covariance. If it is misestimated
the map estimate will still be unbiased, though not any more minimum variance
or maximum likelihood, and its covariance will not be given any more by
Eq.~(\ref{eq:residual_covariance}). 

For example for computational reasons we will find later that using some
other matrix, denoted here as $\mathcal M^{-1}$, rather than the actual
inverse noise covariance, $\mathcal N^{-1}$,  in the calculation of the map
estimates in Eq.~(\ref{eq:min_var_map}) can be helpful. The corresponding
noise correlation matrix for such a map is then given by 
\citep[][ungeneralized case]{Stompor2002}
\begin{equation}
  \label{eq:generalized_residual_covariance}
  \vec N' =
  \langle (\hat{\vec s}'-\vec s')(\hat{\vec s}'-\vec s')^\mathrm T \rangle
  = \left(
    \mathcal R^{-1} + {\vec A'}^\mathrm T\mathcal M^{-1}\vec A'
  \right)^{-1}
  \left(
    \mathcal R^{-1}\mathcal R_\mathcal N\mathcal R^{-1}
    + {\vec A'}^\mathrm T\mathcal M^{-1}\mathcal N\mathcal M^{-1} \vec A'
  \right)
  \left(
    \mathcal R^{-1} + {\vec A'}^\mathrm T\mathcal M^{-1}\vec A'
  \right)^{-1},
\end{equation}
where $\mathcal M$ and $\mathcal R$ define our map-making operator, whereas
$\mathcal N$ and $\mathcal R_\mathcal N$ are the true noise properties.
This expression is significantly more complex and computationally involved
than Eq.~(\ref{eq:residual_covariance}). Fortunately, as we discuss in the
following, in many cases of interest, the latter expression turns out to be
a sufficiently good approximation of the former with $\mathcal N^{-1}$
replaced by $\mathcal M^{-1}$ at least for some of the potential applications.

The {\sc Planck} Working Group 3 (CTP) has performed extensive studies of
different map-making approaches \citep{Poutanen:2005yg,  Ashdown:2007ta,
  Ashdown:2006ey, Ashdown:2008ah}. They have been shown to produce different
residual noise structures in the computed maps studied in detail in those
papers. A map-making method should only be considered complete once the
residual noise covariance associated with it can be understood and
sufficiently well characterized.

The map-making methods considered for {\sc Planck} fall into two general classes
both of which are described by the equations derived above. The first class,
called optimal methods, assumes the sufficient knowledge of the time domain
noise correlations, does not introduce any extra degrees of freedom, $\vec x$,
(or alternately assumes a prior for them with $\mathcal P$ vanishing). In
this case one can derive from Eqs.~(\ref{eq:destripeMap2}) \& (\ref{eq:MmatSS}),
\begin{eqnarray}
 \hat{\vec s} & = &
  \left(  {\vec A}^\mathrm T\mathcal N^{-1}\vec A \right)^{-1}\,
  {\vec A}^\mathrm T\mathcal N^{-1}\vec d,
  \label{eq:optimal_map}
  \\
  \vec N & = & \left(  {\vec A}^\mathrm T\mathcal N^{-1}\vec A \right)^{-1}
  \label{eq:optimal_ncm}
\end{eqnarray}
where the latter is a covariance of the residual noise of the former.

The second class of methods introduces a number of baseline offsets with
a Gaussian correlated (so-called generalized destripers)  or uncorrelated
(standard destripers) prior on them and describe the noise as an
uncorrelated Gaussian process. The destriper maps are evaluated via
Eq.~(\ref{eq:destripeMap1}) or
Eq.~(\ref{eq:destripeMap2}), with $\mathcal N$ assumed to be diagonal.
Clearly on the time-domain level the destriper model is just an approximation,
therefore at least \emph{a priori} we should use the full expression in
Eq.~(\ref{eq:generalized_residual_covariance}) to estimate its covariance.
The CTP papers have shown that for {\sc Planck} the two methods produce
maps which are very close to one another. Moreover, they have shown that using a generalized
destriper, the derived maps eventually become
nearly identical to those obtained with the optimal methods, if an
appropriate length of the baseline and a number of the baseline offsets is
adopted together with a consistently evaluated prior. This motivates using the 
simplified Eq.~(\ref{eq:MmatSS}) as an approximation for the
noise covariance of the destriped maps, i.e., $\vec N = \vec M$. We will
investigate the quality and applicability of this approximation later in
this paper.

In this paper we extend the analysis presented in those earlier CTP papers.
We first study the covariances derived for the different map-making
algorithms using Eq.~(\ref{eq:MmatSS}), compare their properties and test
how well they describe the residual noise in the actual maps. As all those
calculations can not be performed at the full instrumental resolution, we
also discuss  methods of producing the low-resolution version of the maps.

\subsection{Time domain noise} \label{sec:todnoise}

\begin{figure}[!tbp]
  \centering
  \sidecaption
  \resizebox{12cm}{!}{\includegraphics[trim=20 18 16 20,clip]
    {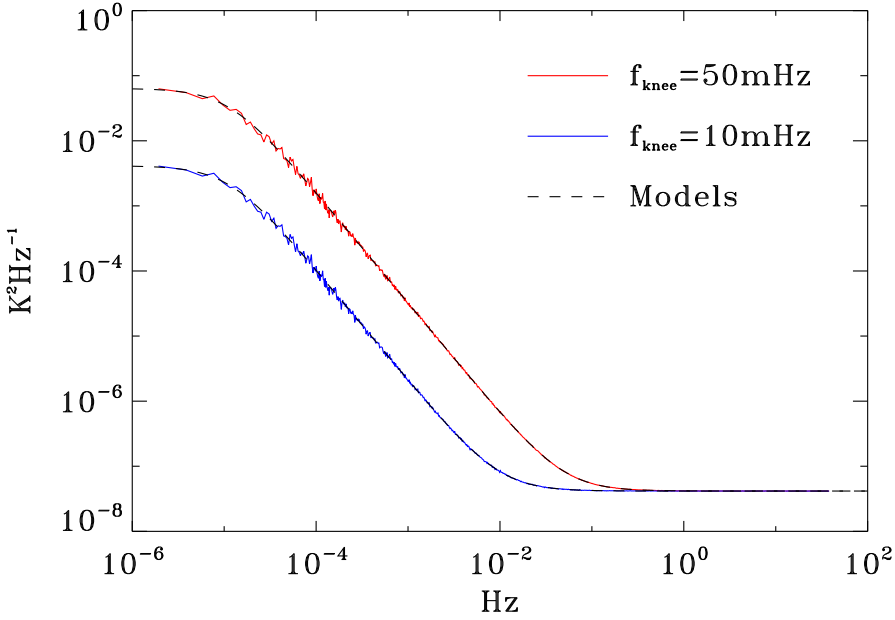}}
  \caption{
    Noise model (dashed lines) and the spectra of simulated noise (solid
    lines).
    The two sets of curves correspond to the two considered knee frequencies
    with $f_\mathrm{min}=1.15\times 10^{-5}$ and $\alpha = 1.7$.
  }
  \label{fig:noise_model}
\end{figure}

We assume that the time domain noise is a Gaussian process and for the
simulations we take the noise power spectral density to have the form
\begin{equation}
  \label{eq:psd}
  P(f) = \frac{\sigma^2}{f_\mathrm{sample}}\cdot
  \frac{f_\mathrm{min}^\alpha+f^\alpha}{f_\mathrm{knee}^\alpha+f^\alpha},
\end{equation}
where the shape is defined by the slope, minimum and knee frequencies 
($\alpha$, $f_\mathrm{min}$ and $f_\mathrm{knee}$ respectively) and the
scaling by the white-noise sample variance and sampling frequency ($\sigma$
and $f_\mathrm{sample}$). Two examples of the theoretical and simulated
noise spectra can be seen in Fig.~\ref{fig:noise_model}.

In the calculations of the maps using the optimal algorithms or generalized
destripers we will assume that noise 
power spectrum is known precisely. As the noise simulated in the cases analyzed
here is piece-wise stationary, with no correlations allowed between the data in 
the different pieces (see Sect.~\ref{sec:simulation}) the respective noise
correlation  matrix, $\mathcal N$,
is block Toeplitz with each of the blocks, describing the noise correlations
of one of the stationary pieces, defined by the noise power spectrum.
Given that we will approximate the inverse
of $\mathcal N$ as also a block Toeplitz matrix with each blocks given by an
inverse noise power spectrum. Though this is just an approximation it has
been demonstrated in the past that it performs exquisitely well
in particularly in the cases with long continuous pieces of the stationary
noise \citep{Stompor2002}, as it is a case in all simulations considered here.
 
 
\subsection{Low-resolution maps} \label{downgrading}

{\sc Planck}
will produce maps with resolution of $\sim 5$ arc minutes at frequencies of
$217\,$GHz and above, and $\leq 13$ arc minutes from $70$--$143\,$GHz. 
The sky maps pixelized at the
full available resolution will therefore include as many as
${\mathcal O}\left(10^7\right)$ pixels per Stokes parameter. Though it
has been demonstrated in previous CTP papers that a calculation of such maps is
feasible, the computations of the covariances of such maps is clearly well
beyond the reach of the current and near-future supercomputers. At the same
time production of low-resolution maps from data of a high-resolution
experiment is not a straightforward task, which in the CMB context is made
even more difficult due to a disparity in amplitudes of the total intensity
anisotropies on the one hand and Q and U Stokes parameters (or E and
B polarization modes) on the other. As we emphasized earlier the map-making
methods described in the previous sections work very well but only in the
limit of sufficiently small pixels. Those need to be much smaller than the
typical variability scale of the considered sky signal, which is usually set
by the experimental resolution. Such an assumption is clearly not fulfilled
in the case of the low-resolution map-making. We therefore expect that the
pixel effects are non-negligible in this latter case. Moreover as we solve
simultaneously for all three Stokes parameters even relatively mild pixel
effects present in the total intensity maps may have significant consequences
for the Q and U Stokes parameter maps.

In this Section we define three alternative methods of producing low-resolution
maps from high-resolution observations. The first two, direct and (inverse)
noise weighting methods, have already been used in the WMAP analysis
\citep{Jarosik:2006ib}. As the third option we  consider at the end
smoothed (low-pass-filtered) maps and their noise covariance.

\subsubsection{Direct method}

The most straightforward method to produce a low-resolution map is to project
the detector observations directly to the pixels of the final target
resolution. Hereafter we will refer to it as the direct method.
  
The direct method is clearly the best choice as far as the described noise
covariance is concerned. However, it does not pay any particular attention
to minimizing the pixel effects. In particular, it may lead to a
position-dependent signal smoothing due to a non-uniform sampling of the
low-resolution pixels --- an
effect which may further cause aliasing problems at the, for example, power
spectrum estimation stage. Moreover, for the destripers the direct method
means that the baseline offsets are solved at the low target resolution.
If the subpixel structure of the pixels can be neglected, this will lead
to a better determination of the baseline offsets, and less residual noise,
as the number of crossing points between baselines increases
\citep{Ashdown:2007ta}. If, however, sub-pixel power is present, it may
affect adversely the offset estimation, with magnitude of the effect
increasing with the pixel resolution. None of the discussed map-making
methods is designed to correct for subpixel structure. Therefore the direct
method can be taken to regard the sky as already smoothed to eliminate the
subpixel structure within the large, low-resolution pixels.

In Sect.~\ref{sec:bandwidth} we quantify the signal error for the different
low-resolution maps and map-makers.

The noise covariance matrices  for such low-resolution maps can be computed
directly using formalism presented in Sect.~\ref{sec:maps}, for example,
Eqs.~(\ref{eq:residual_covariance}) and
(\ref{eq:generalized_residual_covariance}).

Hereafter, we will use the direct method as a reference with respect to which
we compare the other approaches.

\subsubsection{Inverse noise weighting (INW)} \label{subsec:noise_weighted}

In the case of nested pixelization schemes, such as HEALPix
\citep{Gorski:2004by} used in this paper,
to downgrade a temperature only map, one may compute a weighted average
of the subpixel temperatures. A natural choice are the optimal
weights, where the temperature of a small pixel is weighted with the
inverse of its noise variance (or with its hit count provided that
the detectors have equal noise equivalent temperatures). This
weighting leads to the lowest noise of the large pixel in the absence of
pixel-pixel correlations. We will refer to these maps as inverse noise
weighted (INW) maps.

A similar weighting scheme exists for the polarized data as well. The
procedure goes as follows: first the estimated high-resolution maps
are noise-weighted, then their resolution is downgraded, and the resulting
low-resolution, noise-weighted maps are subsequently multiplied by
the low-resolution noise covariance, which needs to be estimated
in parallel. Algebraically, the entire procedure can be summarized
succinctly on the map level as,
\begin{equation}
  \label{eq:noise_weighting}
  \hat{\vec s}' = \vec W''\,\vec X \,\vec W \,\hat{\vec s},
\end{equation}
where $\vec W$ and $\vec W''$ are weight matrices for the high and low
resolution maps, respectively. They depend on $\vec A$ and $\vec A''$
--- the pointing matrices at high and low resolution. $\vec X$
simply sums the pixels in resolution $N_\mathrm{pix}$ to resolution
$N_\mathrm{pix}'$,
\begin{equation}
  \label{eq:X}
  \vec X_{qp} = \left\{
  \begin{array}{cc}
    1,&p\;\mathrm{subpixel}\;\mathrm{of}\;q\\
    0,&\mathrm{otherwise}\\
  \end{array} \right.
\end{equation}
In the following we will assume either block-diagonal or diagonal weighting.
In the former case the weights are given by,
\begin{eqnarray}
  \vec W'' & = &
  \left({\vec A''}^\mathrm T\mathcal N_\mathrm u^{-1}\vec A''\right)^{-1},\\
  \vec W & = & \left(\vec A^\mathrm T\mathcal N_\mathrm u^{-1}\vec A\right)^{-1},
\end{eqnarray}
while in the latter case they are made of the diagonal elements of the
above matrices. Matrix $\mathcal N_\mathrm u$ is the time domain covariance
matrix of the uncorrelated part of the noise, $\vec n$. In the block-diagonal
case the noise weighting mixes different Stokes parameters, while in the
diagonal one each Stokes parameter map resolution is downgraded independently.
Throughout this work we will use only the block diagonal weighting which,
in the cases studied here, turns out to be very close to the diagonal one.

The covariances for the maps obtained via such a procedure can be derived
from Eq.~(\ref{eq:noise_weighting}) and the expressions described in
Sect.~\ref{sec:maps}.


For the destriper technique there is one more extra factor which makes this
manner of resolution downgrading differ from the direct method of the
previous Section. As the maps outputted directly by a destriper code are
of a high resolution, the baseline offsets are also determined at that
resolution. If the block-diagonal weighting is then used to downgrade the
map, the result is equivalent to the direct calculation of the low-resolution
map with the baselines determined from the high-resolution analysis, 
Eq.~(\ref{eq:destripeMap2}).

Noise weighting reduces signal errors by first solving the map at
a resolution where subpixel structure is weak. In comparison to the direct
method, the more accurate signal is gained at the cost of higher noise.
Like the direct method,  INW also disregards any subpixel structure but at
high resolution. In this case the instrument beam naturally smoothes out
small scale structure causing the approximation to hold.

\subsubsection{Harmonic smoothing}

Both methods described above may result in the signal smoothing
(or its band-width) being position-dependent as it is achieved via
averaging of the \emph{observed} high-resolution pixel amplitudes contained
within each low-resolution pixel. This may result in the aliasing of sky power. 

Applying a smoothing operator to each of the high-resolution maps prior
to resolution downgrading could alleviate such a problem. The smoothing
operation needs to take care properly of the high frequency power contained
in the maps avoiding thus its being aliased to the power
at the scales of interest. As the smoothing operation is usually performed 
in the harmonic domain it requires that the high-resolution map is first
expanded in spherical harmonics. If the map has unobserved pixels, they
will induce undesired mode coupling. For sufficiently complete sky
coverage we can ``patch'' the high-resolution map by adding averages
of the neighbouring pixels into blank pixels. If the coverage is more
incomplete, the missing pixels can be replaced by a constrained realization
of signal and noise, such methods are used for example in the so-called 
sampling techniques \citep[e.g.,][]{Jewell2004}, which have been successfully
applied to simulated {\sc Planck} data.
 Simple patching will clearly affect only very small
scale statistical properties. If a constrained realization is applied, 
the spherical expansion will depend on the input model. 
In this work we only deal with a complete sky coverage leaving an
investigation of those effects to the future work.

To suppress small angular scale power the expansion is convolved with
an  axially symmetric window function (e.g. a symmetric Gaussian
window function \citep{Challinor:2000df}),
\begin{eqnarray}
  \label{eq:beam_window}
  \tilde a^\mathrm T_{\ell m} = W_\ell a^\mathrm T_{\ell m},
  & \quad & W_\ell = e^{-\frac{1}{2}\ell(\ell+1)\sigma^2} \\
  \tilde a^\mathrm E_{\ell m} = {}_2W_\ell a^\mathrm E_{\ell m},
  & \quad & {}_2W_\ell = e^{-\frac{1}{2}[\ell(\ell+1)-4]\sigma^2},
\end{eqnarray}
chosen to leave only negligible power at angular scales that
are not supported by the low target resolution.
Finally the regularized
expansion is synthesized into a low-resolution map by sampling the expansion
values at pixel centers. We conduct most of our studies using a beam having
a full width at half maximum (FWHM) of twice the average pixel side. For 
the $N_\mathrm{side}=32$ resolution this is approximately $220'$
($3\fdg 7$). Whenever transforms between harmonic and pixel space are
conducted, it is important to consider the range of multipoles included
in the transformation. We advocate using such an aggressive smoothing that
the harmonic expansion has negligible power beyond $\ell=3N_\mathrm{side}$
and results are stable for any $\ell_\mathrm{max}$ beyond this. For
completeness we have set $\ell_\mathrm{max}=4N_\mathrm{side}$
but stress that any residual power beyond $\ell=3N_\mathrm{side}$
will lead to aliasing.

The smoothing window does not need to be a Gaussian but it is preferable
to avoid sharp cut-offs that may induce ``ringing'' phenomena.
\citet{Benabed:2009af} suggest a window function that preserves
the signal basically unchanged until a chosen threshold and then smoothly
kills all power quickly above that angular resolution. Their window is
\begin{equation}
  \label{eq:teasingbeam}
  W_\ell = \left\{
    \begin{array}{cl}
      1, & \ell \leq \ell_1 \\
      \frac{1}{2}\left[
        1+\cos\left((\ell-\ell_1)\pi/(\ell_2-\ell_1)\right)\right], &
      \ell_1 < \ell \leq \ell_2 \\
      0, & \ell > \ell_2
    \end{array}
  \right.
\end{equation}
with the typical choice $\ell_1=5N_\mathrm{side}/2$ and
$\ell_2=3N_\mathrm{side}$.

This method can be considered optimal from the (large-scale) signal viewpoint;
however it may be suboptimal as far as the noise is concerned, in particular in 
cases with a strongly inhomogeneous noise distribution on the observed sky.
The noise
covariance matrices described in Sec.~\ref{sec:maps} need to be amended to
accurately characterize the residual noise of the smoothed
maps and thus we need to smooth the matrices as well.

Smoothing of a map is a linear operation. For any linear
operator, $\vec L$, acting on a map, $\vec m$, we can compute its
covariance as
\begin{equation}
  \langle(\vec L\vec m)(\vec L\vec m)^\mathrm T\rangle
  = \vec L\langle\vec m \vec m^\mathrm T\rangle\vec L^\mathrm T
  = \vec L\vec N\vec L^\mathrm T
  = \sum_i \lambda_i\,\vec L\hat{\vec e}_i\cdot(\vec L\hat{\vec e}_i)^\mathrm T,
 \label{eq:covSmoothing}
\end{equation}
where $\lambda_i$ and $\vec e_i$ stand for eigenvalues and eigenvectors of the 
noise covariance, $\vec N$, i.e.,
$\vec N =\sum_i\lambda_i\widehat{\vec e}_i\widehat{\vec e}_i^\mathrm T$,
and $\vec m$ is understood here to contain the noise only.
We note that in general one should replicate the same processing steps as 
are to be applied to maps and therefore the smoothing operation should be
applied to the noise covariance of the high-resolution map and its resolution
downgraded later. All these steps are described by the operator $\vec L$ 
introduced above. In this case $\vec L$ is a rectangular matrix with many
fewer rows (given by the number of low-resolution pixels) than columns (the number of high-resolution pixels). In such a case rather than performing 
the eigenvalue decomposition as suggested by the right-most term of
Eq.~(\ref{eq:covSmoothing}), which would require as many operation as the cube of the
number of high-resolution pixels, it may be more efficient to
perform the matrix-matrix multiplications in Eq.~(\ref{eq:covSmoothing})
explicitly. In fact in the latter approach one could rephrase the problem
as a series of PCG solutions of a map-making type, each of which would result
in the computation of the high-resolution covariance, $\vec N$, times one of
the columns of the smoothing operator, $\vec L^\mathrm T$.
This could bring the cost of the covariance smoothing down to that comparable
with actual map-making operation repeated for each of the low-resolution
pixels. Though this may be in a realm of capabilities of the present day
supercomputers it is certainly a huge effort not warranted at the present
stage of this investigation. 

Alternately, one may choose to commute the order of the smoothing and
downgrading operations as highlighted above. Though these two operations
are clearly not exchangeable on the map level, due to the potential presence
of the sub-pixel power, such an approach can be more justifiable for the noise
covariances. In this case we can explicitly compute the low-resolution
unsmoothed covariance matrices directly and subsequently smooth them with
the signal smoothing kernel. One side advantage of this approach is that
the low-resolution maps are more likely to be genuine full-sky maps than
their high-resolution counterparts. Applying the smoothing is therefore
less likely to require any additional pre-processing.

In the following we will apply the smoothing technique to both high and low
resolution maps already downgraded using some of the other approaches. We
will demonstrate that such a combined approach results in controllable
properties of the residual noise on the one hand and well defined sky
signal bandwidth on the other. Unlike both the direct and INW methods, the
low-resolution maps are actually solved from a signal that lacks subpixel
structure.

\section{Numerical calculations of residual noise covariance} 

\label{sec:estimating}

This section presents numerical methods to compute the residual noise
covariance matrix and describes briefly their implementations 
corresponding to three different map-making methods, the optimal method 
(MADping and ROMA implementations) and the generalized (Madam) and classical
destriping (Springtide) methods.

\subsection{Optimal map covariance} \label{sec:optimalmm}

The noise covariance for the maps computed by optimal algorithms using true 
time domain correlations is given by Eq.~(\ref{eq:optimal_ncm}). The
calculation of such a matrix proceeds in two steps and two different
implementations have been developed in the course of the work described here.
During the first step the inverse covariance matrix, ${\vec A}^\mathrm T
\mathcal N^{-1}\vec A$ needs to be assembled and subsequently inverted. 
Given that the matrix can be singular the latter step needs to be taken with
care and a pseudo-inverse may need to be computed. The computation of the
latter involves a eigenvalue decomposition of the inverse noise matrix.
Because the noise matrix is symmetric and in principle non-negative definite,
its eigenvalues are real and non-negative ($\lambda_i\geq 0$), and its
eigenvectors form a complete orthogonal basis. This allows us to expand
the matrix as
\begin{equation}
  \label{eq:eigevaluedecomposition}
  \vec N^{-1}
  = \vec U \vec \Lambda \vec U^\mathrm T
  = \sum \lambda_i\,\hat{\vec e}_i \hat{\vec e}_i^\mathrm T.
\end{equation}
Here $\lambda_i$ are the eigenvalues and $\hat{\vec e}_i$ are the
corresponding eigenvectors of the matrix $\vec N^{-1}$. We can now
invert $\vec N^{-1}$ by using its eigenvalue decomposition,
\begin{equation}
  \label{eq:eigendecomp}
  \vec N = \sum_i \lambda_i^{-1}\hat{\vec e}_i\hat{\vec e}_i^\mathrm T,
\end{equation}
and controlling the ill-conditioned eigenmodes. Any ill-conditioned
eigenmode will have an eigenvalue several orders of magnitude
smaller than the largest eigenvalue. By including in the sum only
the well-conditioned eigenmodes we effectively project out the
correlation patterns that our methods cannot discern. This way of calculating
the noise covariance is implemented in the ROMA and MADping codes.

MADping is one of the codes of the MADCAP\footnote{
  \tt http://crd.lbl.gov/$\sim$borrill/cmb/madcap/} suite of CMB analysis
tools. The code is parallelized using MPI and all the operations are
distributed across multiple processors \citep{madcap}. It uses the M3
library mentioned in \citep{Cantalupo:2009} for data reading and time-domain
noise correlation generation. Load balancing is performed based on both the
number of pixels per processor and the number of time samples falling into those pixels.  Each processor scans through its sections of time-ordered
data (correctly handling overlap with other processors' data) and
accumulates its local piece of the inverse noise covariance. These pieces
are then gathered and written to disk. The scaling of this technique is
\begin{equation}
  \label{eq:madping_scaling}
  N_\mathrm{flops} \sim \mathcal{O}\left(
    {n_\mathrm{samples}\,\cdot\,n_\mathrm{correlation}}
  \right),
\end{equation}
where $n_\mathrm{correlation}$ is the filter length set by the noise
autocorrelation length.

For a {\sc Planck}-sized, full-sky dataset and using a reasonable pixel
resolution (half a degree), the construction of the inverse noise covariance
dominates over the computational cost of inverting this matrix.
Nevertheless, inversion methods as well as the eigendecomposition scale as
$\mathcal O(N_\mathrm{pix}^3)$.

In order to correctly treat the signal component of the data map in
Eq.~(\ref{eq:min_var_map}), we must apply our low-resolution noise
covariance to a noise-weighted map which has been downgraded from higher
resolution. This downgrading process is equivalent to the technique discussed
in Sect.~\ref{subsec:noise_weighted}, and ensures that signal variations
inside a low-resolution pixel are accounted for. The high-resolution noise-weighted map consistent with the above formalism is constructed
as the first step of the map-making carried out by the MADmap program
\citep{Cantalupo:2009}. The matrix eigenvalue decomposition is done using
a ScaLAPACK\footnote{\tt
  http://www.netlib.org/scalapack/scalapack\_home.html}
interface that allows efficient parallel eigenvalue decomposition of
large matrices using a divide and conquer algorithm.

The ROMA code \citep{natoli01, deg05} is an implementation of the optimal GLS
iterative map making specifically designed for {\sc Planck}, but also
successfully used on suborbital experiments such as BOOMERanG \citep{masi06}.
To estimate noise covariance as in
Eq.~(\ref{eq:residual_covariance}) we start by calculating row $i$ of its inverse, $N^{-1}$, by computing its action on the unit vector
along axis $i$:
\begin{equation}
  (N^{-1})_{i:} = (\vec A^\mathrm{T} \mathcal N^{-1}\vec A) \cdot \vec e_i\;.
\end{equation}

This calculation is implemented as follows: i) projecting the unit vector
into a TOD by applying $\vec A$ on $\hat{\vec e}_i$; ii) noise-filtering the
TOD in Fourier space; iii) projecting the TOD into a map by applying $\vec
A^\mathrm{T}$. By computing each column independently we reduce memory
usage because we allocate memory only for one map instead of allocating
memory for the full matrix. The computational cost of the full calculation
is dominated by FFTs that are repeated as many times as the number of
columns, hence the scaling can be expressed as:
\begin{equation}
N_\mathrm{flops} \sim
  \mathcal O\left(
    n_\mathrm{samples} \cdot\log_2 (n_\mathrm{filter})\cdot n_\mathrm{pix}
  \right).
\end{equation}
Once the inverse noise matrix is assembled it is inverted in a fashion
similar to the one described above. We note that the resulting covariance
matrix has to be symmetric. Though this is not \emph{a priori} ensured by the
algorithm it is the case within expected numerical errors. To ameliorate any effect
we symmetrize the result by averaging the matrix with its transpose.

Alternately, we can compute the NCM column-by-column with help of 
multiple map-making-like operations, i.e.,
\begin{equation}
  \vec{y}_{i:}
  = (\vec A^\mathrm{T} \mathcal N^{-1}\vec A)^{-1} \cdot \vec e_i,
\end{equation}
where $\hat {\vec e}_i$ is a unit vector as defined above and $\vec y_i$
stands for a column of $\vec N$. We rewrite the above equation as
\begin{equation}
(\vec A^\mathrm{T} \mathcal N^{-1}\vec A) \cdot {\vec y}_i = \vec e_i,
\end{equation}
and solve it using the standard PCG map-making solver. We note that in such
a case there is no need to store a full inverse noise covariance matrix in
a memory of a computer at any single time as the operations on the left
hand side can be performed from right to left. As a result this approach
can be applied also for high-resolution cases for which the direct method
described above would not be any more feasible.

We note that unlike in the previous approaches based on the direct matrix
inversion in the latter case there is no special care taken of potential
singularities. Though the presence of those does not hamper the PCG
procedure \cite[e.g.,][]{Cantalupo:2009}, nonetheless care must be
taken while interpreting its results.

\subsection{Destriped map covariance}

In the destriping approach to map-making, we model all noise correlations
by baseline offsets. Thus we write Eq.~(\ref{eq:tod}) as
\begin{equation} \label{eq:ds_tod}
  \vec d = \vec A\vec s + \vec B\vec x + \vec n_\mathrm u,
\end{equation}
where $\vec n_{\mathrm u}$ is a vector of uncorrelated white noise
samples. Accordingly, we must replace the time-domain noise covariance
matrix, $\mathcal N$, by a diagonal matrix, $\mathcal N_\mathrm u$. All
noise correlation is then included in the prior baseline offset covariance
matrix, $\mathcal P$.

If we now apply the destriping approximation to Eqs.~(\ref{eq:MmatSS0}) 
we find for the pixel-pixel residual noise covariance matrix:
\begin{equation}
  \label{eq:madam_corr}
  \vec M^{-1} = \vec A^\mathrm T\mathcal N_\mathrm u^{-1}\vec A -
  \vec A^\mathrm T\mathcal N_\mathrm u^{-1} \vec B
  \left(
    \mathcal P^{-1}+\vec B^\mathrm T\mathcal N_\mathrm u^{-1}\vec B
  \right)^{-1}
  \vec B^\mathrm T\mathcal N_\mathrm u^{-1}\vec A\;.
\end{equation}
The first term on the {\sc rhs} is the binned white noise contribution (a
diagonal or block-diagonal matrix for temperature-only and polarized cases
respectively) and the second term describes the pixel-pixel correlations
due to errors in solving for the baselines, i.e., the difference between
the solved and actual baselines \citep{Kurki-Suonio:2009}.

When making a map using destriping, one can use high resolution to solve
for baselines and still bin the map at low resolution. Since this
is equivalent to producing first a high-resolution map and then downgrading
through inverse noise weighting, we will always assume the same pixel size
for both of these steps.

\subsubsection{Conventional destriping}

Springtide~\citep{Ashdown:2006ey} is an implementation of the
conventional destriping approach which solves for one baseline per
pointing period.  Since the baselines are so long, it allows for a
number of optimizations in the handling of the data.  During one
pointing period, the same narrow strip of sky is observed many times,
so the time-ordered data are compressed into rings before doing the
destriping.  Another effect of the long baselines is that the prior
covariance matrix of the baselines, $\mathcal P$, is strongly
diagonal-dominant, so to a very good approximation can be assumed
to be diagonal.  As a consequence, the matrix $\left(\mathcal
  P^{-1}+\vec B^\mathrm T\mathcal N_\mathrm u^{-1}\vec B\right)^{-1}$
that appears in the expression for the inverse map covariance matrix
(\ref{eq:madam_corr}) is also diagonal. Thus the number of operations
taken to compute (\ref{eq:madam_corr}) is
\begin{equation}
N_\mathrm{flops} \sim
  \mathcal{O}(n_{\mathrm{base}}(n_{\mathrm{pix/base}})^2).
\end{equation}

The number of baselines is small compared to the generalized
destriping approach, so another method of computing the noise
covariance matrix of the map becomes feasible.  It is possible to
compute the inverse posterior covariance matrix of the baseline
offsets explicitly, to invert it and use it to compute the map
covariance matrix.  This method has the advantage that the resolution
at which the destriping is performed is not constrained to be the same
as the resolution of the final map covariance matrix.  The destriping
can instead be done at the natural resolution of the data to avoid
subpixel striping effects. The inverse of the posterior baseline error
covariance matrix can be calculated using Eq.~(\ref{eq:MmatXX})
and the corresponding map covariance matrix is given by
Eq.~(\ref{eq:MmatSS}).
However, the pointing matrices, $\vec A$, need not be for the same
resolution in both steps.

The structure of the inverse posterior baseline covariance matrix,
Eq.~(\ref{eq:MmatXX}), depends on the scanning strategy, but it is in general
a dense matrix. Inverting the matrix and using it to compute
Eq.~(\ref{eq:MmatSS}) involves dense matrix operations, so this
method is computationally more demanding than the other approach
described above.  However, the posterior baseline covariance matrix
needs only to be computed once and stored, and then can be used many
times to compute the map covariance matrix for any desired resolution.
It is also possible to use Eq.~(\ref{eq:MmatSS}) to compute the
residual noise covariance matrix for a subset of the pixels in the map.

\subsubsection{Generalized destriping}

Madam \citep{Keihanen:2004yj, Keihanen:2009} is an implementation of
the generalized destriping principle. It is flexible in the choice
of baseline length and makes use of prior information of baseline covariance
($\mathcal P$ is not approximated to be diagonal).

Even for a generalised destriper, all the matrices in
Eq.~(\ref{eq:madam_corr}) are extremely sparse.  Most of the
multiplications only require operations proportional to the number of
pixels, baselines or samples.  The matrices $\mathcal P^{-1}$ and
$\left(\mathcal P^{-1}+\vec B^\mathrm T\mathcal N_\mathrm u^{-1}\vec
  B\right)^{-1}$ are approximately circulant, band-diagonal matrices
whose width is determined by the noise spectrum. For all cases studied in
this paper, the latter matrix is limited to the order of $10^3$
non-negligible elements per row. We call this width the
\emph{baseline correlation length}, $n_\mathrm{corr}$, and includes
the white-noise contribution as well. $n_\mathrm{corr}$ corresponds to a
few hours of samples and is inversely proportional to baseline length,
$n_\mathrm{bl}$.

We evaluate the prior baseline offset matrix from the power spectral density
of the correlated noise, $P(f)$, by Fourier transforming the baseline
PSD, $P_\mathrm x(f)$, into the autocorrelation function.
The baseline PSD is evaluated as \citep{Keihanen:2009}:
\begin{equation}
  \label{eq:psd_x}
  P_\mathrm x(f) = \frac{1}{t_\mathrm{base}}\sum_{m=-\infty}^{\infty}
  P(f+m/t_\mathrm{base})g(ft_\mathrm{base}+m),
  \quad
  \mathrm{where}
  \quad
  g(x) = \frac{\sin^2 \pi x}{(\pi x)^2}.
\end{equation}
The sum converges after including only a few $m$ around the origin. For
stationary noise, any row of $\mathcal P^{-1}$ can then be evaluated as
a cyclic permutation of $\mathcal F^{-1}[1/P_\mathrm x(f)]$.

We evaluate (\ref{eq:madam_corr}) after computing (\ref{eq:psd_x}) by
approximating the inner matrix, $\left(\mathcal P^{-1}+\vec B^\mathrm
  T\mathcal N_\mathrm u^{-1}\vec B\right)^{-1}$, as circulant. This allows us
to invert the $n_\mathrm{base}\times n_\mathrm{base}$ band diagonal matrix
by two short Fourier transforms. It turns out that the matrix multiplications
are most conveniently performed by first evaluating the sparse
$\vec A^\mathrm T\vec N_\mathrm u^{-1}\vec B$ matrix and then operating with
it on the inner matrix from both sides.

In effect the inverse covariance matrix gains a contribution from all
quadruplets $(\vec x_i, \vec x_j, p, q)$, where baselines $\vec x_i$
and $\vec x_j$ are within baseline correlation length and hit pixels
$p$ and $q$.  The number of operations required to complete the
estimate is then proportional to
\begin{equation}
  \label{eq:madam_scaling}
  N_\mathrm{flops} \sim
  \mathcal{O}\left(
    n_\mathrm{base} \cdot n_\mathrm{corr} \cdot (n_\mathrm{pix/base})^2
  \right),
\end{equation}
where $n_\mathrm{base}$ and $n_\mathrm{corr}$, the number of baselines
per survey and correlation length respectively, are inversely
proportional to the length of a baseline.  In contrast,
$n_\mathrm{pix/base}$, pixels per baseline, is proportional to
baseline length. For short baselines and low-resolution maps, the
magnitude of $n_\mathrm{pix/base}$ is close to unity.  Table
\ref{tab:pix2base} lists low-resolution $N_\mathrm{side}$ parameters
and the baseline lengths that correspond to average pixel sizes.  It
can be used to estimate $n_\mathrm{pix/base}$ and shows that, for
example, $1.25\,$s baseline offset at $N_\mathrm{side}=32$ resolution
covers approximately $4$ pixels.  The success of this approach is to
replace $n_\mathrm{bl}^2$ in the computation complexity by
$n_\mathrm{pix/base}^2$.

\begin{table}
  \caption{Pixel side to baseline length at 1 rpm spin rate}
  \label{tab:pix2base}
  \centering
  \begin{tabular}{rrrl}
    \hline \hline
    $N_\mathrm{side}$ & $\sqrt{\mathrm{pixel}\,\mathrm{area}}$ & $N_\mathrm{pix}$ & $t_\mathrm{base}$ \\
    \hline
    4   & $14\fdg 658$ &     192 & 2.443s\\
    8   & ~$7\fdg 329$ &     768 & 1.222s\\
    16  & ~$3\fdg 665$ &   3,072 & 0.611s\\
    32  & ~$1\fdg 832$ &  12,288 & 0.305s\\
    64  & ~$0\fdg 916$ &  49,152 & 0.153s\\
    128 & ~$0\fdg 458$ & 196,608 & 0.076s\\
    \hline
  \end{tabular}
\end{table}

\subsection{Smoothed covariance matrices} \label{sec:smoothNCM}

In order to apply the smoothing operator to the low-resolution noise
covariance matrices, $\vec N$, (Eq.~(\ref{eq:covSmoothing})), we assume
that low-resolution maps, and thus also their covariance matrices, are
expected to cover the entire sky. We can also use the eigen-decomposition
of the noise covariance matrices as it is available from the matrix
inversion procedure described earlier. Consequently, we perform the
smoothing to the eigenvectors of the noise covariance matrix,
Eq.~(\ref{eq:covSmoothing}).

Each eigenvector is a $3N_\mathrm{pix}$ map itself (I, Q, U
map) and it has therefore an expansion in the spherical harmonic
domain with a set of coefficients $\vec a_i$
\begin{equation}
  \label{eq:eigen2spherical}
  \vec a_i = \vec Y\hat{\vec e}_i.
\end{equation}
Here $\vec Y$ is the matrix that performs the transformation. It is
made of spin-0 and spin-2 spherical harmonic functions. The
coefficients shall now be smoothed with the same window function
that we used on the high-resolution map:
\begin{equation}
  \label{eq:smoothed_alm}
  \widetilde{\vec a}_i = \vec W \vec a_i.
\end{equation}
Next we turn the smoothed coefficients back to a map
\begin{equation}
  \label{eq:smoothed_eigenmodes}
  \widetilde{\vec e}_i = \vec Y^{-1} \widetilde{\vec a}_i
\end{equation}
and compose the smoothed matrix
\begin{equation}
  \label{eq:eigevaluerecomposition}
  \widetilde{\vec N}
  = \sum \lambda_i\:
  \widetilde{\vec e}_i\cdot\widetilde{\vec e}_i^\mathrm T.
\end{equation}
We note that because $\widetilde{\vec N}$ is symmetric, its eigenvalues are
real and its eigenvectors make an orthogonal system. However, $\lambda_i$
and $\widetilde{\vec e}_i$ are not in general the eigenvalues and
eigenvectors of the smoothed matrix $\widetilde{\vec N}$. This is
particularly important whenever the unsmoothed noise covariance matrix,
$\vec N$, is singular and therefore its calculation
needs to be regularized, as described in Sect.~\ref{sec:optimalmm}. In
such cases, special care may need to be taken to account for effects of
such singularities on the smoothed covariance.
As we point out in the next Section, the unsmoothed covariance is indeed
commonly expected to be singular or nearly so and therefore a general
procedure of treating singular cases is needed. We will discuss a proper
way of dealing with such an issue in Sect.~\ref{sec:ncmval}.

In addition, the smoothing procedure on its own will often lead to singular
eigenvectors of the smoothed covariance matrix, with the eigenvalues
corresponding to those close to zero. Though at a first glance such
eigenvectors may look like being strongly constrained by the data, their
actual value in the analysis is negligible as the sky signal in those modes
is also smoothed. The nearly vanishing variance of those modes will often
spuriously exaggerate the smoothing and map-making artifacts likely present
whenever any inverse noise weighting needs to be applied. To avoid such
problems hereafter we compute the inverse of the smoothed covariance via
its eigenvalue decomposition and set the eigenvalues of all the nearly singular
modes to zero. The criterion for selecting the nearly singular modes will in
general depend on the case at hand.

In some cases the eigenvalue decomposition of the smoothed matrix may 
not be readily available or its computation not desirable. We can then
regularize the inversion of $\vec {\tilde N}$ by adding some low level of
the pixel-independent uncorrelated noise. For consistency, a random
realization of such noise should also be added to the corresponding maps.
We note that both approaches are effectively equivalent and that the choice
of the singularity threshold needed to select the singular eigenmodes 
corresponds roughly to the choice of the noise level to be added. We will
commonly use the latter approach in some of the power spectrum
tests discussed later.

%

\subsection{Singularities} \label{sec:singularities}

As we have pointed out in Sect.~\ref{sec:optimalmm}, the inversion of the
inverse noise covariance matrix, $\vec N^{-1}$,
Eqs.~(\ref{eq:optimal_ncm}) \& (\ref{eq:MmatSS0}), often needs to be
regularized due to the presence of singular or numerically singular modes.
In this Section we discuss the origin of such modes.

We first note that in all cases considered here the inverse covariance
can be expressed as
\begin{equation}
  \vec N^{-1} =  \vec A^\mathrm T \, \mathcal M^{-1} \, \vec A,
\end{equation}
where $\mathcal M^{-1}$ is defined to be,
\begin{equation}
  \mathcal M^{-1} \equiv
  \left\{
    \begin{array}{l c l}
      \mathcal N^{-1}, & \quad & \hbox{\rm for the optimal maps;}\\
      \mathcal N_\mathrm u^{-1} - \mathcal N_\mathrm u^{-1}\vec B 
      \left(
        \mathcal P^{-1} + \vec B^\mathrm T\mathcal N_\mathrm u^{-1}\vec B
      \right)^{-1}\vec B^\mathrm T\mathcal N_\mathrm u^{-1}, & &
    \hbox{\rm for the destriped maps.}\\
    \end{array}
  \right.
\end{equation}
We assume that the pointing matrix, $\vec A$, has full column rank, and
thus $\vec A\vec x = \vec 0\Rightarrow\vec x=\vec 0$. This is equivalent
to an assumption that the sky signal can indeed be estimated from a given
data set. Though this may not be always the case, in particular for the
polarization sensitive experiments, it can usually be achieved if some of
the ill-constrained pixels are removed from consideration. Given this
assumption, the problem of the singular modes of $\vec N^{-1}$ becomes that
of the matrix, $\mathcal M^{-1}$, defined above.

Let us consider the optimal map case first. The matrix $\mathcal M^{-1}$ is
equal to the inverse of the time-domain noise covariance, $\mathcal N^{-1}$.
The latter, Sect.~\ref{sec:todnoise}, is a block Toeplitz matrix with each
block defined by an inverse of the noise power spectrum, Eq.~(\ref{eq:psd}).
Each of those blocks describes the noise properties of one of the stationary
data segments assumed in the simulations. For each block the eigenmodes 
corresponding to the lowest frequencies as permitted by the length of
the segment have eigenvalues vastly smaller than the high frequency modes.
These modes can therefore lead to near singularities. This is specifically
true for zero-frequency modes corresponding to an offset of each of the
stationary data segments. The (near) null space of the full matrix will be 
therefore spanned by all such vectors corresponding to each of the segments. 

Due to projection effects, not all of those modes result in singular modes
of the final pixel domain noise covariance. However, if for a mode, $\vec t$, 
from the null space of $\mathcal N^{-1}$, there exists a pixel domain vector,
$\vec x$, such as $\vec t = \vec A \, \vec x$, the inverse noise covariance
will be singular with eigenvector equal to $\vec x$.

In the studied case, the scanning strategy is such that the sky areas
observed in each of the stationary periods overlap, which efficiently removes
most of the potential degenerate vectors. In fact only a single
pixel-domain $IQU$ vector of which the $I$ part is one and all the others
zero, called hereafter a global offset, can  potentially be singular. We
will indeed confirm these expectations via numerical results later~\footnote{
  We note that in the argument presented here the global offset mode
  is only nearly singular. This is  due to our assumed noise power spectrum,
  which is finite at zero frequency. In a more realistic case the
  offsets of the stationary segments will be however unknown, corresponding
  to an infinite amplitude at zero frequency. The noise covariance in such
  cases should be therefore considered to be strictly singular with the
  global offset being the singular eigenvector. In such cases the noise
  weighting on the right term of the map making equation,
  Eq.~\ref{eq:optimal_map}, will force the offset of each of the
  stationary data segments to be strictly zero. This may not be a
  sufficiently good approximation in particular for short stationary
  time segments. This could, however, be alleviated by introducing the
  segment offsets as extra degrees of freedom contained in the vector,
  $\vec x$, \citep[e.g.][]{Stompor2002}.
}.

In the case of the maps produced with the destriper codes in the absence
of any priors, i.e., $\mathcal P^{-1} = 0$  the $\mathcal M^{-1}$ matrix has
as many singular  vectors as the baseline offsets defined by the
columns of the offset `pointing'  matrix, $\vec B$. However, as long as all
of the offsets cross on the sky the only singular vector of the pixel-domain
covariance will again correspond to the global offset vector as in the
optimal map case. We note however that unlike in that case, this time this
vector is exactly singular. If a prior is employed, as is the case in
both the classical and generalized implementations of the destriper technique
discussed here, the columns of the matrix $\vec B$ are no longer
singular vectors of the matrix $\mathcal M^{-1}$, nor is the global offset
vector a singular vector of $\vec N$. Nevertheless, at least for some
common choices of the prior the global offset vector remains nearly singular.

\section{Numerical tests and comparison metric}

This paper has two main goals. On the one hand we propose and  compare various 
methods devised to produce the low-resolution maps, searching for the map-making
method which leads to low-resolution maps virtually free of artifacts 
such as those due to sub-pixel power aliasing. In parallel we develop
the tools to estimate residual noise covariance for such maps, which properly
describe the error of the estimated maps due to residual noise.
The numerical results presented in the subsequent sections of the paper aim
therefore at comparing and validating the algorithms which we have presented
earlier. The discussed comparisons involve standard statistical tests, such
Kolmogorov-Smirnov, $\chi^2$, etc.\ and ones which are specifically devised
in the light of the anticipated future applications of the maps and their
covariances. We describe such tests in this Section.

\subsection{Quadratic maximum likelihood power spectrum estimation}
\label{sec:QML}

One of the main applications for the low-resolution noise covariance matrices
we envisage is to the estimation of power spectra, $C_\ell$. This is often
separated into the estimation of large and small angular scales, usually
associated with high and low signal to noise regimes. The methods discussed
here are relevant for large angular scales. The successful estimation of
the underlying true power spectrum of the sky signal sets demanding
requirements for the quality of the maps produced for such a purpose as well
as the consistency of the estimated noise covariance and the actual noise
contained in the map.  For this reason we will use hereafter power spectrum
estimation as one of the metrics with which to evaluate the quality of the
proposed algorithms.

We will use the Quadratic Maximum Likelihood (QML) method for the power
spectrum estimation as introduced in \citep{Tegmark:1996qt} and later
extended to polarization in \citep{Tegmark:2001zv}. Given a map in
temperature and polarization $\vec m=(\vec T,\vec Q,\vec U)$,
the QML provides estimates of the power spectra, that read,
\begin{equation}
  \hat{C}_\ell^X 
  = \sum_{\ell' \,, X'} (\vec F^{-1})^{X \, X'}_{\ell\ell'}
  \left[
    \vec m^\mathrm T\vec E^{\ell'}_{X'}\vec m
    -\mathrm{tr}(\vec N{\vec E}^{\ell'}_{X'})
  \right].
  \label{eq:QMLestim}
\end{equation}
Here, $\hat {C}_\ell^X$ is an estimated power spectrum, $X = \mathrm{TT}, 
\mathrm{EE}, \mathrm{TE}, \mathrm{BB}, \mathrm{TB},$ or $\mathrm{EB}$,
and $\vec F_{X X'}^{\ell \ell '}$ is the Fisher matrix defined as
\begin{equation}
  \label{eq:QMLfisher}
  \vec F^{\ell\ell'}_{X X'}
  =\frac{1}{2}\mathrm{tr}\Big[{\vec C}^{-1}\frac{\partial
    {\vec C}}{\partial
    C_\ell^X}{\vec C}^{-1}\frac{\partial {\vec C}}{\partial
    C_{\ell'}^{X'}}\Big] \,.
\end{equation}
The $\vec E$ matrix is given by
\begin{equation}
  \label{eq:QMLelle}
  \vec E^\ell_X
  =\frac{1}{2}{\vec C}^{-1}\frac{\partial\vec C}{\partial C_\ell^X}\vec C^{-1},
\end{equation}
where  $\vec C=\vec S(C_{\ell})+\vec N$ is the covariance matrix (signal plus
noise contribution) of the map, $\vec m$. Here $C_\ell$ is a fiducial power
spectrum needed for the calculation of the signal part of the covariance.
In this paper we will take it to be given by the true power spectrum as used
to produce the simulated skies. Though this would be an unfair assumption,
while testing the performance of this power spectrum estimation technique,
it is justified in our case, where the fact that it leads to the minimal
estimation uncertainties increases the power of our test. (Indeed the QML estimator is in fact also known to be equivalent to a single iteration of a quasi-Newton-Raphson procedure to search for the true likelihood maximum \citep{Bond:1998zw}.)

More details about the QML method can be found elsewhere
\citep[e.g.][]{Tegmark:1996qt, Tegmark:2001zv, Efstathiou:2006wt}. 
\citet{Gruppuso2009} describes the specific implementation of the method,
nicknamed Bolpol, as used in this work and discusses its performance in the
application to the WMAP 5 year data. 

Hereafter, we neglect any systematic effects of either instrumental
and/or astrophysical origins. Nevertheless we note that if the final CMB
map is obtained via some linear cleaning procedure involving maps computed
either for different detectors and/or frequency channels, the results of
this paper will be still relevant and the noise covariance of the `cleaned'
CMB map can be computed via a linear combination of the single map
covariances calculated in turn with help of the procedure discussed here.

\begin{table}[!tbp]
  \caption{Frequently used symbols in this paper.}
  \label{tab:notation}
  \centering
  \begin{tabular}{cl}
    \hline \hline
    Symbol & Definition \\
    \hline
    $\vec C$          & pixel-pixel covariance matrix\\
    $\vec N$          & noise covariance, map domain\\
    $\mathcal N$      & noise covariance, time domain\\
    $\mathcal N'$     & correlated ($1/f$) noise covariance, time domain\\
    $\vec M^{-1}$      & white noise covariance, map domain\\ 
    $\mathcal N_\mathrm u$ & white noise covariance, time domain\\
    $\vec M_\mathrm p$ & $3\times 3$ observation matrix\\
    $\mathcal P$      & prior baseline offset covariance \\
    $\vec A$          & detector pointing matrix\\
    $\vec B$          & offset-to-TOD matrix \\
    $\vec x$          & baseline offset vector\\
    $\vec m$          & $3N_\mathrm{pix}$ Stokes I,Q,U map vector\\
    $\hat{\vec s}$    & map estimate\\
    $\vec n$          & noise vector\\
    $\vec n'$         & correlated noise vector\\
    $\vec n_\mathrm u$ & white noise vector\\
    $\vec d$          & TOD vector\\
    $\vec s$          & sky map\\
    $\vec e_i$  & Cartesian unit vector along the $i$:th axis\\
    \hline
  \end{tabular}
\end{table}

\subsection{Noise bias} \label{sec:noise_bias}

In the map-making methods considered in this paper residual noise in the maps 
is independent of the sky properties and completely defined by the time-domain
noise properties and the scanning strategy. The noise present in the maps
contributes to the power spectrum estimates of the map signal.
We will therefore refer to this contribution as the noise bias and use it to 
quantify the noise level expected in the maps of our different methods in a
manner more succinct and manageable than the full noise covariance.

The noise bias
is defined to be the expectation value for angular power spectrum in the
noise-only case
\begin{equation}
  \label{eq:noise_bias}
  N_\ell^{XY} = \langle C^{XY,\mathrm{noise}}_\ell\rangle,
\end{equation}
where $X$ and $Y$ stand for T, E and B. In the general case of a quadratic
estimator, such as QML, Eq.~(\ref{eq:QMLestim}), \citep{Tegmark:2001zv}, 
the power spectrum estimates are given as a quadratic form of the input
map, $\vec m$,
\begin{equation}
  \label{eq:qml_bias}
  C_\ell^{XY} = \vec m^\mathrm T \vec Q_\ell^{XY}\vec m
  = \mathrm{tr}\:\vec Q_\ell^{XY}\vec m\vec m^\mathrm T.
\end{equation}
By taking $\vec m$ to be noise-only, the noise bias can be expressed as
\begin{equation} \label{eq:noise_bias1}
  N_\ell^{XY} = \langle C^{XY,\mathrm{noise}}_\ell\rangle
  = \mathrm{tr}\:\vec Q_\ell^{XY}\vec N.
\end{equation}
Given the eigenvalue decomposition of the noise covariance matrix,
$\vec N = \sum_i\lambda_i\hat{\vec e}_i\hat{\vec e}_i^\mathrm T$,
we can evaluate the noise bias as
\begin{equation}
  \label{eq:noise_bias2}
  N_\ell^{XY} 
  = \sum_i\lambda_i\hat{\vec e}_i^\mathrm T\vec Q_\ell^{XY}\hat{\vec e}_i.
\end{equation}
In the following we will validate the noise covariance matrices, estimated
for the recovered sky maps, with the help of their respective noise biases,
which we will compare to results of Monte Carlo simulations. In this
context it is particularly useful to consider a pseudo-$C_\ell$ estimator
that assumes uniform pixel weights and full sky. For this estimator, the
operator $\vec Q_\ell^{XY}$ is
\begin{equation}
  \label{eq:pseudo_cl}
  \vec Q_\ell^{XY} = \frac{1}{2\ell+1}
  \left(\vec Y_\ell^X\right)^\dagger\vec Y_\ell^Y,
\end{equation}
where $\vec Y_\ell^X$ has $2\ell+1$ rows that are maps of the appropriate
spherical harmonics. It maps a map vector into a vector of spherical
harmonic expansion coefficients $\{a_{\ell m}^X\}$ where $m=-\ell\ldots\ell$.

The procedure we implement here involves two steps. First for every
estimated noise covariance we compute analytically the noise bias using
Eqs.~(\ref{eq:noise_bias2}) and~(\ref{eq:pseudo_cl}). Second, we compute
the bias using Monte Carlo realizations of the noise-only maps produced
using the corresponding map-making procedure. The multiplications
$\vec Y_\ell^X\vec u_i$ are conveniently implemented using the HEALPix
\citep{Gorski:2004by} Fortran 90 subroutine {\tt map2alm}. We note that as
we consider hereafter only full-sky cases, the noise biases we compute as 
described above  would be equal to those expected in the Maximum Likelihood
estimates of the sky power spectrum, were it not for the imperfection of the
sky quadrature due to pixelization effects \citep{Gorski:2004by}.

\section{Simulation} \label{sec:simulation}

\subsection{Scanning strategy}

In this study the {\sc Planck} satellite orbits around the second Lagrangian
point (L2) of the Earth-Sun system \citep{Dupac:2004pg}. The spin axis lies
near the ecliptic plane, precessing around the anti-Sun direction once
every six months with an amplitude of $7\fdg 5$.
The telescope line-of-sight forms an
$85^\circ$ angle with the spin axis. In addition to these modes, we include
a nutation of the spin axis and slight variations to the $1\,$rpm spin rate.
Details of the scanning simulation can be found from \citet{Ashdown:2008ah}
where it was used in a map-making study.

\subsection{{{\sc Planck}} detectors}

In this Paper we study residual noise in the {\sc Planck} $70\,$GHz
frequency maps. {\sc Planck} has twelve detectors at $70\,$GHz.
In the focal plane they are located behind six horn antennas, a pair of
detectors (``Side'' and ``Main'' detectors\footnote{Side and Main
refer to two detector branches downstream from the orthomode
transducer that separates the two orthogonal linear polarizations.})
sharing a horn. A pair of detectors measures two orthogonal linear
polarizations. The horns are split in two groups (three horns in a
group). The Side and Main polarization sensitive axes of a group are
nearly aligned and the polarization directions of the second group
differ from the first group nominally by $45^\circ$. Two horns from
the different groups make a polarization pair that follows the same
scan path in the sky (three pairs in total with slightly different
scan paths). As a minimum the observations of a polarization pair
are required to build a polarization map. Due to implementation
restrictions the Side and Main polarization axes are not fully
orthogonal and the polarization direction differences between the
groups are not exactly $45^\circ$, but the deviations from these
nominal values are small ($\lesssim 0\fdg 2$). The Side polarization
axes of the two groups differ by $+22\fdg 5$ and $-22\fdg 5$ from
the scan direction.

The beams of the detectors were assumed circularly symmetric with a
$14'$ FWHM (full width half maximum) beam width. The beams do not
impact the residual noise maps or covariance matrices. None of the
map-making methods studied here make an attempt to correct for beam
effects in the maps.

\subsection{Time ordered data}

We computed the NCM's of our three map-making methods using our
noise model spectrum and the one year pointing data of twelve
$70\,$GHz detectors. We produced NCM's for both $N_\mathrm{side}=8$
and $N_\mathrm{side}=32$ pixel sizes. We wanted to compare these
NCM's to the noise maps made by the same map-making methods. For
that purpose we simulated $50$ noise-only timelines and made maps
from them. Our correlated noise streams were simulated in six day
chunks by inverse Fourier transforming realizations of the noise
spectrum \citep{Natoli:2001ca}. We assumed an independence between
the chunks and between the detectors. Fig.~\ref{fig:noise_model}
contains a comparison between the power spectra of the generated
noise streams and the model spectra.

Twenty-five of the surveys featured a relatively high $1/f$ contribution
having the knee frequency, $f_\mathrm{knee}$, set to $50\,$mHz. The
other half was simulated to have a more favorable $f_\mathrm{knee}=10\,$mHz.
It should be noted that these frequencies have been chosen above and
below the satellite spin frequency, $1\,\mathrm{rpm}\approx 17\,$mHz.
The slope of the $1/f$ noise power
spectrum was $\alpha=-1.7$. The correlation timescale of the $1/f$ noise was
restricted to about one day. This made our noise spectrum flat at
low frequencies (below a minimum frequency $f_\mathrm{min}=1.15\times
10^{-5}\,\mathrm{Hz}\thickapprox 1/24\,$h). As we described earlier,
it is the minimum frequency that determines the correlation length
of the noise filter in the optimal map-making. In the noise covariance
matrix of the generalized destriping, the baseline correlation length is,
however, determined by the knee frequency.

We used a uniform white noise NET of $204\,\mu K\sqrt{\mathrm s}$ for all
detectors\footnote{$\sigma = \mathrm{NET}\sqrt{f_\mathrm{sample}}$, where
$\sigma$ and $f_\mathrm{sample}$ were defined in Eq.~(\ref{eq:psd}).
The $70\,$GHz detectors had $f_\mathrm{sample}=76.8\,$Hz.}. We chose
this NET because we wanted to produce noise maps and covariance
matrices whose noise levels are compatible with another CTP study
\citep{Jaffe:2009}. In all map-making and NCM computations we
assumed a perfect knowledge of the detector noise spectrum.

The noise timelines were processed directly into both low-resolution
($N_\mathrm{side}=32$) and high-resolution ($N_\mathrm{side}=1024$)
HEALPix maps using the discussed map-making
codes. In the $N_\mathrm{side}=1024$ temperature and
polarization maps the mean standard deviations of white noise per
map pixel were $44$ and $63\,\mu$K (Rayleigh-Jeans $\mu$K). For
$N_\mathrm{side}=32$ maps the corresponding values were $1.4$ and
$2.0\,\mu$K.

The high-resolution maps were in turn downgraded to the low target resolution
using the schemes detailed in Sect.~\ref{downgrading}.

For the signal error studies described in Sect.~\ref{sec:bandwidth}, we
scanned simulated foreground maps into signal-only timelines. These we
processed into low-resolution ($N_\mathrm{side}=8$) maps using the same
methods as for the $N_\mathrm{side}=32$ (both directly and through high
resolution). We then extracted the signal error part by subtracting a
binned map from the destriped map. The foreground signal errors were
summed with a CMB map to provide a worst case scenario of signal striping
in otherwise perfectly separated CMB map.

\subsection{Input maps} \label{sec:input_maps}

To study bandwidth limitation with respect to downgrading we simulated
$117$ high-resolution $N_\mathrm{side}\mathrm =1024$ CMB skies
corresponding to the same theoretical spectrum, $C_\ell$. These
maps were smoothed and downgraded to $N_\mathrm{side}\mathrm =8$ using
three different Gaussian beams of widths $5^\circ$, $10^\circ$, $20^\circ$
and three apodized step functions with the choices of $(\ell_1,\ell_2)$
being $(20,24)$, $(16,24)$ and $(16,20)$. A seventh set of downgraded
maps was produced by noise weighting according to the scanning strategy.
To comply with this last case, the smoothing windows include also the
$N_\mathrm{side}\mathrm =8$ pixel window function from the HEALPix package.

For the signal error exercise we used the {\sc Planck} sky model,
PSM\footnote{see
  {\tt http://www.apc.univ-paris7.fr/APC\_CS/Recherche/Adamis/PSM/psky-en.php}}
version $1.6.3$,
to simulate the full microwave sky at $70\,$GHz. For diffuse galactic
emissions we included thermal and spinning dust, free-free and synchrotron
emissions. We then added a Sunyaev-Zeldovich map and finally completed
the sky with radio and infrared point sources. The combined 
$N_\mathrm{side}\mathrm =2048$ map was smoothed with a symmetric Gaussian
beam and scanned into a timeline according to the scanning strategy.

For final validation the noise covariance matrices were tested in
power spectrum estimation. Each noise map was added to a random CMB
map drawn from the theoretical distribution defined by a fixed
theoretical CMB spectrum. The theoretical spectrum is the WMAP
first-year best fit spectrum and has zero BB mode.

All maps in this work are presented in the ecliptic coordinate system. This
choice is useful for {\sc Planck} analysis since the scanning circles and
many map-making artifacts form circles that connect the polar regions of the
map. In this coordinate system the galaxy is not positioned in the ecliptic
plane but forms a vertical horse shoe shape around the center of the map
(see Fig.~\ref{fig:striping}).

\section{Results} \label{sec:results}

In this Section we first focus on the noise covariance matrices computed 
using different map-making techniques. We discuss and compare the overall
noise patterns implied by such matrices and test the quality of the destriper
approximation as applied to the noise covariance predictions.
In the second part of the Section we discuss the low-resolution maps
and evaluate their quality in the light of their future potential applications,
such as those to the large angular scale power spectrum estimation.
Due to the computational resource restrictions the low-resolution results 
presented here are obtained either with the HEALPix $N_\mathrm{side}=8$
--- in particular tests in Sect.~\ref{sec:bandwidth} and power spectrum
estimation tests in Sect.~\ref{sec:by_pse} --- or $N_\mathrm{side}=32$ ---
in most of the other Sections.

\subsection{Noise covariance matrices}

First we discuss the noise covariance matrices computed for the low
resolution maps of the direct method. As explained in Sect.~\ref{downgrading}
we compute from these matrices the noise covariances of the other
downgrading techniques. In the following we consider noise
covariance calculated using 4 different ways. In the first way we compute
the noise covariance using the optimal algorithm. For this purpose we
have developed two codes MADping or ROMA, which are described in
Sect.~\ref{sec:optimalmm}. However, as they are just two different
implementations of the same algorithm, we derive most of the results
presented in the  following and involving the optimal covariance using
MADping. We note that whenever results from the both codes are available
they have turned out to be virtually identical within the numerical
precision expected from this kind of calculations. The optimal noise
covariance matrices are expected to provide an accurate description of the
noise level found in the actual optimal maps. We will test this expectation
in the following and use the optimal results as a reference with which to
compare the destriping results.

The three remaining computations of the noise covariance are based on the
destriping approach and correspond to different assumptions about the
offset prior as well as baseline length. We consider the following specific
cases: a classical destriper calculation with a baseline of $3600\,$s
(Springtide) and two generalized destriper computations with a baseline of
$1.25,$s and $60\,$s (Madam). For each of these cases we will compare the
covariance matrices with each other, with the optimal covariances and then
test their consistency with the noise found in the simulated maps. We note
again that this last property is not any more ensured given the approximate
character of the destriper approach.

\begin{figure}[!tbp]
  \centering
  \sidecaption
  \resizebox{10cm}{!}{
    \includegraphics[trim=30 13 20 20,clip]{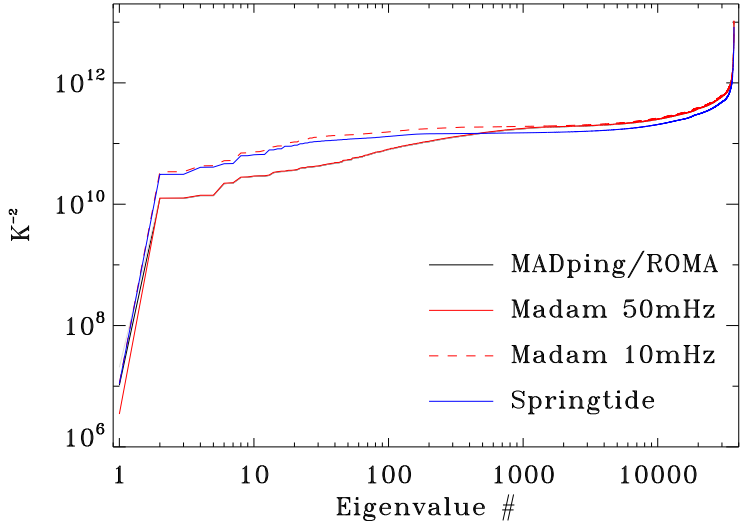}
  }
  \caption{
    Eigenspectra of the inverse covariance matrices $\vec N^{-1}$. MADping,
    ROMA and Madam results for $f_\mathrm{knee}\mathrm=50\,$mHz overlap
    completely. Springtide results are for $10\,$mHz.
  }
  \label{fig:eigenspectra}
\end{figure}

Fig.~\ref{fig:eigenspectra} shows the eigenvalue spectra of some inverse
NCMs. We note that all matrices possess a positive semi-definite
eigenspectrum as is required for any covariance matrix, yet at the same
time they all have one nearly ill-conditioned eigenmode\footnote{
  A double precision ($64$ bit) matrix is numerically ill-conditioned
  when the condition number exceeds approximately $10^{12}$ \citep{nr}.
  For our matrices the situation is not as dire but the first eigenmode
  still deserves special attention.}
, which renders the condition number, i.e., the ratio
of the largest and smallest eigenvalue, very large. This is in agreement
with our expectations as described in Sect.~\ref{sec:singularities}.
Indeed the peculiar eigenmodes corresponding to the
smallest eigenvalues of the inverse matrices are also found to be non-zero
and constant for the $I$ part of the vector and nearly zero for its polarized
components, and thus close to the global offset vector discussed in
Sect.~\ref{sec:singularities}. The small deviations, on the order of $10^{-3}$,
exist, as expected, as none of the peculiar modes is in fact truly singular.
We note that the MADping and ROMA results, both computed in this test, are
seen in the figure to be indistinguishable. They also coincide very closely
with the Madam results computed for the same $f_\mathrm{knee}=50\,$mHz.
The Springtide results, computed with $f_\mathrm{knee}=10\,$mHz are close to,
though not identical with, the Madam results for the very same value of
$f_\mathrm{knee}$ when a longer ($60\,$s) baseline is used for Madam.

\begin{figure}[!tbh]
  \centering
  \resizebox{\hsize}{!}{
    \includegraphics[width=0.2\textwidth,angle=90]{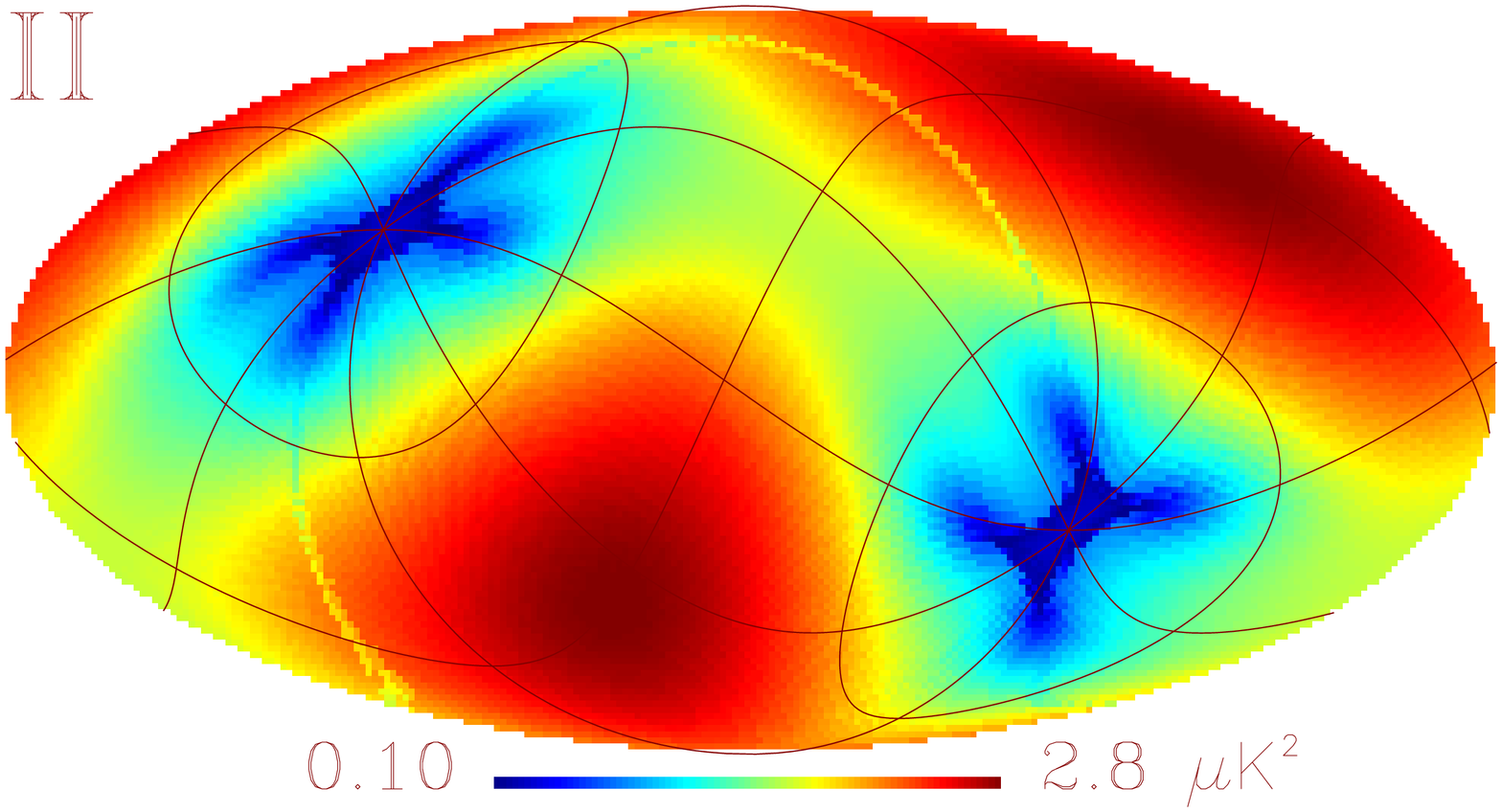}
    \includegraphics[width=0.2\textwidth,angle=90]{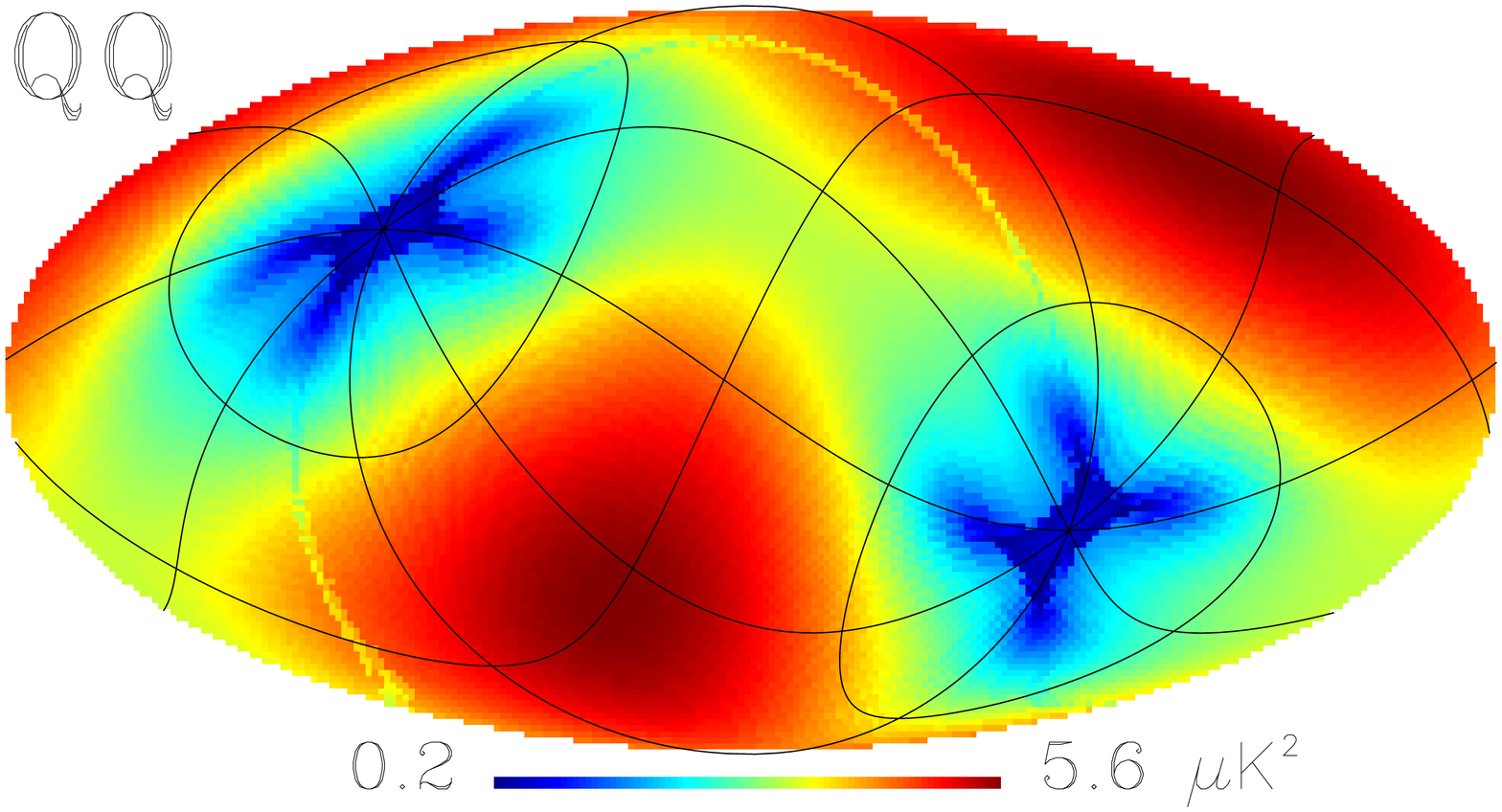}
    \includegraphics[width=0.2\textwidth,angle=90]{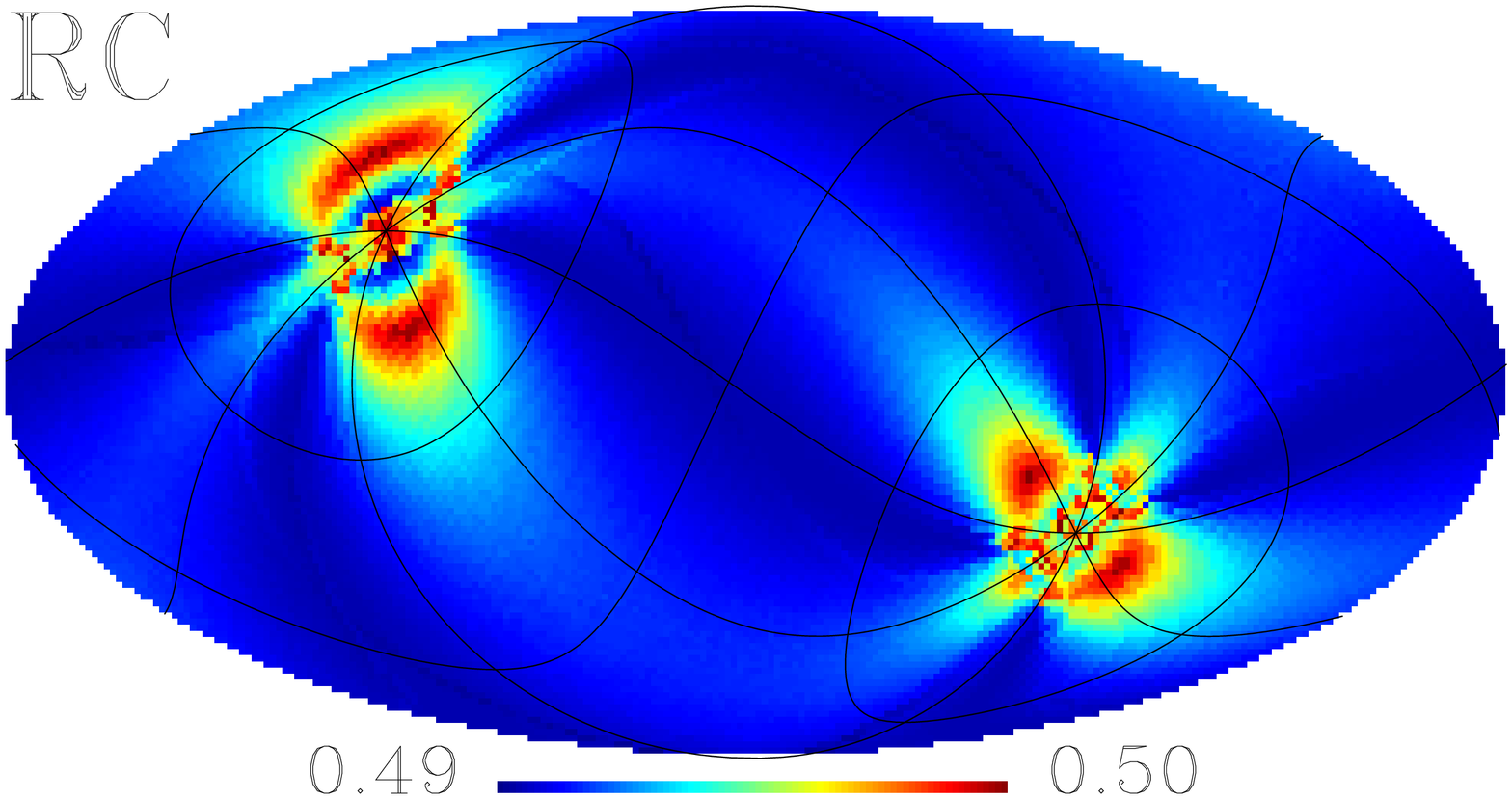}
  }
  \resizebox{\hsize}{!}{
    \includegraphics[width=0.2\textwidth,angle=90]{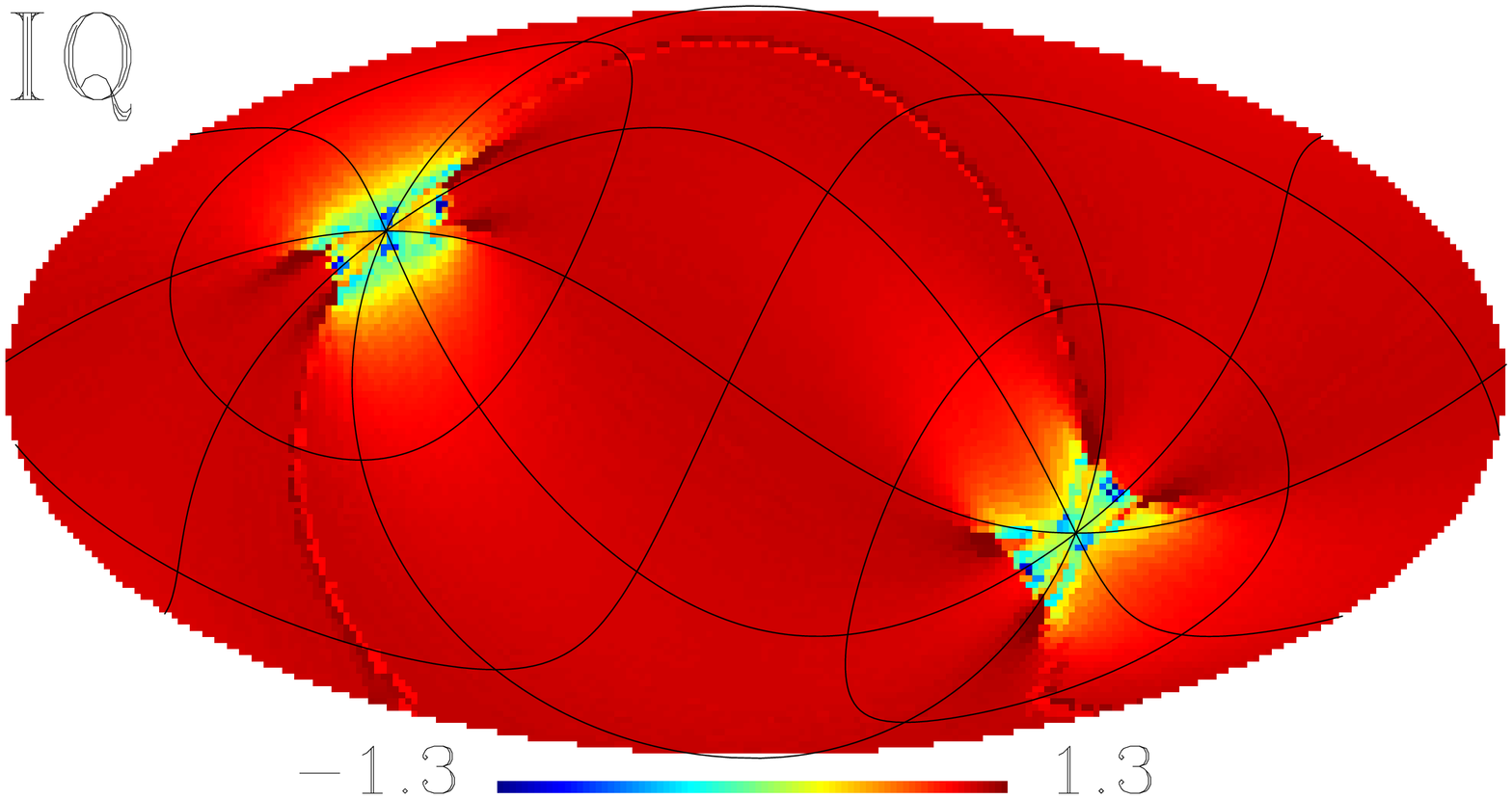}
    \includegraphics[width=0.2\textwidth,angle=90]{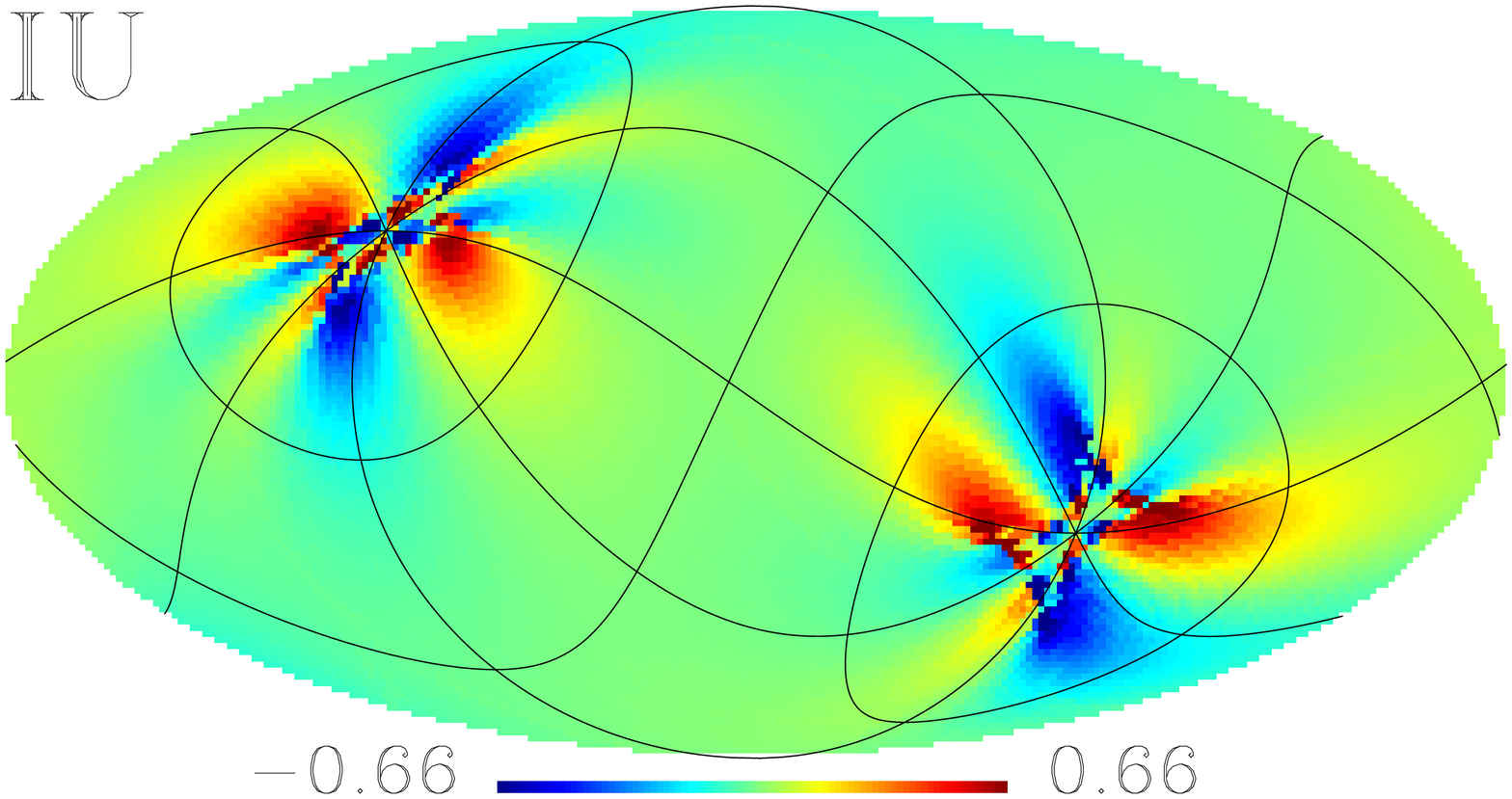}
    \includegraphics[width=0.2\textwidth,angle=90]{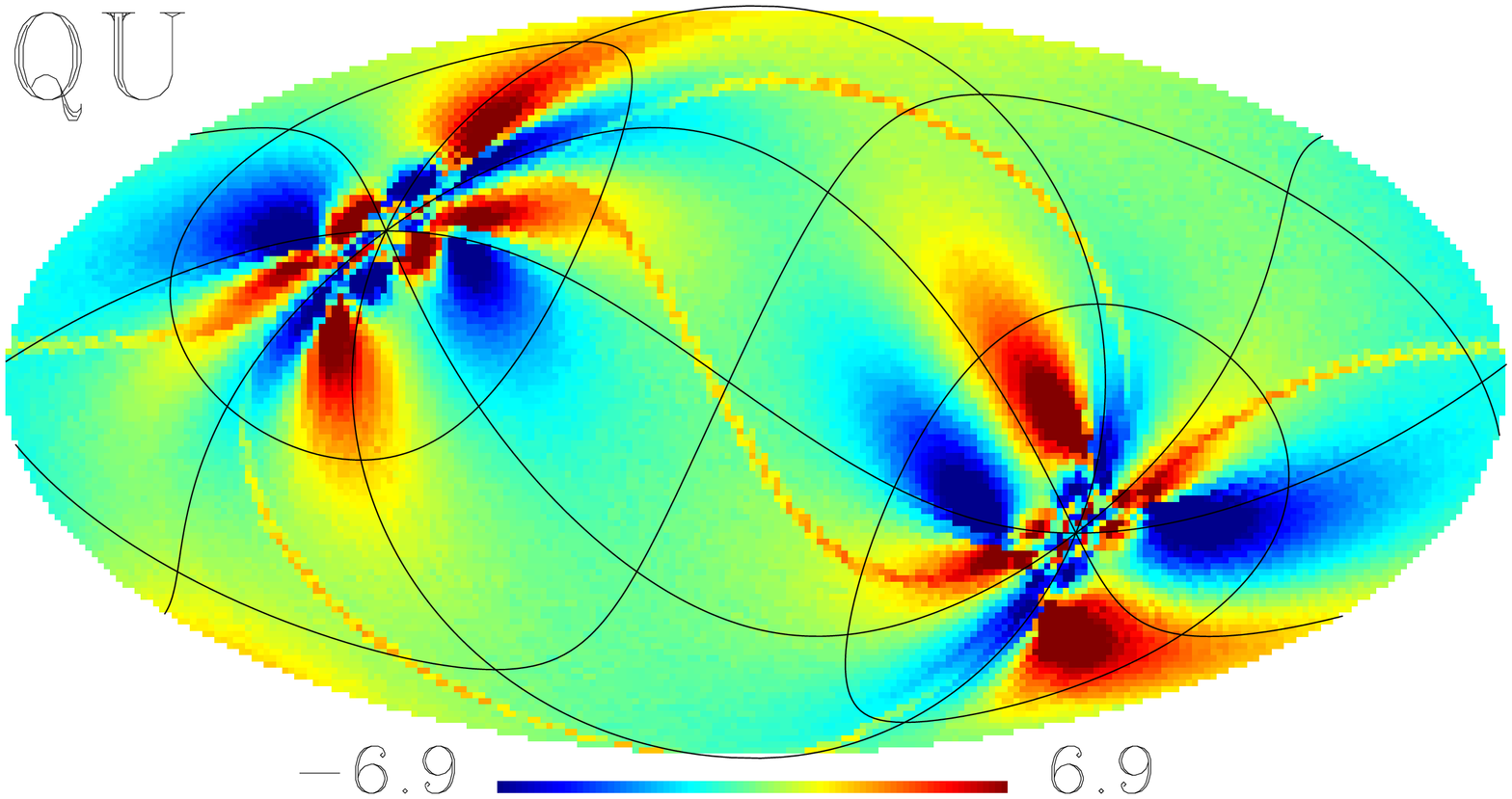}
  }
  \caption{
    \emph{Top:} MADping pixel variances for temperature and polarization and
    the reciprocal condition number of the pixel observation matrices.
    \emph{Bottom:} Correlation coefficient $\times 10^3$ between
    I-Q, I-U and Q-U pixels.
    This part of the noise covariance is dominated by white noise which is
    modelled equivalently in all three paradigms. Hence,
    MADmap, ROMA, Madam and Springtide results are nearly identical. Maps
    are rotated into galactic coordinates to show the structure near the
    ecliptic poles.
  }
  \label{fig:MadpingNCVM_variances}
\end{figure}

Fig.~\ref{fig:MadpingNCVM_variances} depicts the estimated Stokes
I, Q and U pixel variances as well as the covariance between them. These
quantities are dominated by the white noise contribution and all methods
describe white noise in the same manner. The top right-most panel shows
the reciprocal condition number (1/condition number) of the $3\times 3$
blocks of the matrix $\vec A^\mathrm T\mathcal N^{-1}_\mathrm u\vec A$ for
each of the sky pixels. These numbers define our ability to disentangle
the three Stokes parameter for each of the pixels. Whenever they are
equal to $1/2$ the parameters can be not only determined but their
uncertainties will not be correlated. If the reciprocal condition number
for a selected pixel approaches $0$, the Stokes parameters can not be
constrained. In the cases considered here, the Stokes parameters can
clearly be determined for all the pixels.

Fig.~\ref{fig:MadpingNCVM_variances} shows a strong asymmetry
between the IQ and IU blocks. The fact that the polarization axes of the
two detectors of a horn are not fully perpendicular makes the I
noise of a pixel correlate with the Q and U noises of the same
pixel. We can imagine an instrument basis, where one group of three
horns measures the Q of this basis and the other group measures the
U of the same basis. Because Side and Main polarization axis
deviations from the orthogonality have similar magnitudes in the two
groups, we expect the diagonals of IQ and IU blocks to be similar
(symmetric in IQ and IU) in the instrument basis. In the map we use a
different polarization basis\footnote{
  In a HEALPix map the Stokes parameters Q and U at a
  point in the sky are defined in a (x,y,z) reference coordinate,
  where the x-axis is along the meridian and points to south, the
  y-axis is along the latitude and points to east, and the z-axis
  points to the sky~\citep{Gorski:2004by}.}.
Building the noise weighted map (in the map-making) rotates the Q
and U from the instrument basis to the map basis. The IQ and IU
blocks become asymmetric in this rotation.

\begin{figure}[!tbp]
  \centering
  \resizebox{\hsize}{!}{
    \includegraphics[width=0.2\textwidth,angle=90,trim=25 0 45 0,clip]
    {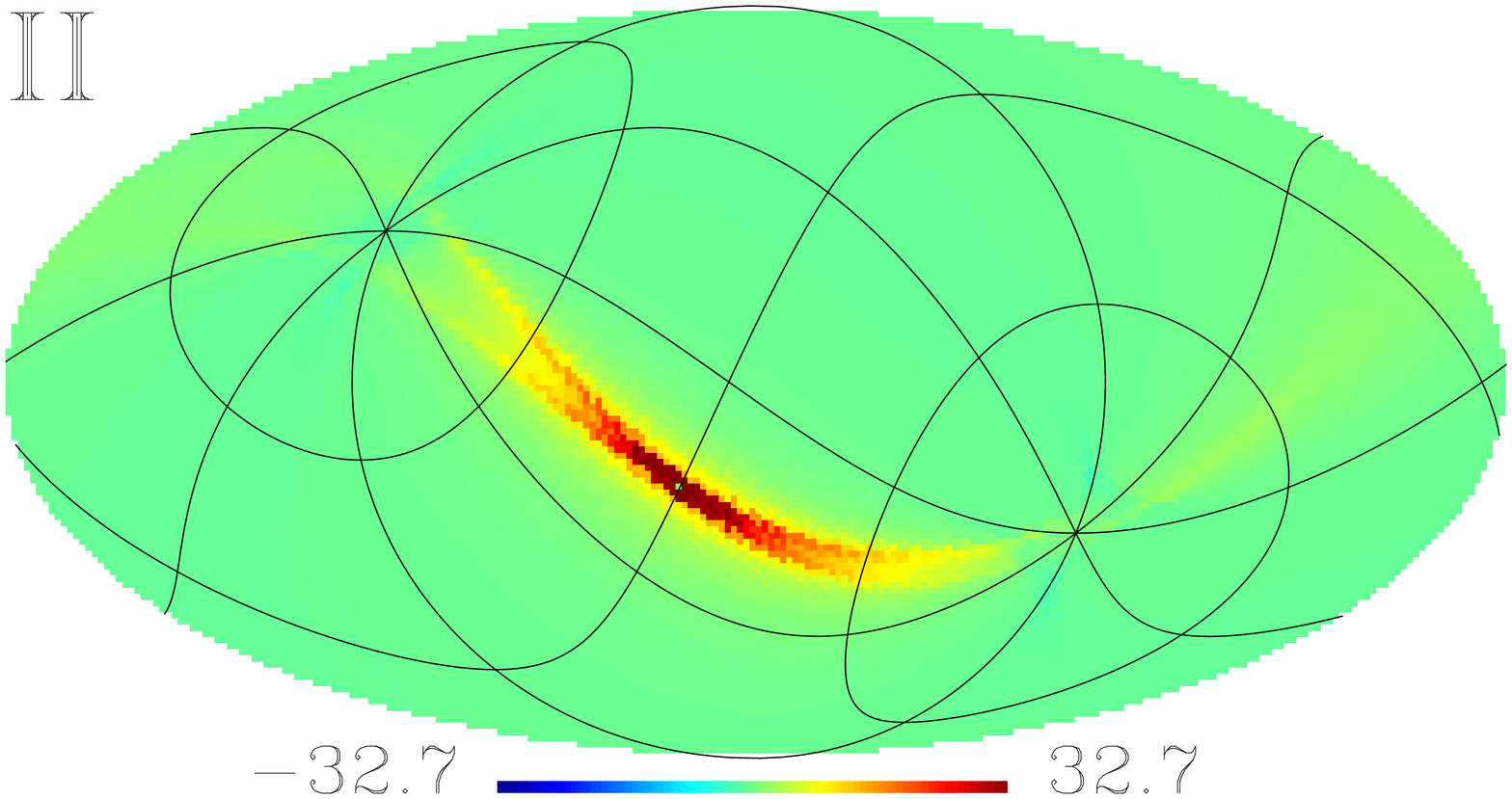}
    \includegraphics[width=0.2\textwidth,angle=90,trim=25 0 45 0,clip]
    {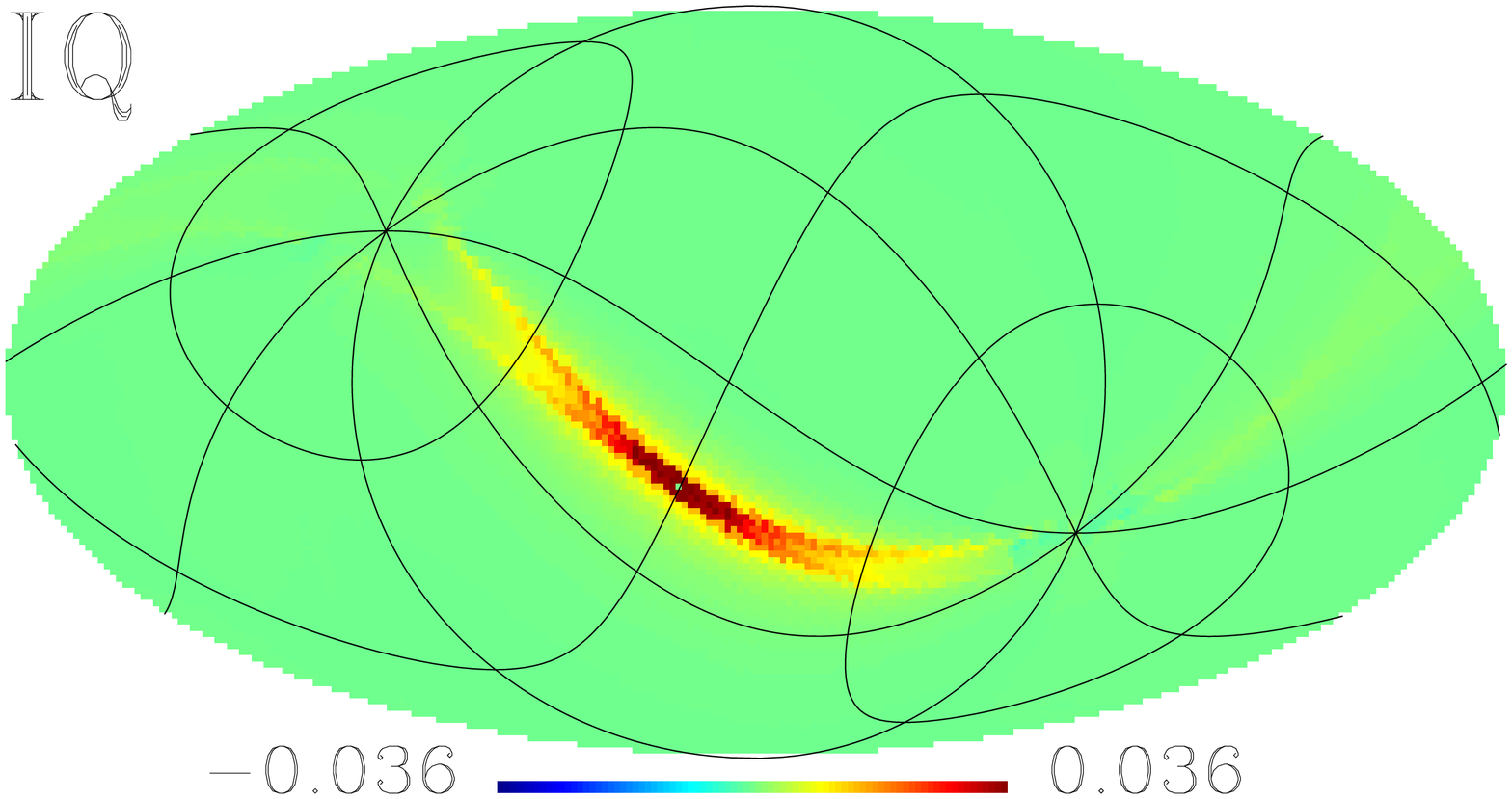}
    \includegraphics[width=0.2\textwidth,angle=90,trim=25 0 45 0,clip]
    {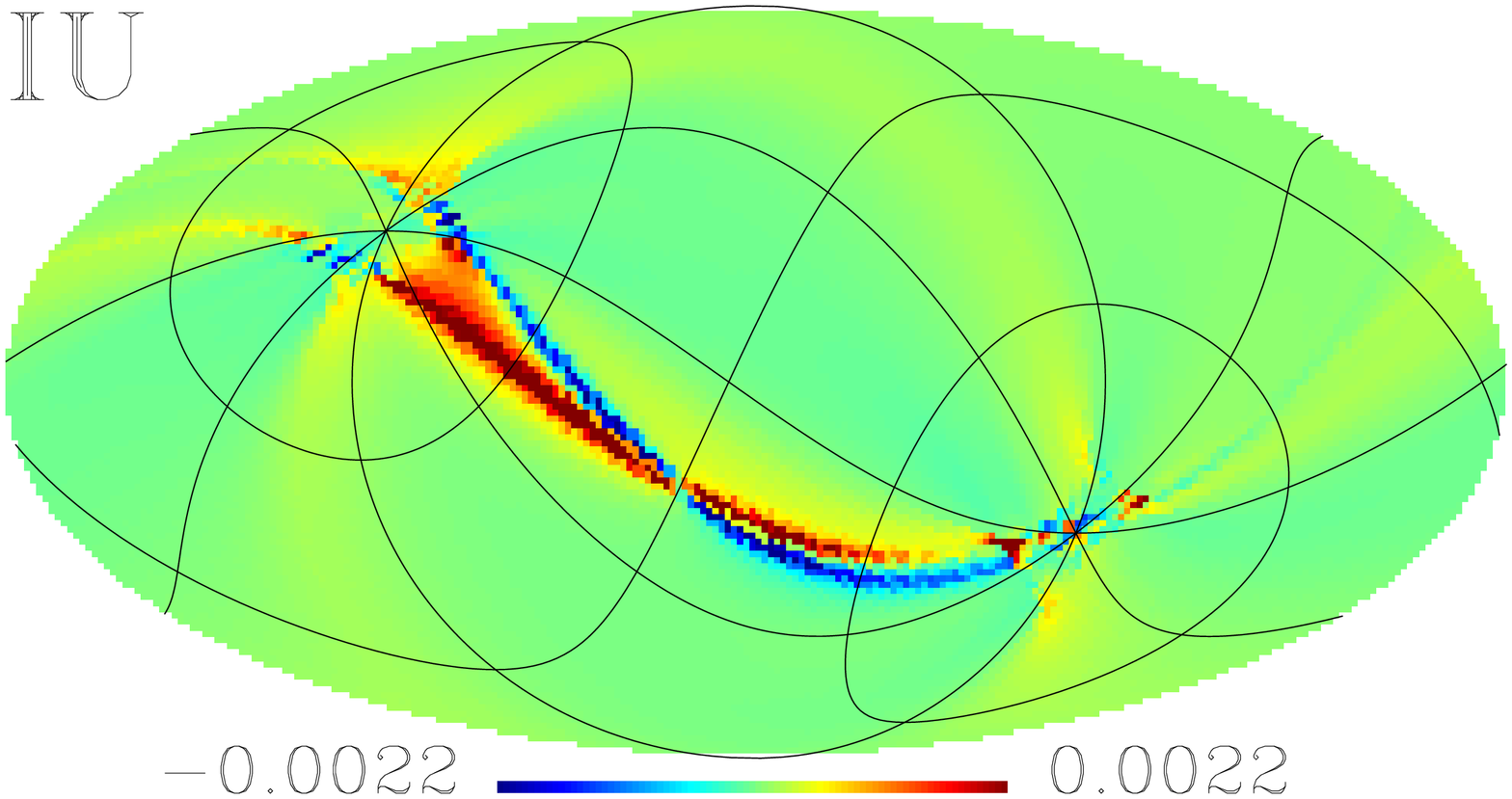}
  }
  \resizebox{\hsize}{!}{
    \includegraphics[width=0.2\textwidth,angle=90,trim=25 0 45 0,clip]
    {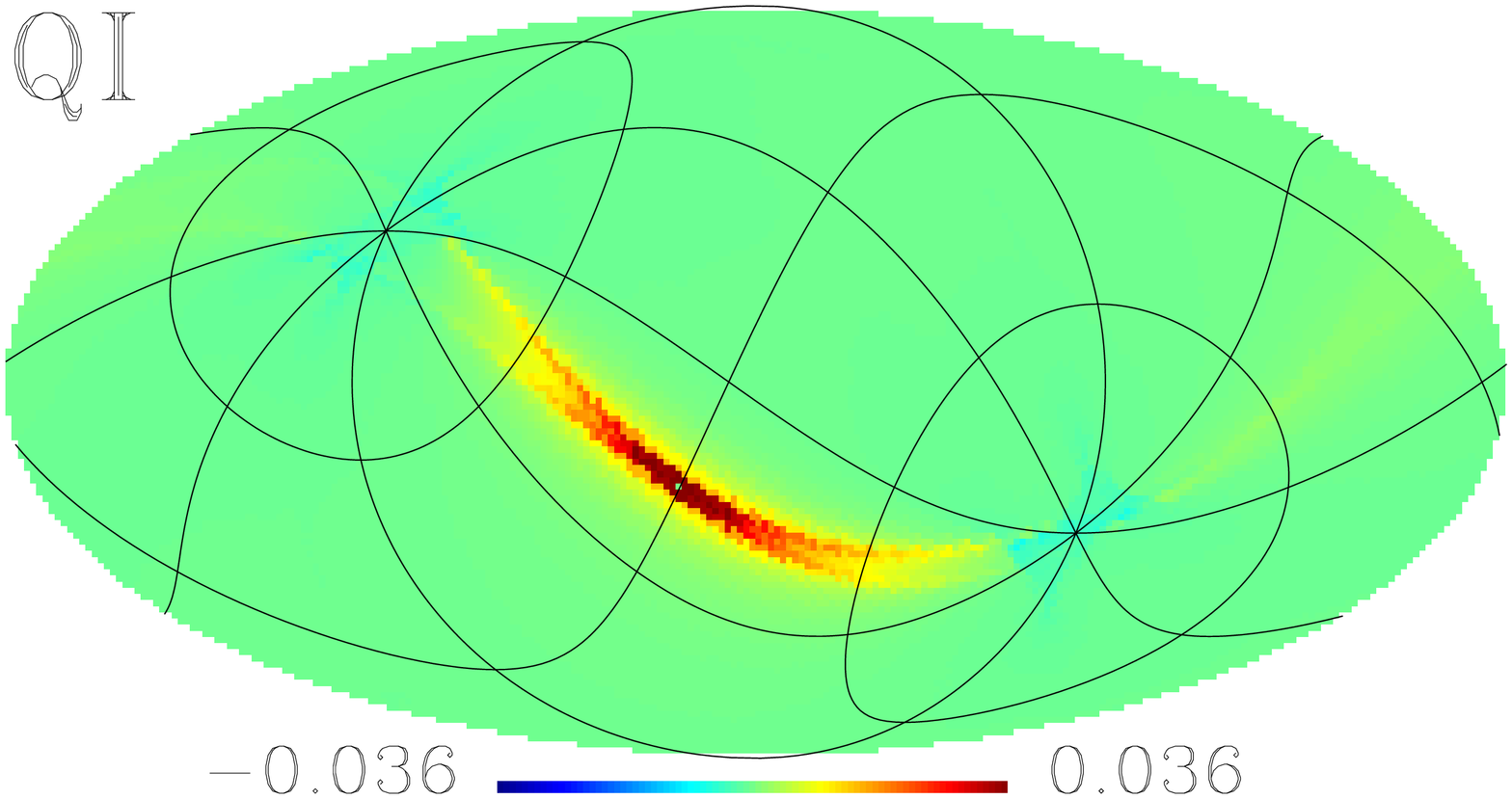}
    \includegraphics[width=0.2\textwidth,angle=90,trim=25 0 45 0,clip]
    {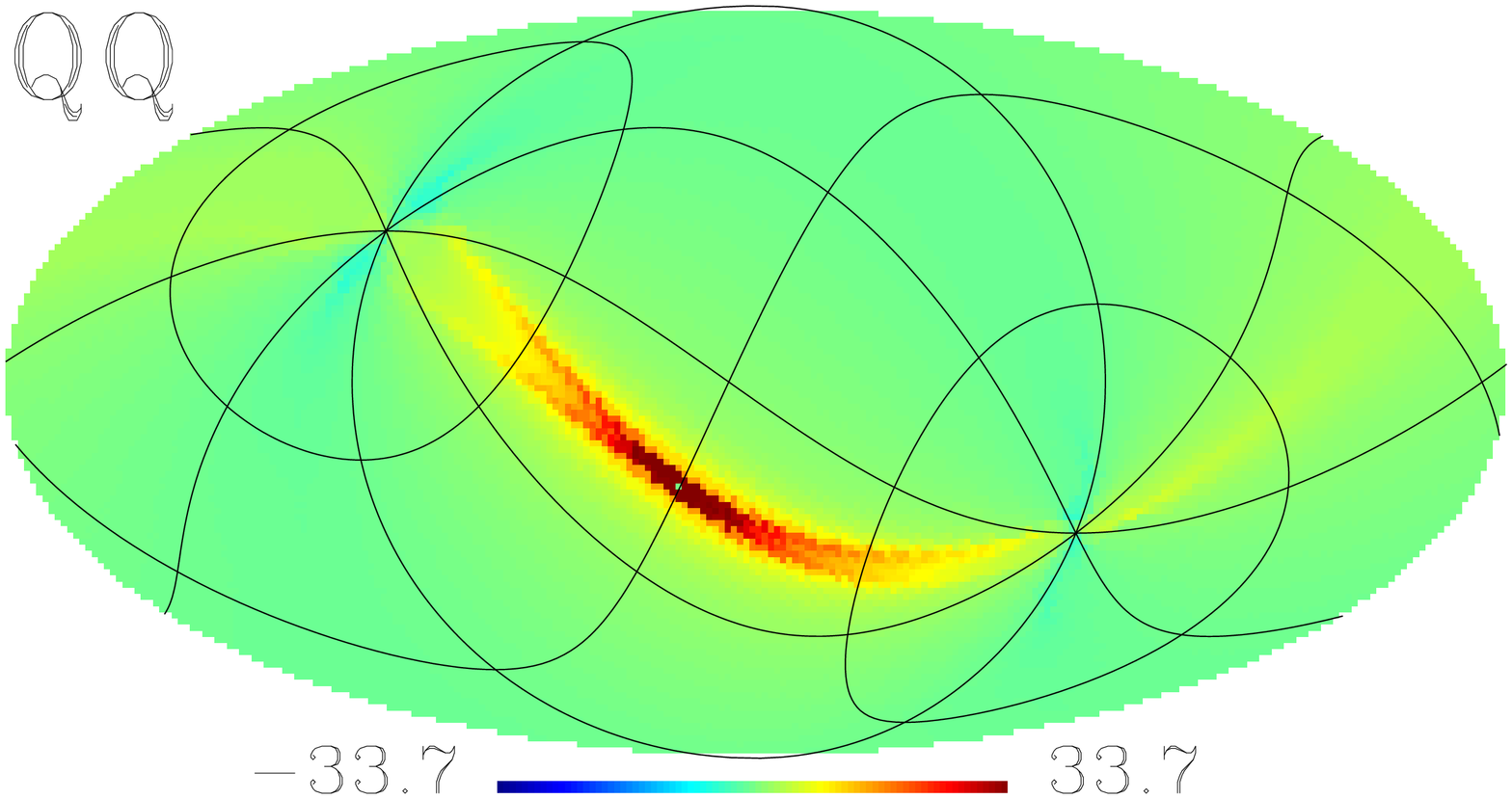}
    \includegraphics[width=0.2\textwidth,angle=90,trim=25 0 45 0,clip]
    {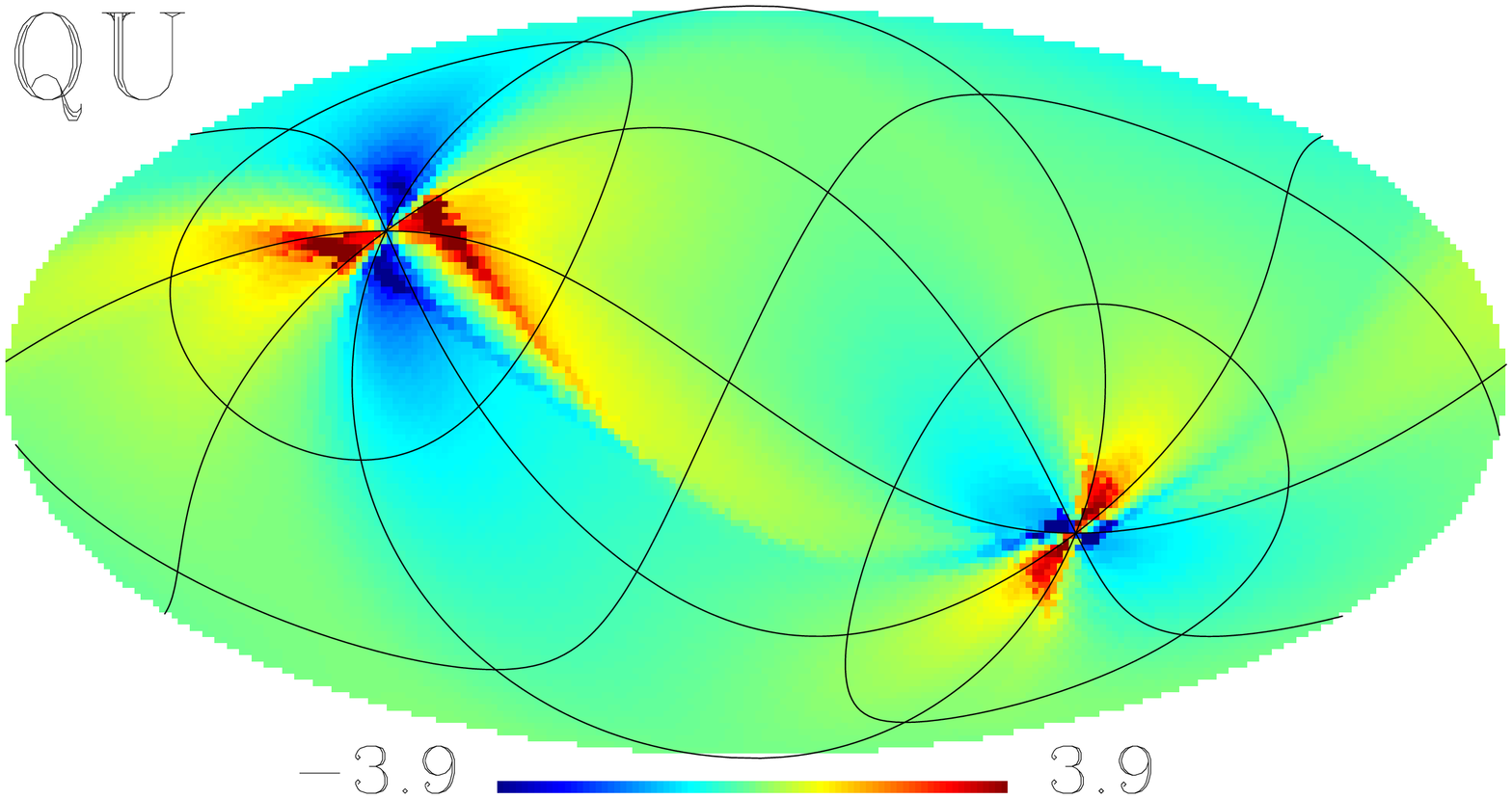}
  }
  \resizebox{\hsize}{!}{
    \includegraphics[width=0.2\textwidth,angle=90,trim=25 0 45 0,clip]
    {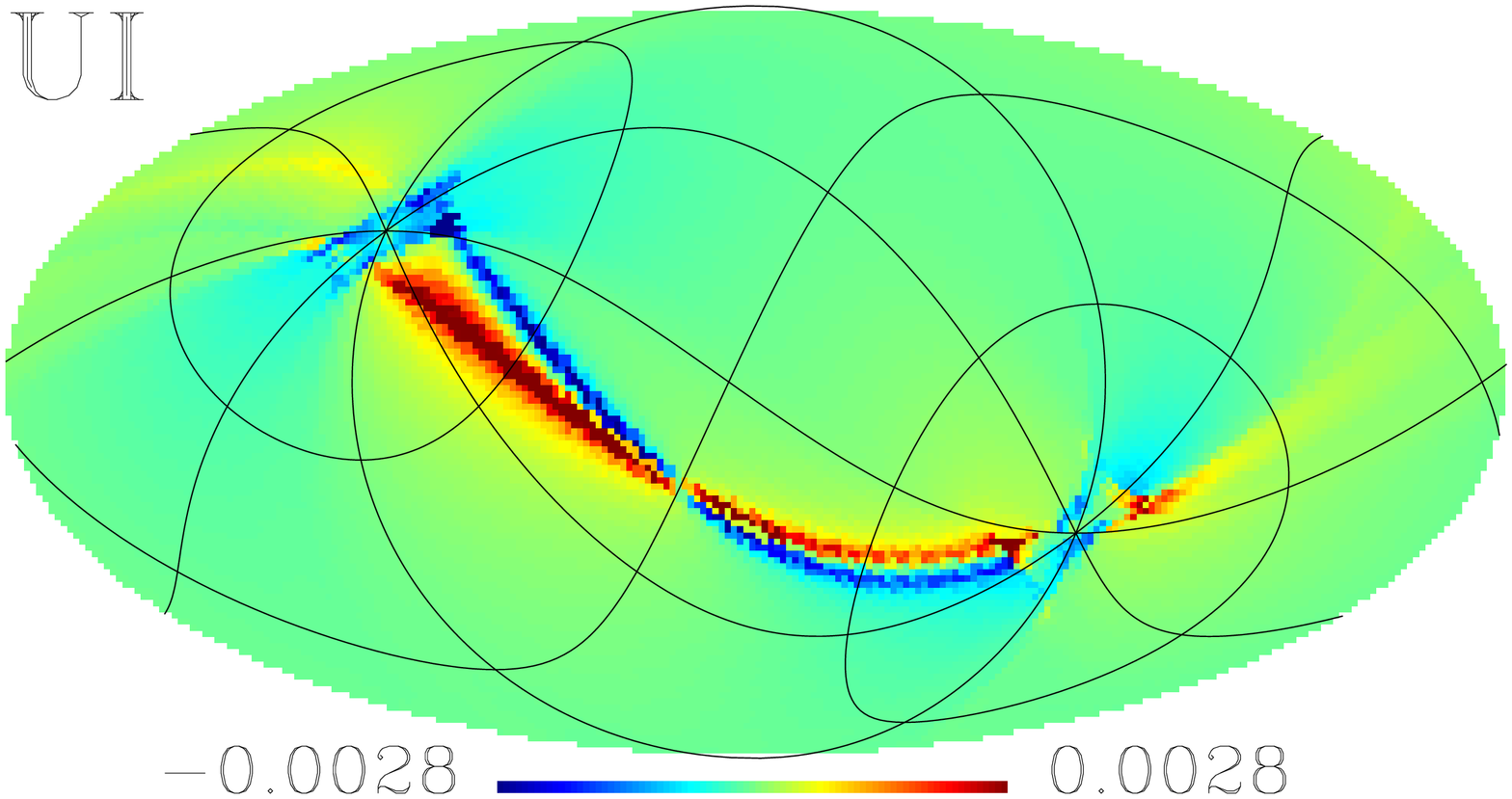}
    \includegraphics[width=0.2\textwidth,angle=90,trim=25 0 45 0,clip]
    {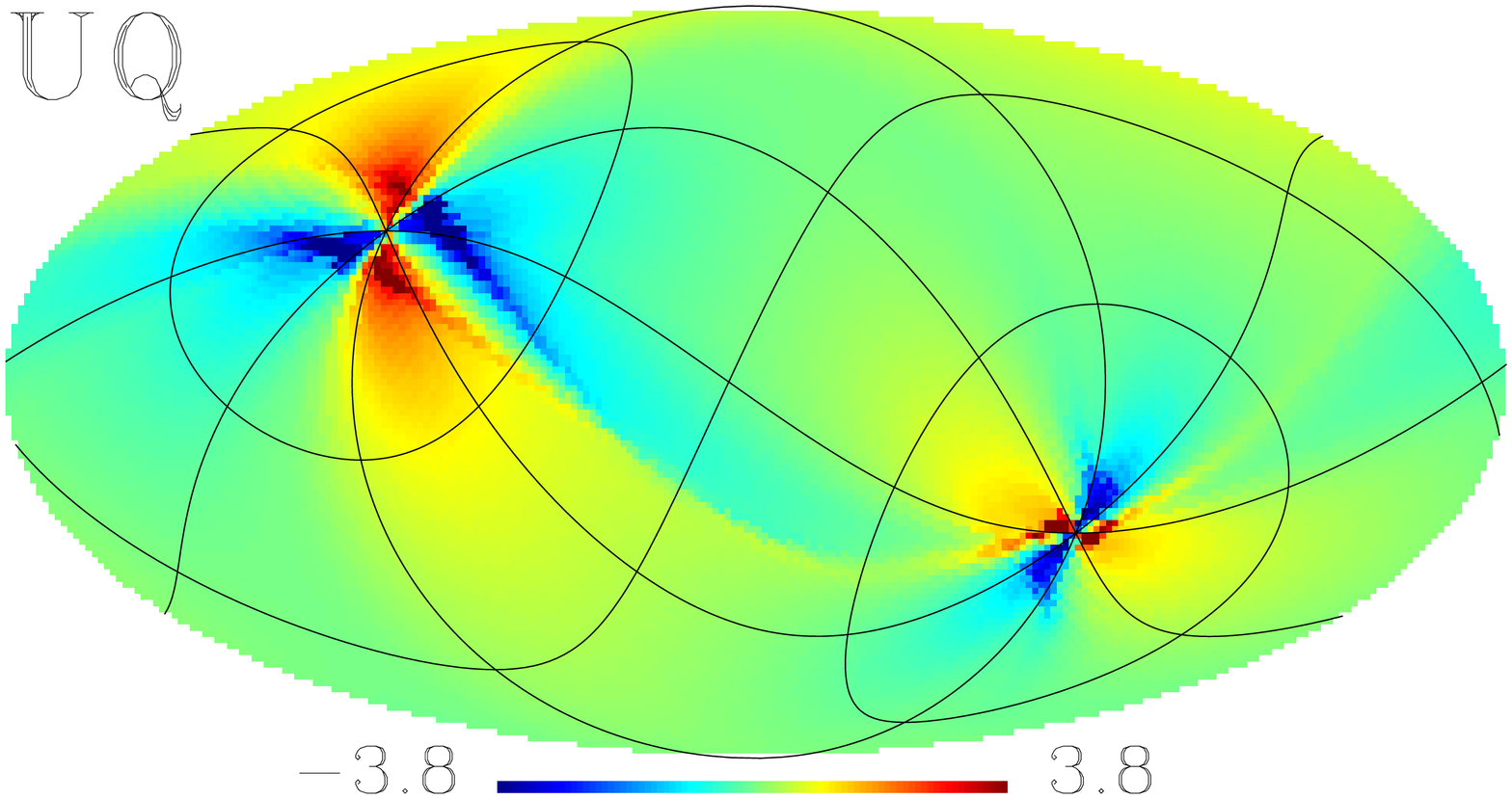}
    \includegraphics[width=0.2\textwidth,angle=90,trim=25 0 45 0,clip]
    {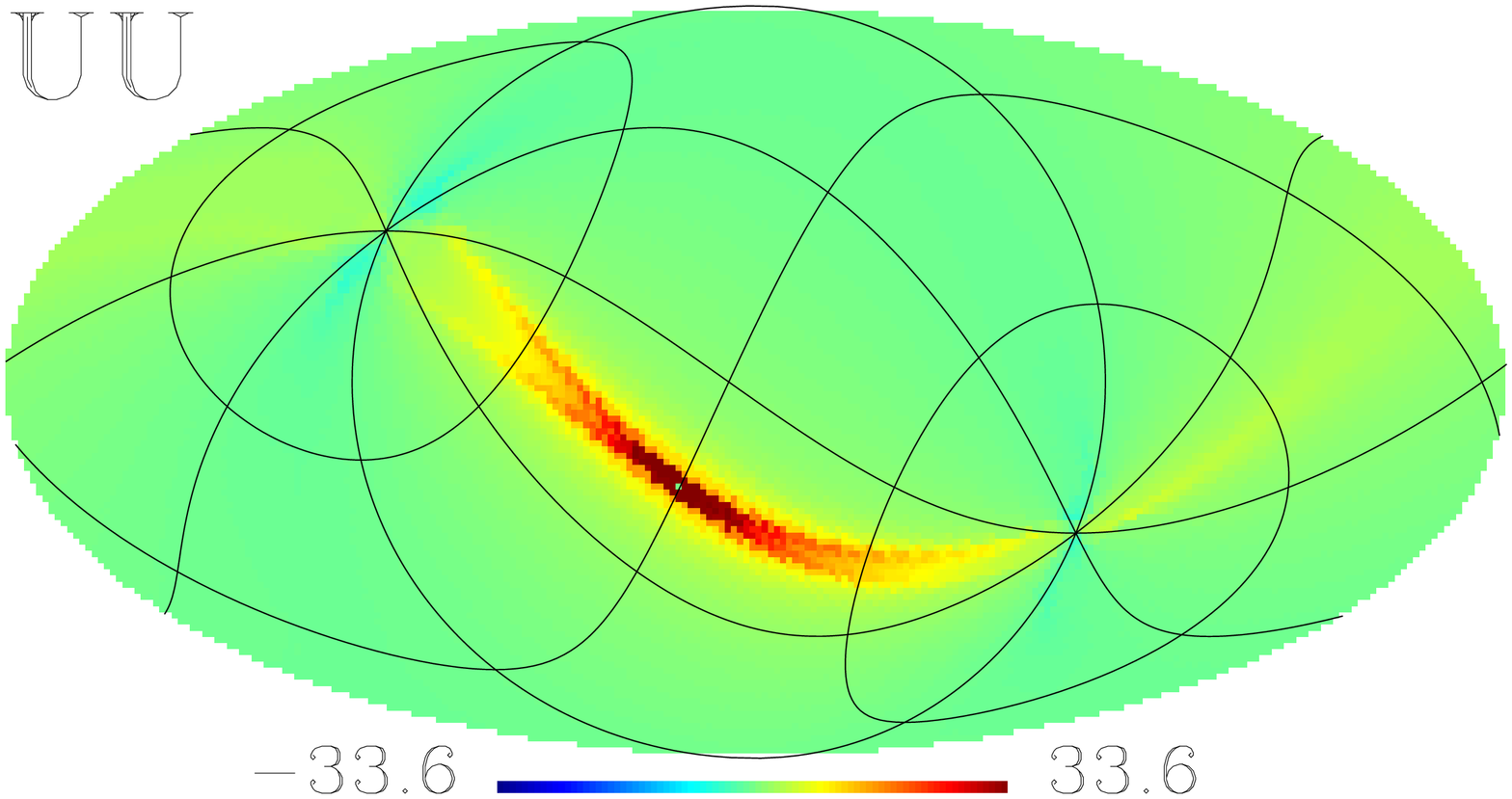}
  }
  \caption{
    A single column of the MADping covariance matrix corresponding to a
    pixel at the ecliptic equator. For both ROMA and Madam $1.25\,$s baseline
    counterparts all visible characteristics remain unchanged.
    We plot the value of the correlation coefficient, $R$, multiplied by
    $10^3$. In order to enhance the features, we have halved the range of
    the color scale.
  }
  \label{fig:MadpingNCVM_equator}
  \centering
  \resizebox{\hsize}{!}{
    \includegraphics[width=0.2\textwidth,angle=90,trim=25 0 45 0,clip]
    {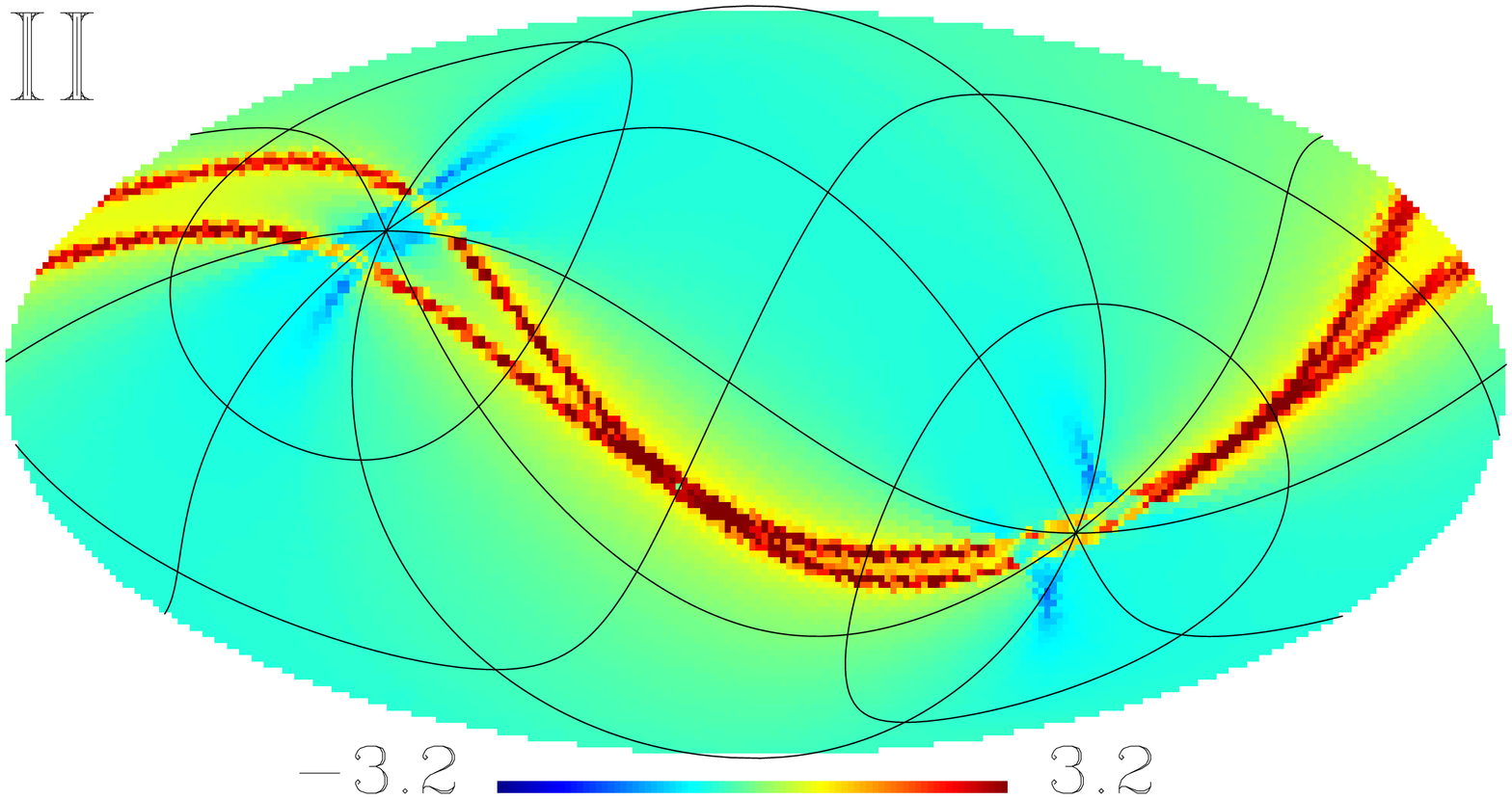}
    \includegraphics[width=0.2\textwidth,angle=90,trim=25 0 45 0,clip]
    {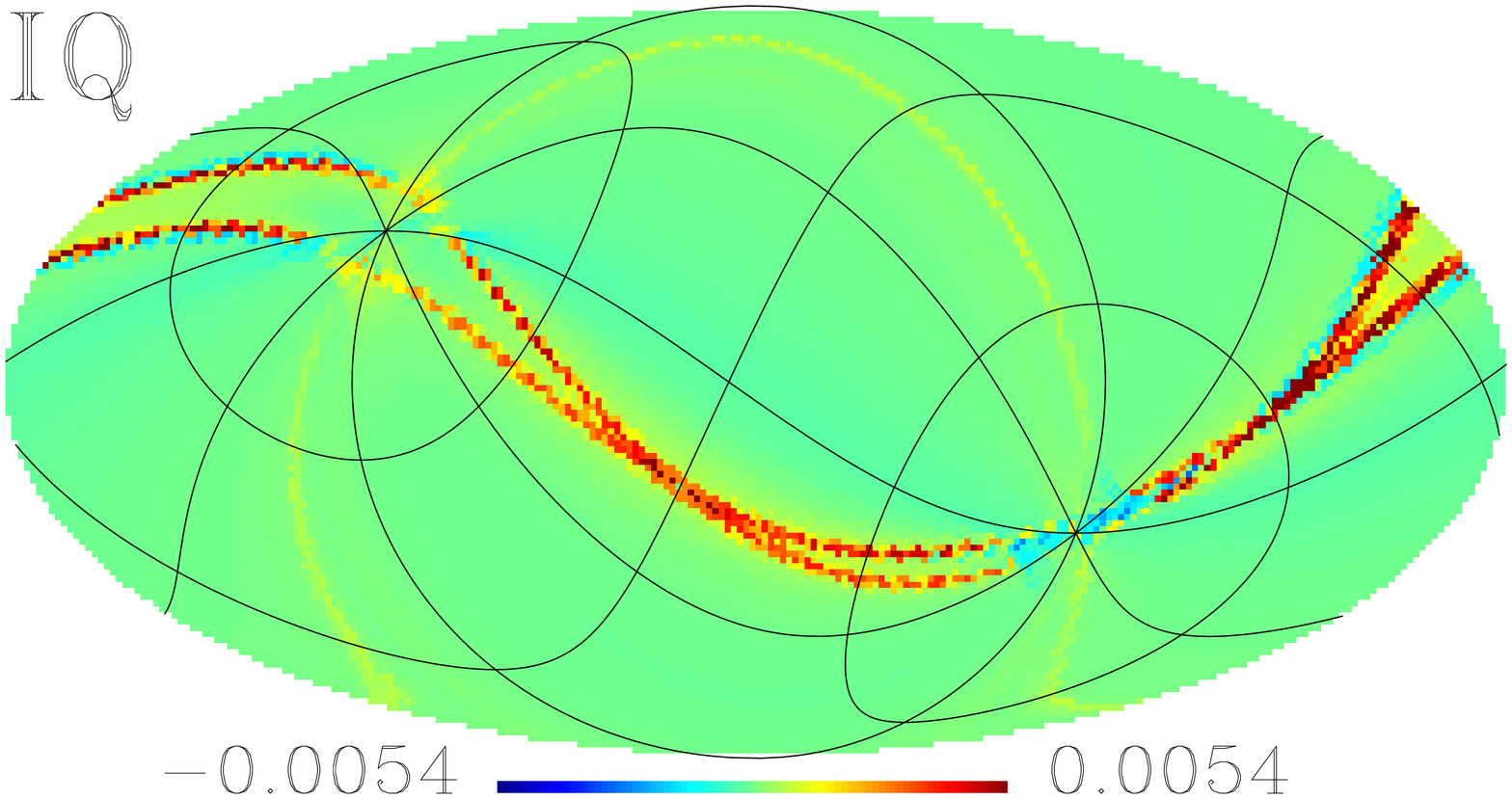}
    \includegraphics[width=0.2\textwidth,angle=90,trim=25 0 45 0,clip]
    {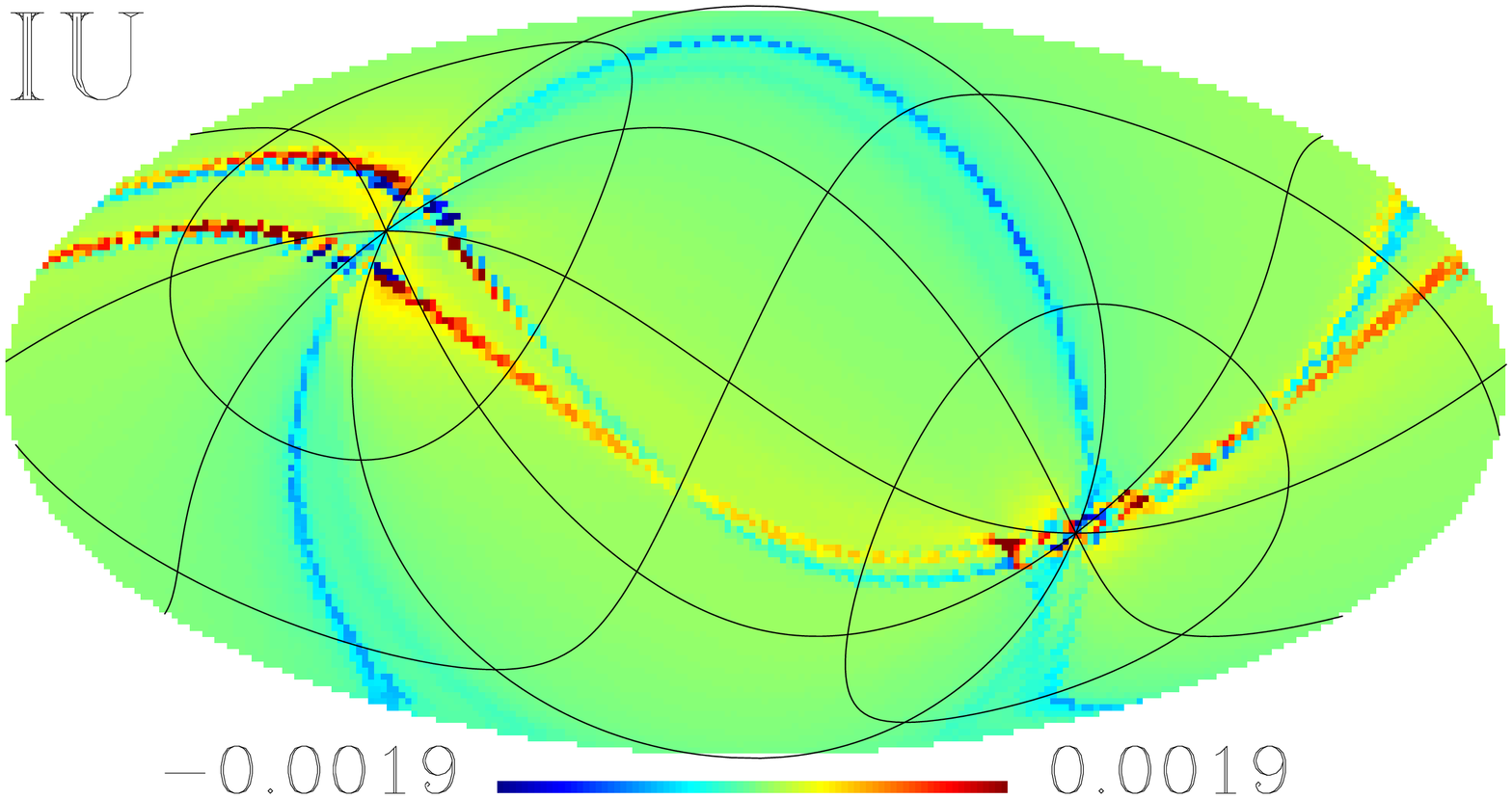}
  }
  \resizebox{\hsize}{!}{
    \includegraphics[width=0.2\textwidth,angle=90,trim=25 0 45 0,clip]
    {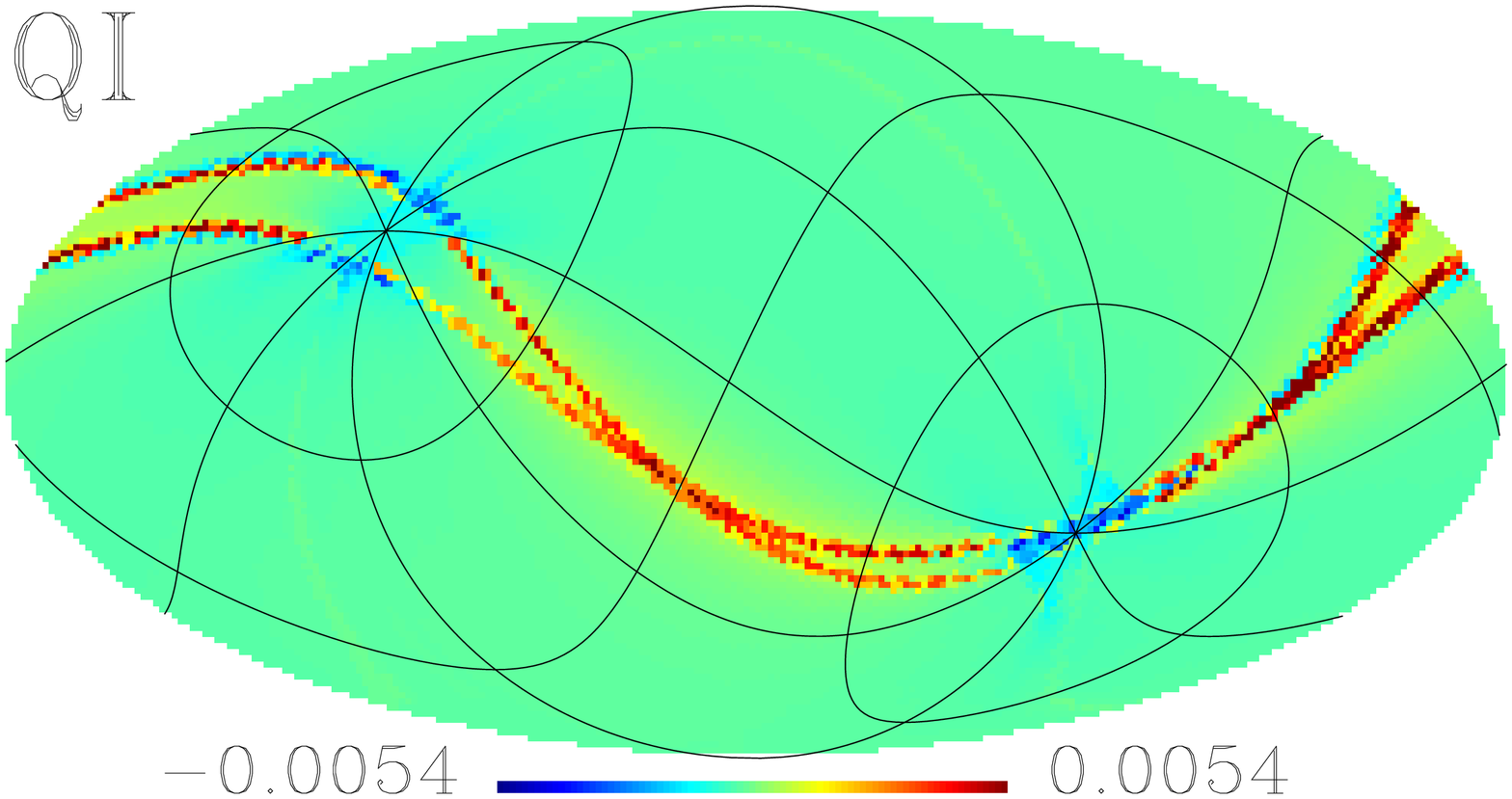}
    \includegraphics[width=0.2\textwidth,angle=90,trim=25 0 45 0,clip]
    {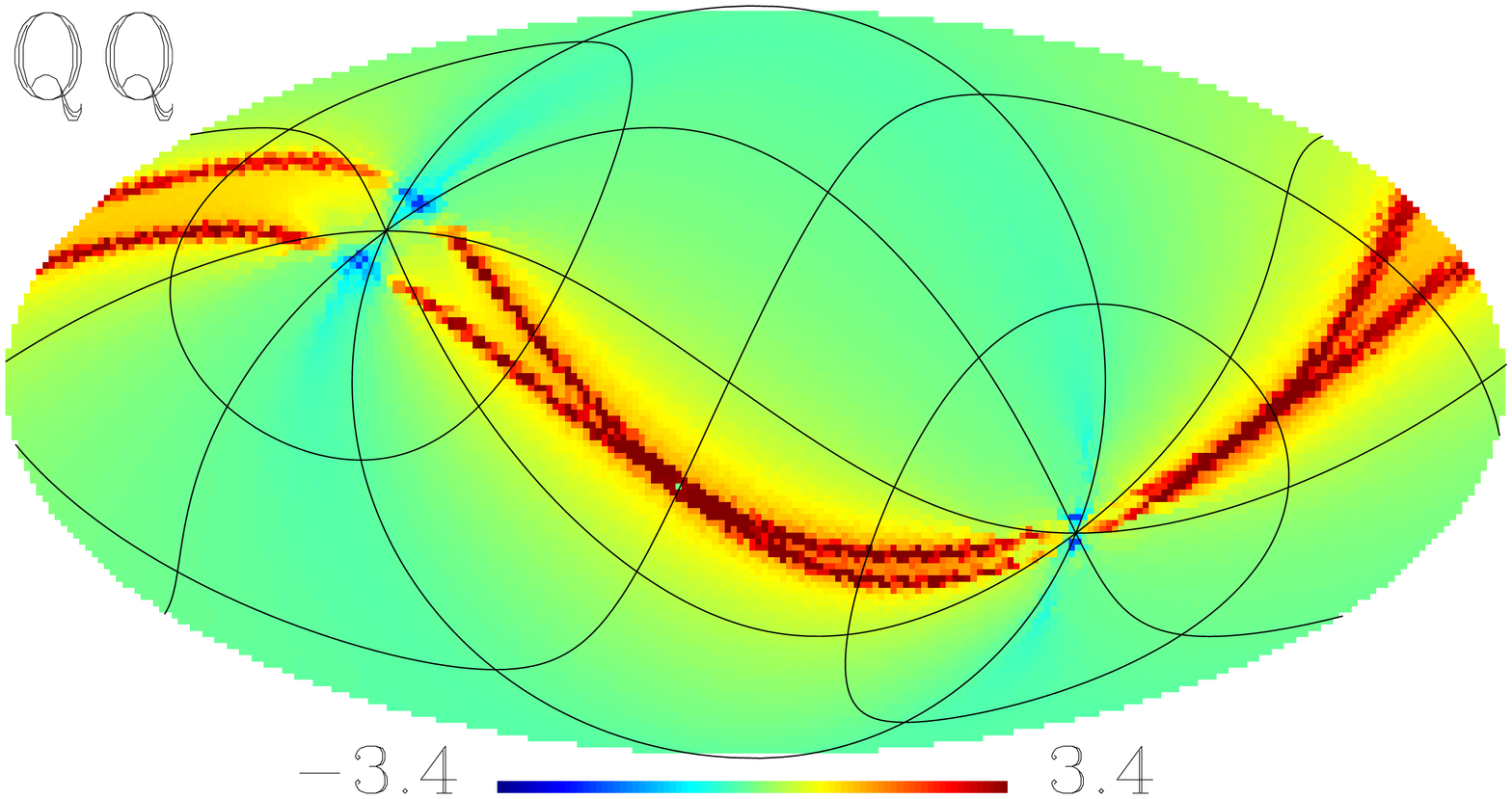}
    \includegraphics[width=0.2\textwidth,angle=90,trim=25 0 45 0,clip]
    {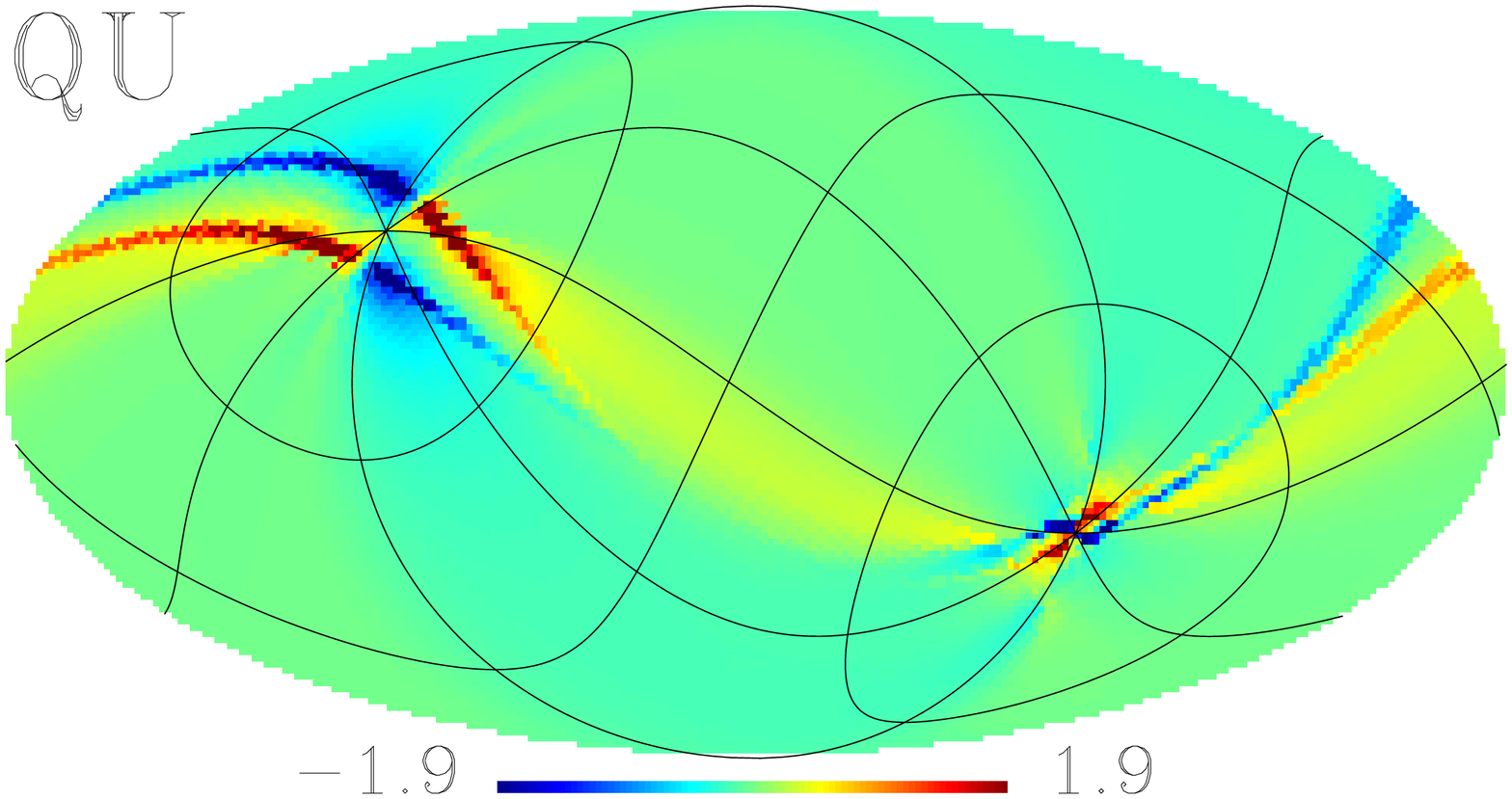}
  }
  \resizebox{\hsize}{!}{
    \includegraphics[width=0.2\textwidth,angle=90,trim=25 0 45 0,clip]
    {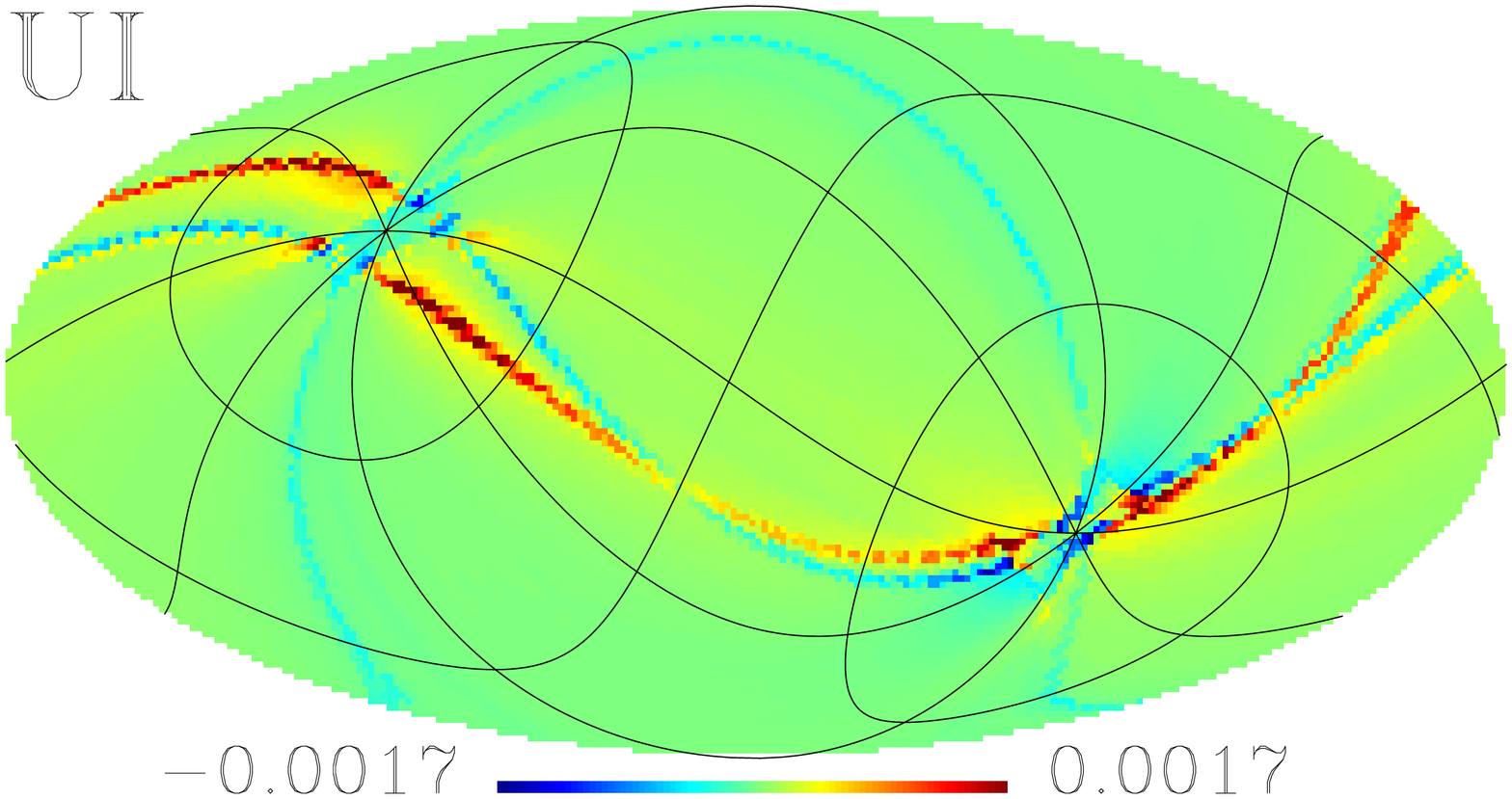}
    \includegraphics[width=0.2\textwidth,angle=90,trim=25 0 45 0,clip]
    {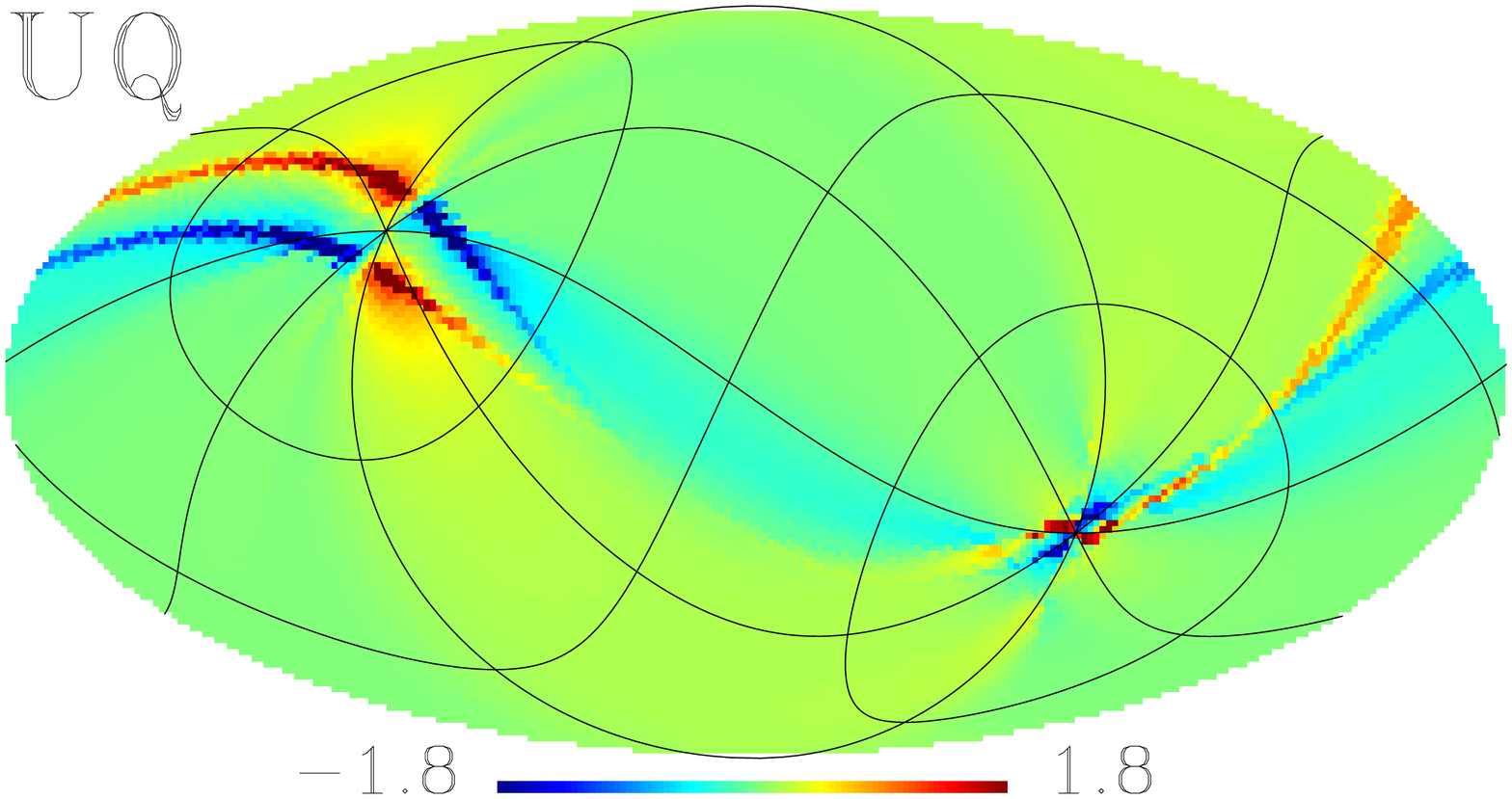}
    \includegraphics[width=0.2\textwidth,angle=90,trim=25 0 45 0,clip]
    {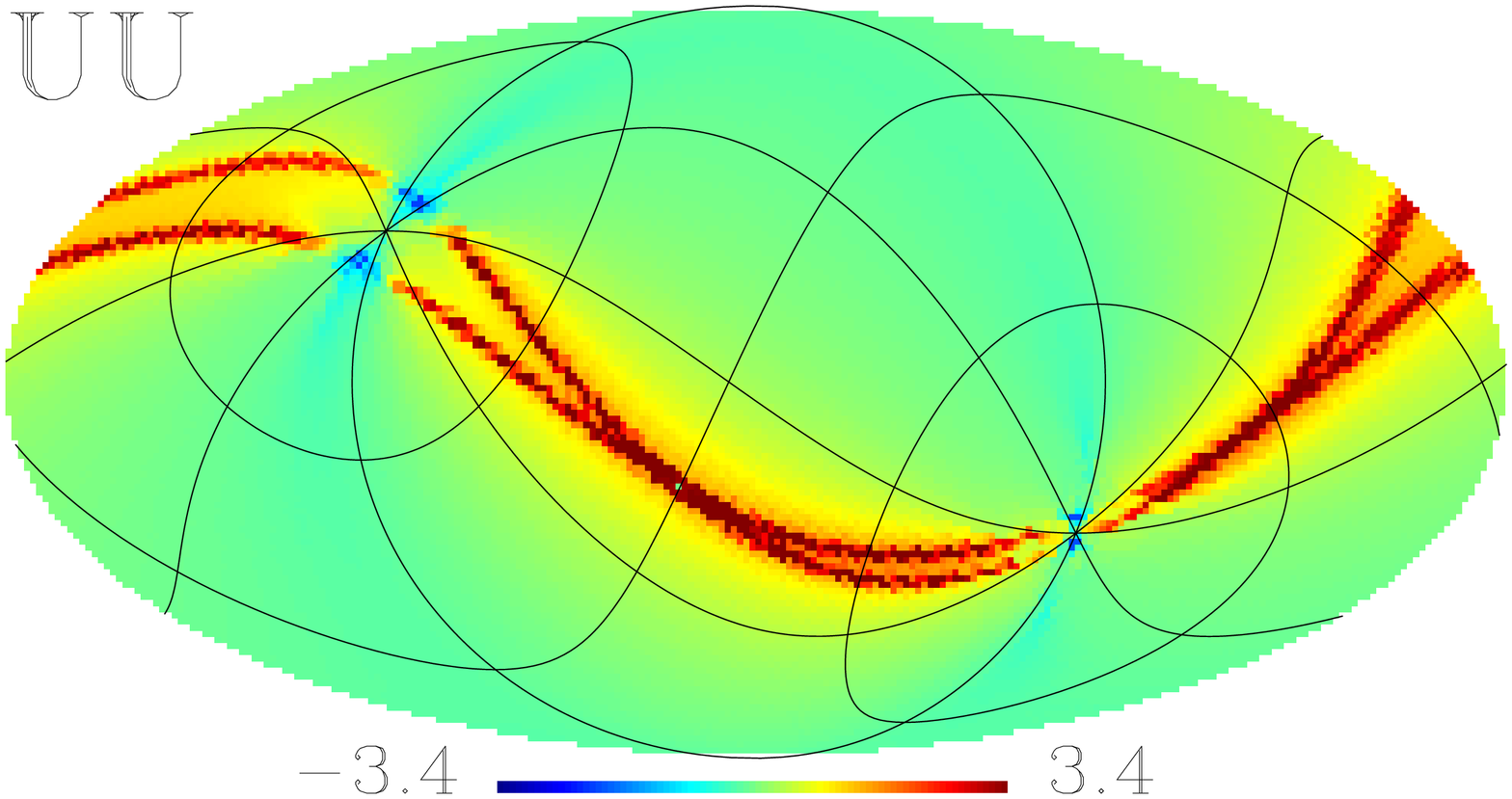}
  }
  \caption{
    A single column of the Springtide covariance matrix corresponding to a
    pixel near the ecliptic equator. For description of the normalization,
    see text.
  }
  \label{fig:SprintideNCVM_equator}
\end{figure}

Figs.~\ref{fig:MadpingNCVM_equator}--\ref{fig:SprintideNCVM_equator} show
plots of a single column of the noise covariance matrix. The column
corresponds to reference pixel number 0.
In the HEALPix nested pixelization scheme for
$N_\mathrm{side}=32$ resolution,  pixel 0 is located at the equator.
In the plots each pixel has the value $\langle m_p m_q\rangle$ normalized by
$\sqrt{\langle m_p^2 \rangle \langle m_q^2  \rangle}$.
Thus the pixel values of the plots represent correlation coefficients.
Due to this normalization, the reference pixel automatically gets unit value
and is later set to zero in order to bring out smaller
features of the other elements of the columns.


The sky is scanned from one ecliptic pole to the other. The NCM
column maps are characterized by bands of correlation along the
scanning rings.
Pixels near the equator,
such as the reference pixel 0, are only observed during a few-hour window
as the satellite scanning ring is rotated over the course of the survey.
The two crossing bands of higher correlation correspond to two pointing
periods half a year a part that observe the reference pixel 0.

For both generalized destriping and optimal map-making, there is a visible
gradient in the correlation along the scanning ring. Pixels that are
observed immediately before or after the reference pixel have the strongest
correlation. The conventional destriping with its hour long baselines
assumes constant correlated noise over the scanning ring and does not,
therefore, show this feature.

Figs.~\ref{fig:MadpingNCVM_equator}
and \ref{fig:SprintideNCVM_equator} show that the strongest
cross-correlations ($\sim 1\%$) exist between Q and U noise maps.
Side and Main polarization sensitive directions differ slightly from
$90^\circ$ and, as a result, small ($\sim 10^{-4}\%$) IQ and IU
correlations remain.

\begin{figure}[!tbp]
  \centering
  \resizebox{\hsize}{!}{
    \includegraphics[width=0.2\textwidth,angle=90,trim=25 0 45 0,clip]
    {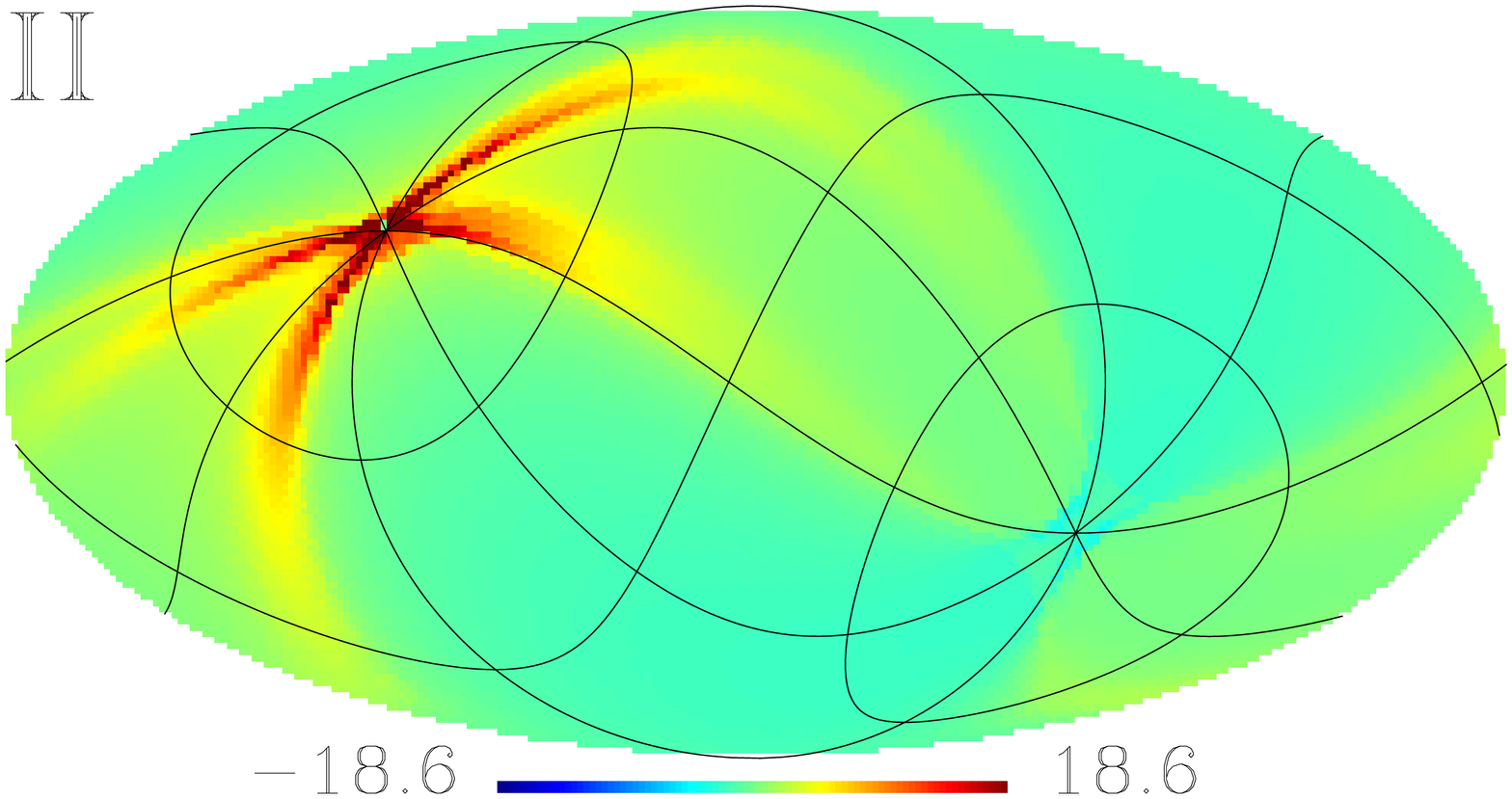}
    \includegraphics[width=0.2\textwidth,angle=90,trim=25 0 45 0,clip]
    {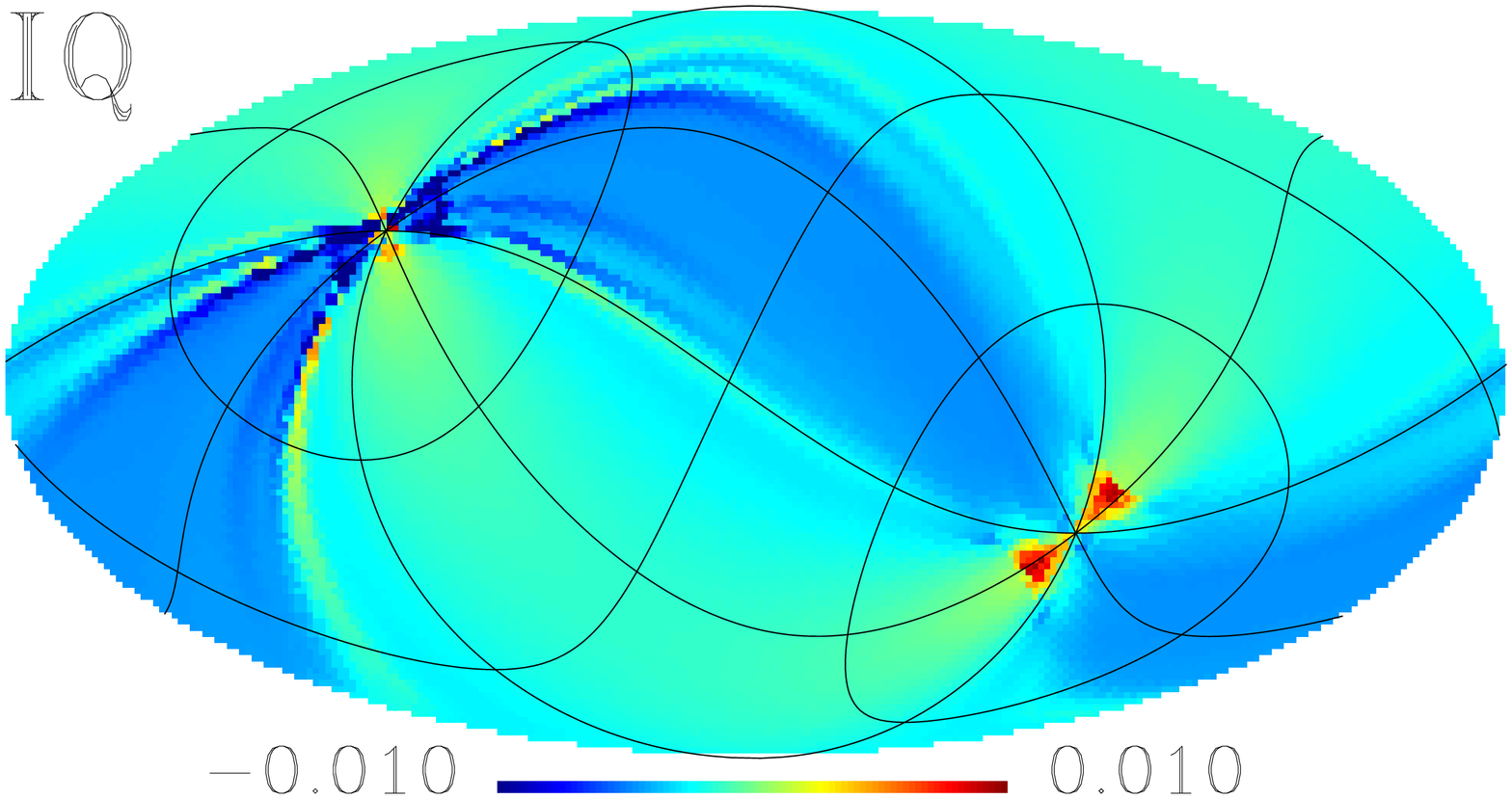}
    \includegraphics[width=0.2\textwidth,angle=90,trim=25 0 45 0,clip]
    {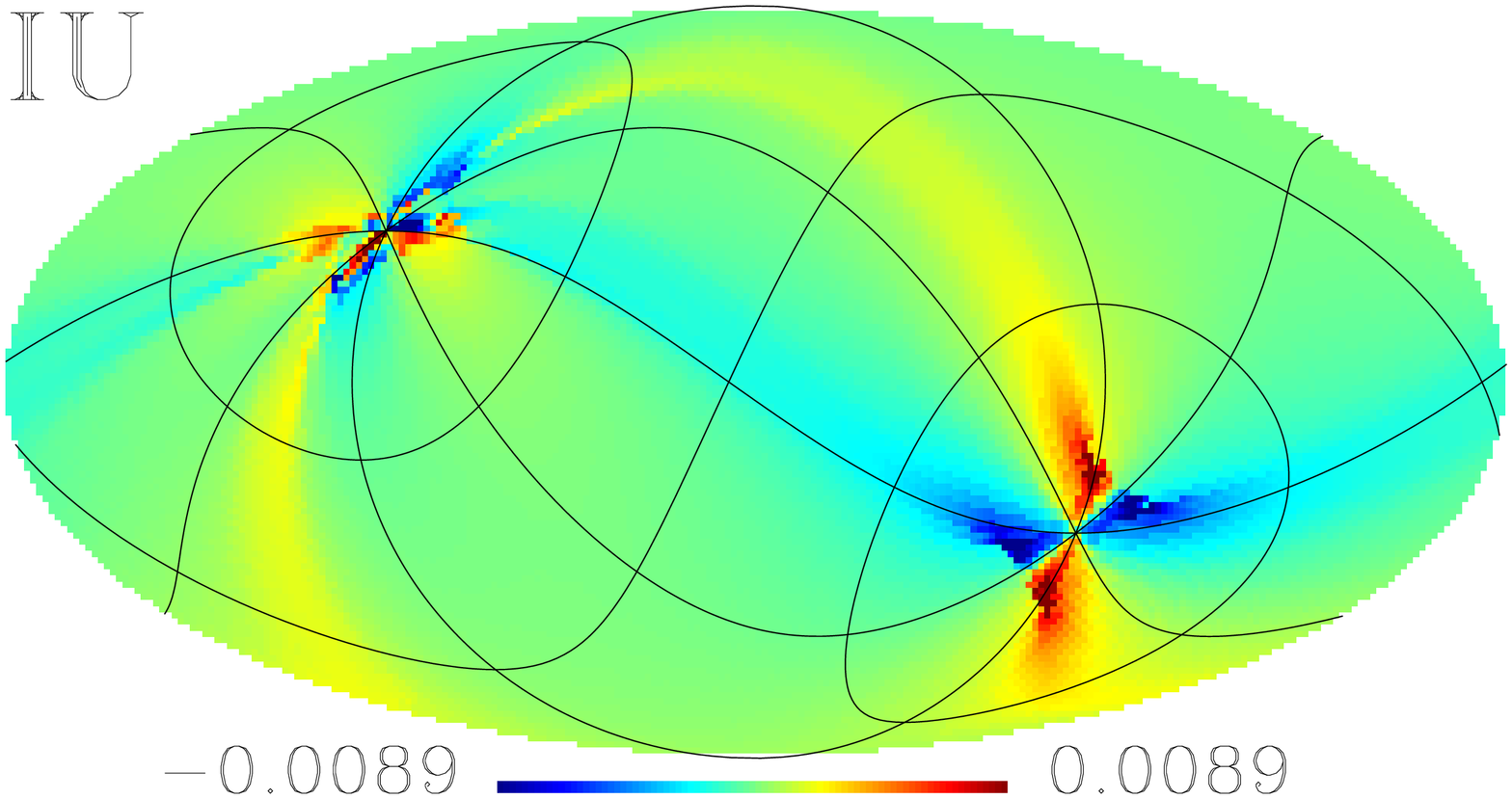}
  }
  \resizebox{\hsize}{!}{
    \includegraphics[width=0.2\textwidth,angle=90,trim=25 0 45 0,clip]
    {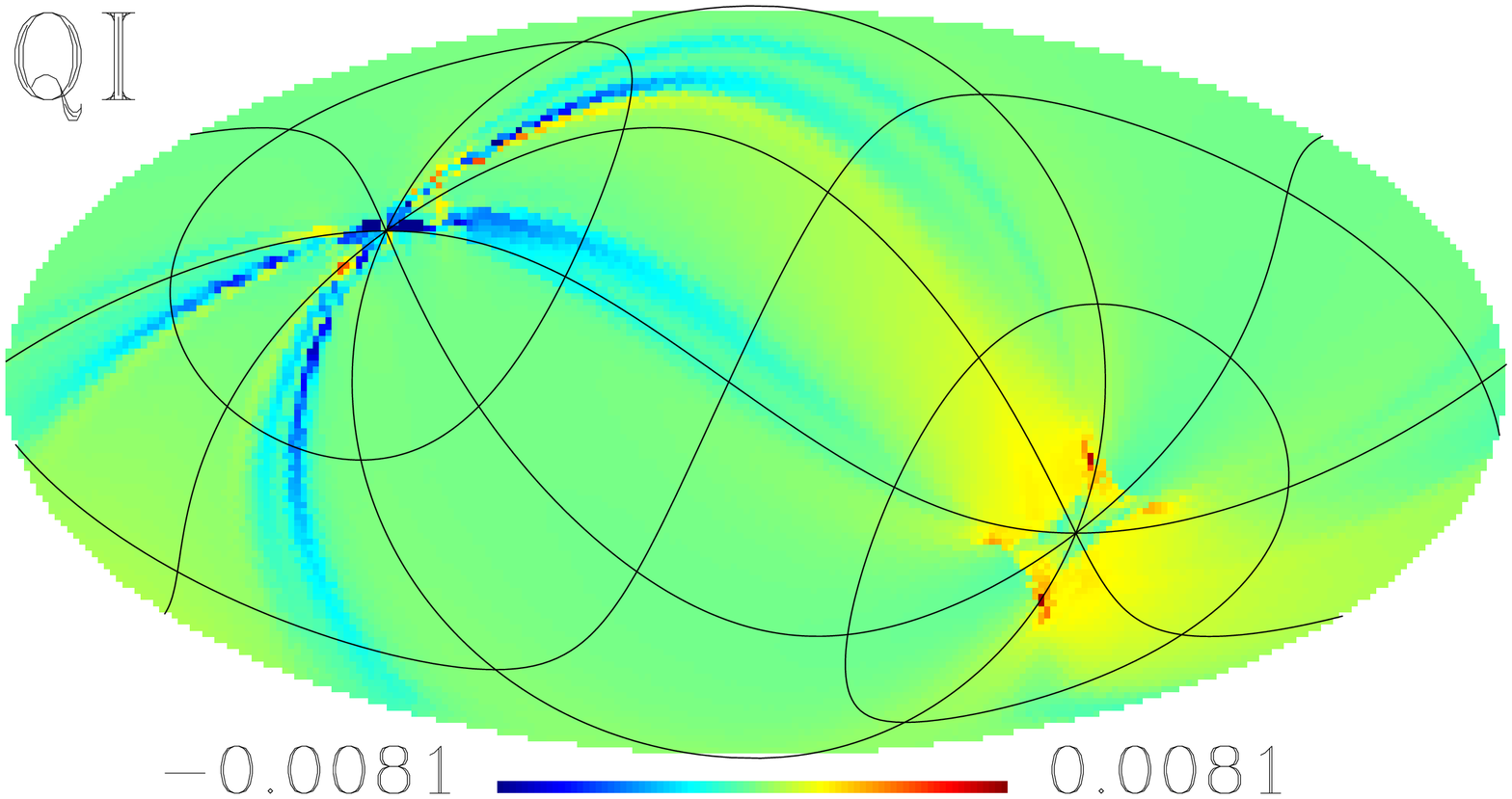}
    \includegraphics[width=0.2\textwidth,angle=90,trim=25 0 45 0,clip]
    {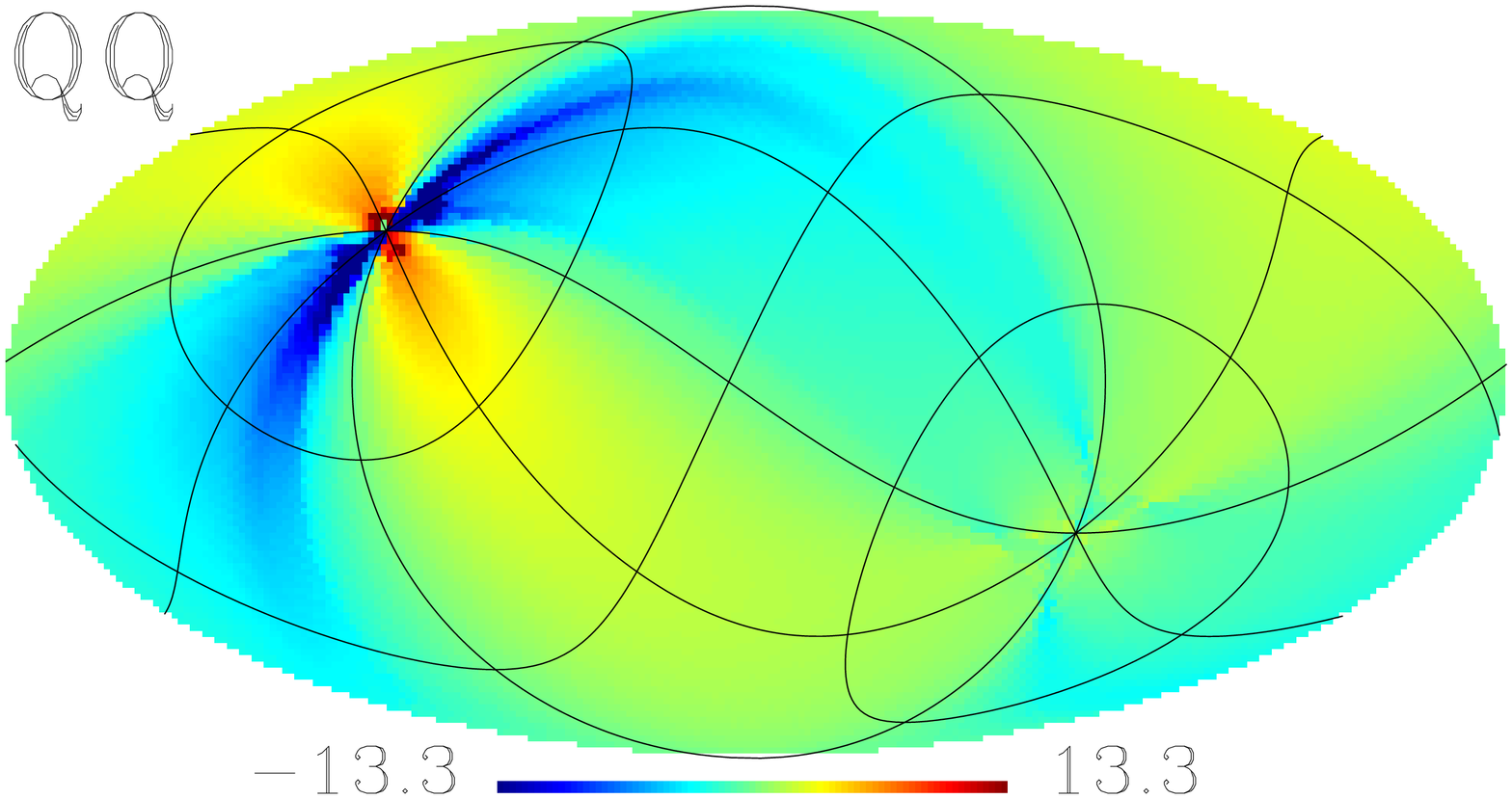}
    \includegraphics[width=0.2\textwidth,angle=90,trim=25 0 45 0,clip]
    {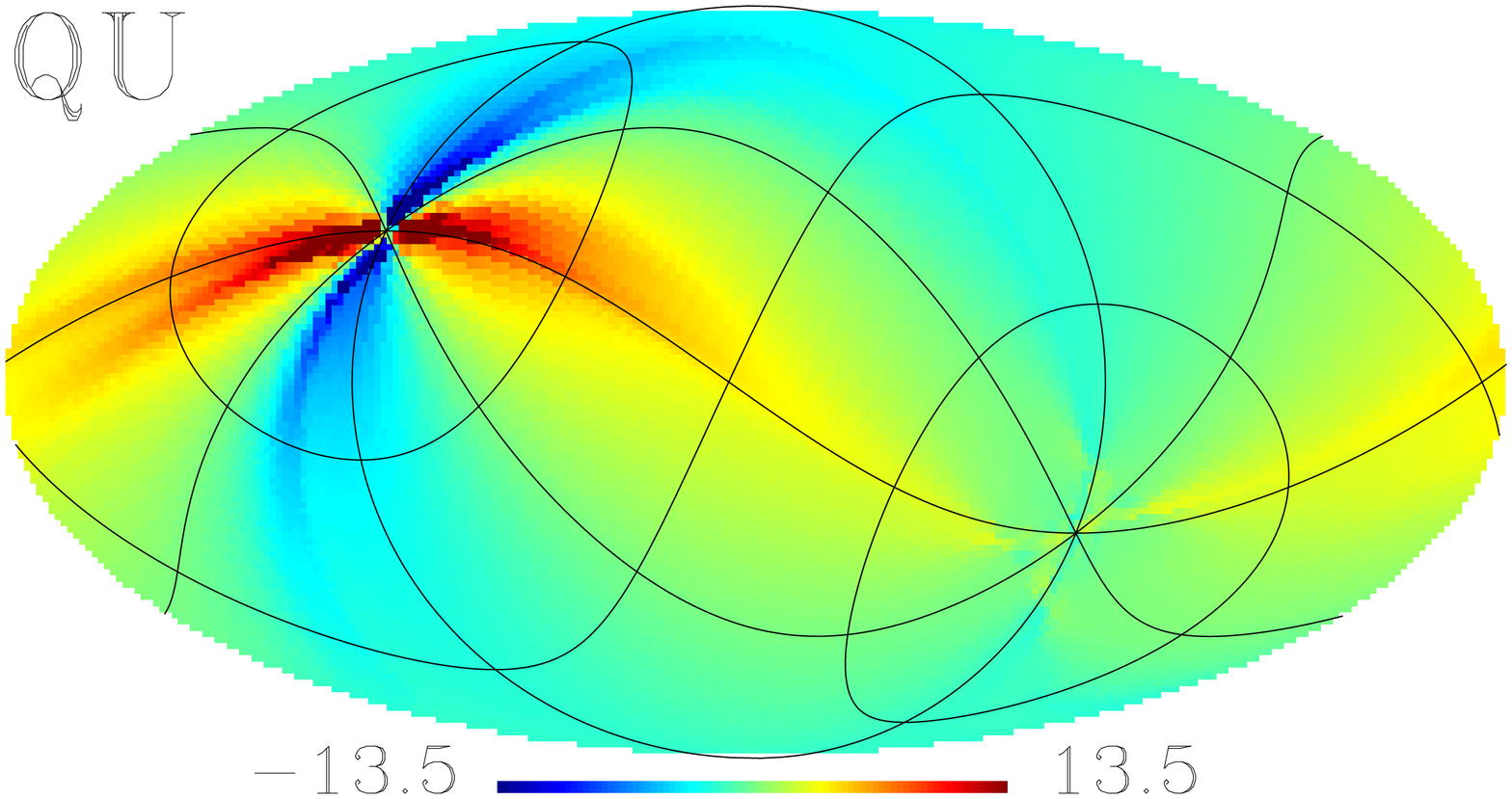}
  }
  \resizebox{\hsize}{!}{
    \includegraphics[width=0.2\textwidth,angle=90,trim=25 0 45 0,clip]
    {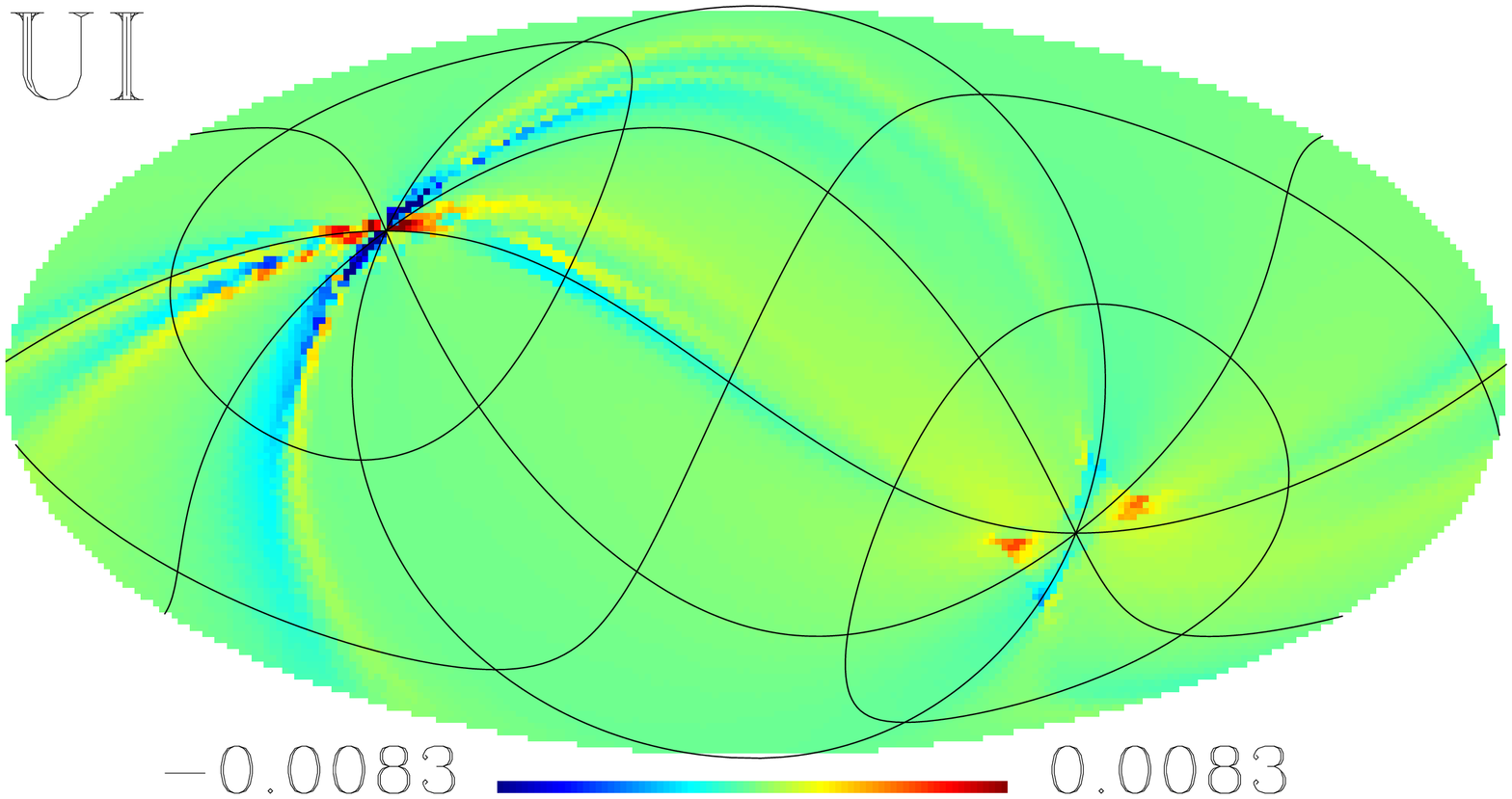}
    \includegraphics[width=0.2\textwidth,angle=90,trim=25 0 45 0,clip]
    {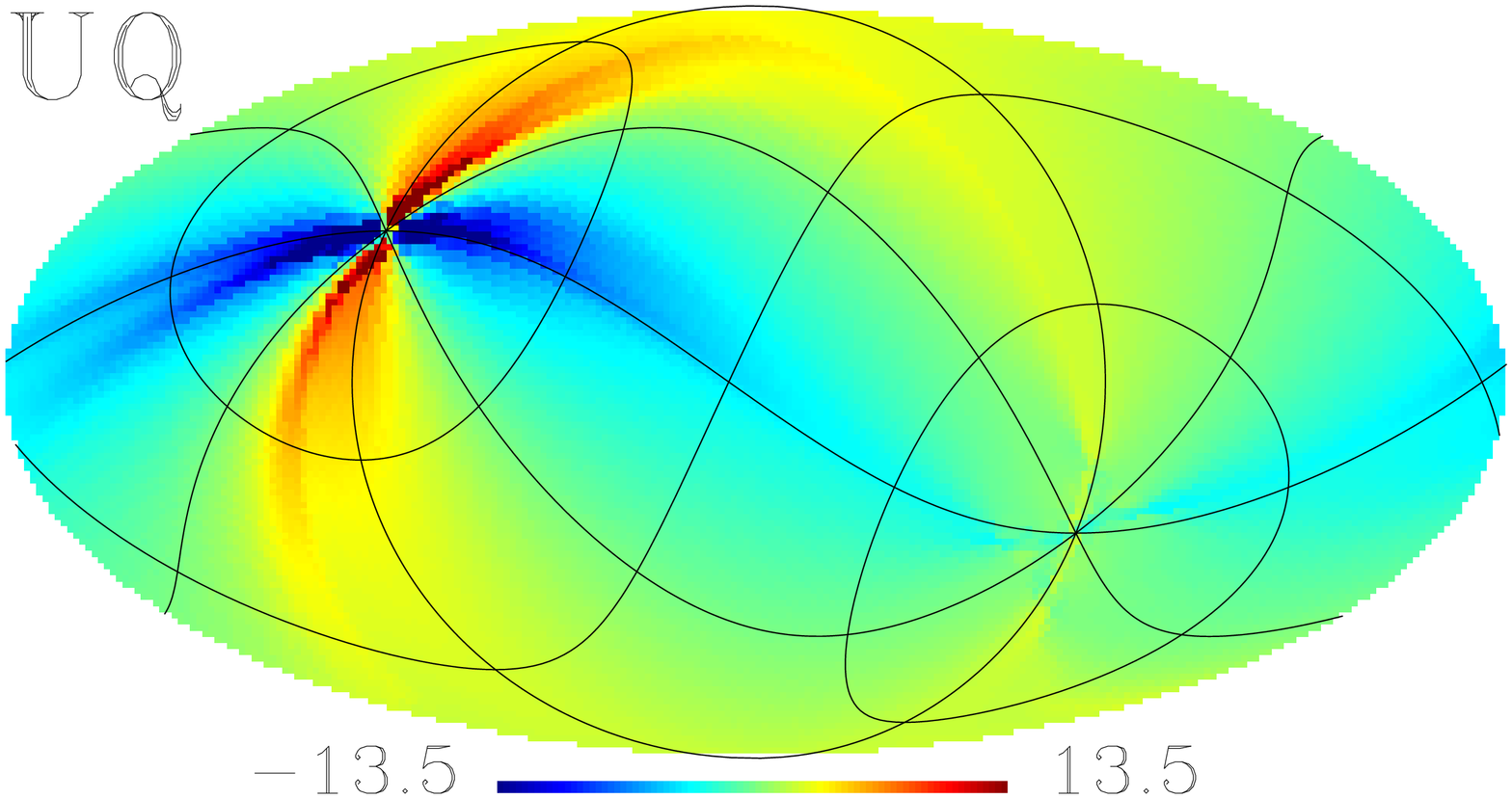}
    \includegraphics[width=0.2\textwidth,angle=90,trim=25 0 45 0,clip]
    {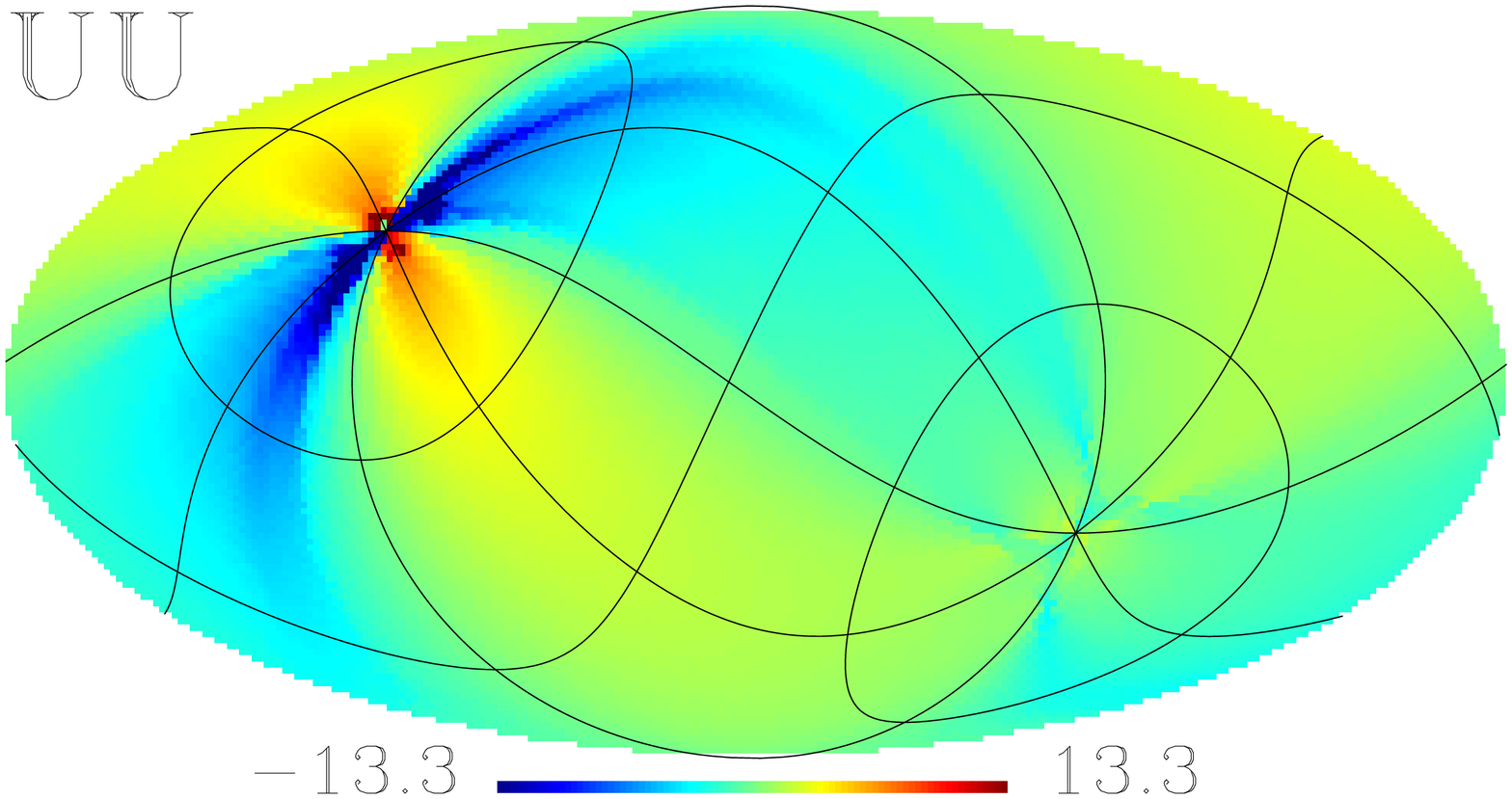}
  }  \caption{
    A single column of the MADping covariance matrix corresponding to a
    pixel at the ecliptic pole. For both ROMA and Madam $1.25\,$s baseline
    counterparts all visible characteristics remain unchanged.
    We plot the value of the correlation coefficient, $R$, multiplied by
    $10^3$. In order to enhance the features, we have halved the range of
    the color scale.
  }
  \label{fig:MadpingNCVM_pole}
  \resizebox{\hsize}{!}{
    \includegraphics[width=0.2\textwidth,angle=90,trim=25 0 45 0,clip]
    {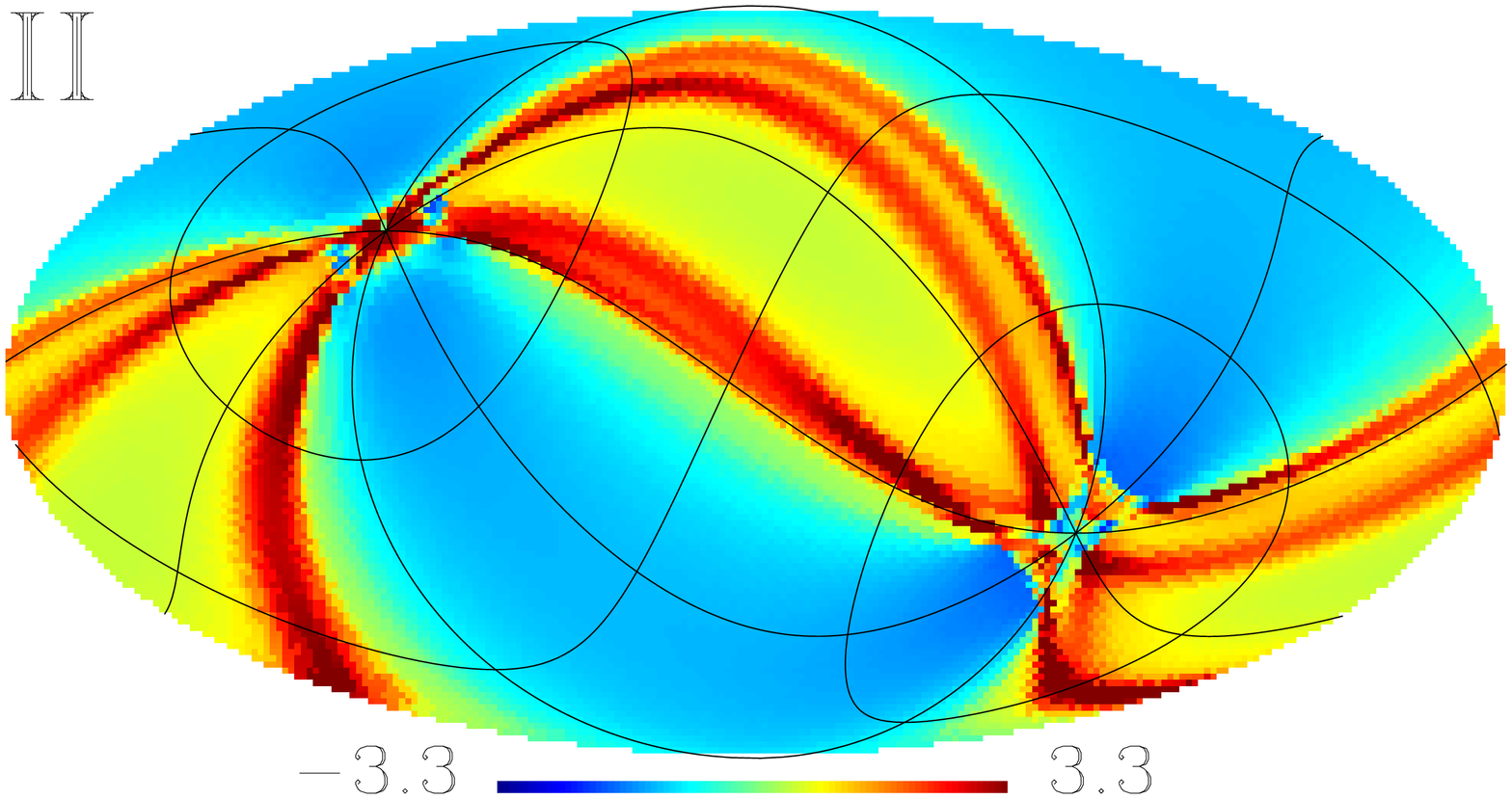}
    \includegraphics[width=0.2\textwidth,angle=90,trim=25 0 45 0,clip]
    {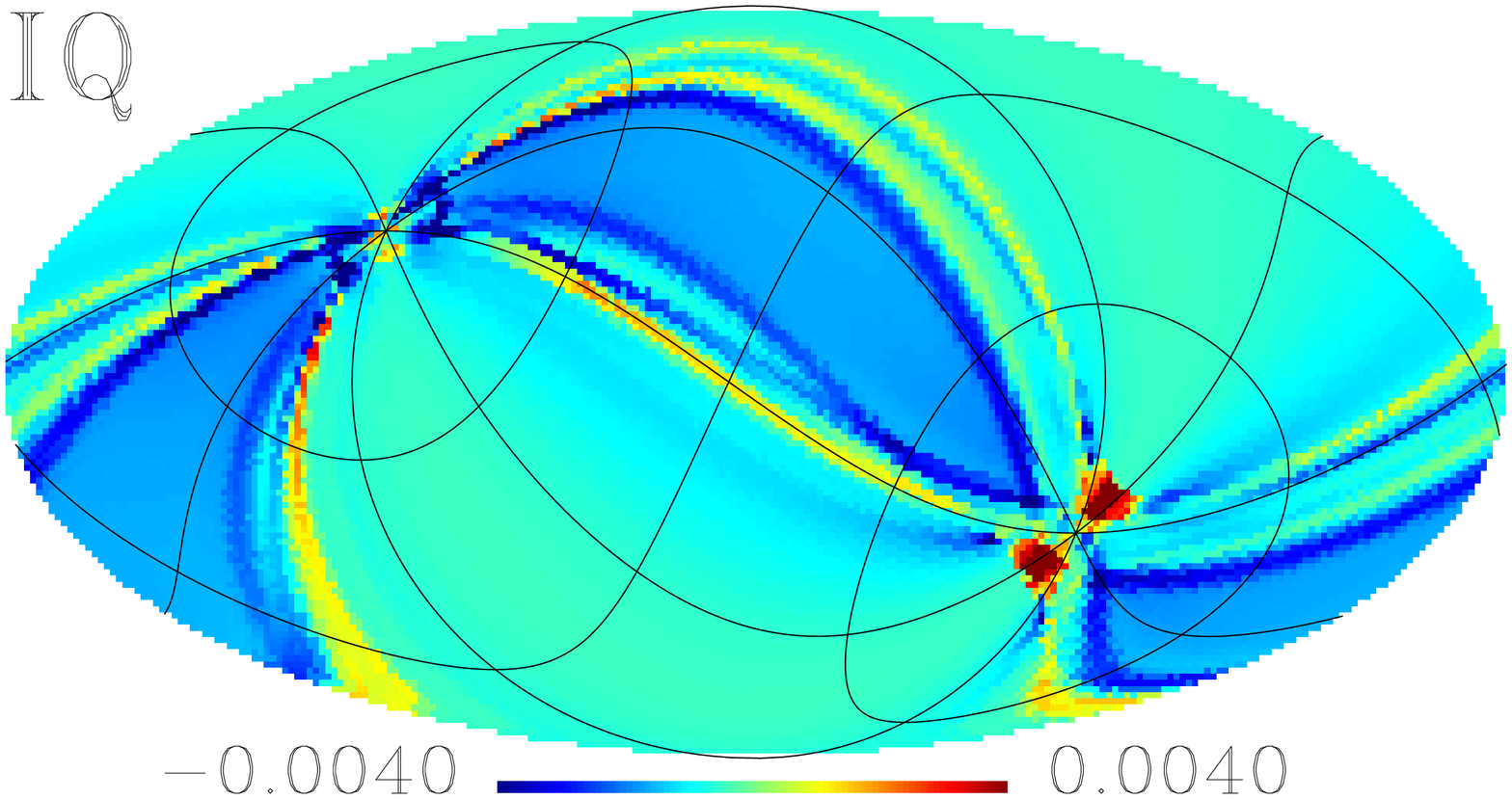}
    \includegraphics[width=0.2\textwidth,angle=90,trim=25 0 45 0,clip]
    {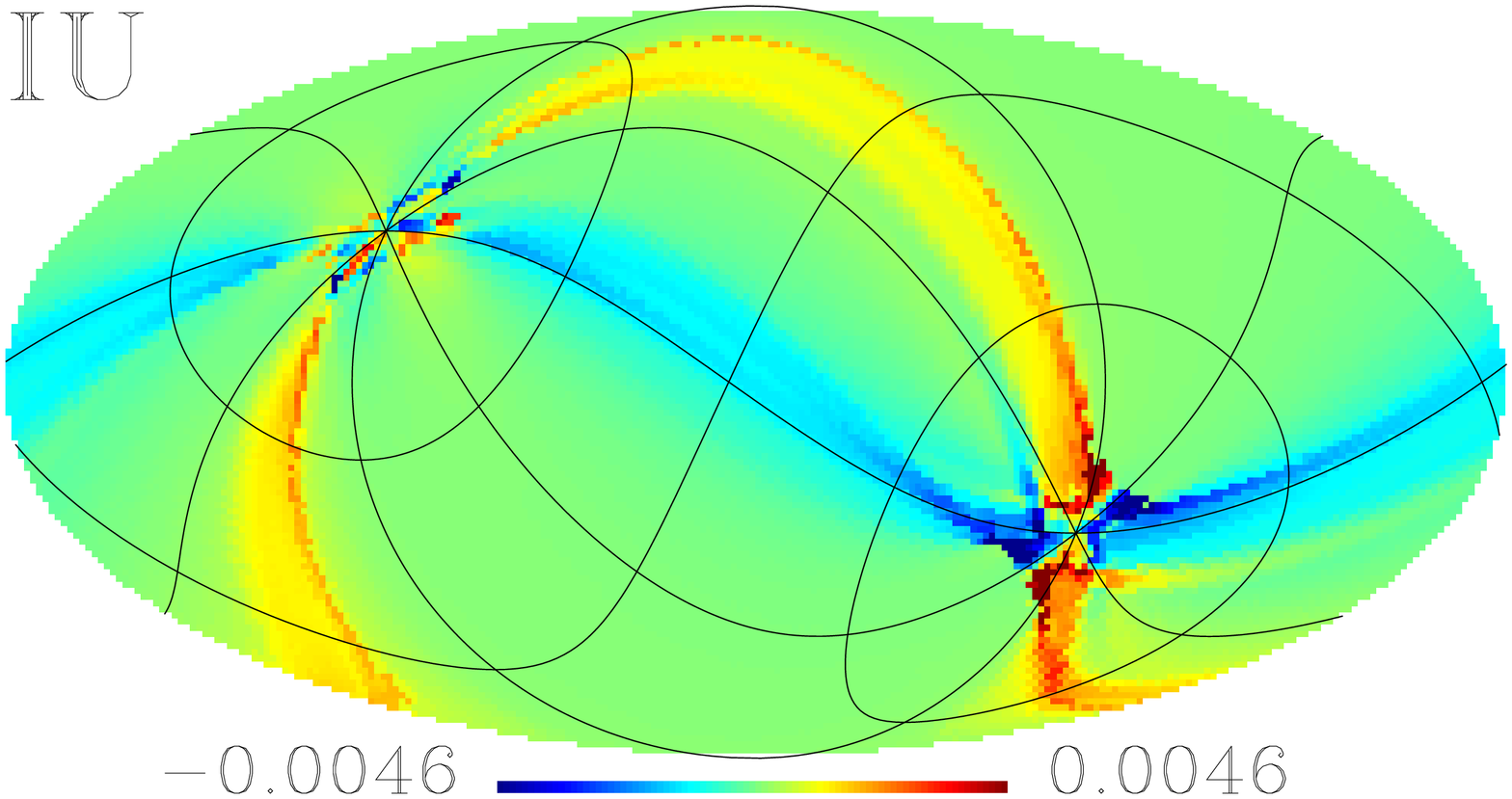}
  }
  \resizebox{\hsize}{!}{
    \includegraphics[width=0.2\textwidth,angle=90,trim=25 0 45 0,clip]
    {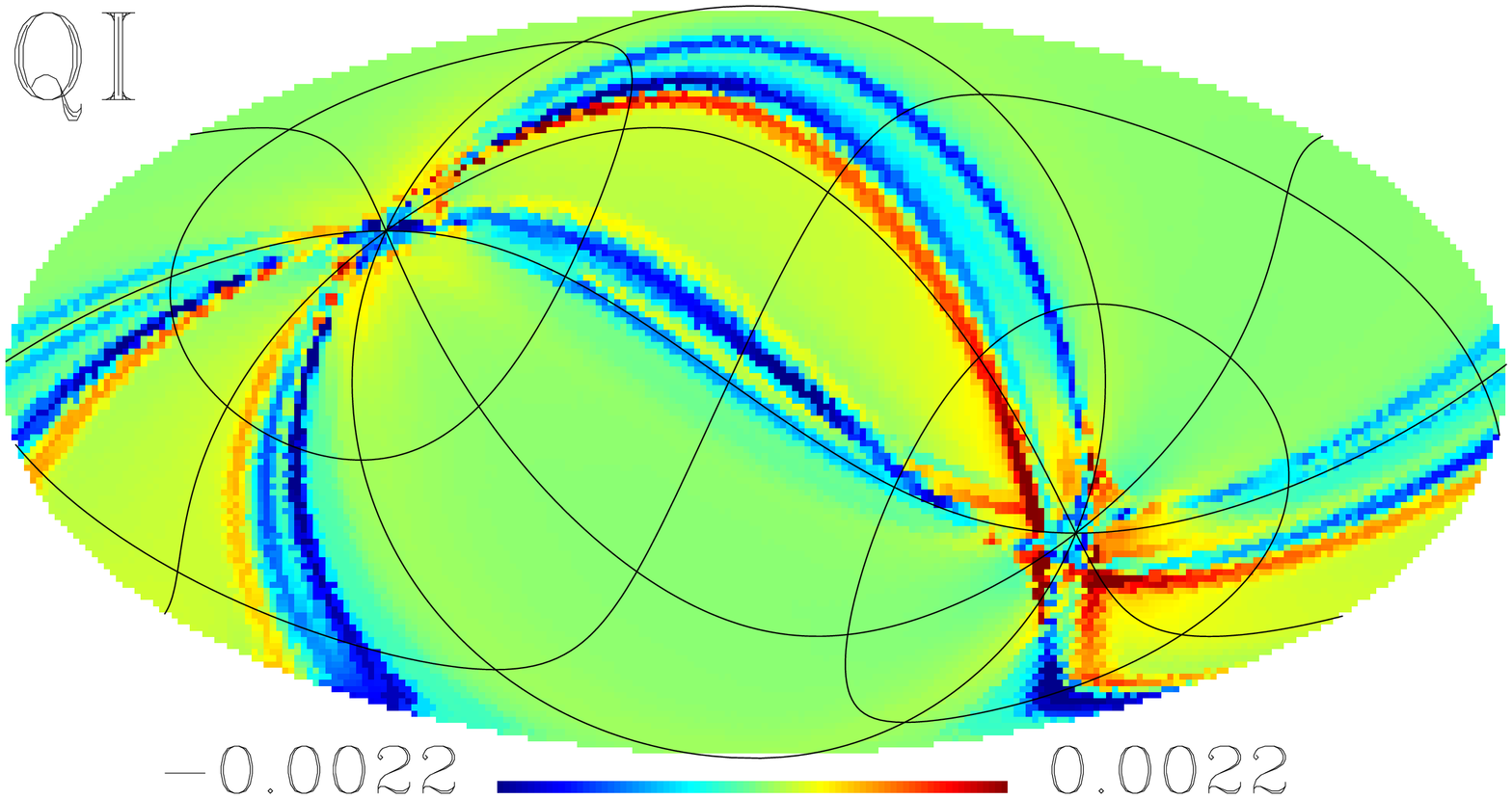}
    \includegraphics[width=0.2\textwidth,angle=90,trim=25 0 45 0,clip]
    {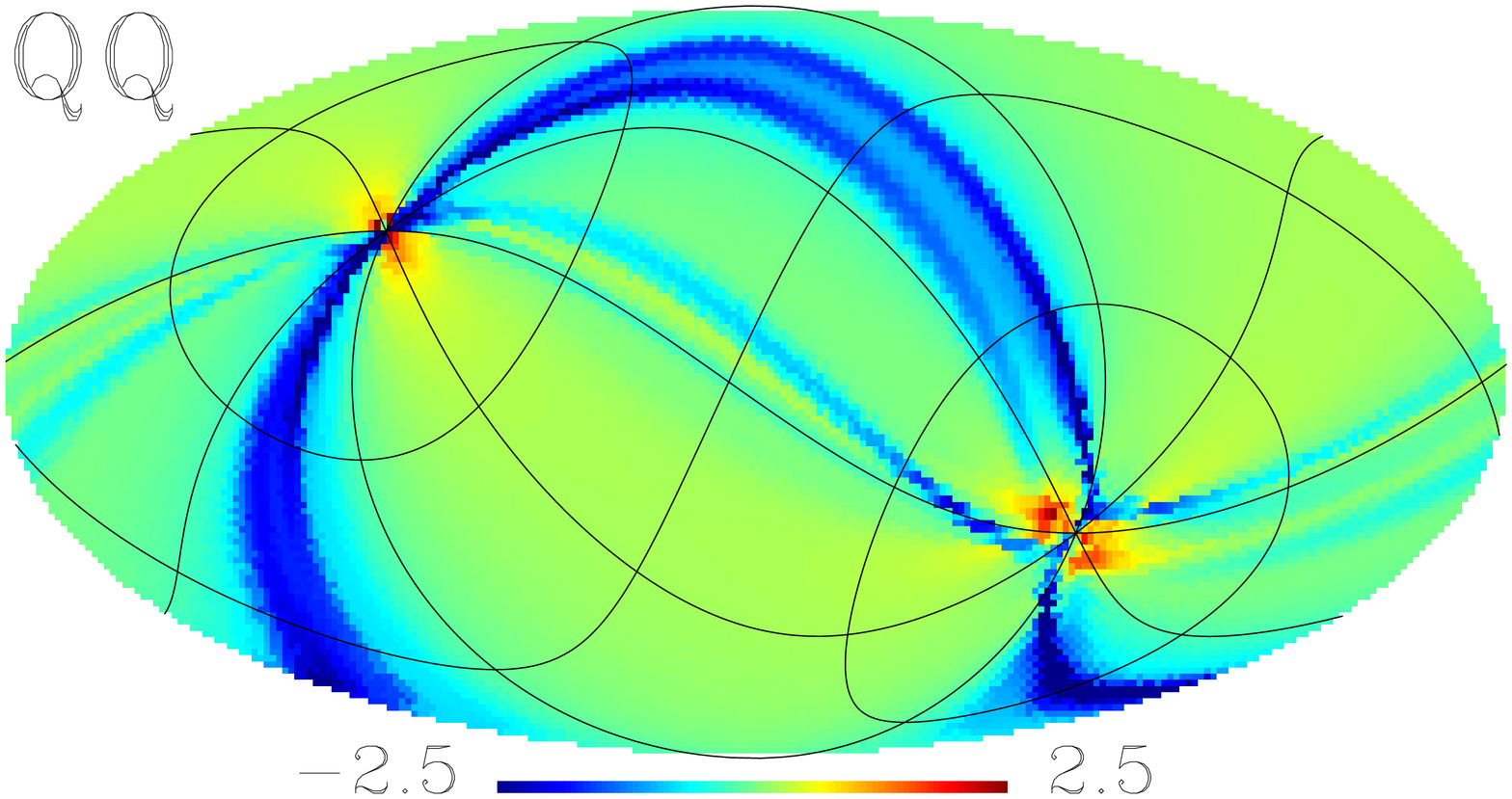}
    \includegraphics[width=0.2\textwidth,angle=90,trim=25 0 45 0,clip]
    {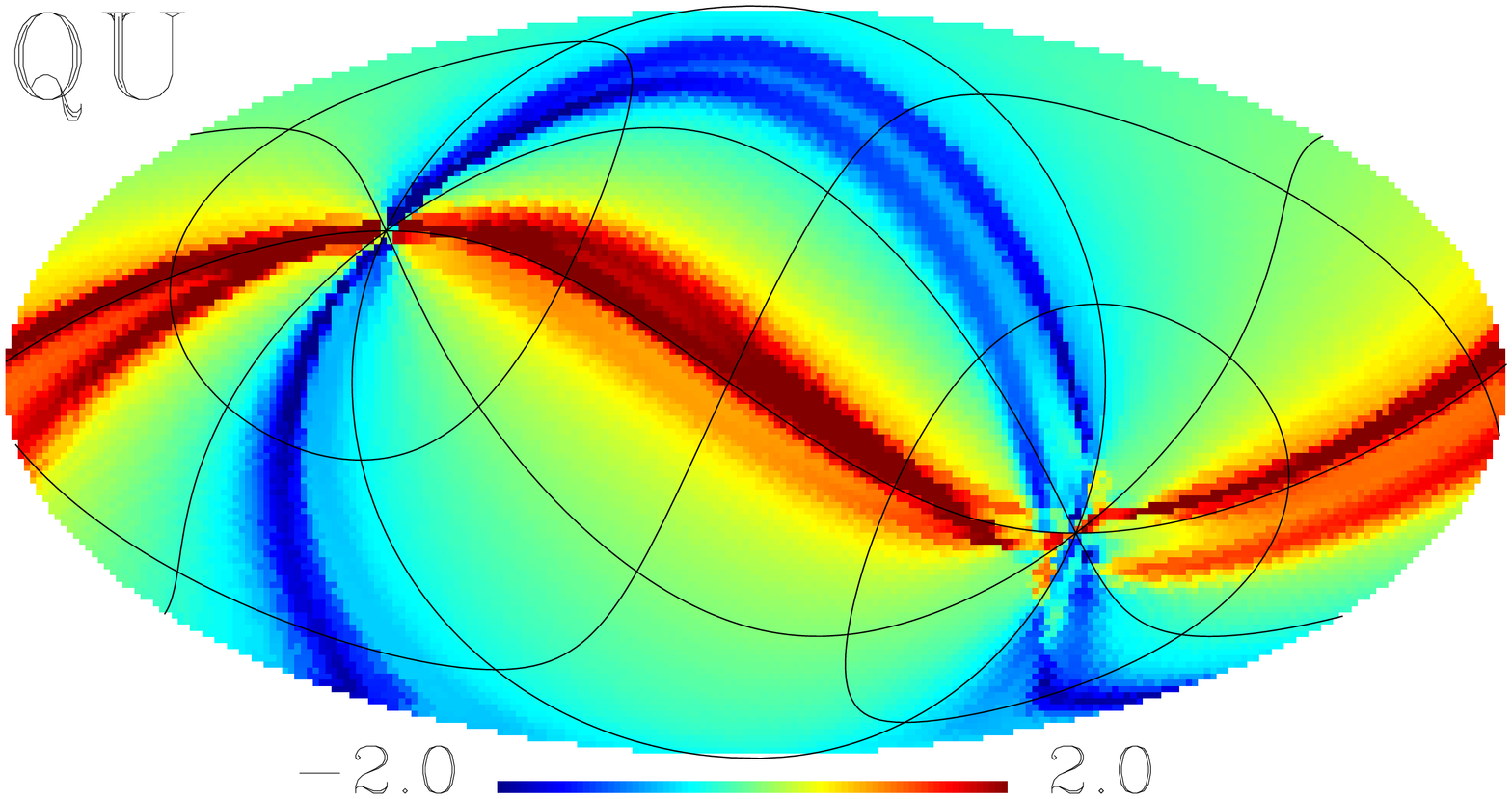}
  }
  \resizebox{\hsize}{!}{
    \includegraphics[width=0.2\textwidth,angle=90,trim=25 0 45 0,clip]
    {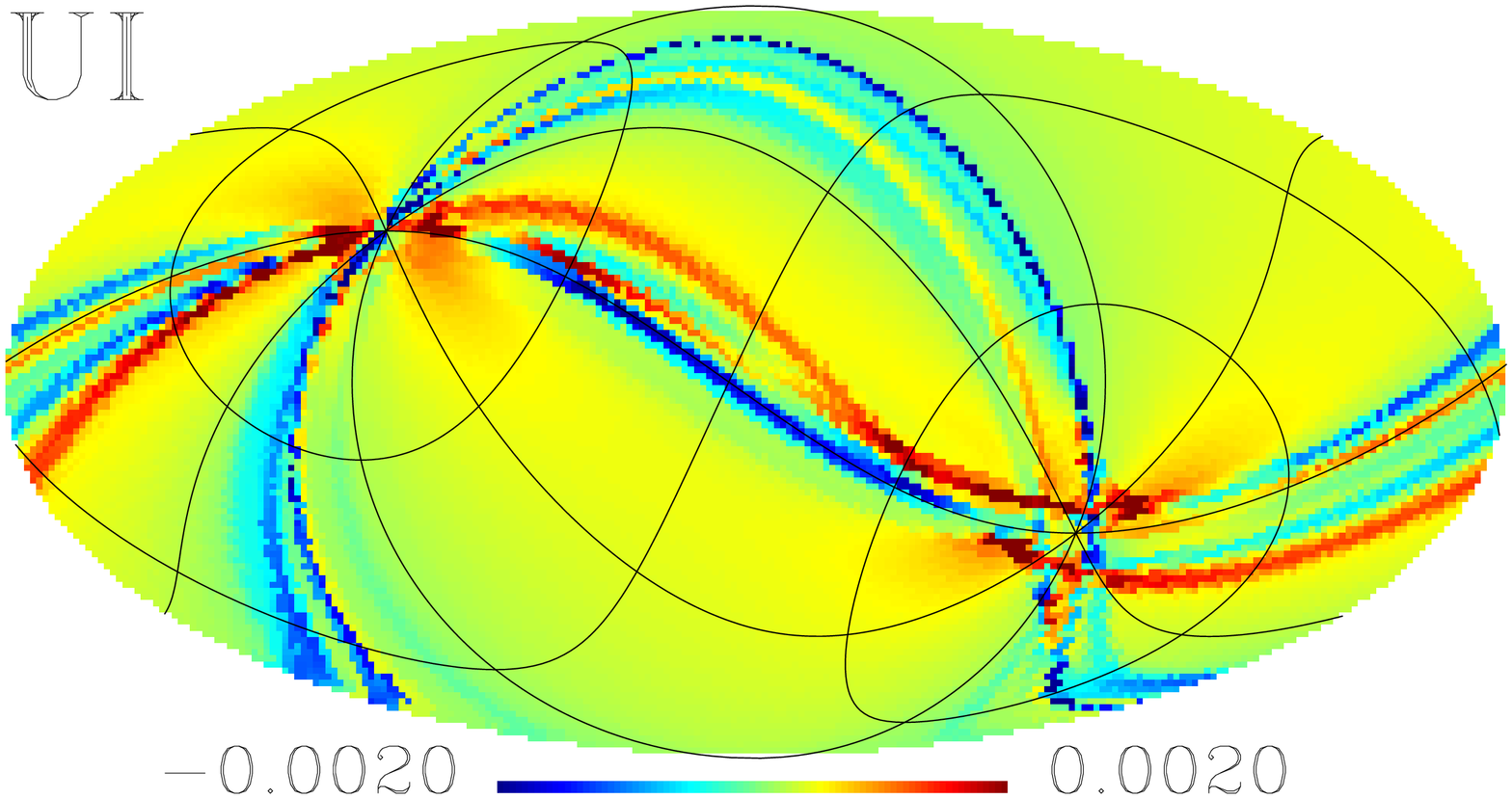}
    \includegraphics[width=0.2\textwidth,angle=90,trim=25 0 45 0,clip]
    {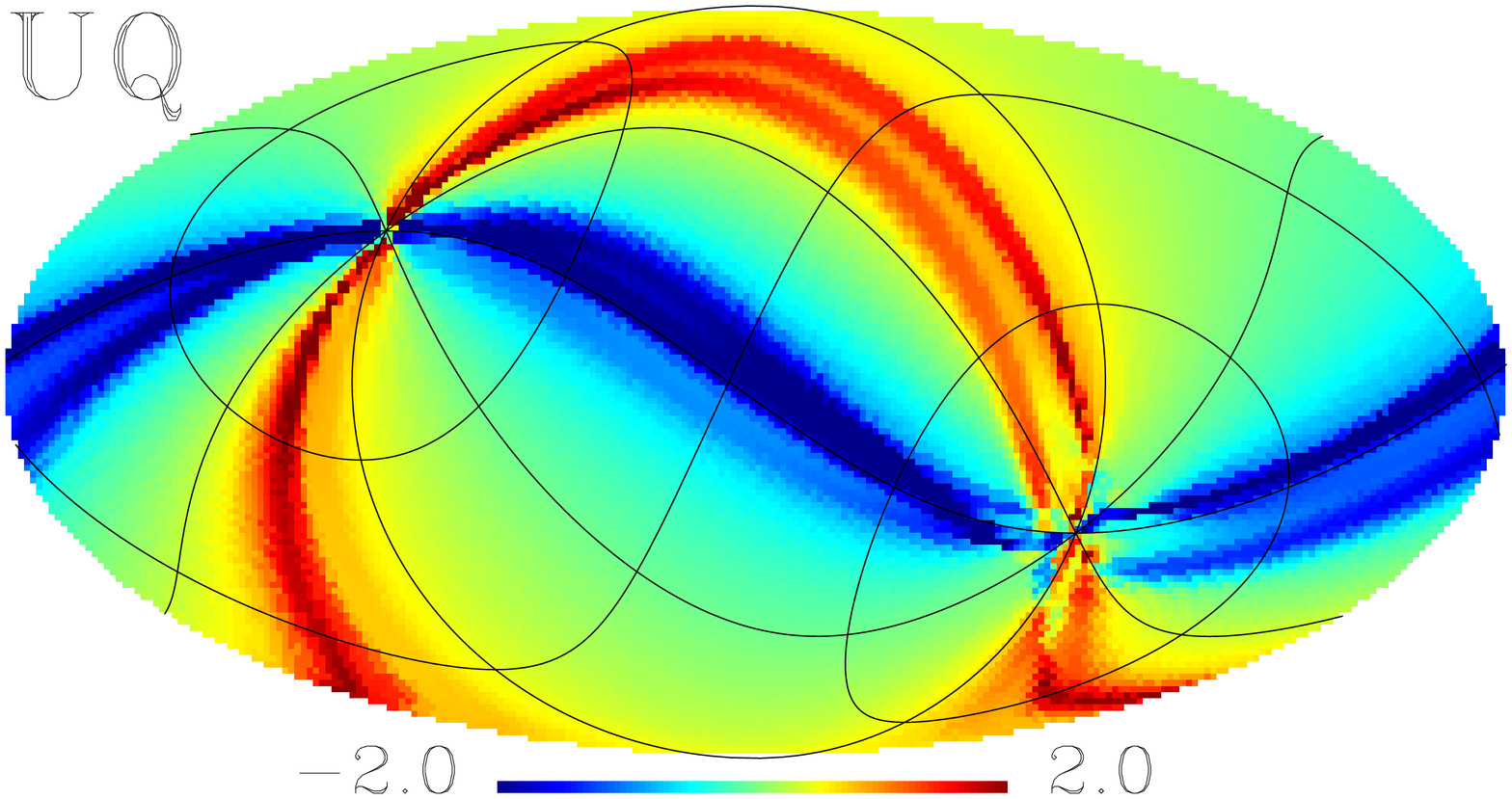}
    \includegraphics[width=0.2\textwidth,angle=90,trim=25 0 45 0,clip]
    {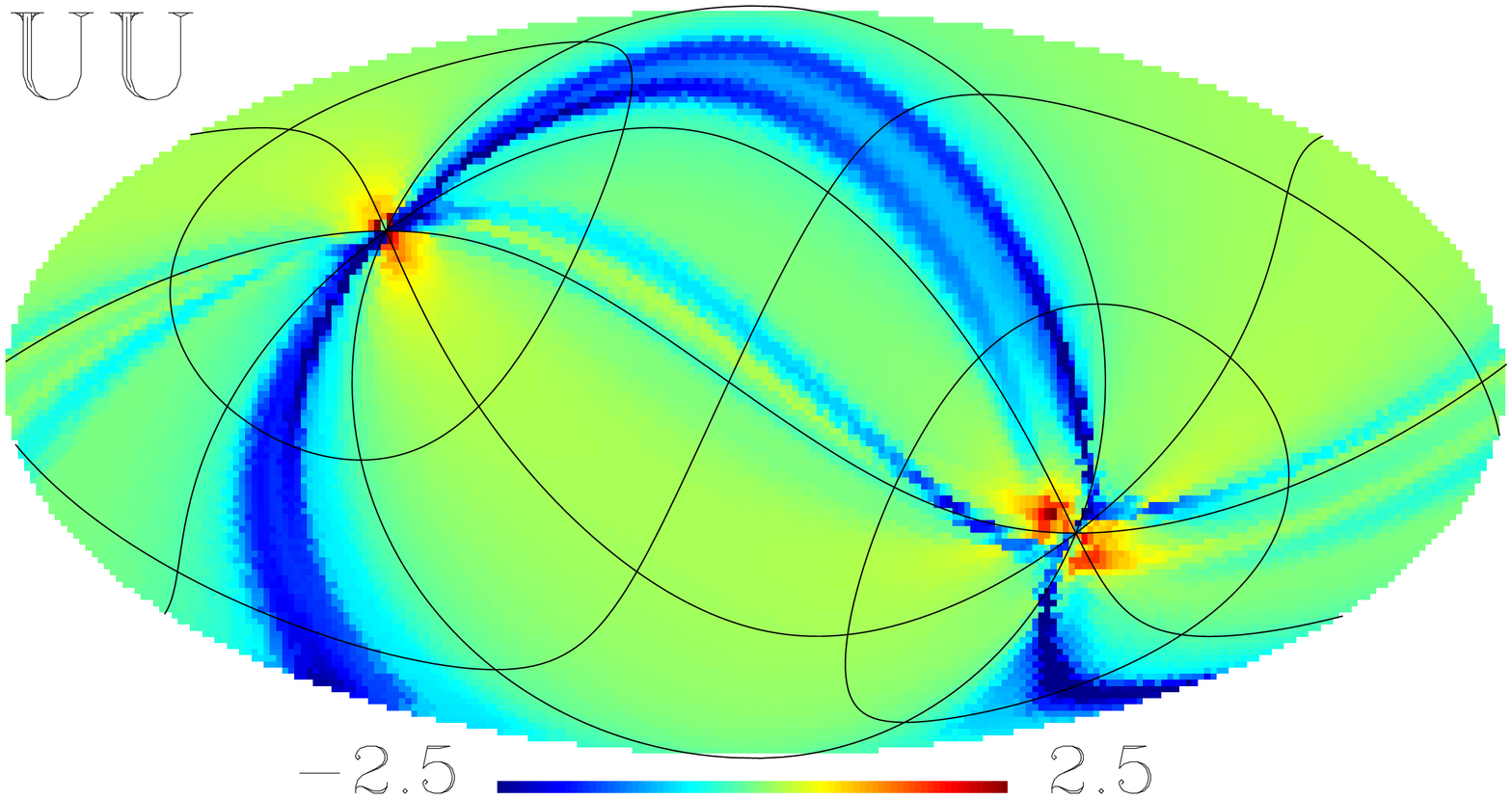}
  }
  \caption{
    A single column of the Springtide covariance matrix corresponding to a
    pixel near the ecliptic north pole. For description of the normalization,
    see text.
  }
  \label{fig:SpringtideNCVM_pole}
\end{figure}

Figs.~\ref{fig:MadpingNCVM_pole}-\ref{fig:SpringtideNCVM_pole} show
plots of the NCM columns of another reference pixel (number
2047). This pixel is located at the northern ecliptic pole
region and it exhibits a very different correlation pattern
compared to the previous case. Since the pole is visited
frequently through the course of the survey, it becomes correlated
with the rest of the map as a whole. Correlation amplitude is
increased by a factor of 3 from the equator pixel case (the increase can
be seen from the color bar ranges) and there is now a
distinct asymmetry between northern and southern hemispheres.
As one expects, the asymmetry only appears in optimal and
generalized destriping estimates.

\subsection{Noise covariance validation}\label{sec:ncmval}

In this Section we report the results of three different validation tests.
We performed a $\chi^2$ test, compared the noise biases computed from the
matrices and corresponding Monte Carlo maps and finally used the matrices
and Monte Carlo maps as inputs to angular power spectrum estimation.

\subsubsection{By $\chi^2$}

Residual noise expected in the recovered maps is Gaussian due to the linear
character of all the map-making methods considered here. Thus the noise is
completely described by its covariance matrix. More specifically, in the
absence of any singular modes of the estimated residual covariance,
$\vec N$, the residual maps, $\vec m = \hat{\vec s} - \vec s$, 
are drawn from a multivariate Gaussian distribution defined
by $\vec N$. Therefore, the $\chi^2$ statistic, defined as,
$\chi^2 = \vec m^\mathrm T \vec N^{-1} \vec m$, is drawn from a $\chi^2$
distribution with $3N_\mathrm{pix}$ degrees of freedom (dof). 

If any singular mode is present we simply replace the matrix, $\vec N^{-1}$,
in the definition of $\chi^2$, by a matrix, $\vec N'^{-1}$, which is like
$\vec N^{-1}$ in all respects but has the eigenvalue corresponding to the
singular vector set to $0$. We note that if the eigenvalue decomposition
of the matrix $\vec N^{-1}$ is not available or too costly to compute, we
can achieve numerically the same effect by defining $\vec N' =
\vec N +\eta^2\,\vec v \vec v^t$, where $\vec v$ is a singular vector we
want to project out and $\eta^2$ is a large positive  number for which
however inversion of $\vec N'$ is still stable \citep[e.g.][]{Bond:1998qg}.
As we subtract one degree of freedom corresponding to the excluded,
ill-conditioned eigenmode we expect that there are in total
$3N_{\mathrm pix}-1$ degrees of freedom left in our maps

We can apply the analogous test to the smoothed noise covariance. The $\chi^2$
statistic is defined as before with the smoothed covariance matrix as
well as residuals used now in place of the respective unsmoothed objects. 
As we commented on that in Sect.~\ref{sec:smoothNCM}, the inverse of the
smoothed covariance has to be appropriately regularized, to avoid the
results of the test being biased by the artifacts potentially present at
the scales smaller than the smoothing kernel and therefore not containing
any cosmologically useful information. The effective number of degrees of
freedom left in the data will coincide then with the number of eigenvalues
which have not been set to zero in the regularization process. Alternately,
if the preferred regularization approach involved adding some low level
of the white noise, Sect.~\ref{sec:smoothNCM}, the number of the degrees
of freedom is equal to $3N_{\mathrm pix}$.

In addition, one needs to take care of the singularity of the unsmoothed
noise covariance. This has to be done explicitly if the smoothed version
of the ill-conditioned eigenmode, $\vec v$, does not belong to the null
space of the inverse smoothed NCM, i.e., 
$\vec {\tilde N}^{-1}\left(\vec L\vec v\right) \nsimeq 0$.
To do so, we employ the same approach as before, replacing the regularized
inverse of the smoothed covariance matrix, $\vec {\tilde N}^{-1}$, by,
\begin{eqnarray}
  \tilde{\vec N}^{-1} \rightarrow 
  \left[
    \vec {\tilde N} +\eta^2\,\vec L \vec v \left(\vec L \vec v\right)^\mathrm T
  \right]^{-1} &
  \stackrel{\eta^2 \rightarrow \infty}{\longrightarrow} &
  \vec {\tilde N}^{-1} -  \left(
    \vec {\tilde N}^{-1}\,\vec L \vec v
  \right)  \left[
    \left(\vec L \vec v\right)^t \vec {\tilde N}^{-1} \left(\vec L\vec v\right)
  \right]^{-1}
  \left( \vec {\tilde N}^{-1}\,\vec L \vec v \right)^\mathrm T
\end{eqnarray}
where $\vec L$ is a smoothing operator, Eq.~(\ref{eq:covSmoothing}), and the
last expression follows from the Sherman-Morrison-Woodbury
formula~\citep{Woodbury:1950}. This last operation additionally reduces
the number of degrees of freedom by 1 (or whatever number of
modes, $\vec v$ is to be projected out).

The Kolmogorov-Smirnov test can be used to test whether a set of samples
conforms to some theoretical distribution. The test estimates the probability
of the maximal difference between the \emph{empirical distribution function},
\begin{equation}
  \label{eq:edf}
  F_n(x) = \frac{1}{n}\sum_{i=1}^n \Theta(x_i),
\end{equation}
of the observations $x_i$ (in our case the individual $\chi^2$)
and the theoretical cumulative distribution function. $\Theta(x)$ is
the Heaviside step function. We note that in this work we take an advantage
of the fact that we can simulate the residual noise directly. Though this is
clearly not the case when real data are considered, the tests described here
can be  applied to a difference of two sky maps produced by disjoint sets
of detectors operating at the same frequency and can therefore be a useful
test of the real life data processing integrity \citep[e.g.][]{Stompor2002}.

\begin{figure}[!tbh]
  \centering
  \sidecaption
  \resizebox{12cm}{!}{\includegraphics[trim=70 20 20 5,clip]
    {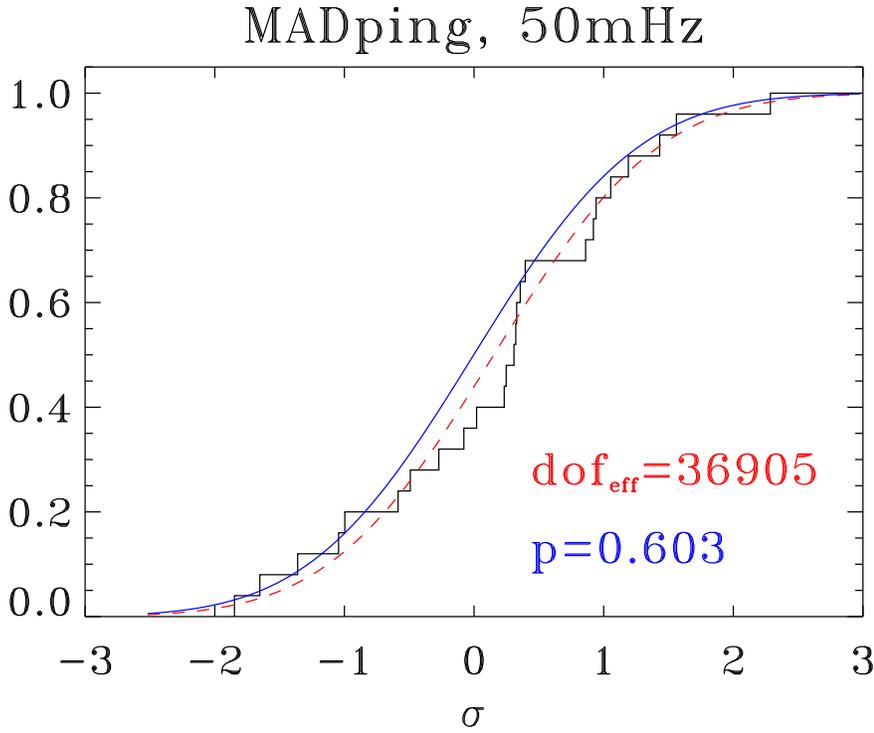}}
  \caption{
    \emph{MADping}
    empirical $\chi^2$ distribution function from the $25$ residual noise
    maps compared with the theoretical cumulative probability density.
    The black stair line is the empirical distribution function,
    the blue solid line is the theoretical $\chi^2$ distribution for 
    $3N_\mathrm{pix}-1=36,863$ degrees of freedom. It is the same for all
    direct method maps in
    Figs.~\ref{fig:MADping_chisq}--\ref{fig:Springtide_chisq}.
    The red dashed line is the least squares fit of the $\chi^2$
    distribution to the experimental distribution (dof being the fitting
    parameter). The horizontal axis
    is translated to the expected center of the distribution,
    $\langle\chi^2\rangle=\mathrm{dof}=36,863$, and scaled by the
    expected deviation, $\sigma_{\chi^2}=\sqrt{2\mathrm{dof}}$.
  }
  \label{fig:MADping_chisq}
\end{figure}

\begin{figure}[!tbh]
  \centering
  \sidecaption
  \resizebox{12cm}{!}{\includegraphics[trim=30 23 0 5,clip]
     {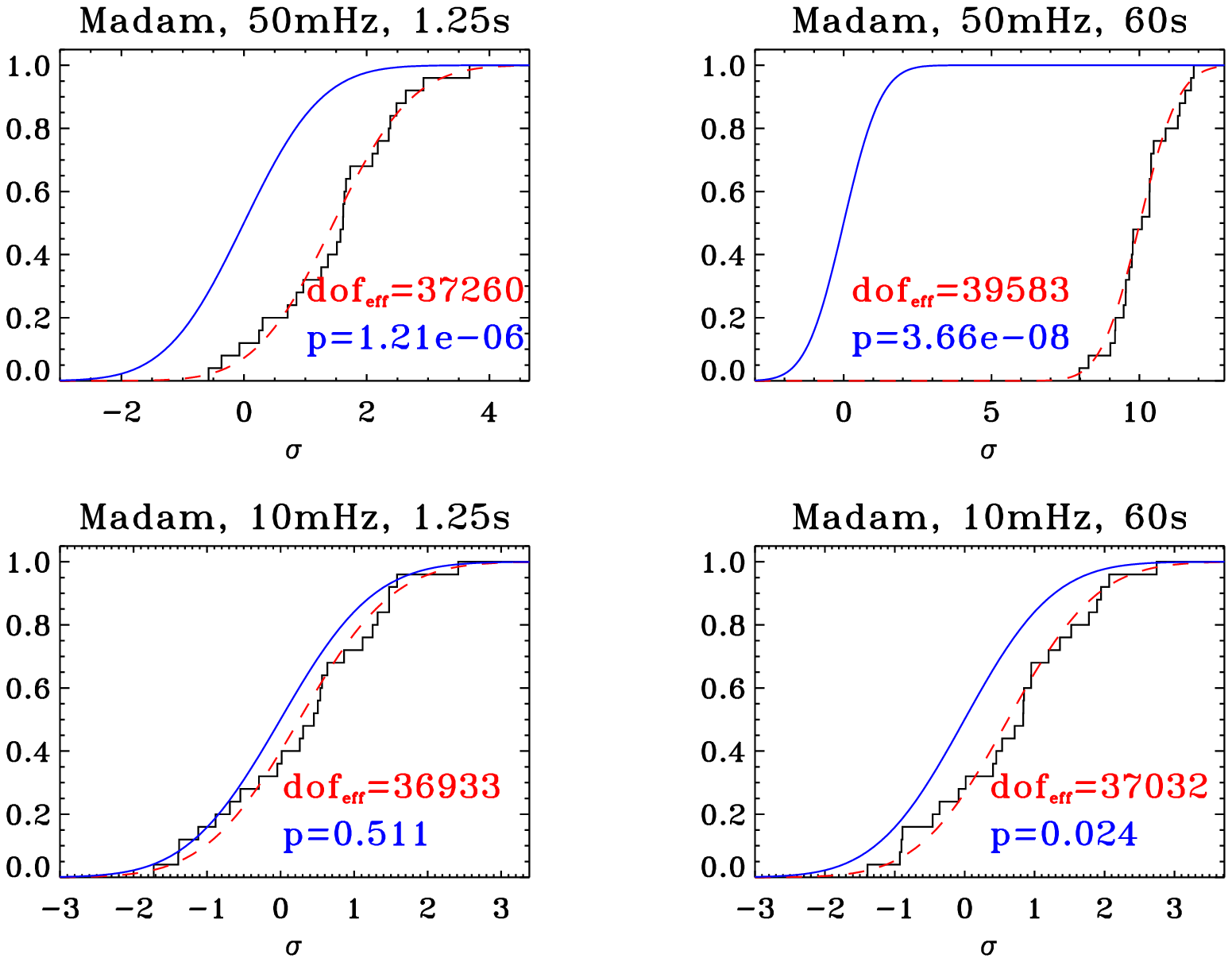}}
  \caption{
    \emph{Madam}
    empirical $\chi^2$ distribution functions from the $25$ residual noise
    maps compared with the theoretical cumulative probability density.
  }
  \label{fig:Madam_chisq}
\end{figure}

\begin{figure}[!tbh]
  \centering
  \sidecaption
  \resizebox{12cm}{!}{\includegraphics[trim = 35 203 0 5, clip]
    {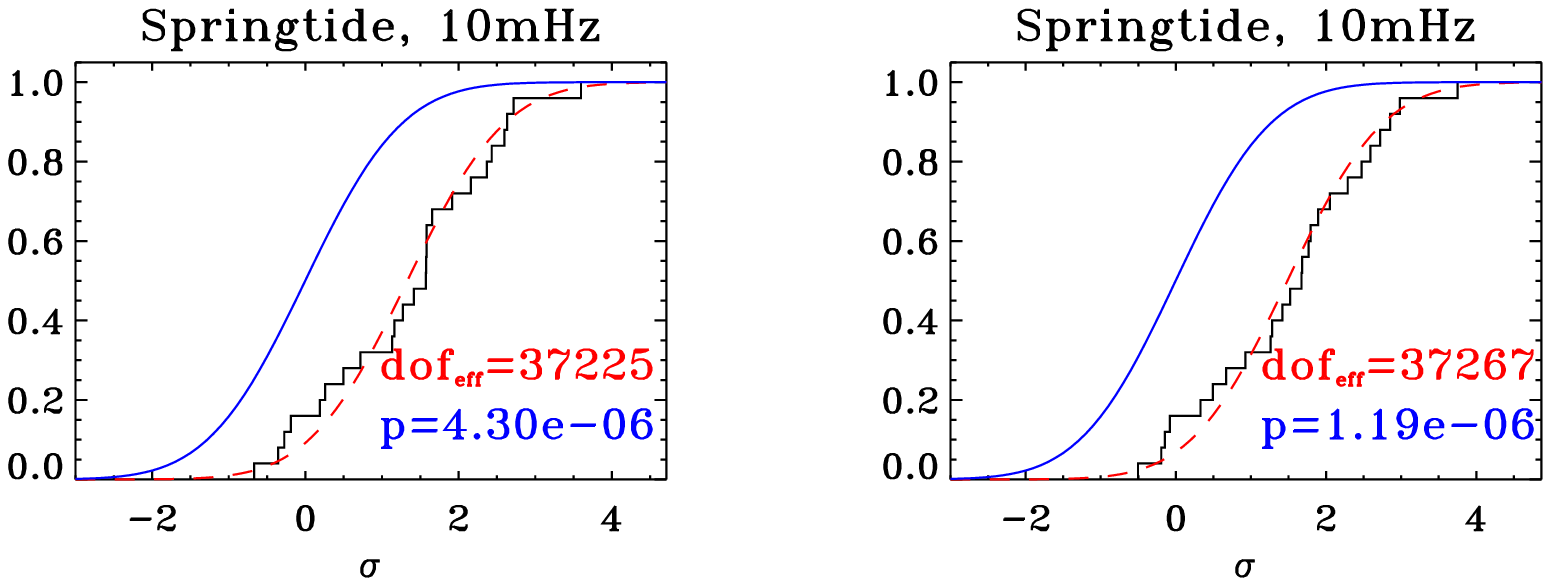}}
  \caption{
    \emph{Springtide}
    empirical $\chi^2$ distribution functions from the $25$ residual noise
    maps compared with the theoretical cumulative probability density.
    \emph{Left:} direct method.
    \emph{Right:} INW.
  }
  \label{fig:Springtide_chisq}
\end{figure}

Figs.~\ref{fig:MADping_chisq}--\ref{fig:Springtide_chisq} show the
two cumulative distribution functions for $25$ noise maps in the case of the
direct method. Reported $p$-values are the
probabilities of observing this level of disagreement even if the noise
description was exact. Conventionally, the level $p<0.05$ is considered to
be enough to reject the null hypothesis that the distributions match.

We then proceed to study the agreement between downgraded noise maps and
the noise covariance matrices. As a test case, we use the Madam NCM
for $10\,$mHz knee frequency, $1.25\,$s baselines. We smoothed the
covariance matrix using an apodized window function, setting the
thresholds to $2N_\mathrm{side}$ and $3N_\mathrm{side}$ respectively. As
expected, the smoothed matrix is extremely singular. We compute its 
inverse by including only the eigenvalues that are greater than
$10^{-2}$ times the largest eigenvalue, including $20,882$ of the $36,864$
available modes.

\begin{figure}[!tbh]
  \centering
  \resizebox{\hsize}{!}{
    \includegraphics[trim=80 20 30 5,clip]{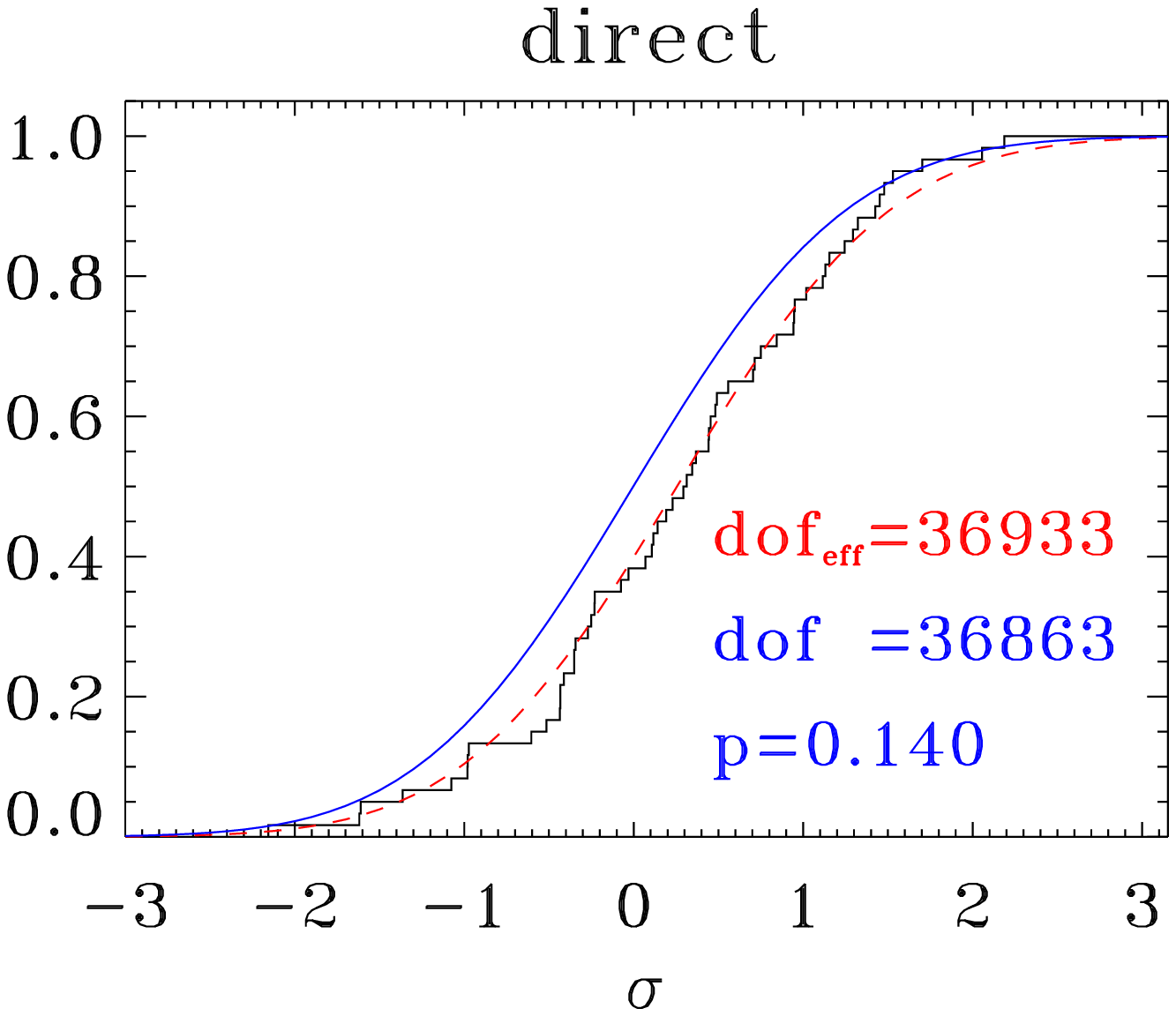}
    \includegraphics[trim=80 20 30 5,clip]{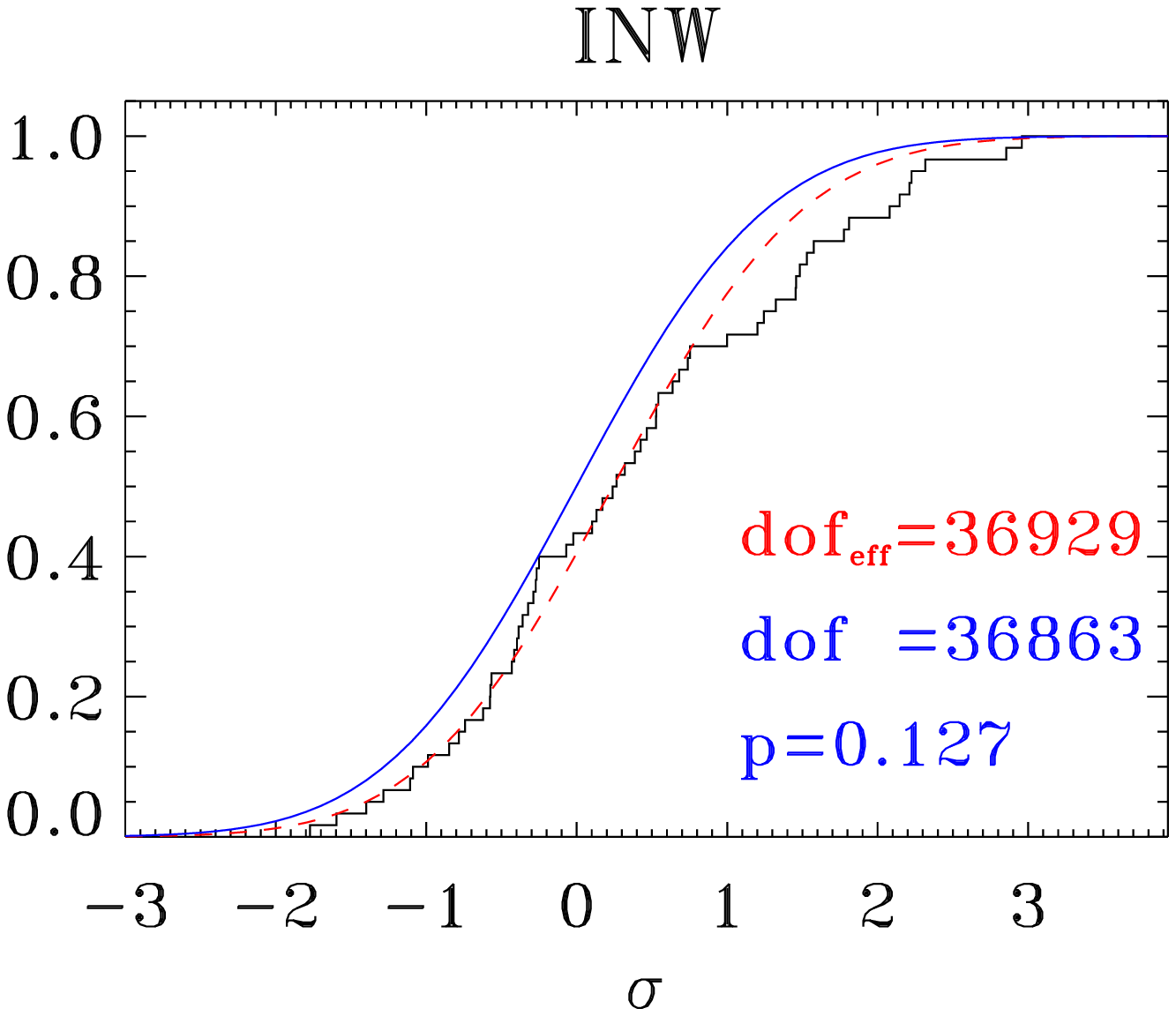}
    \includegraphics[trim=80 20 30 5,clip]{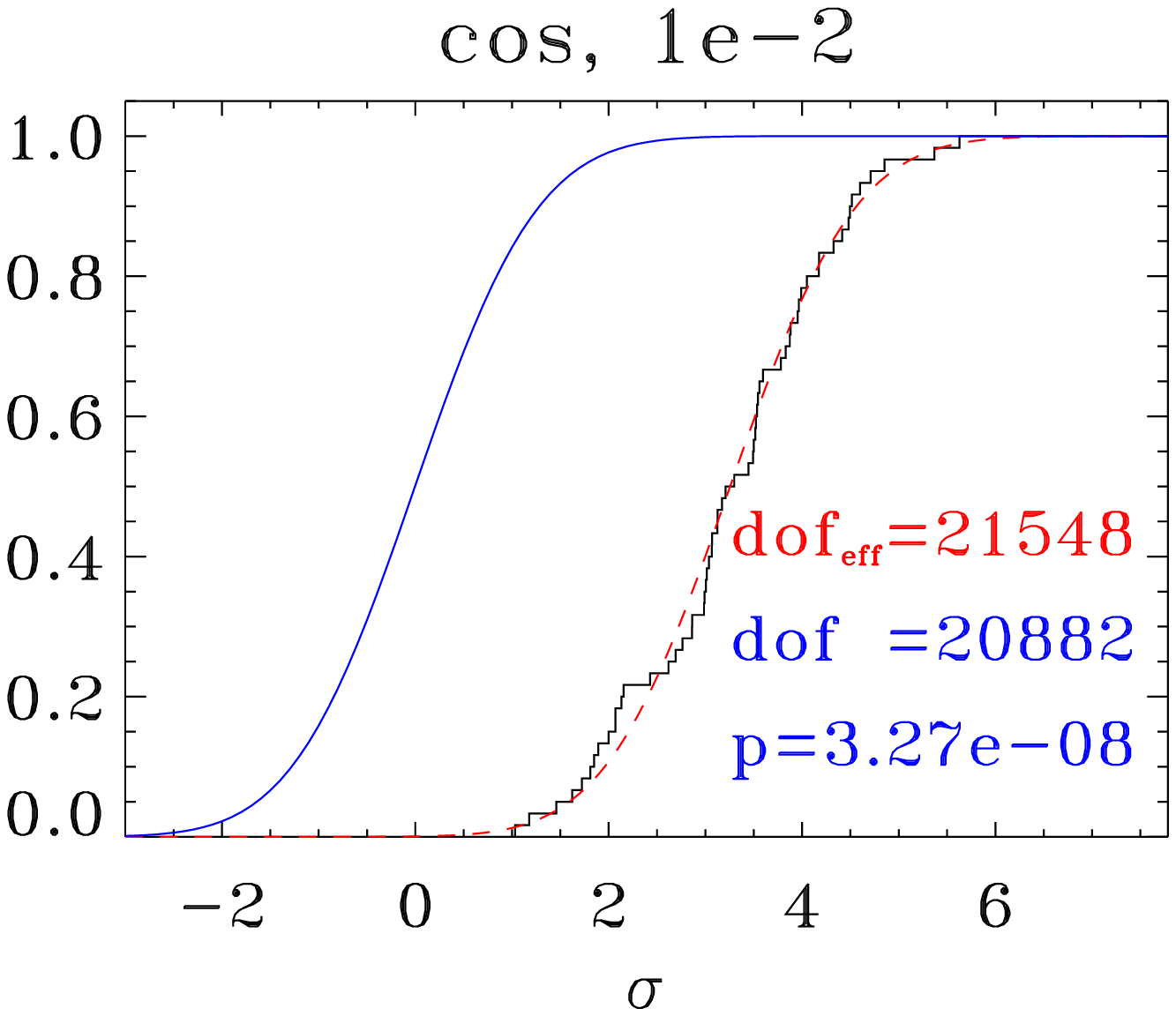}
  }
  \resizebox{\textwidth}{!}{
    \includegraphics[trim=80 20 30 5,clip]
    {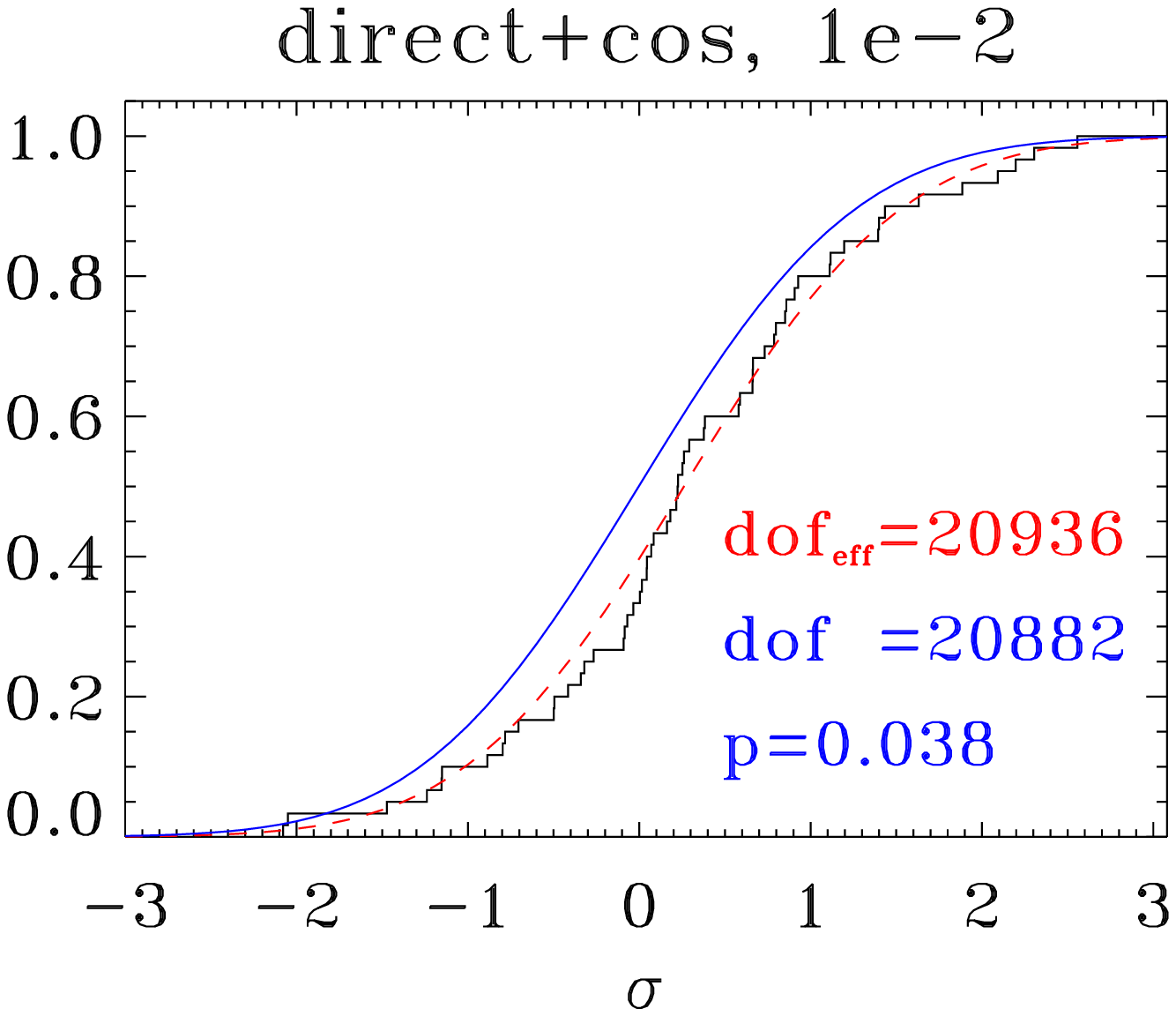}
    \includegraphics[trim=80 20 30 5,clip]
    {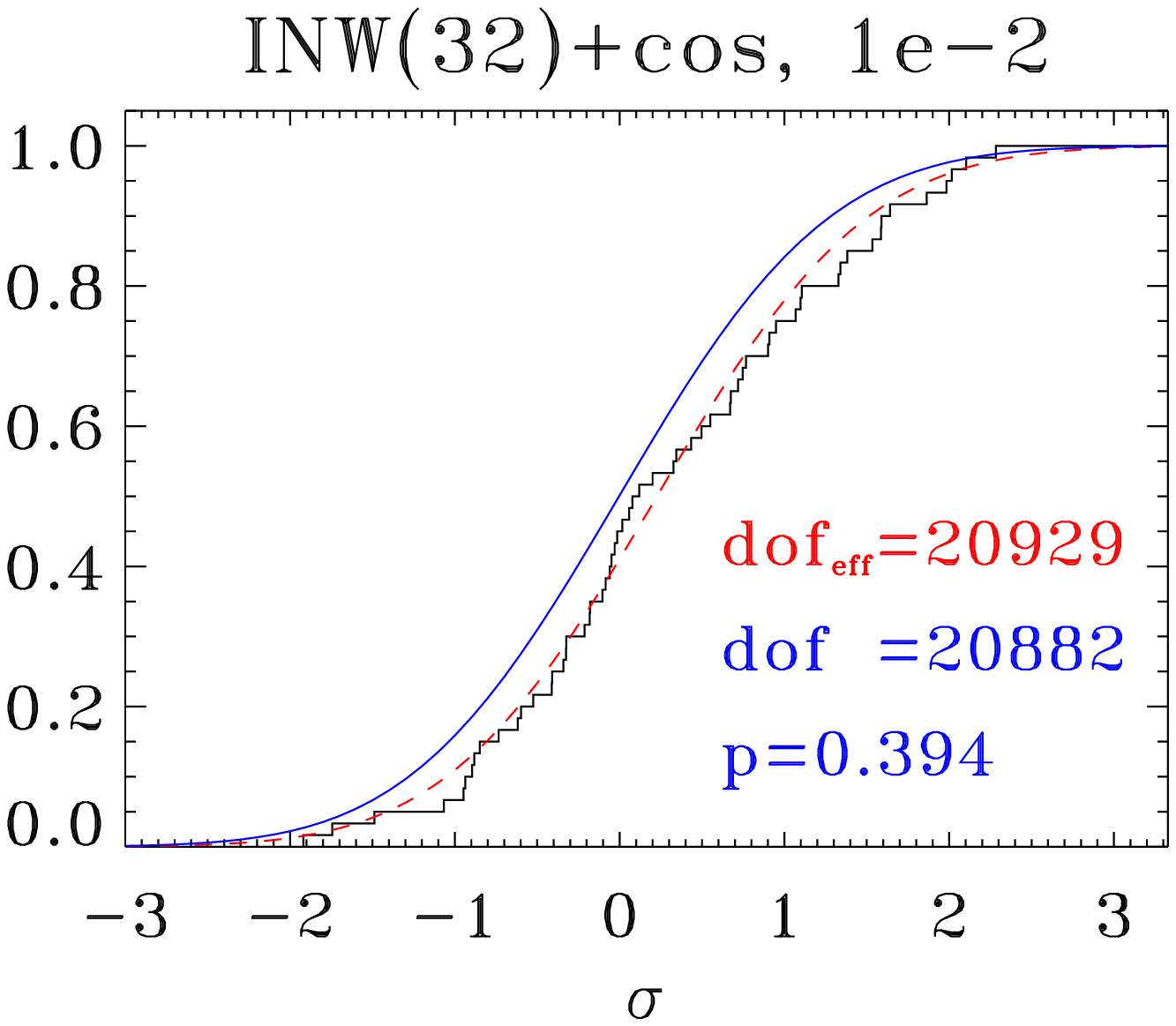}
    \includegraphics[trim=80 20 30 5,clip]
    {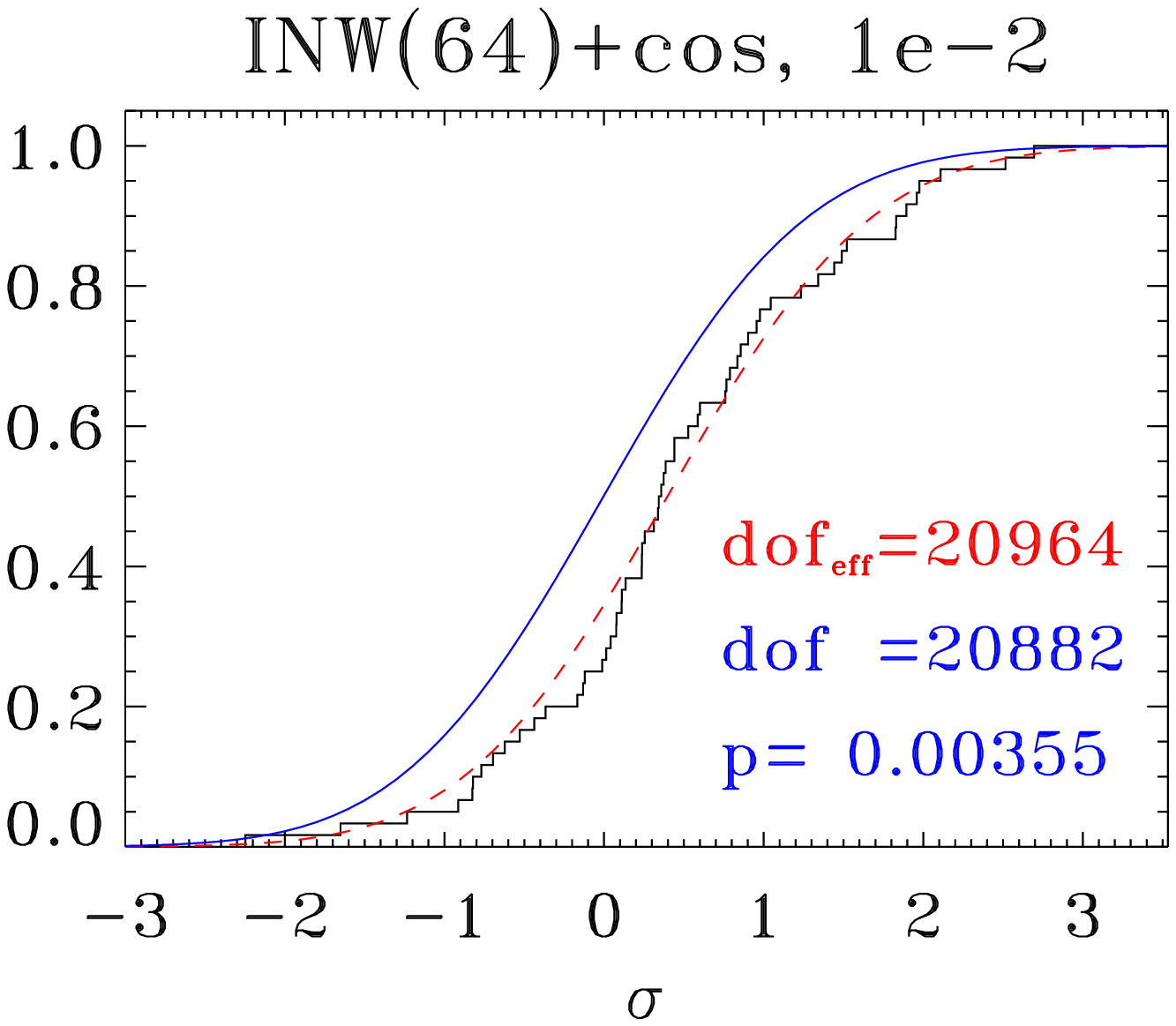}
  }
  \caption{
    \emph{Madam}
    empirical $\chi^2$ distribution functions from $60$ residual noise
    maps compared with the theoretical cumulative probability density. For
    a smoothed map, we count the degrees of freedom as the number of included
    eigenmodes in the inverse NCM.
    \emph{Top:} Three sets of low-resolutions maps using just one of the
    low-resolution methods at a time. The high-resolution maps for
    INW and smoothing methods had $N_\mathrm{side}=1024$.
    \emph{Bottom:} Since it is suboptimal to compute a low-resolution
    spherical harmonic expansion from a noisy high-resolution map, we
    test how well the smoothing approach works \emph{in conjunction} with
    the two pixel-based downgrading methods.
  }
  \label{fig:smoothed_chisq}
\end{figure}

Fig.~\ref{fig:smoothed_chisq} shows the empirical distribution functions
of the $\chi^2$. Even though the matrix is computed for the direct method,
the inverse noise weighted (INW) maps conform well to it. However, when
we apply the smoothing kernel to the high-resolution maps, there is a clear
disagreement. This stems from the fact that in this downgrading the high
resolution pixels are not correctly (inverse noise variance) weighted when
we compute the low-resolution map. If we first produce a low-resolution
direct method or INW map and then smooth it, the agreement is much
better. This is shown in the bottom row of
Fig.~\ref{fig:smoothed_chisq}.

In the case of the direct method maps our results show that only optimal (noise)
maps and their respective noise covariance are mutually consistent in the
light of the $\chi^2$ statistics. The good statistical agreement in this
case does not depend  on the time domain noise characteristics nor map
resolution. This is expected given that the noise covariance
estimator implemented in the optimal codes, Eq.~(\ref{eq:optimal_ncm}),
is an exact expression describing the noise properties in the pixel domain,
and that we have assumed perfect knowledge of the time domain noise.

The level of consistency found in the destriping cases varies depending on
the underlying time domain properties, i.e., $f_\mathrm{knee}$, and on the
assumed baseline length. In the case of $f_\mathrm{knee}=50\,$mHz, we have
not found a satisfactory agreement in any of the considered destriping
cases. For the lower $f_\mathrm{knee}=10$ mHz the results obtained with
the generalized destriper, Madam, are satisfactory for the short, $1.25\,$s,
baseline choice, and marginal for the long one, $60\,$s. The latter result
is consistent also with that obtained using the classical destriper,
Springtide.

In the case of direct low-resolution map-making, the discrepancy between
the noise maps and the
noise covariance matrix stems from the destriping approximations. Both
destriping approaches assume that the correlated part of the noise is
perfectly modelled by a set of baseline offsets. Correlated noise
occurring at frequencies higher than what the baselines can model is
not removed from the TOD and therefore is binned onto the map For short
baselines or low knee frequencies the unmodelled noise manifests as a
relatively small angular scale correlation and does not bias the power
spectrum estimates at low $\ell$.

We conclude that the noise covariance of the low-resolution
maps produced by the destriping algorithm needs to be used with
care. The flexibility of the generalized destripers permitting them
to use different baseline lengths makes them in this context the
preferred choice. We emphasize however that, if the accuracy of
the noise description is the major concern, then only the
optimal techniques are suitable. In the next Sections we will
reconsider all these low-resolution map-making techniques in the
context more specific to the large angular scale power
spectrum estimation work, which is envisaged as the main
application of the low-resolution maps and their covariances.

\subsubsection{By noise bias} \label{subsec:noise_bias}

In this Section we describe the calculation of the average
angular power spectrum of noise maps, i.e., the noise bias --- see 
Sect.~\ref{sec:noise_bias}, 
using a pseudo $C_\ell$ estimator for which both the NCM estimate and the
Monte Carlo map averages are feasible to compute.
Testing the noise covariance matrix by comparing estimated and measured
noise biases can be viewed as complementary to the $\chi^2$ tests described 
in the previous Section. It can certainly provide more information than
the plain $\chi^2$ test, as instead of simple pass or fail indicator,
the noise bias comparison will tell \emph{at which angular resolution}
the noise model agrees with the data. At the same time, the noise
bias is less sensitive to the anisotropic features present in the
residual noise. The noise bias test is clearly more directly relevant for
power spectrum estimation.

\begin{figure}[!tbh]
  \centering
  \resizebox{\hsize}{!}{
    \includegraphics[trim=50 20 5 0,clip]
    {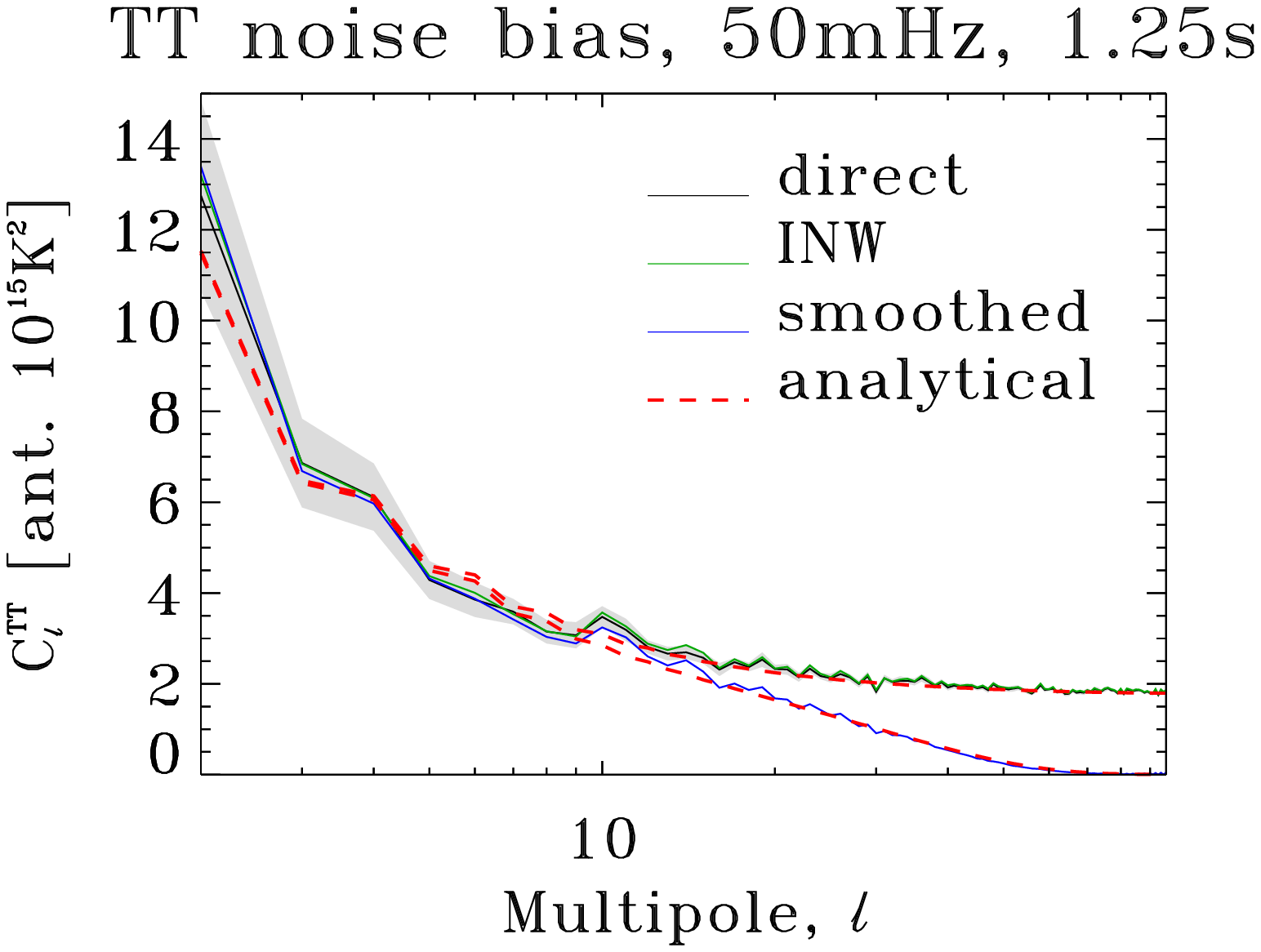}
    \includegraphics[trim=50 20 5 0,clip]
    {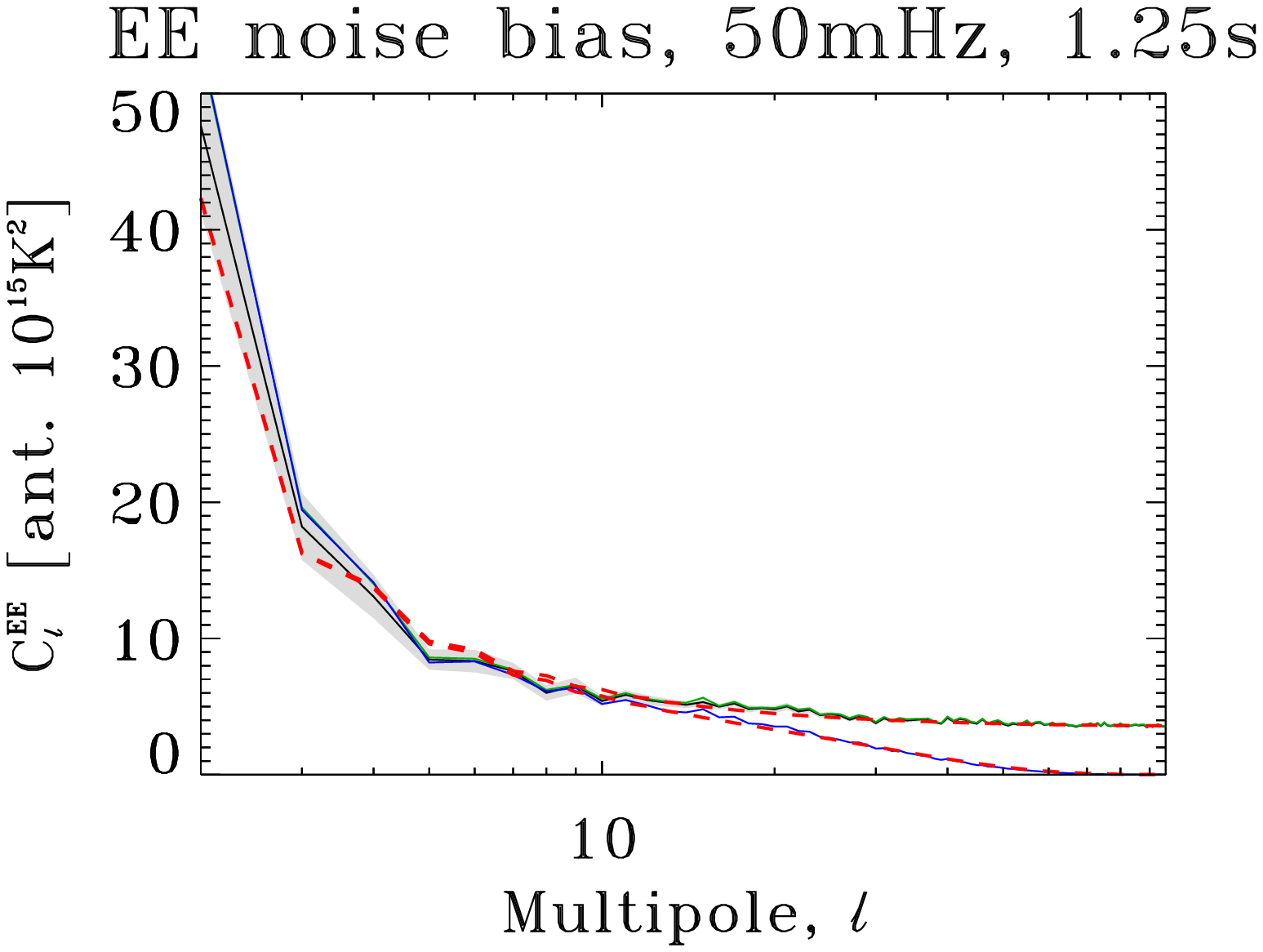}
    \includegraphics[trim=50 20 5 0,clip]
    {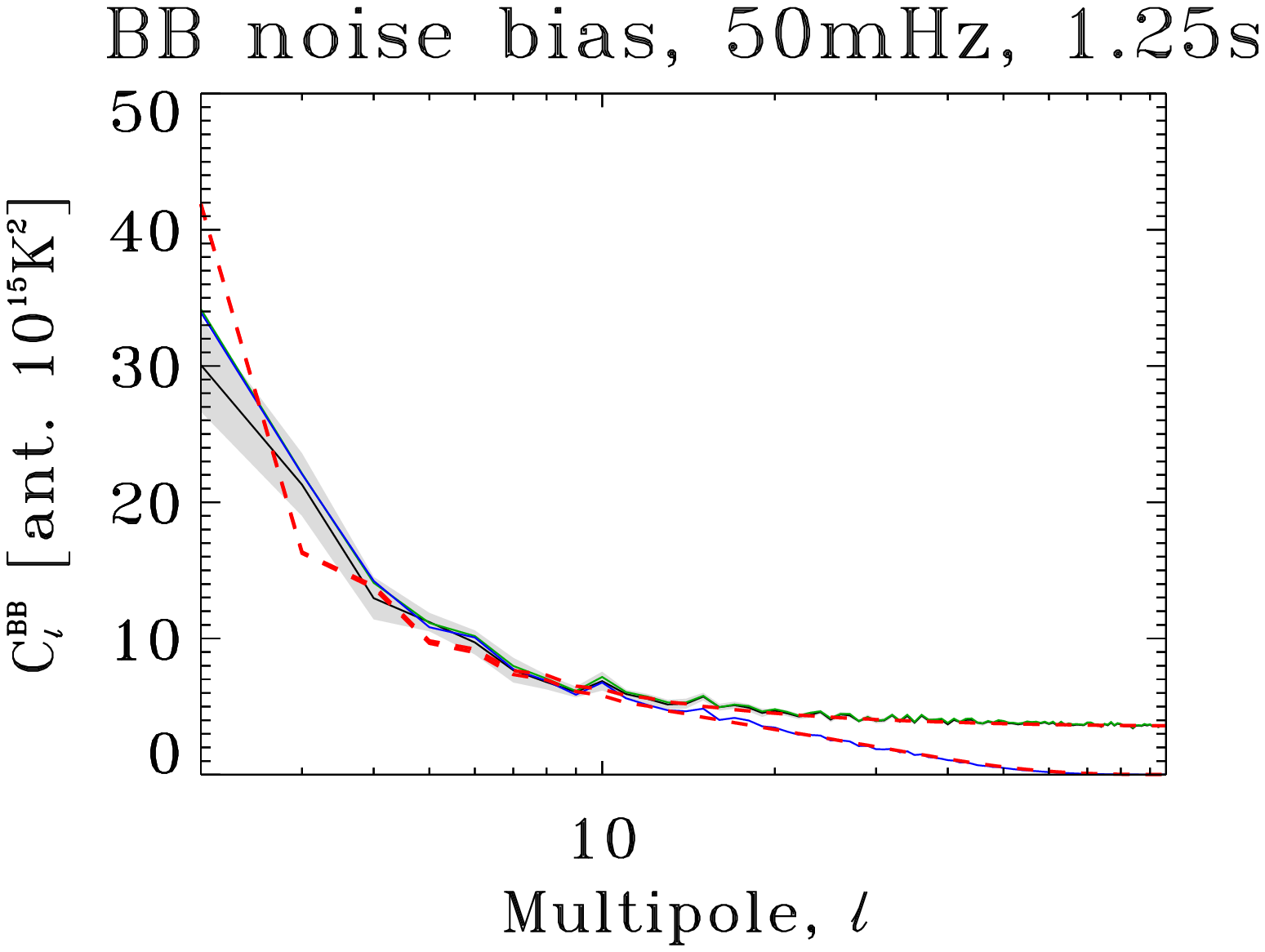}
  }
  \caption{
    Analytical and mean Monte Carlo noise biases from Madam runs at 
    $f_\mathrm{knee}=50$mHz using a short $1.25\,$s baseline. Grey band is the
    $1$-$\sigma$ region for the average, computed by dividing the sample
    variance by $\sqrt{25}$. 
  }
  \label{fig:Madam_mean_bias}
\end{figure}

\begin{figure}[!tbh]
  \centering 
  \resizebox{\hsize}{!}{
    \includegraphics[trim=50 20 5 0,clip]
    {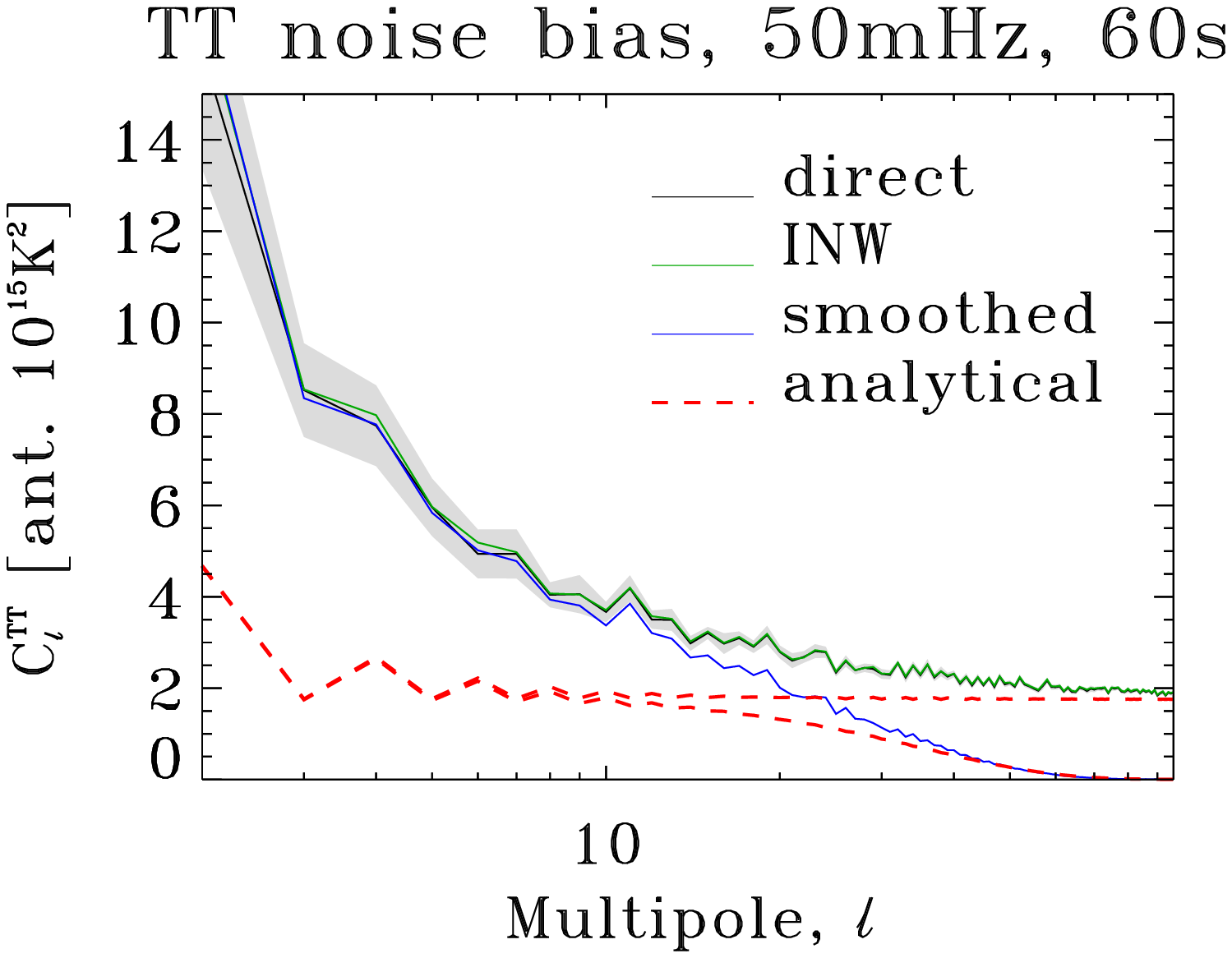}
    \includegraphics[trim=50 20 5 0,clip]
    {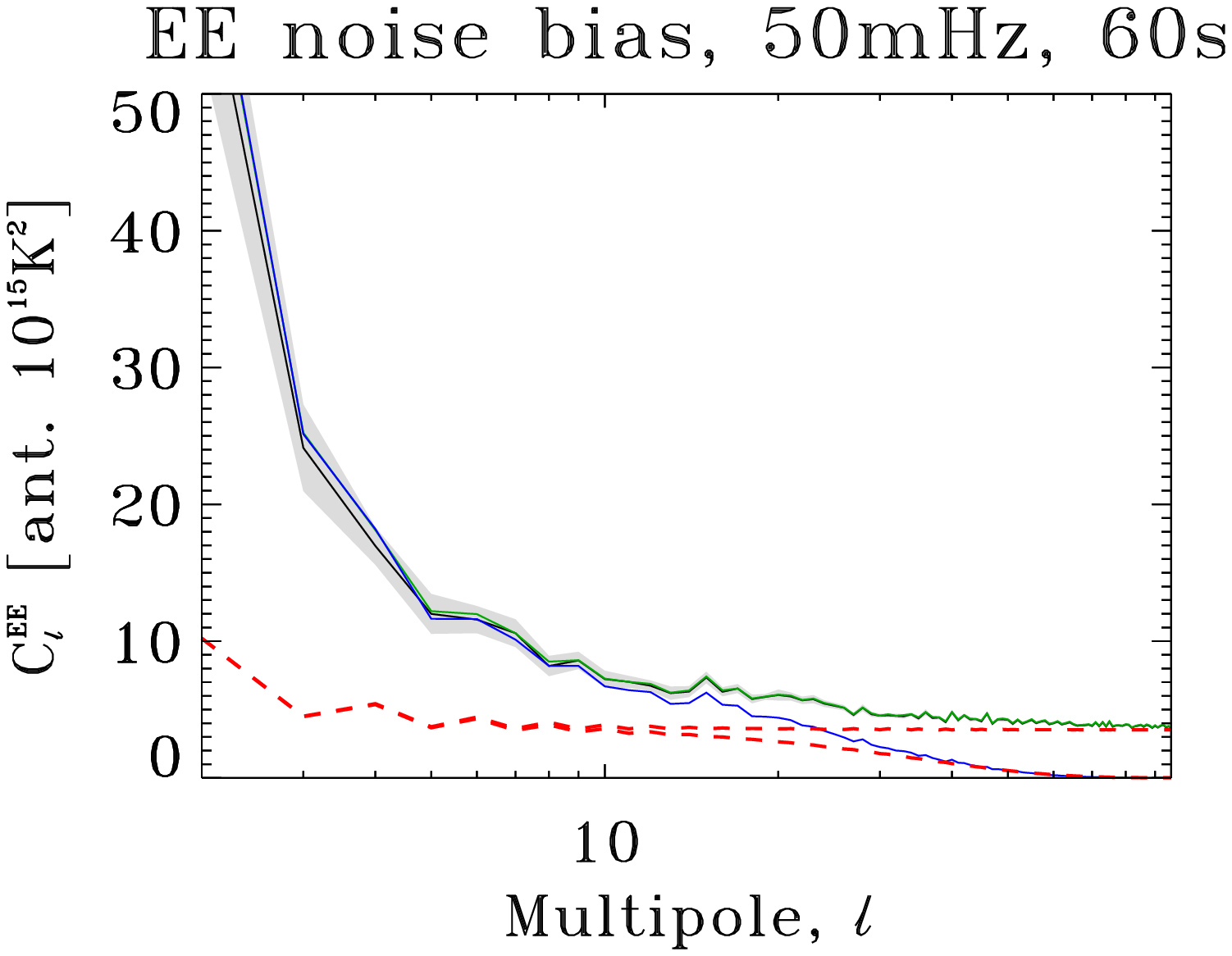}
    \includegraphics[trim=50 20 5 0,clip]
    {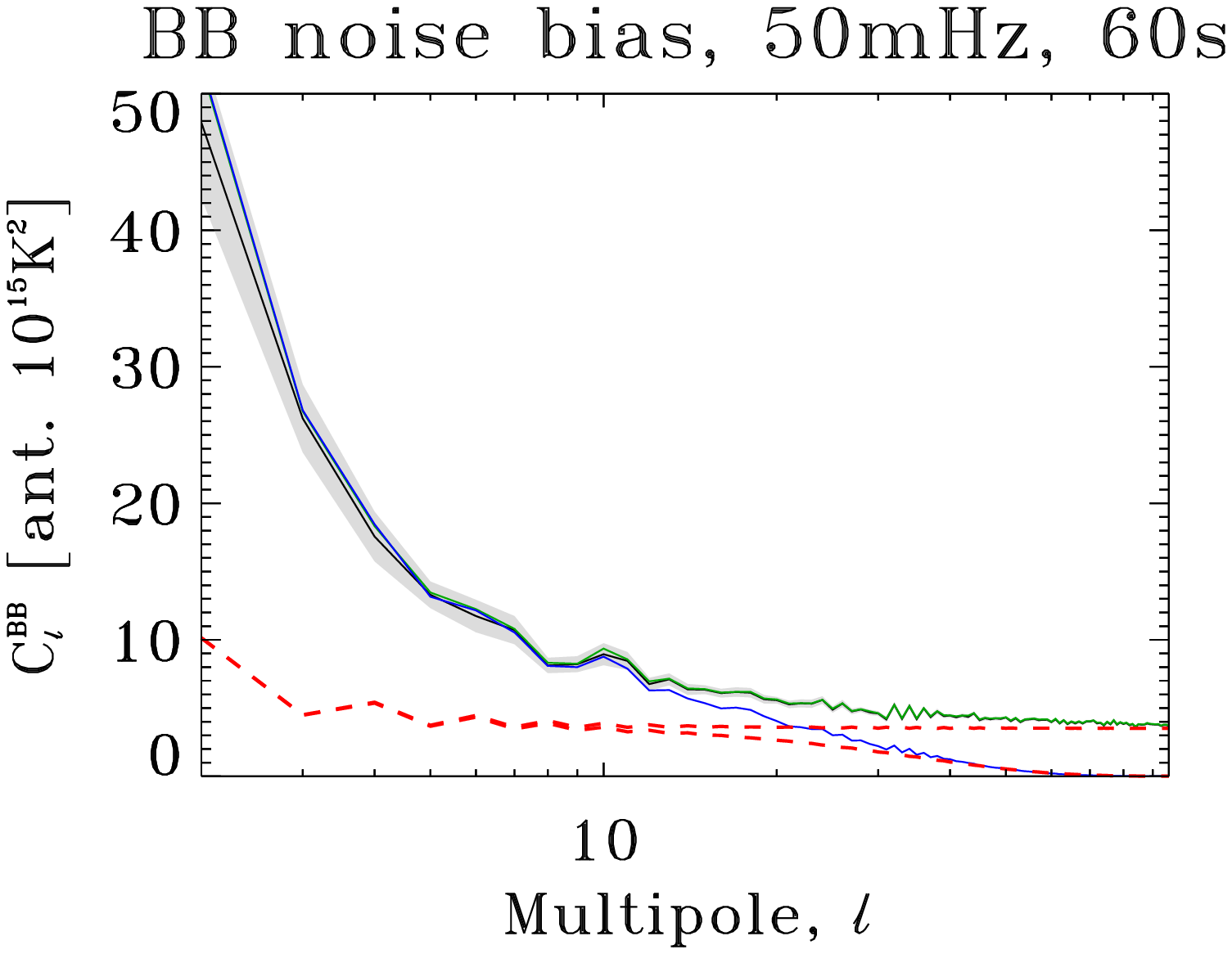}
  }
  \caption{
    Analytical and mean Monte Carlo noise biases from Madam runs at 
    $f_\mathrm{knee}=50\,$mHz using a long $60\,$s baseline. Grey band is the
    $1$-$\sigma$ region for the average, computed by dividing the sample
    variance by $\sqrt{25}$. Long baselines clearly fail to model the
    correlated noise.
  }
  \label{fig:Madam_mean_bias2}
\end{figure}

\begin{figure}[!tbh]
  \centering
  \resizebox{\hsize}{!}{
    \includegraphics[trim=50 20 5 0,clip]
    {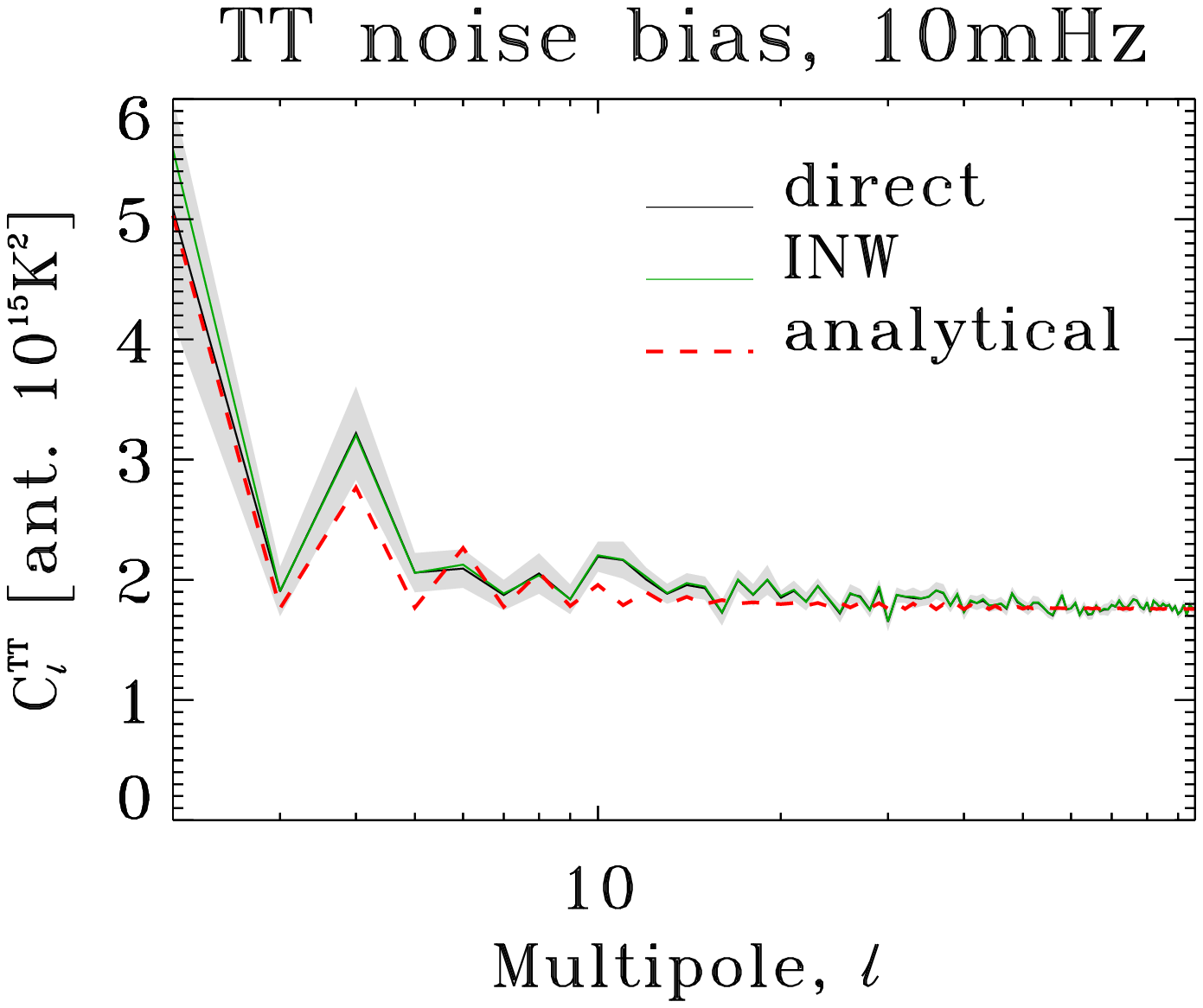}
    \includegraphics[trim=50 20 5 0,clip]
    {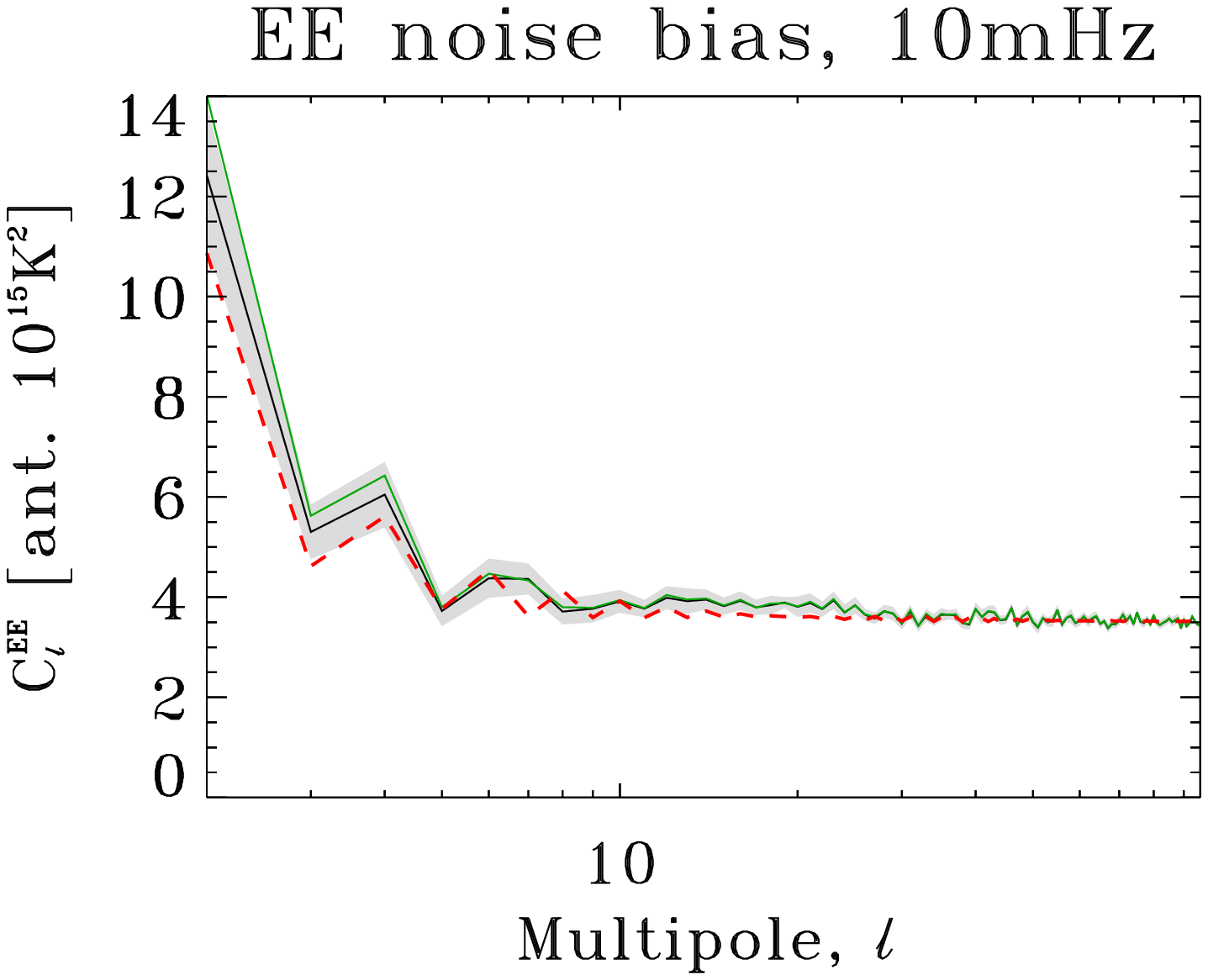}
    \includegraphics[trim=50 20 5 0,clip]
    {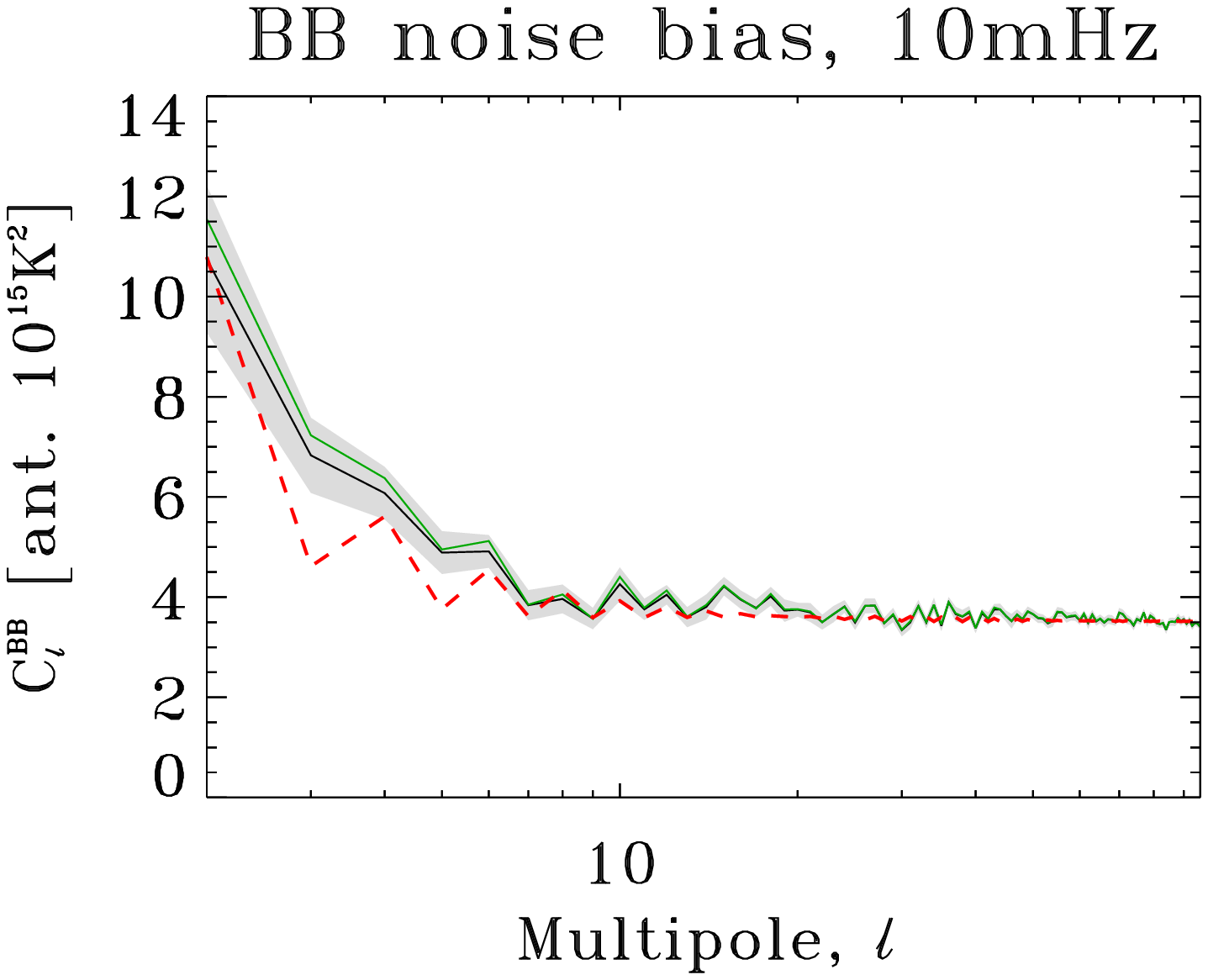}
  }
  \caption{
    Analytical and mean Monte Carlo noise biases from Springtide runs
    at $f_\mathrm{knee}=10\,$mHz.
  }
  \label{fig:Springtide_mean_bias}
\end{figure}

Figs.~\ref{fig:Madam_mean_bias}--\ref{fig:Springtide_mean_bias}
compare noise bias averages from $25$ noise realizations of the maps of the
noise residuals to the analytical estimates, Eq.~(\ref{eq:noise_bias2}),
based on the estimated noise covariance matrix. Map spectra are computed
using the HEALPix {\tt anafast} utility and the estimated noise biases
using the corresponding {\tt map2alm} subroutine. We only show the 
autospectra, TT, EE and BB, as the scanning has decoupled the modes to
large extent and only minimal coupling between the modes exist. 

The error band around the averages is the standard deviation of the individual
$C_\ell$ values divided by $\sqrt{25}$. Each plot exhibits up to five
curves: direct, noise weighted and harmonic smoothed noise
biases, and two analytical estimates.

The results derived for the case of the Madam runs with the short baseline,
$1.25\,$s, and the high knee frequency, $50\,$mHz, as shown in
Fig.~\ref{fig:Madam_mean_bias}, agree now very well with the MADping results.
This is unlike in the $chi^2$ test discussed earlier, indicating that those
were the anisotropic features responsible for the latter disagreement. The
numerical calculations of the noise bias are in this case agree very well
with the analytic predictions, Sect.~\ref{sec:noise_bias}. Both these facts
validate the destriper approximation to the noise covariance in the light
of this test, which is found to describe sufficiently precisely noise in
the low-resolution map. This conclusion agrees with those derived using the
$\chi^2$ test earlier.
 
The long baseline case is shown in Fig.~\ref{fig:Madam_mean_bias2} for
$f_\mathrm{knee}=50\,$mHz and computed using the generalized destriper, and
in Fig.~\ref{fig:Springtide_mean_bias} for the low value of the knee
frequency, $10\,$mHz, and based on Springtide results. We find that in the
high knee frequency case the prediction and numerical results differ rather
dramatically, highlighting the failure of the destriper approximation in
such case already seen with the $\chi^2$ test. We note here that the failure
seems to be affecting the largest angular scales as both the numerical and
analytical results tend to converge at the highest $\ell$ end considered in
this analysis. Similar results are found for the classical destriper maps
and covariances. For the low $f_\mathrm{knee}$ case the agreement is found to be
marginal, with visible deviations seen generally at $\ell \lesssim 5$ and are
the most significant in the case of the BB mode spectrum. The results
obtained using Madam in the analogous case are nearly indistinguishable.
 
We note that our analytic prediction are well in line with
WMAP findings \citep{Hinshaw:2006ia,Page:2006} and studies of the
destriping framework \citep{Efstathiou:2004eu,Efstathiou:2006wt}.

Averaged noise biases conform to the analytical estimates on almost all
accounts and deviations are small compared to the absolute noise bias.
As a warning example we include the results of modelling residual noise
from the $f_\mathrm{knee}=50\,$mHz timelines using $60\,$s baseline offsets.
This leaves more noise in the maps but the corresponding analytical
estimate is actually lower, since noise not modelled by the baseline
offsets is neglected.

At $10\,$mHz the Madam results for long $60\,$s baselines are equivalent
with Springtide results: both succeed to estimate noise bias at low knee
frequency but should not be used for high knee frequencies as such.

Our noise bias estimates for EE and BB spectra are equal, but the averaged
spectra for the Monte Carlo maps appear visually different in this respect.
To ensure that this is only due to Monte Carlo noise we ran Madam in Monte
Carlo mode, simulating noise on the fly and avoiding the costs associated
with storing of the time ordered data. After averaging over 117
$N_\mathrm{side}=8$ noise map spectra we found that the differences between
EE and BB noise map spectra were at most $10\%$.

In the Appendix~\ref{app:noiseBias} we replot some of these figures after
subtracting the analytical bias and dividing by the Monte Carlo sample
deviation to highlight the differences between the analytical and numerical
results.

\subsubsection{By power spectrum estimation} \label{sec:by_pse}

Our final validation procedure for the noise covariance matrices was to
use them in $C_\ell$ estimation. Due to resource constraints this
exercise was conducted at a lower $N_\mathrm{side}=8$
resolution. All three map-making codes produced maps from the $25$ noise
realizations. Each realization was paired with an independent realization
of the CMB sky and the co-added maps were processed using the Bolpol code,
an implementation of the QML estimator described in Sect~\ref{sec:QML}.
The $25$ power spectrum estimates for each multipole were then averaged
over and the Bolpol-determined error bars were accordingly divided by
$\sqrt{25}$. 

\begin{figure}[!tbhp]
  \centering
  \resizebox{11cm}{!}{\includegraphics[trim=5 10 15 0,clip]
    {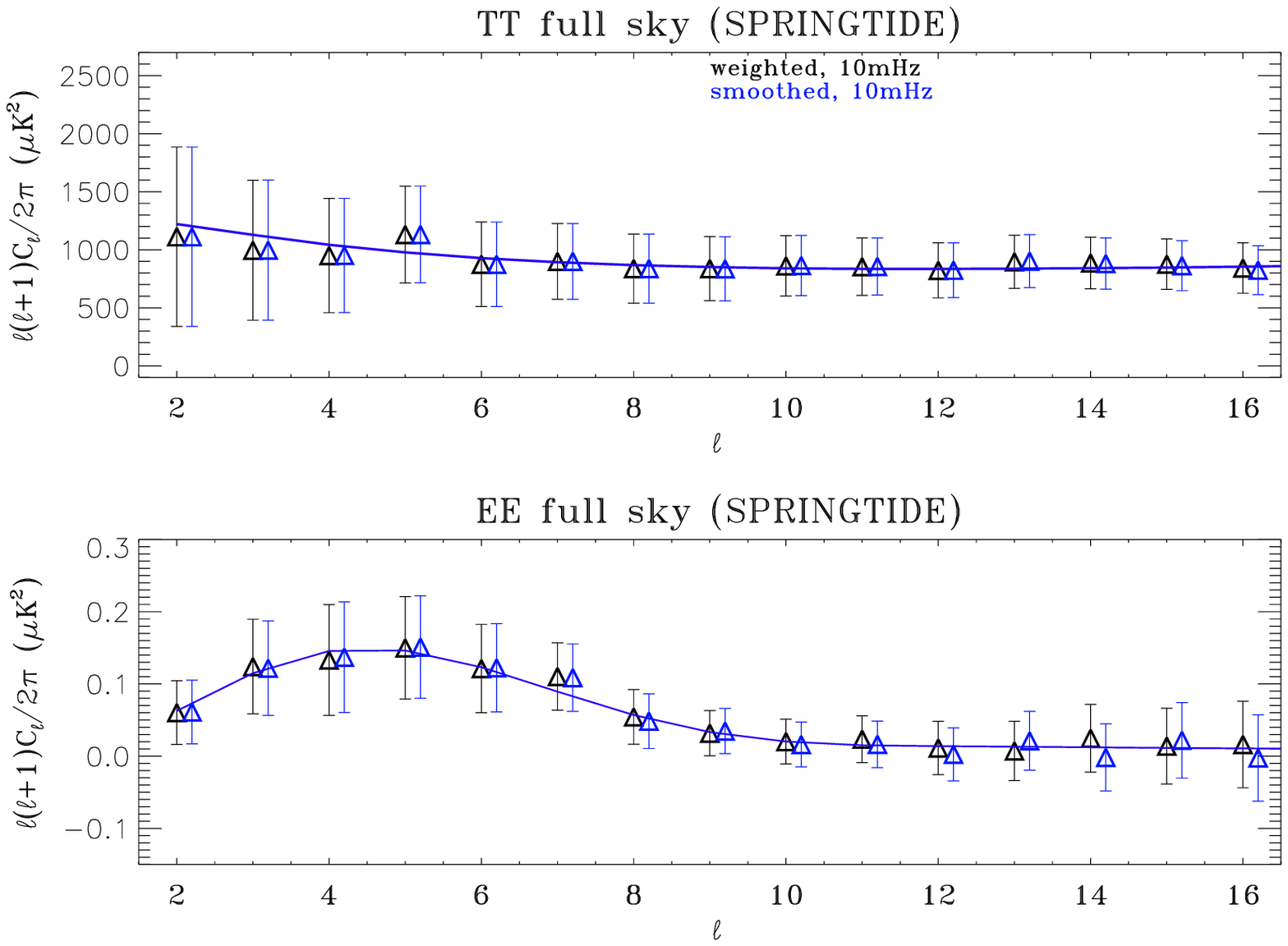}
  }
  \resizebox{11cm}{!}{\includegraphics[trim=5 10 15 0,clip]
    {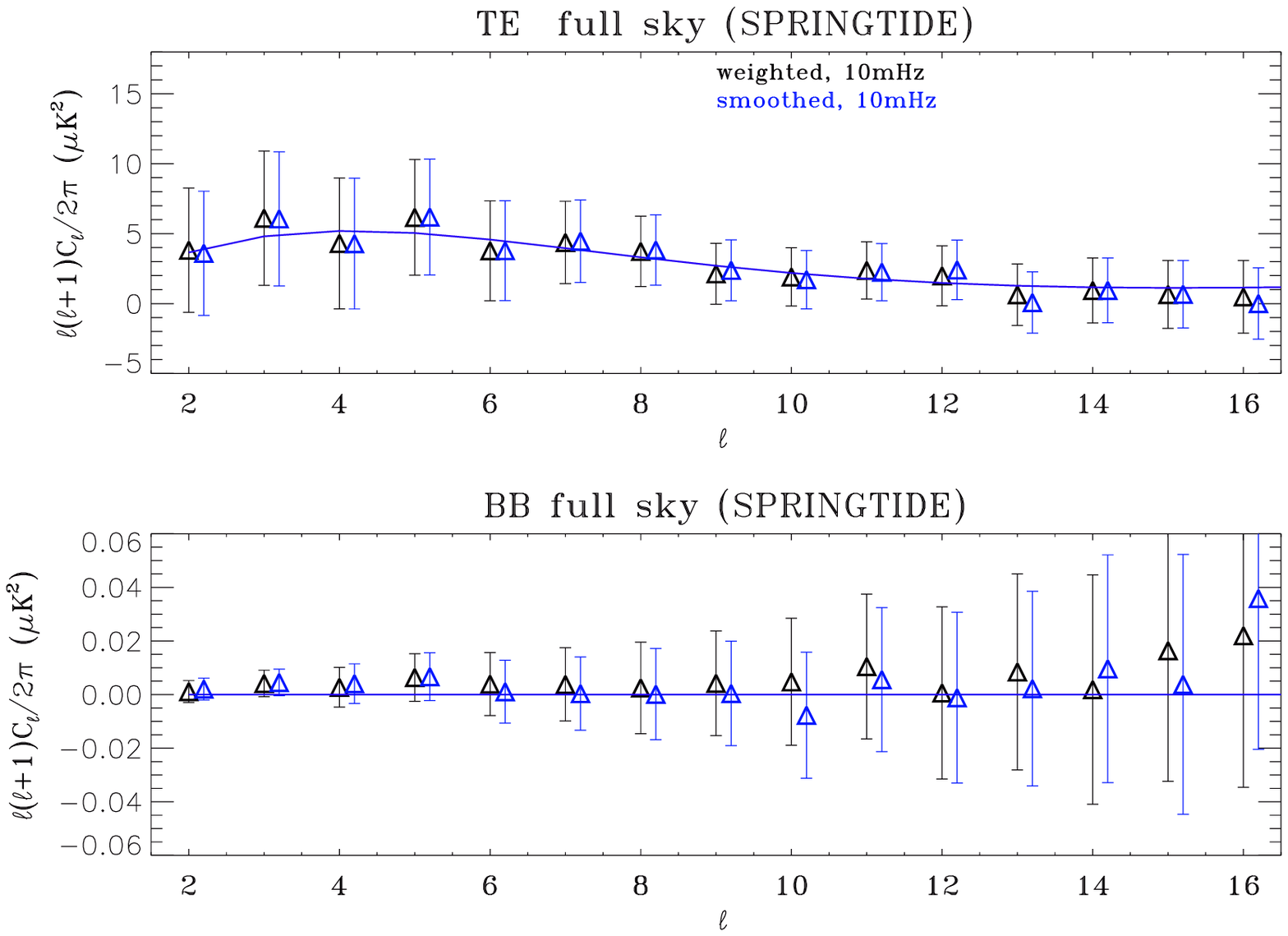}
  }
  \resizebox{11cm}{!}{\includegraphics[trim=5 10 15 0,clip]
    {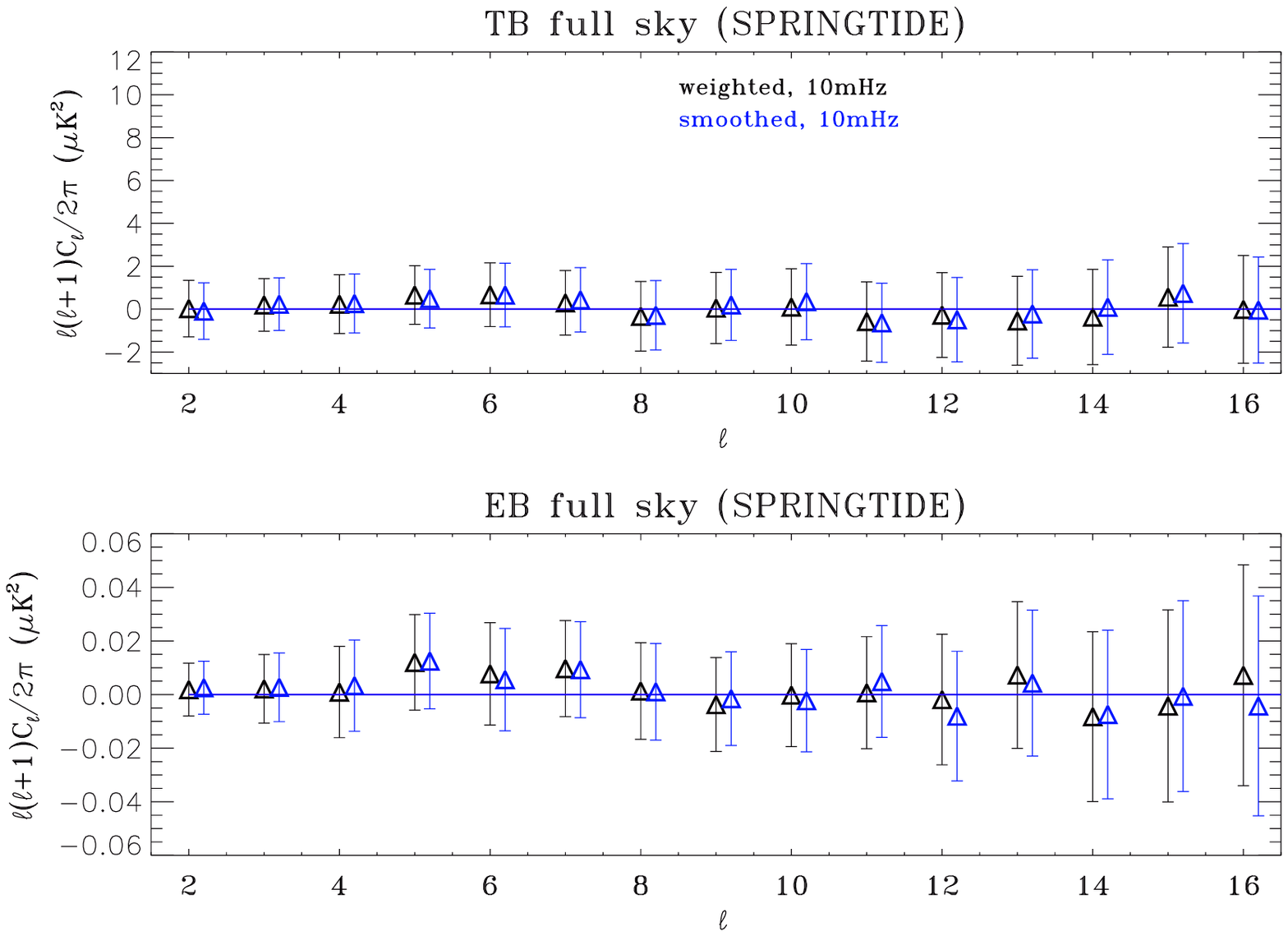}
  }
  \caption{
    Averaged power spectrum estimates over 25 noise and CMB realizations.
    The noise has a $10\,$mHz knee frequency.
  }
  \label{pic:PSE2}
\end{figure}

The example of the results is shown in Fig.~\ref{pic:PSE2}. These estimates
were obtained for the case with the low knee frequency,
$f_\mathrm{knee}=10\,$mHz, using the conventional destriper, Springtide.
We note they do not hint unambiguously at any problem with the estimated
covariance, even if the $\chi^2$ (strongly) and the noise bias (mildly)
tests may indicate otherwise. This is likely in part due to a lower
sensitivity of the power spectrum test on the one hand and on the other due
to the fact that the lower resolution has been used in this last case.

Similar statistically good agreements can also be seen in the case of
the higher value of $f_\mathrm{knee}$, if the covariance is computed using
either the optimal or generalized destriping technique with the short
baselines of $1.25$s. If longer baselines are used, i.e., $60$s, the
estimates of the polarized spectra, both E and B, are visibly
discrepant with the assumed inputs. Similar disagreement can be seen if
the off-diagonal elements of the covariance matrices are neglected.
In both of these cases, no particular effect on the total intensity 
spectrum can be noticed in the range of investigated angular scales.
We illustrate all these statements in Appendix~\ref{app:ps}.
These observations emphasize the importance of precise estimation of
the noise covariance in particular for the polarized power spectra.



\subsection{Low-resolution maps} \label{sec:bandwidth}

In the previous sections we have discussed our ability to estimate correctly
the properties of the noise present in the low-resolution maps. We have
demonstrated that this is indeed the case for all considered resolution
downgrading strategies and both optimal and destriper maps. Though in the
latter case a baseline length needs to be carefully chosen depending on the
time domain noise characteristic.

In this Section we focus on the low-resolution maps themselves. We will
look at them from three different perspectives, evaluating the level of the
map-making artifacts left in the maps, the properties of the sky signal and
the level of the noise.

\begin{figure}[!tbh]
  \centering
  \resizebox{\hsize}{!}{
    \includegraphics[trim=25 0 45 0,clip,angle=90]
    {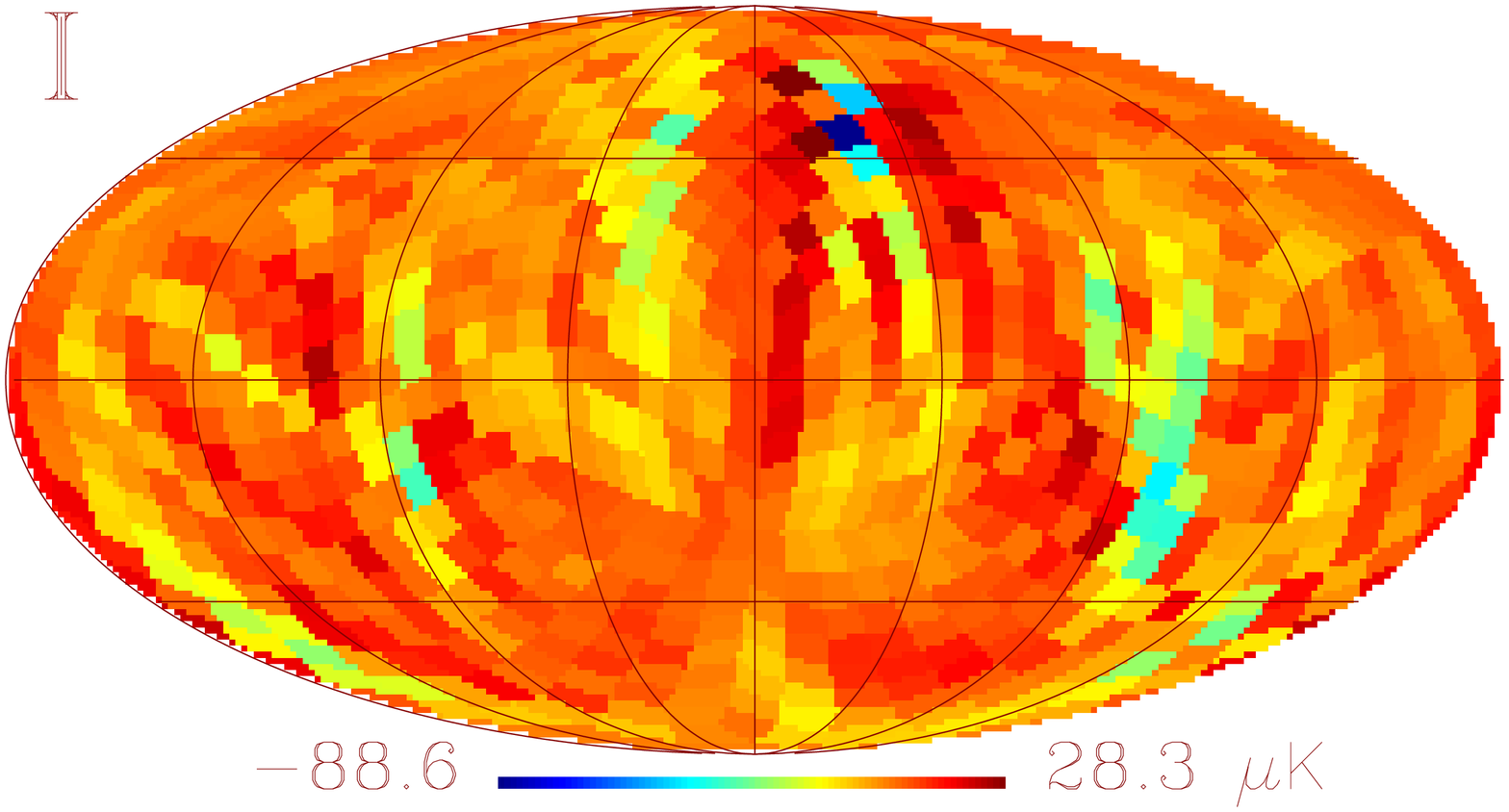}
    \includegraphics[trim=25 0 45 0,clip,angle=90]
    {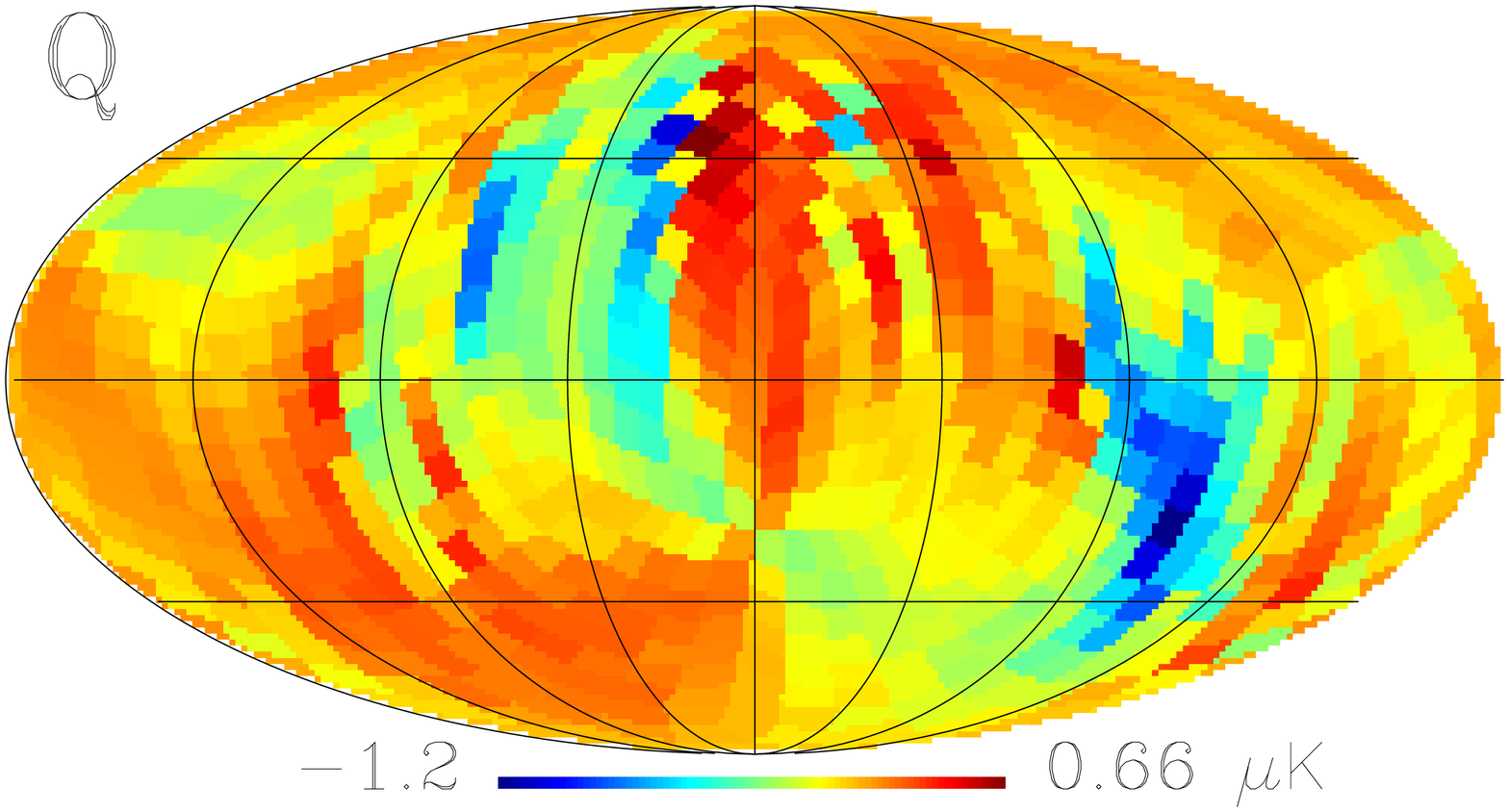}
    \includegraphics[trim=25 0 45 0,clip,angle=90]
    {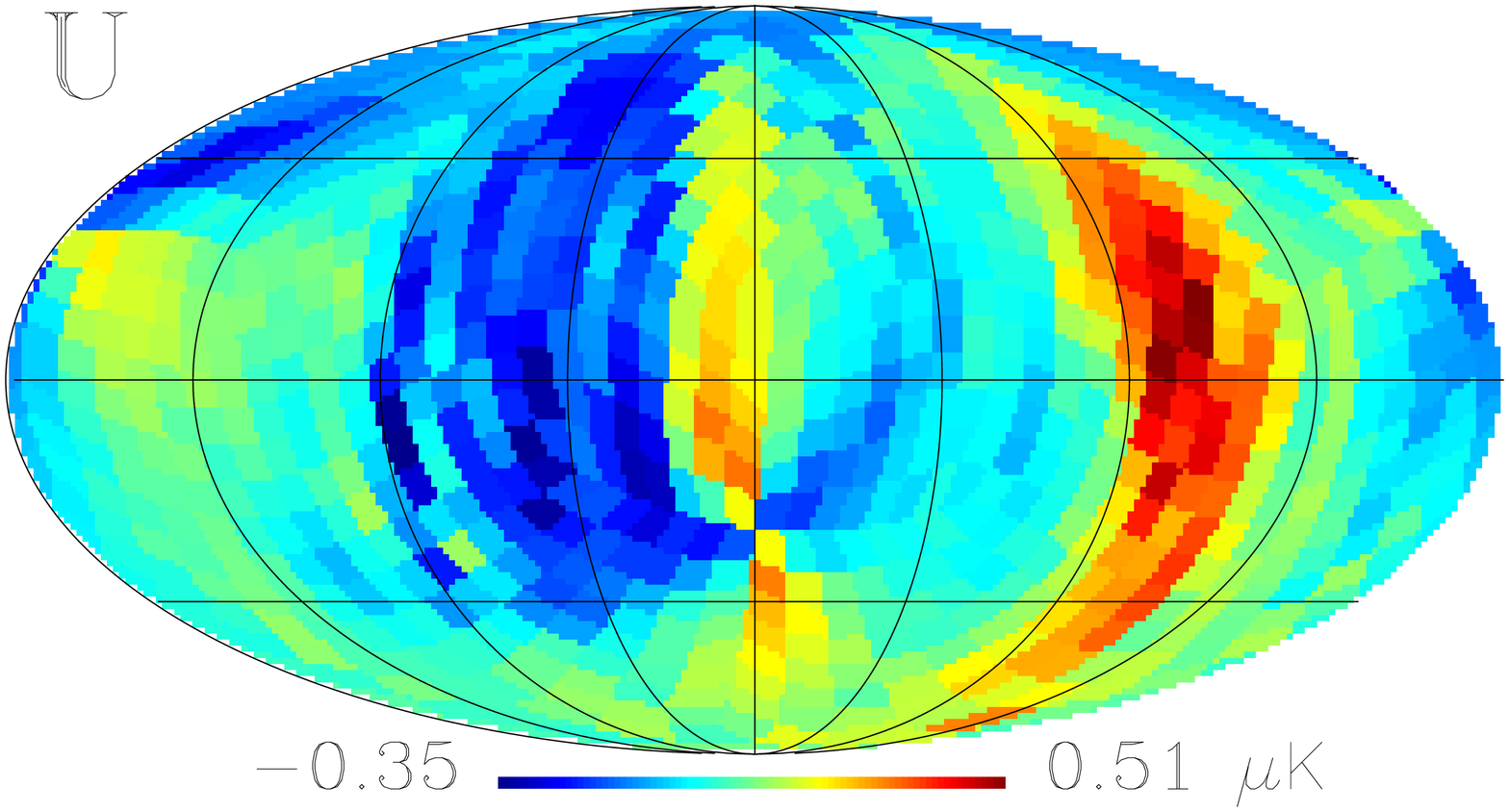}
  }
  \resizebox{\hsize}{!}{
    \includegraphics[trim=25 0 45 0,clip,angle=90]
    {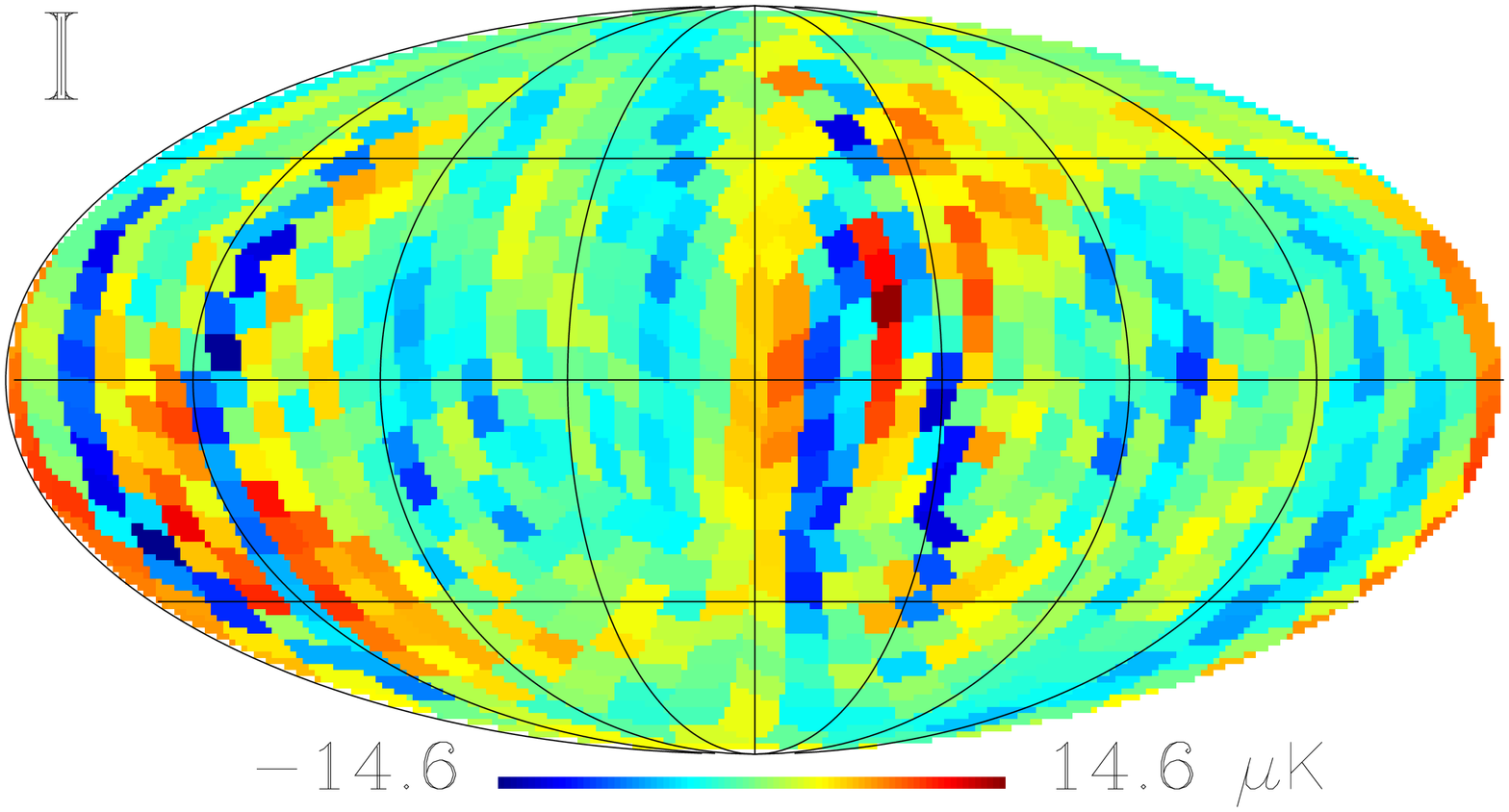}
    \includegraphics[trim=25 0 45 0,clip,angle=90]
    {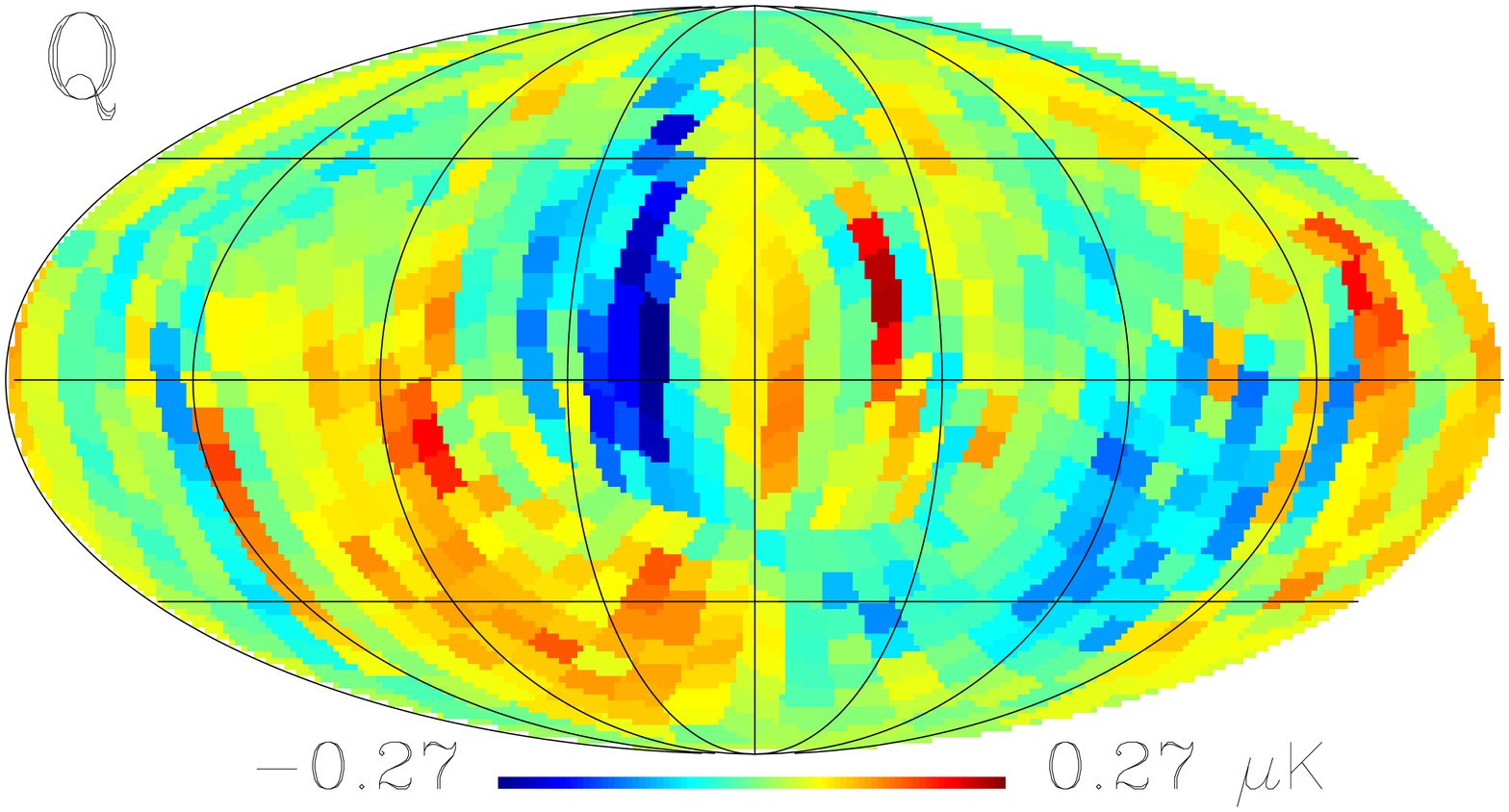}
    \includegraphics[trim=25 0 45 0,clip,angle=90]
    {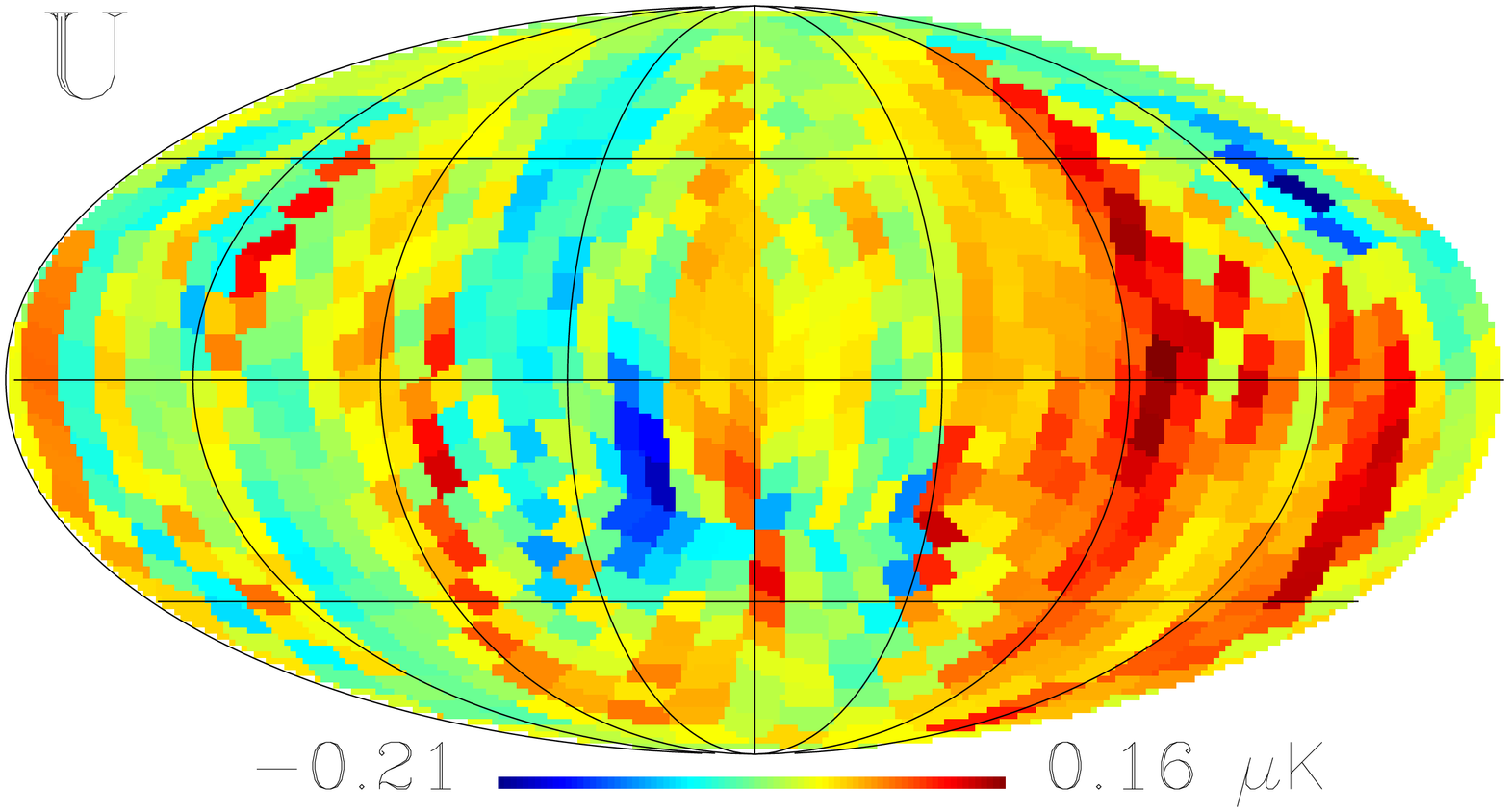}
  }
  \resizebox{\hsize}{!}{
    \includegraphics[trim=25 0 45 0,clip,angle=90]
    {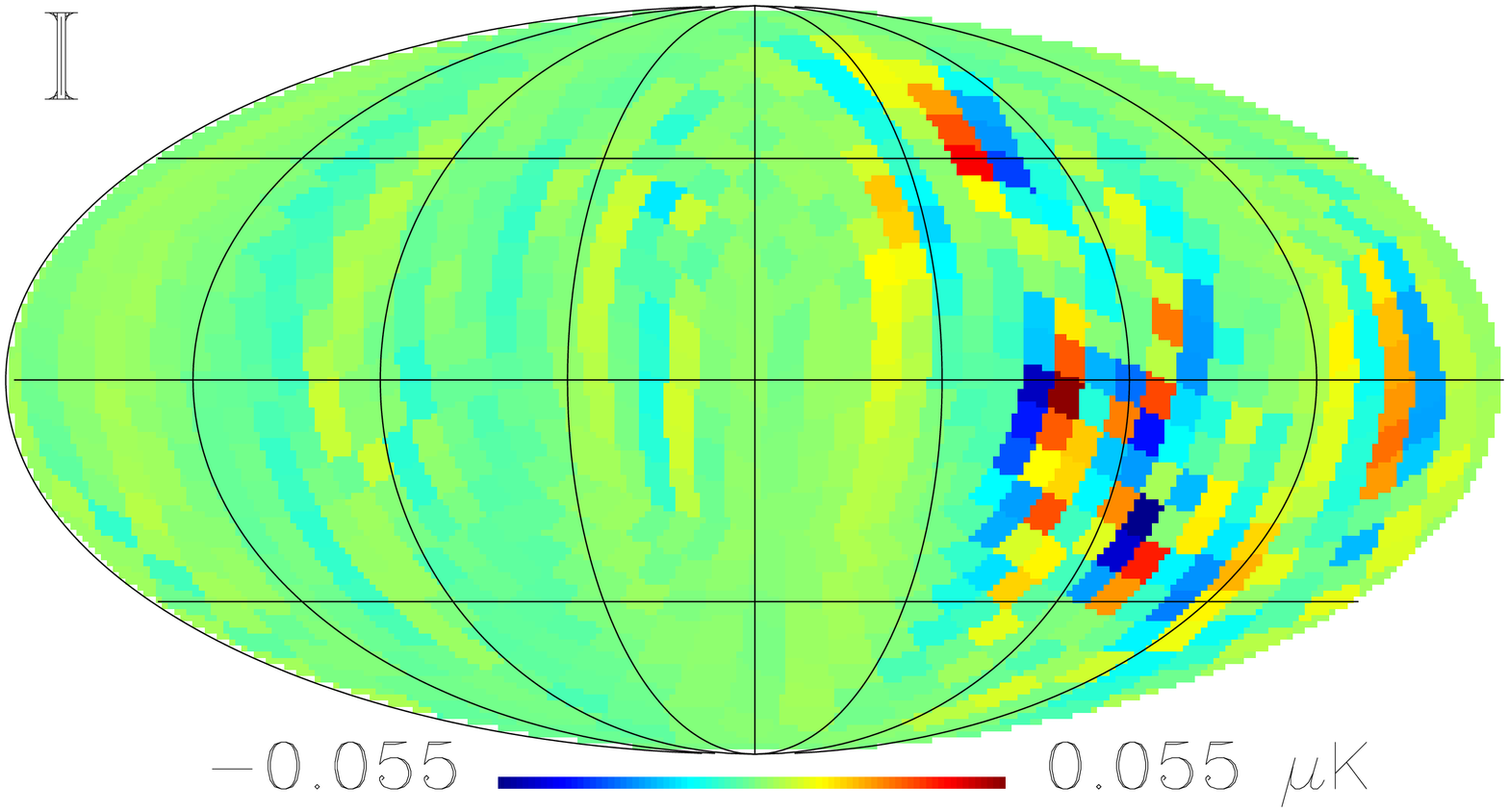}
    \includegraphics[trim=25 0 45 0,clip,angle=90]
    {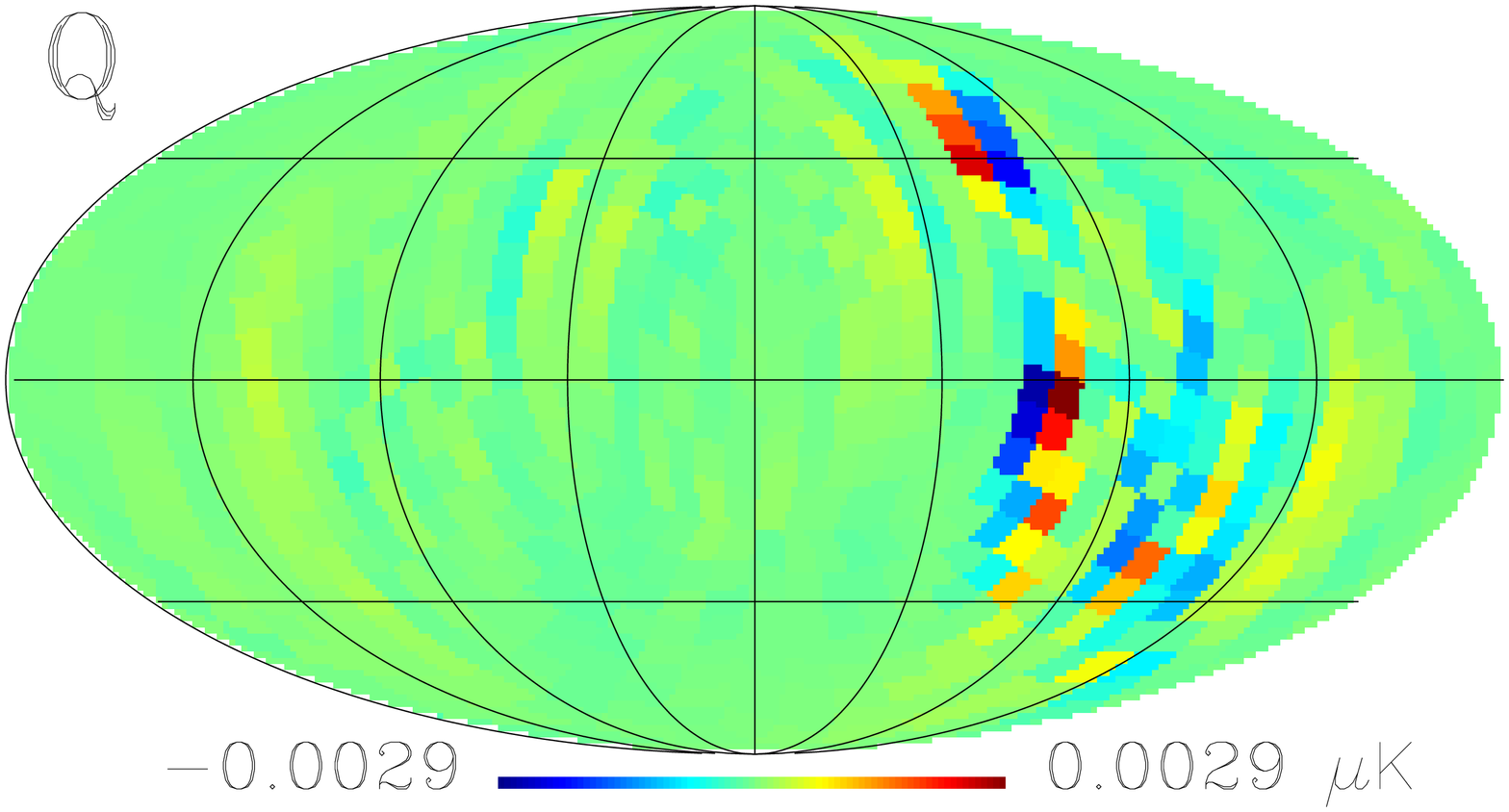}
    \includegraphics[trim=25 0 45 0,clip,angle=90]
    {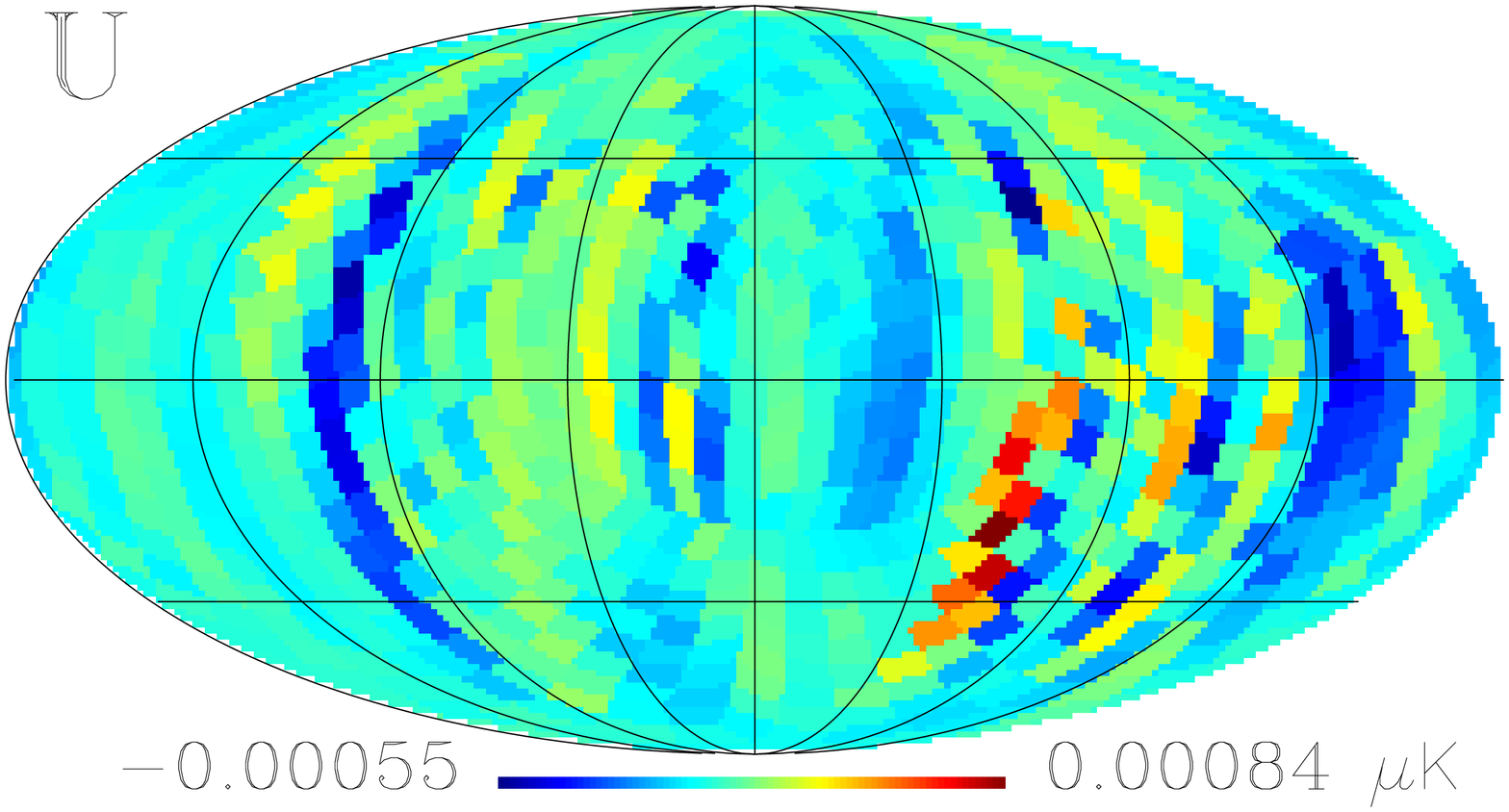}
  }
  \caption{
    Examples of signal striping. We show the difference between a
    binned and destriped signal-only maps.
    Rows correspond to direct Madam results, $1.25\,$s and $60\,$s, and
    inverse noise weighted $1.25\,$s baseline case respectively.
  }
  \label{fig:striping}
\end{figure}

In Fig.~\ref{fig:striping} we show the differences of the noise-free low
resolution maps computed using different downgrading approaches discussed
earlier and the input map used for the simulations. The reference low
resolution version of the input map has been obtained via simple binning
of the sky signal directly into low-resolution pixels.

\begin{figure}[!tbh]
  \centering
  \resizebox{\hsize}{!}{
    \includegraphics[trim=20 10 20 0,clip]{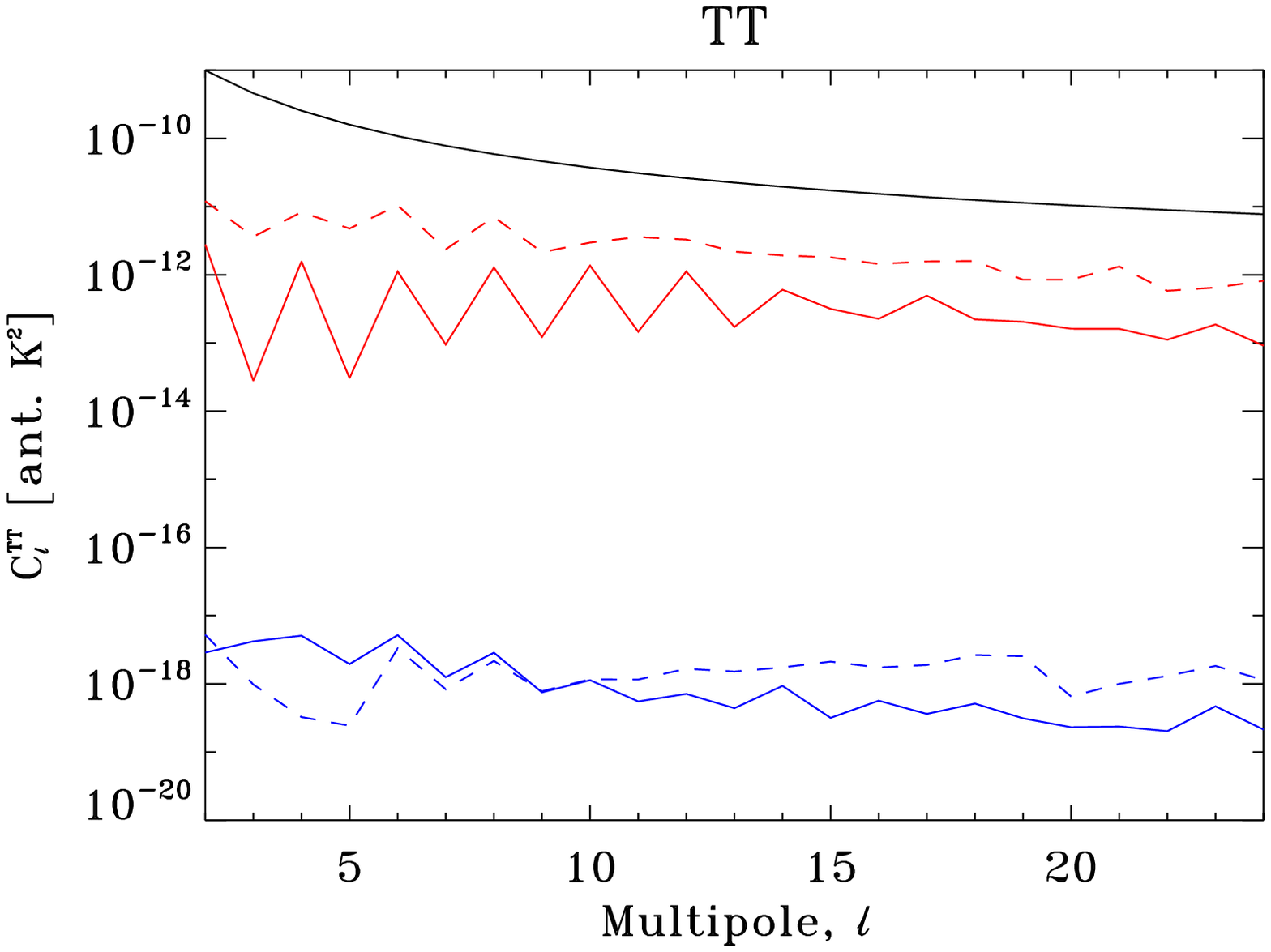}
    \includegraphics[trim=20 10 20 0,clip]{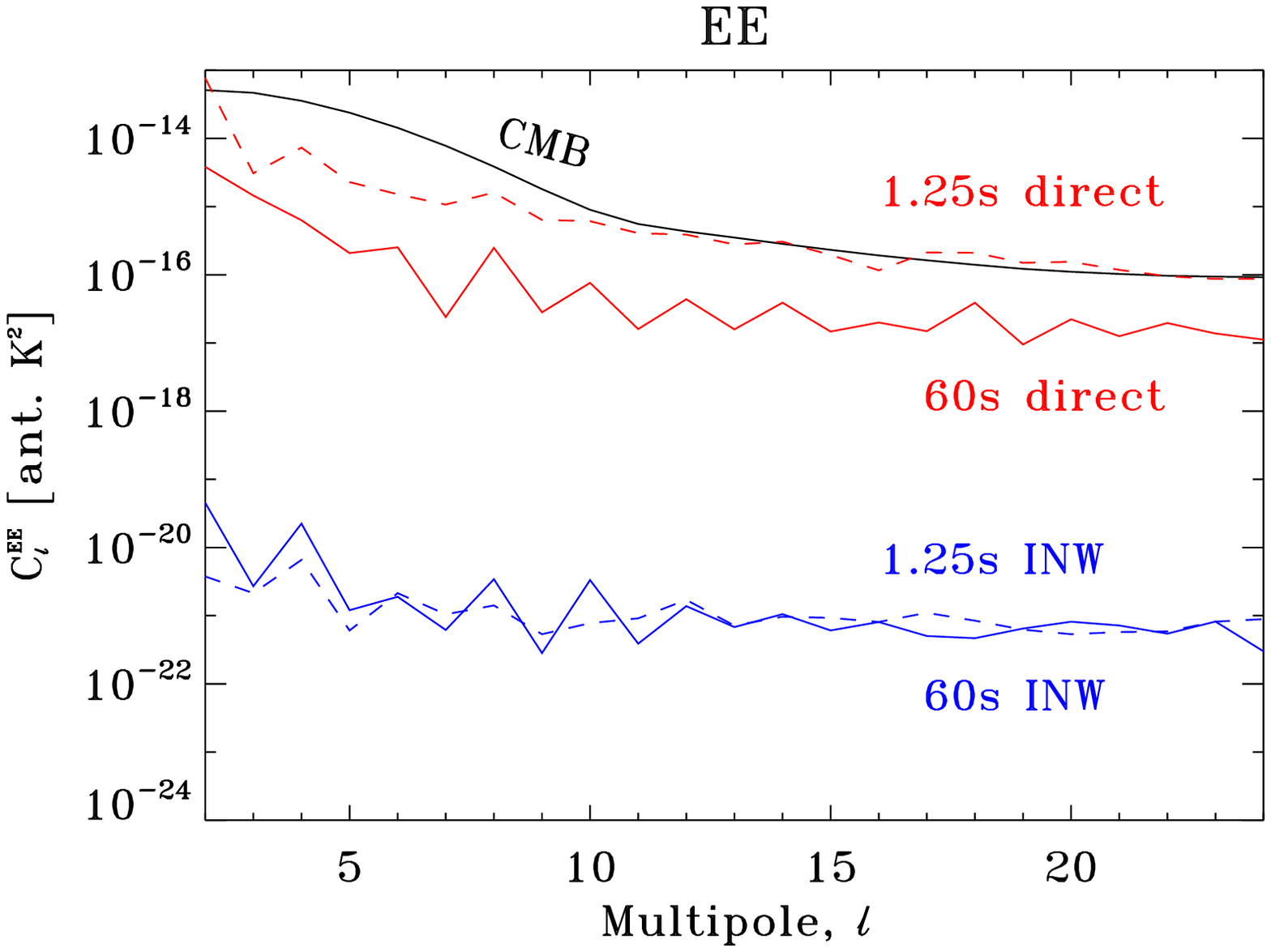}
  }
  \caption{
    Comparison of CMB and stripe power spectra reveals that the striping
    can significantly bias the EE and BB power spectrum estimates.
    Stripe map spectra are computed from maps shown in Fig.~\ref{fig:striping}.
  }
  \label{fig:striping_cl}
\end{figure}

In the case of the direct method we see clearly the extra power
spread all over the sky in all three Stokes parameters. Any other
proposed approach clearly fares much better leading to a substantial
decrease of the level of the observed artifacts. This is quantified with
the help of the pseudo-spectra in Fig.~\ref{fig:striping_cl} and in
Table~\ref{tab:stripe_rms}, where we have collected the root mean square
estimates for the signal residual maps.

\begin{table}
  \caption{
    Comparison of TT power (rms) in the stripe maps at 
    \mbox{$N_\mathrm{side}=8$}
  }
  \centering
  \begin{tabular}{l l l l l l l}
    \hline
    \hline
    & \multicolumn{3}{l}{CMB$^\mathrm a$}
    & \multicolumn{3}{l}{Foregrounds$^\mathrm b$} \\
    Map-maker   & Direct & Averaged & Smoothed & Direct & Averaged & Smoothed\\
    \hline
    Madam, $1.25$s & $5.44\,\mu$K & $7.29\,$nK & $5.20\,$nK & $9.46\,\mu$K & $4.67\,$nK & $2.14\,$nK\\
    Madam, $60$s   & $2.07\,\mu$K & $3.31\,$nK & $1.98\,$nK & $3.81\,\mu$K & $3.21\,$nK & $2.23\,$nK\\
    \hline
  \end{tabular}
  \flushleft
  $^\mathrm a$ Binned CMB map rms is $37.34$ $\mu$K
  \newline
  $^\mathrm b$ Binned foreground map rms is $196.3$ $\mu$K
  \label{tab:stripe_rms}
\end{table}

\begin{figure}[!tbhp]
  \centering
  \resizebox{11cm}{!}{
    \includegraphics[trim=5 10 15 0,clip]
    {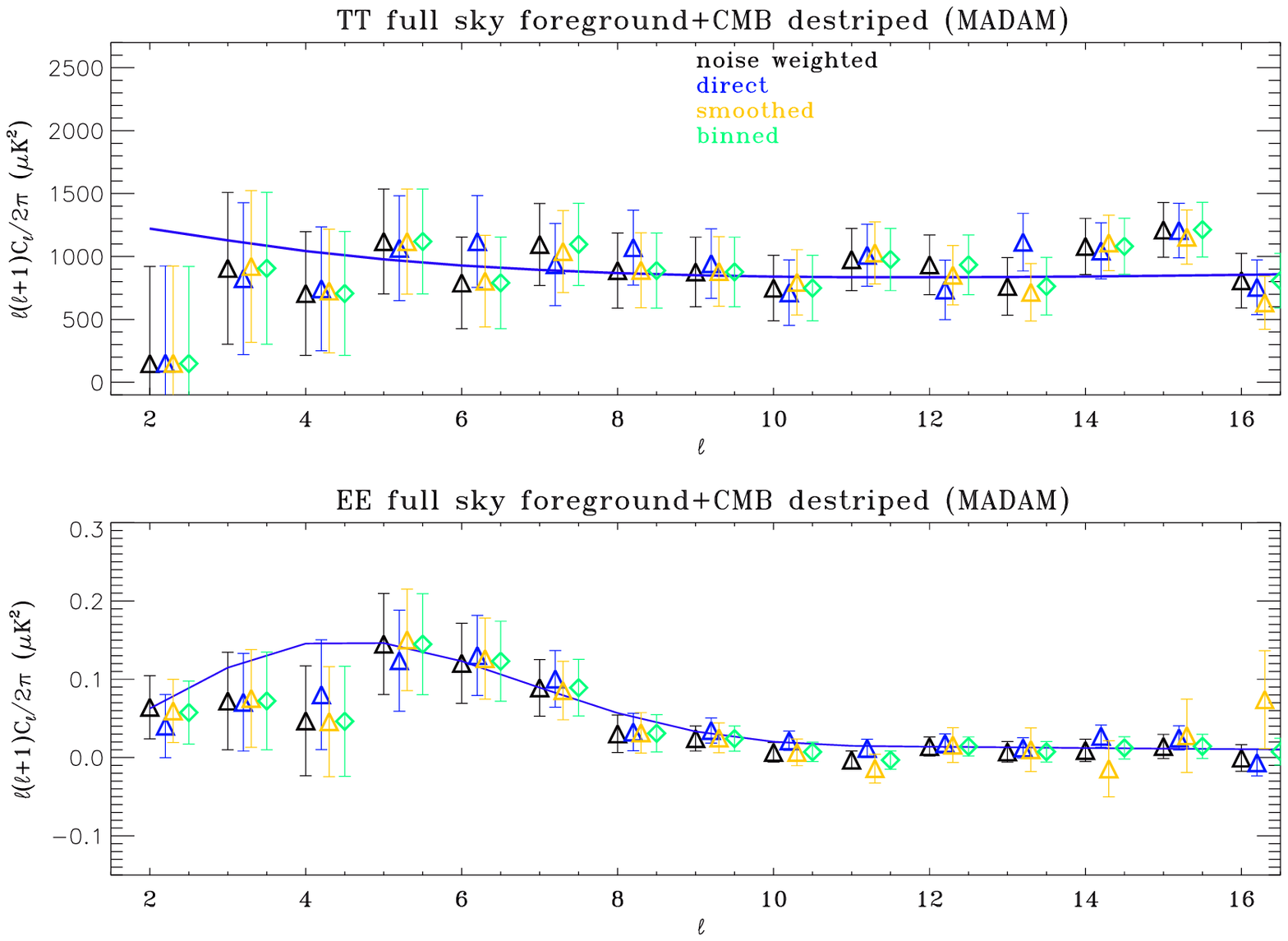}}
  \resizebox{11cm}{!}{
    \includegraphics[trim=5 10 15 0,clip]
    {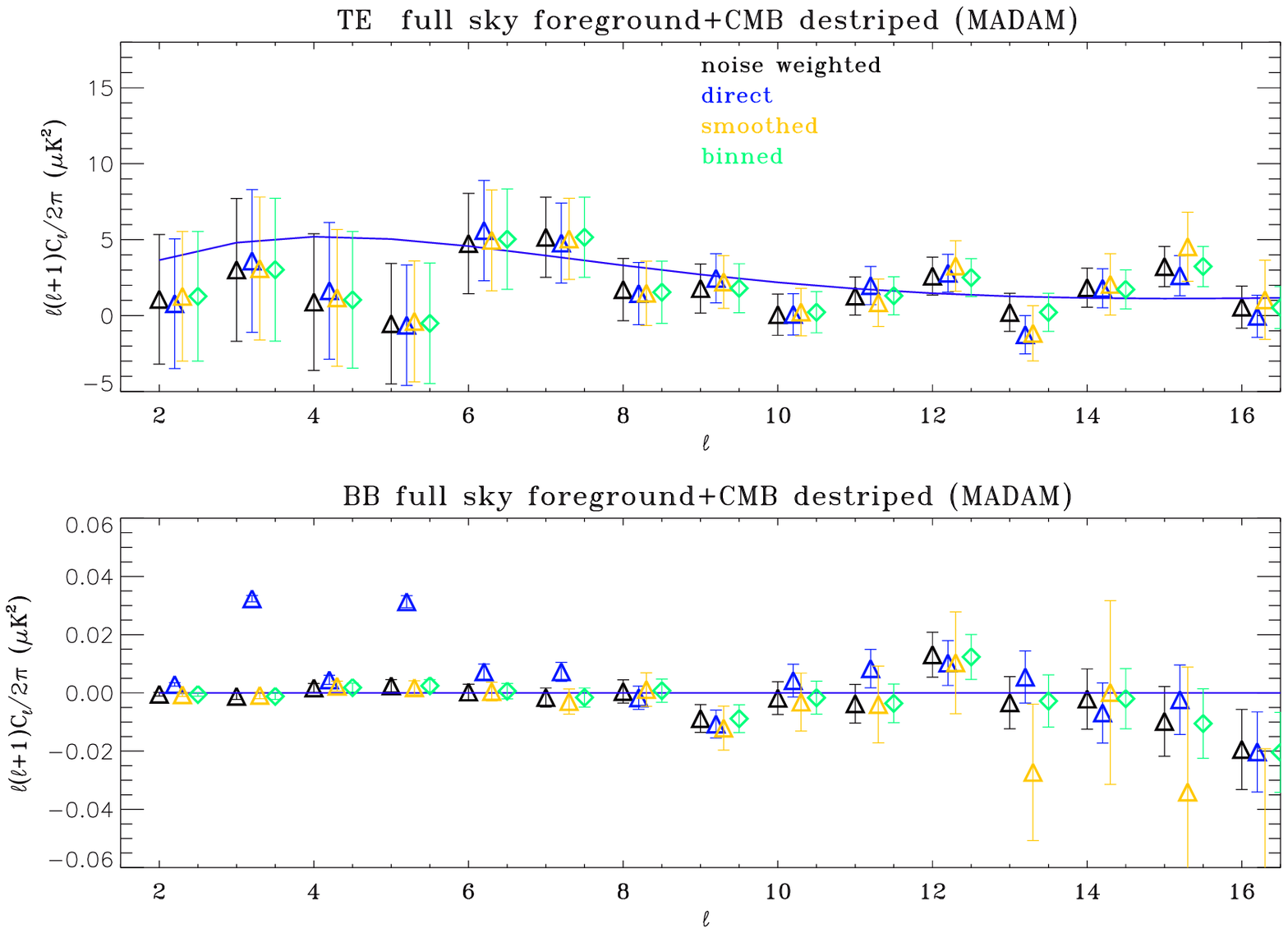}}
  \resizebox{11cm}{!}{
    \includegraphics[trim=5 10 15 0,clip]
    {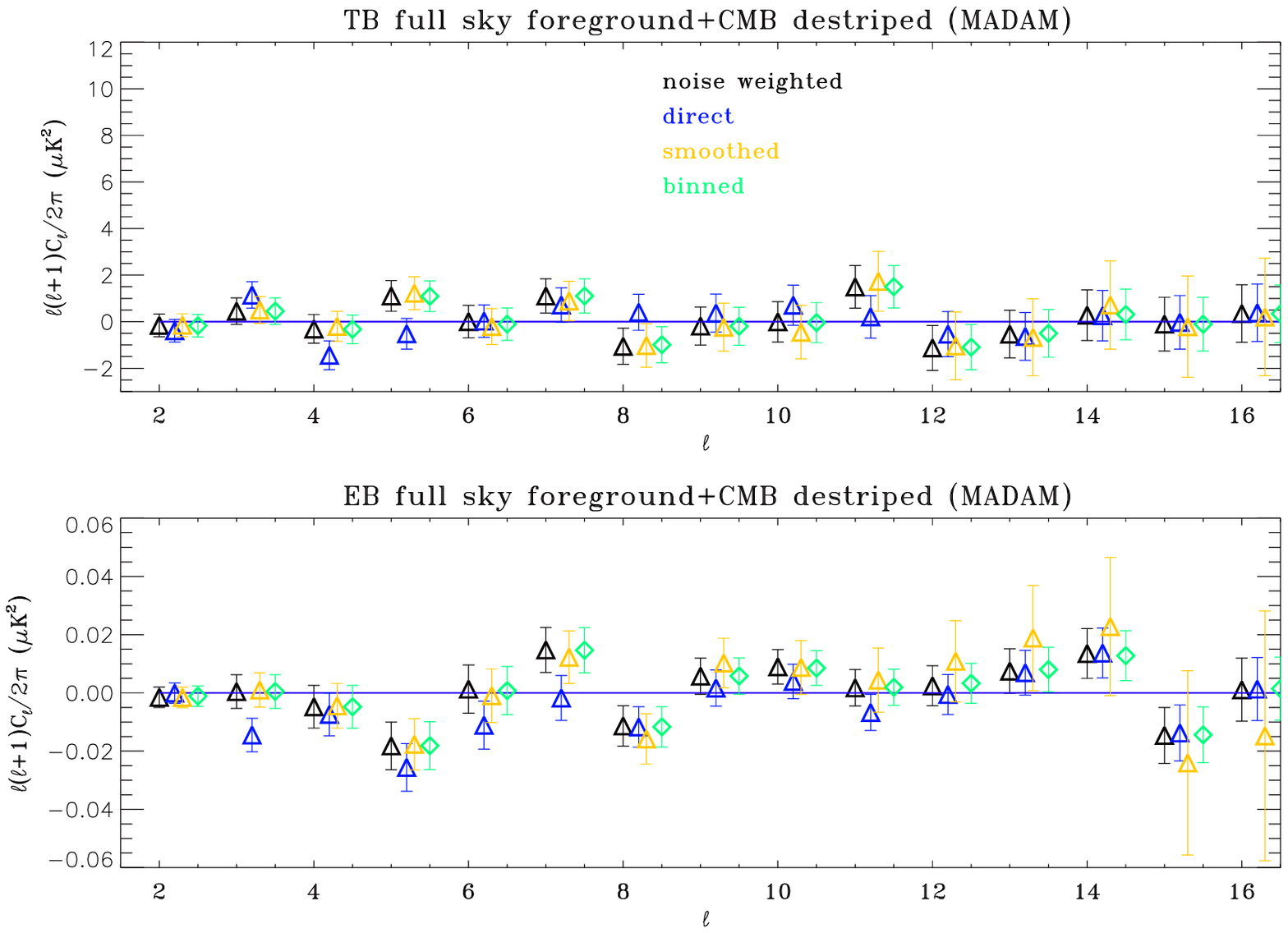}}
  \caption{Band power estimates of CMB with CMB and foreground stripes.}
  \label{fig:PSE_stripes}
\end{figure}

\begin{figure}[!tbh]
  \begin{flushleft}
    \hspace{1.1cm}
    \mbox{Raw $C_\ell$}
    \hspace{5.0cm}
    \mbox{Deconvolved $\ell(\ell+1)C_\ell$}
    \hspace{2.9cm}
    \mbox{Fractional difference}
  \end{flushleft}
  \vspace{-1.5mm}
  \centering
  \resizebox{\hsize}{!}{
    \includegraphics[trim= 0 15 20 28,clip]
    {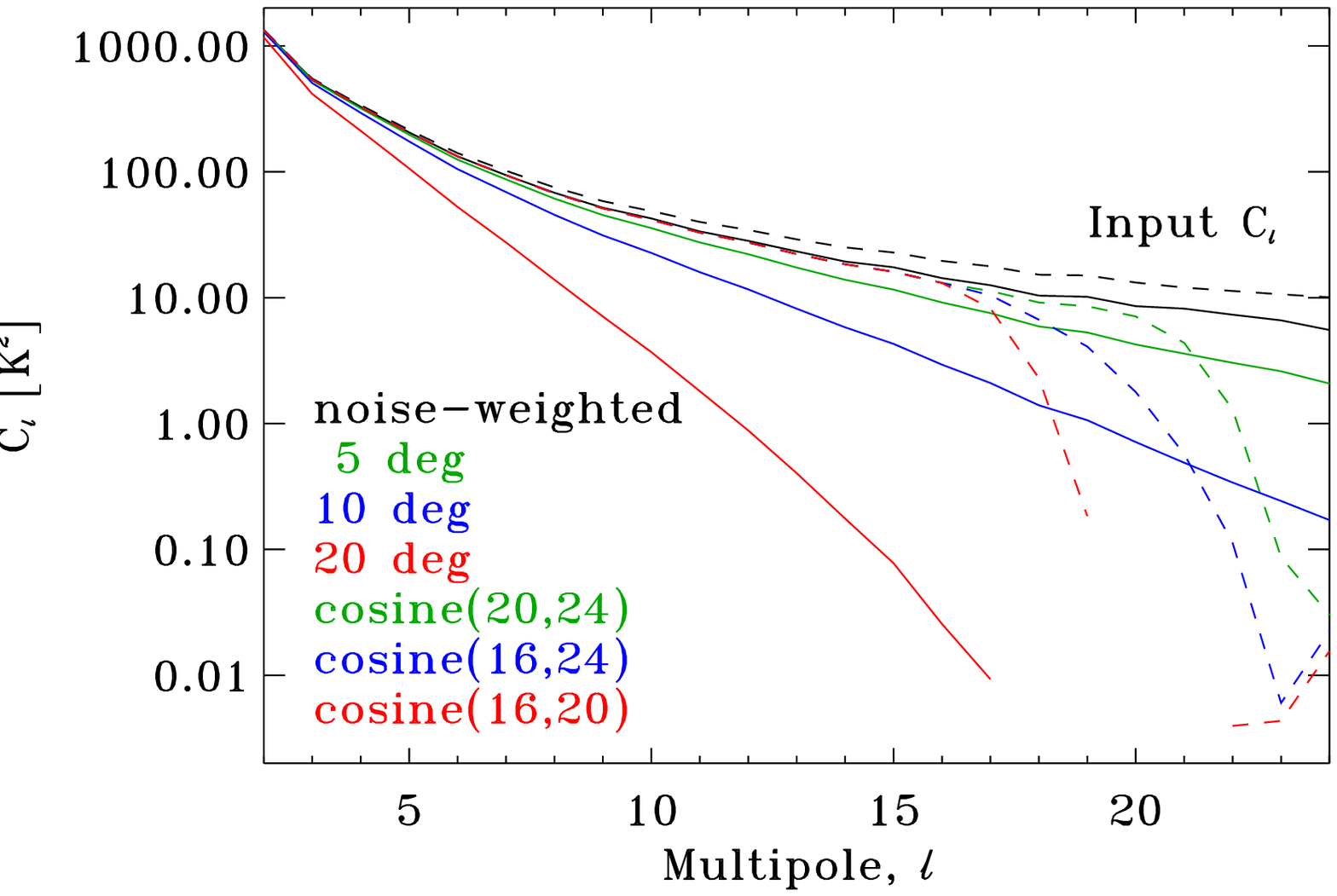}
    \includegraphics[trim=10 15 20 28,clip]
    {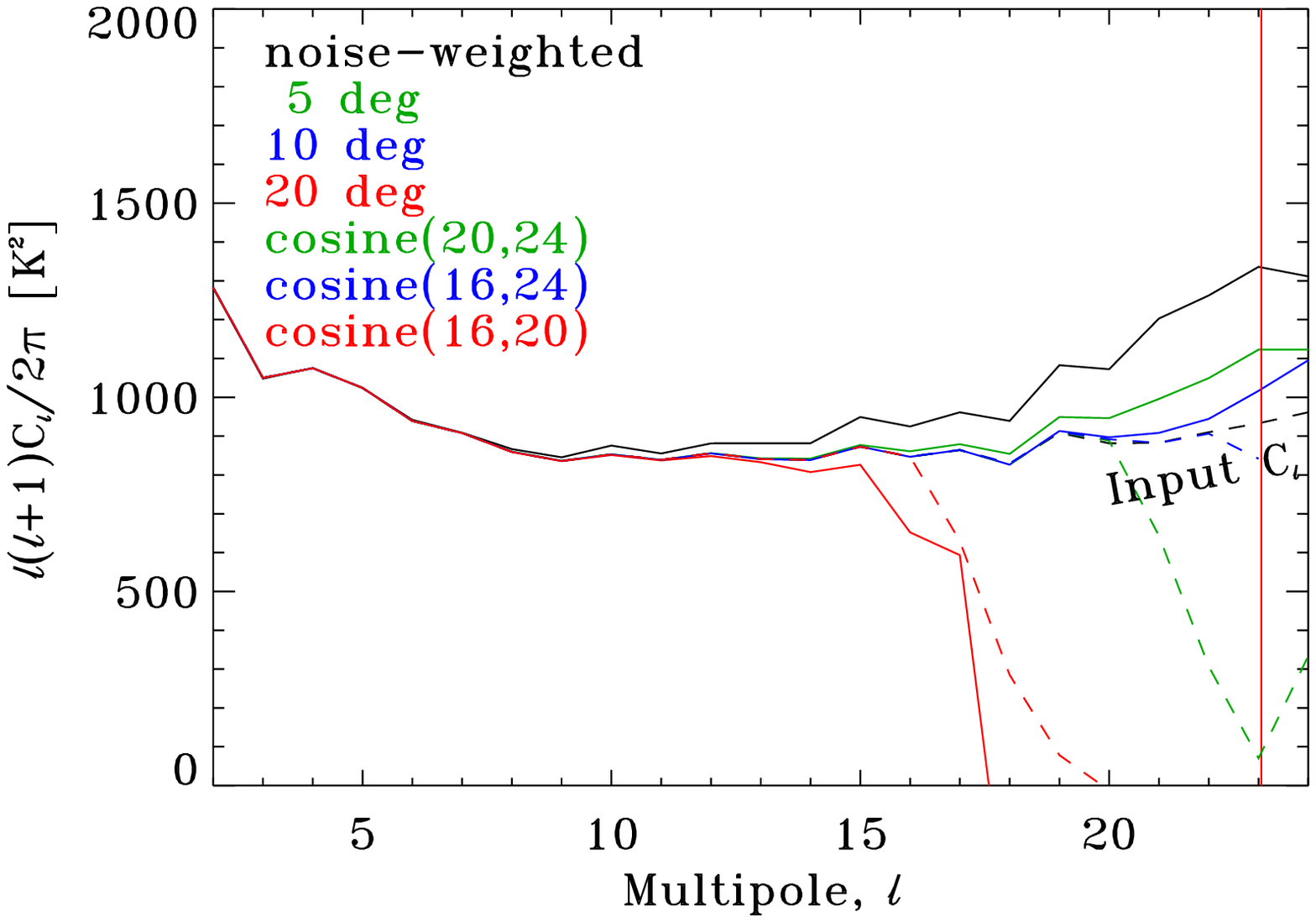}
    \includegraphics[trim=10 15 20 28,clip]
    {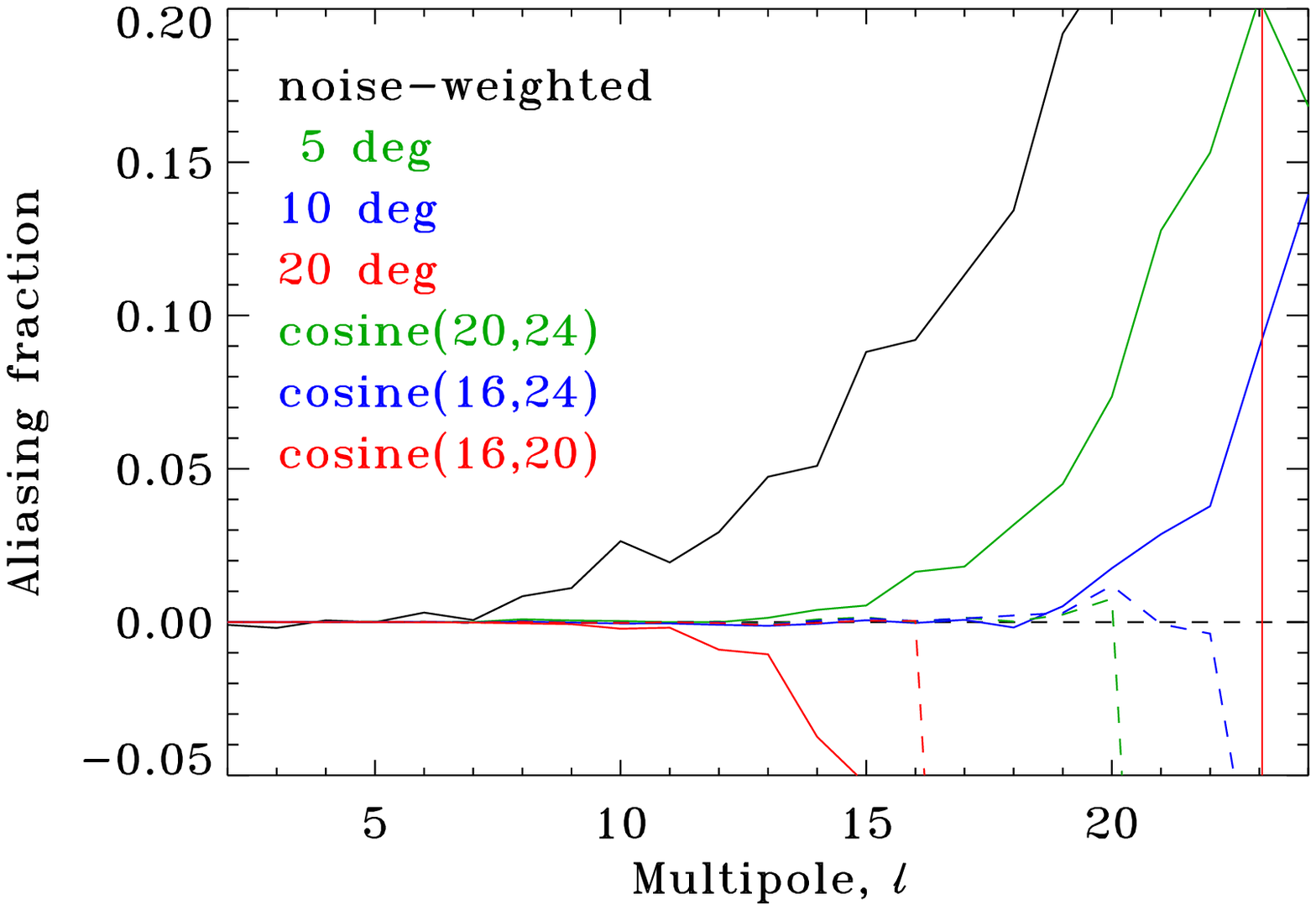}
  }
  \caption{
    Bandwidth limiting the signal using various window functions.
    In the QML method the quadratic map function is multiplied with the
    inverse Fisher matrix to produce the QML power spectrum
    estimate (see Sect.~\ref{sec:QML}). The inverse Fisher matrix can correct
    some of the aliasing effects that cause bias in the power
    spectra. For this figure the pseudo-$C_\ell$ spectra were computed
    from the full sky CMB maps. To simulate the effect of the inverse
    Fisher in QML we deconvolved our pseudo spectra with a mode coupling
    kernel that we computed from a map of ones.
    The three panels show the same curves first
    as raw estimates, then after deconvolving the smoothing and pixel
    windows and finally after subtracting and dividing by input model.
    The apodized step window functions (``cosine'') correspond to choices
    of the thresholds
    $(\ell_1,\ell_2)$ as $(20,24)$, $(16,24)$ and $(16,20)$ respectively.
    Here, solid lines are for Gaussian windows and dashed lines for the
    apodized step functions. Note that the noise weighting and direct
    low-resolution map-making produce similar aliasing effects.
  }
  \label{pic:downgrading}
\end{figure}

The CMB part of the low-resolution maps also depends on the
downgrading technique. These attempt to suppress the high-$\ell$
(subpixel) power and therefore can potentially affect the CMB angular
power spectrum even within the band of interest.
Fig.~\ref{pic:downgrading} shows the full-sky
pseudo-$C_\ell$ spectra averaged over $117$ CMB realizations,
downgraded using inverse noise weighting and a number of smoothing kernels.
Comparison of the spectra shows that it is hard to attain sub-percent
bias even at $\ell=2N_\mathrm{side}$. If these were estimates from an actual 
power spectrum estimation code with correlated noise and a sky cut,
the smoothed covariances would, however, be regularized by adding white
noise effectively leading to a considerable uncertainty already at that
multipole due to white noise and a sky cut. Methods that produce less
than $5\%$ bias for $C_\ell$ estimates at
$\ell=2.5N_\mathrm{side}$ are the Gaussian $10^\circ$ symmetric beam
and the apodized step function for
$(\ell_1,\ell_2)=(2\textrm{ or }2.5N_\mathrm{side},3N_\mathrm{side})$.

In Fig.~\ref{fig:PSE_stripes} we show the actual sky signal spectra estimated
for the low-resolution, noiseless maps. It can be seen that the estimated
band powers are not drastically affected with respect to the estimates
coming from the binned maps. Nevertheless, the case of estimates coming
from the direct low-resolution maps show some deviation from the binned
case. The bias is most prominent in the BB low multipole estimates.
We note that since the the signal covariance matrix is ill-conditioned it was
regularized by adding a small white NCM ($\sigma \simeq 1\,\mu$K) and
each signal map received a noise realization consistent with this
white NCM. 

\begin{figure}[!tbh]
  \centering
  \resizebox{\hsize}{!}{
    \includegraphics[trim=40 15 20 20,clip]{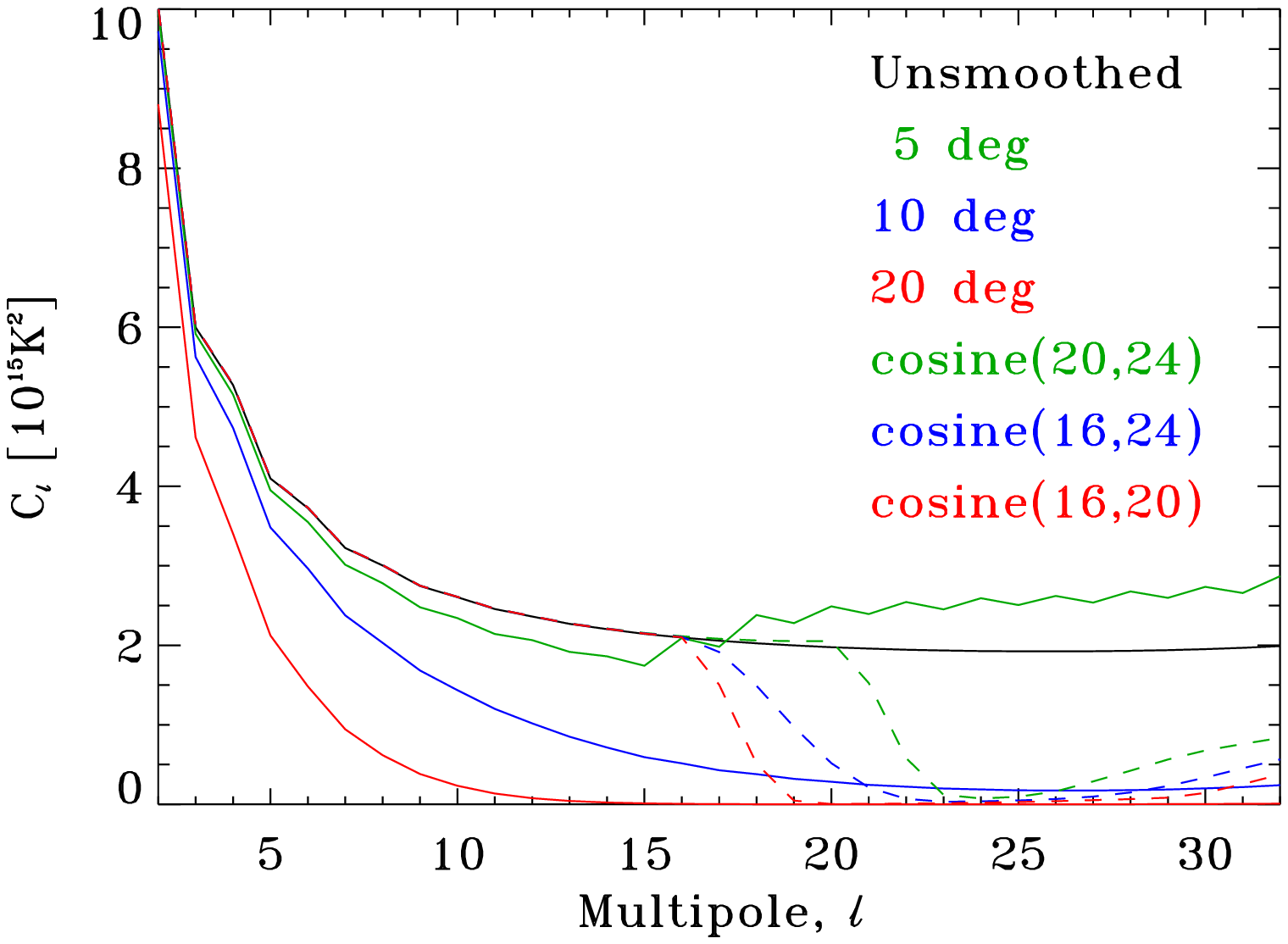}
    \includegraphics[trim=20 15 20 20,clip]{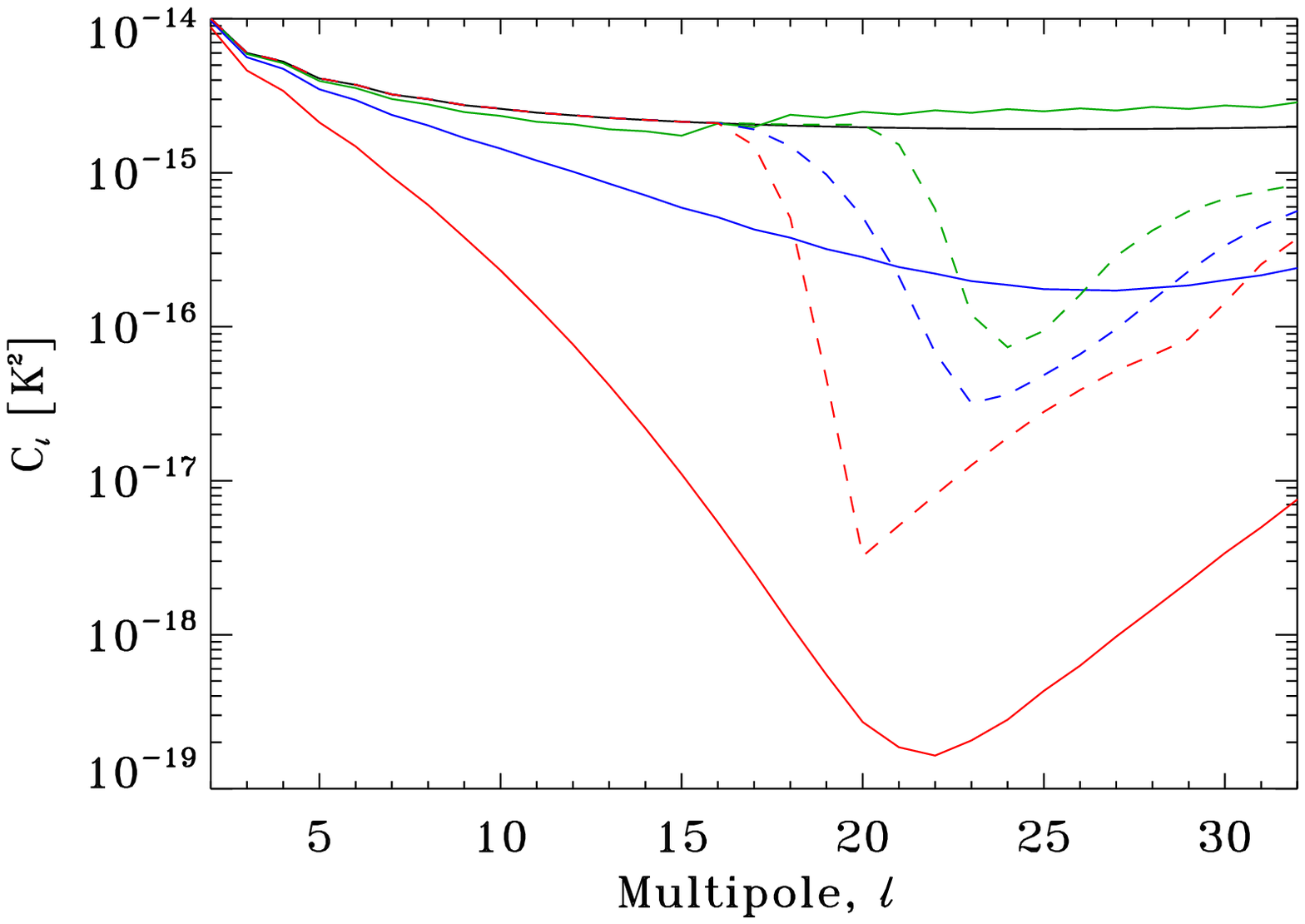}
  }
  \caption{
    TT noise bias computed from smoothed covariance matrices. Solid lines
    correspond to the Gaussian window functions and the dashed ones to
    apodized step functions. \emph{Left:} Linear vertical scale.
    \emph{Right:} Logarithmic vertical scale.
  }
  \label{pic:downgrading2}
\end{figure}

None of the proposed downgrading approaches can yield a noise level
better than the direct method. This is because noise-weighted downgrading
or smoothing both introduce departures from the optimal weighting of the noise
present in the data. The expected level of the noise is therefore an important
metric with which to compare the different downgraded maps. 
Fig.~\ref{pic:downgrading2} shows the analytical noise biases evaluated
from an $N_\mathrm{side}\mathrm =8$ Madam noise covariance matrix after
smoothing using the $6$ different beam window functions defined in
Sect.~\ref{sec:input_maps} and the unsmoothed case that is a close
match to the noise weighted maps. The
narrowest Gaussian window function having an FWHM equal to $5^\circ$,
slightly less than average pixel width, appears pathological due to
aliasing effects. The rest of the test cases are more stable but feature
a non-negligible amount of aliased power for multipole moments beyond
$\ell=3N_\mathrm{side}$ (the $C_\ell$ are not normalized with the conventional
$\ell(\ell+1)/2\pi$).

\subsection{Resource requirements}

Table~\ref{tab:resources} lists some CPU time costs for various 
$N_\mathrm{side}$, baseline length and knee frequency combinations.
Being a considerably expensive operation, we have not tested the scaling
of the optimal calculation for this particular exercise. See
\cite{madcap} for more discussion of scaling. The Madam resource cost
scales roughly linearly with respect to the number of pixels and the
so-called baseline correlation length. 

Resource requirements of the three approaches vary. As with map-making,
the destriping problem size is related to the chosen baseline offset length.
The same consideration applies also for noise covariance estimation.
Both the Madam generalized destriper and the MADping covariance calculations
scale by the length of the noise filter. ROMA does the calculation in
Fourier space and as a results scales as the the logarithm of the noise
filter length. The MADping algorithm has no dependency (ignoring communication
and final file writing) on the number of pixels being used. Although the
computation prefactors are not specified, we can see that at
$N_\mathrm{side}=32$, MADping is already more resource efficient than ROMA
(cf. rows two and three).

\begin{table}
  \begin{minipage}[t]{\columnwidth}
    \caption{
      CPU time costs for 12 70GHz detector years $=3\cdot10^{10}$ samples.
      Calculations are done on 2.6GHz Quad-core Opterons.
    }
    \label{tab:resources}
    \centering
    \renewcommand{\footnoterule}{} 
    \begin{tabular}{lcrrr}
      \hline \hline
      Case&$N_\mathrm{side}$&$w_\mathrm{corr}$\footnote{
        The filter or baseline correlation length}&PEs&CPUh\\
      \hline
      MADping         & 32 &131,073&14,000 & $80,000$\footnote{performed on an earlier, dual-core version of the machine} \\
      MADping         & 32 &  4,097 &14,000& 7,525 \\
      ROMA            & 32 & ~4,097 & ~512 &25,000 \\
      ROMA            & ~4 & 16,385 & ~256 &~~~480 \\
      ROMA            & ~4 & ~4,097 & ~256 &   410 \\
      Madam 1.25s, 50\footnote{10 and 50 refer to $1/f$ knee frequencies in mHz
      } & 64 & 1,759 & 1,024 & ~~901 \\
      Madam 1.25s, 50 & 32 & ~1,759 & ~512 &   114 \\
      Madam 1.25s, 10 & 32 & ~8,285 & ~512 &   438 \\
      Madam ~~10s, 50 & 32 & ~~~289 & ~128 &   ~42 \\
      Madam ~~60s, 50 & 32 & ~~~~84 & ~~64 &   ~34 \\
      Madam 1.25s, 50 & 16 & ~1,759 & ~256 &   ~34 \\
      Springtide, 10  & 32 & ~~~~~1 &1,024 &   256 \\
      Springtide, 10  & 16 & ~~~~~1 &1,024 &   ~51 \\
      \hline
    \end{tabular}
  \end{minipage}
\end{table}

\section{Conclusion}

We have presented the formalism and tools to compute the residual noise
covariance matrix for three map-making paradigms studied for
{\sc Planck} (an optimal method and two destriping methods).
The structure of these matrices follows from the scanning strategy but
is modulated by the underlying noise model that defines the map-making
method. The matrices were tested against Monte Carlo noise maps that were
processed from correlated noise streams into maps using MADmap, Madam and
Springtide map-making codes.

The most accurate correspondence between the covariance
matrix and the noise maps is, as expected, between the optimal map-makers,
MADmap and ROMA, and their covariance matrices and the two codes produce
nearly identical matrices. Both the generalized
(Madam) and classical (Springtide) destripers are shown to disregard some
medium frequency correlated noise that cannot be modelled by the chosen
baseline offset length. It is shown that for a low knee frequency, $10\,$mHz,
the Springtide baseline length of 1 hour is sufficient to model the correlated
noise and compute the residual noise covariance. For a high knee frequency,
$50\,$mHz, even the Madam $60\,$s baselines are too long to suffice. However,
using a short $1.25\,$s baselines (just 96 samples) the Madam results are
extremely close to optimal results even for the high knee frequency.

As a concluding test we used the matrices in actual power spectrum estimation
and verified that all methods model residual noise adequately when the
noise approximation (baseline length) is short enough to model the
correlated noise.

Resource costs of the methods vary greatly. Although both MADping and ROMA
arrive at the same result, the implementations differ and the ROMA result
scales with the resolution of the map. Both optimal implementations are
extremely resource intensive. The Madam method can be used to
produce good approximations of the optimal covariance matrices at a
fraction of their cost. The covariance matrices for these tests were
evaluated for two low resolutions, $N_\mathrm{side}=8$ and
$N_\mathrm{side}=32$. It is possible to compute the matrices up to
$N_\mathrm{side}=64$ (already $162$ gigabytes) or even up to
$N_\mathrm{side}=128$ ($2.6\,$TB) but the computational scaling of the
methods using the matrices will likely set limits to the usefulness of
such resolutions.

We studied two classes of downgrading strategies, those that make
an attempt to limit the signal bandwidth and those that do not. 
The choice of the best downgrading approach depends on both the accuracy
of the resulting noise and signal models. First measuring our ability to
compute an accurate noise covariance matrix and the second describing our
ability to control signal effects such as striping and aliasing.

All methods to produce low-resolution maps have their drawbacks.
Direct map-making at low resolution produces an unacceptable level of
signal striping that is caused by subpixel structure. Downgrading by noise
weighting biases the power spectrum through aliasing effects. The frequently
used Gaussian beam smoothing has a significant drawback of suppressing the
signal at otherwise useful angular resolutions. We find that an apodized step
function is able to retain a great deal of signal power up to
$\ell_\mathrm{max}=2N_\mathrm{side}$ but even then the power spectrum estimates
will be biased beyond $\ell=2.5N_\mathrm{side}$. However, to accurately
evaluate the noise covariance matrix for a smoothed map, we would need
to compute an usmoothed covariance matrix at the high map resolution and then
apply the same smoothing kernel to both the matrix and the map. Disregarding
this requirement leads to disagreement between the map and the matrix that
can be alleviated by combining two or more of the downgrading methods.

Of the  downgrading methods considered, we consider smoothing, with a suitable choice of the window function and possibly
a intermediate downgrading step by inverse noise weighting, to produce
the best possible low-resolution maps for power spectrum analysis.

We presented in this work a method to compute the residual noise
covariance of a smoothed, bandwidth-limited map. The method was shown to
produce an accurate description of the noise in the smoothed maps when
both the map and the matrix agree prior to smoothing. Our
method of smoothing the covariance matrix makes it possible to consider
bandwidth limited low-resolution maps and produce sub-percent level 
unbiased power spectrum estimates up to $\ell=2.5N_\mathrm{side}$.

In this work we have assumed a single frequency channel, uncorrelated noise
between detectors, noise that is white at high frequencies and full sky
coverage. In further work any of these constraints can be lifted.
The only application that we used the covariance matrix was in power spectrum
estimation. The downgrading methods that suit power spectrum analyses best
may not be optimal for different low-resolution analysis, e.g.,
study of large scale topology. A relevant future direction to explore is the
use of the covariance matrices as inputs in a likelihood code for cosmological
parameters.

\begin{acknowledgements}
  The work reported in this paper was done by the CTP Working Group of the
  {\sc Planck} Consortia.
  {\sc Planck} is a mission of the European Space Agency.
  We acknowledge the use of the CAMB (http://camb.info) code for generating
  theoretical CMB spectra.
  Some of the results in this paper have been derived
  using the HEALPix package \citep{Gorski:2004by}.
  We acknowledge the use of version $1.6.3$ of the {\sc Planck} sky model,
  prepared by the members of {\sc Planck} Working Group 2.
  This work has made use of the {\sc Planck} satellite simulation package
  (Level-S), which is assembled by the Max Planck Institute for Astrophysics
  {\sc Planck} Analysis Centre (MPAC).
  This research used resources of the National Energy Research Scientific
  Computing Center, which is supported by the Office of Science of the
  U.S. Department of Energy under Contract No. DE-AC02-05CH11231.
  We thank CSC (Finland) for computational resources.
  RK wishes to thank the Jenny and Antti Wihuri Foundation and the
  V\"ais\"al\"a Foundation for financial support. 
  This work was supported by the Academy of Finland grants 121703 and 121962.
  HKS thanks Waldemar von Frenckells stiftelse and Magnus Ehrnrooth
  Foundation for financial support.
  We acknowledge support from ASI, contract Planck LFI Activity of Phase E2.
  RS acknowledges partial support of the European Commission Marie Curie IR
  Grant, MIRG-CT-2006-036614.
\end{acknowledgements}

\bibliographystyle{aa}
\bibliography{randsvangen.bib}

\appendix

\section{Additional material} \label{app:extra}

\subsection{Noise biases} \label{app:noiseBias}

\begin{figure}[!tbh]
  \centering
  \resizebox{\hsize}{!}{
    \includegraphics[trim=50 20 5 0,clip]
    {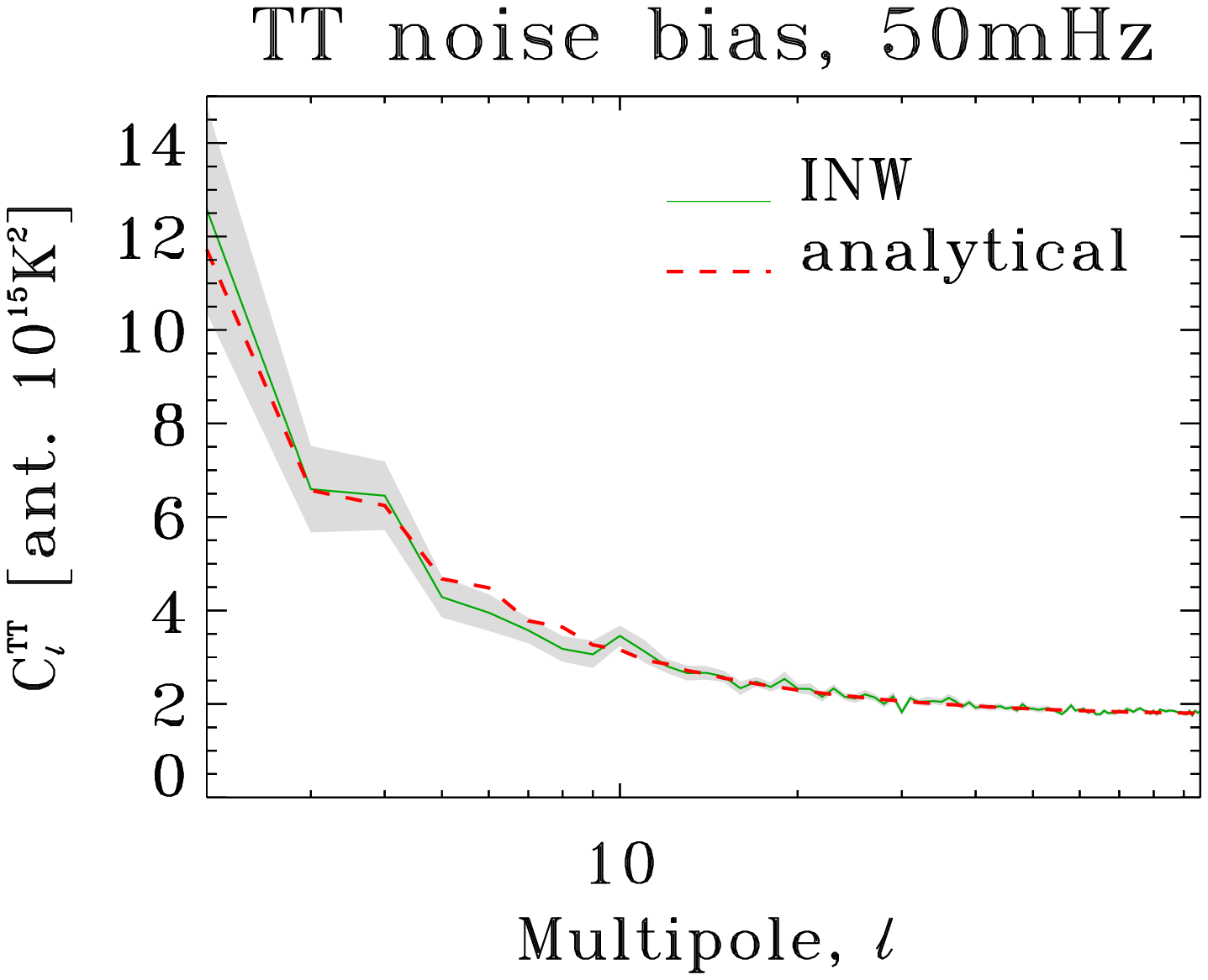}
    \includegraphics[trim=50 20 5 0,clip]
    {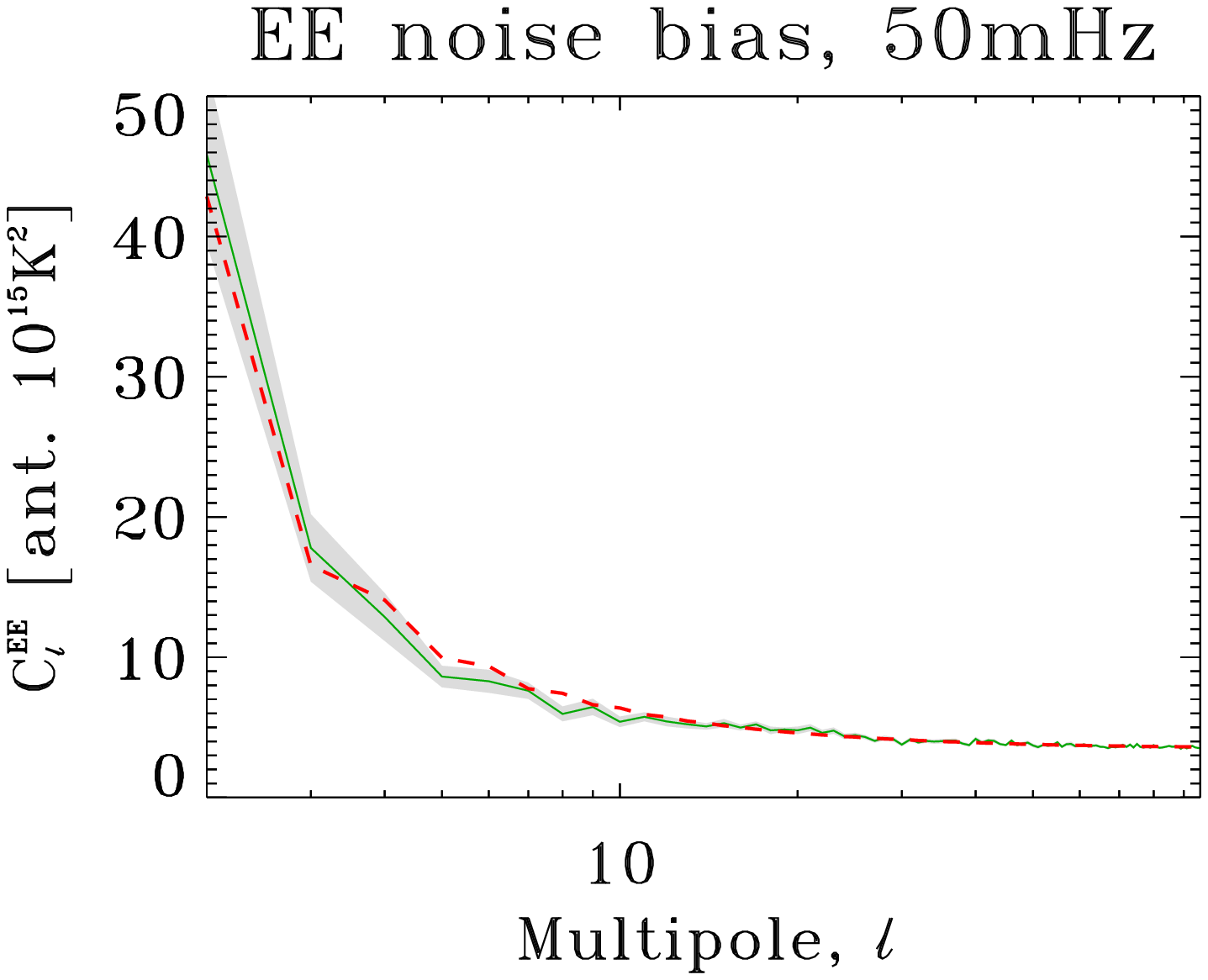}
    \includegraphics[trim=50 20 5 0,clip]
    {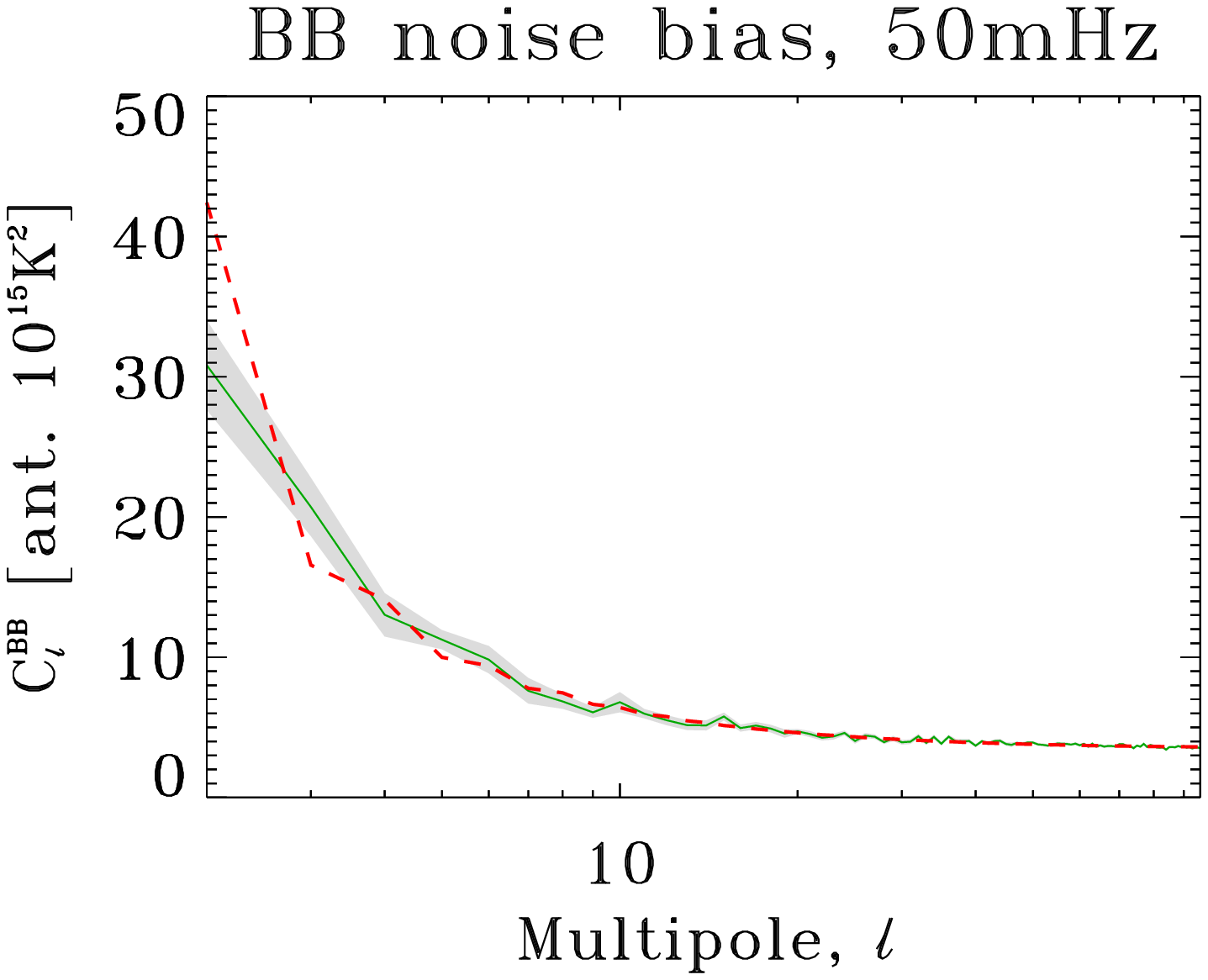}
  }
  \caption{    
    Analytical and mean Monte Carlo noise biases from MADmap runs at 
    $f_\mathrm{knee}=50$mHz. Grey band is the $1$-$\sigma$ region for the
    average, computed by dividing the sample variance by $\sqrt{25}$.
    Like the TE, TB and EB biases are both consistent with zero and
    are not shown here.
    The analytical bias corresponds to the unsmoothed case presented in
    Sect.~\ref{sec:bandwidth}.
  }
  \label{fig:madmap_mean_bias}
\end{figure}

\begin{figure}[!tbh]
  \centering
  \resizebox{\hsize}{!}{
    \includegraphics[trim=50 20 5 0,clip]
    {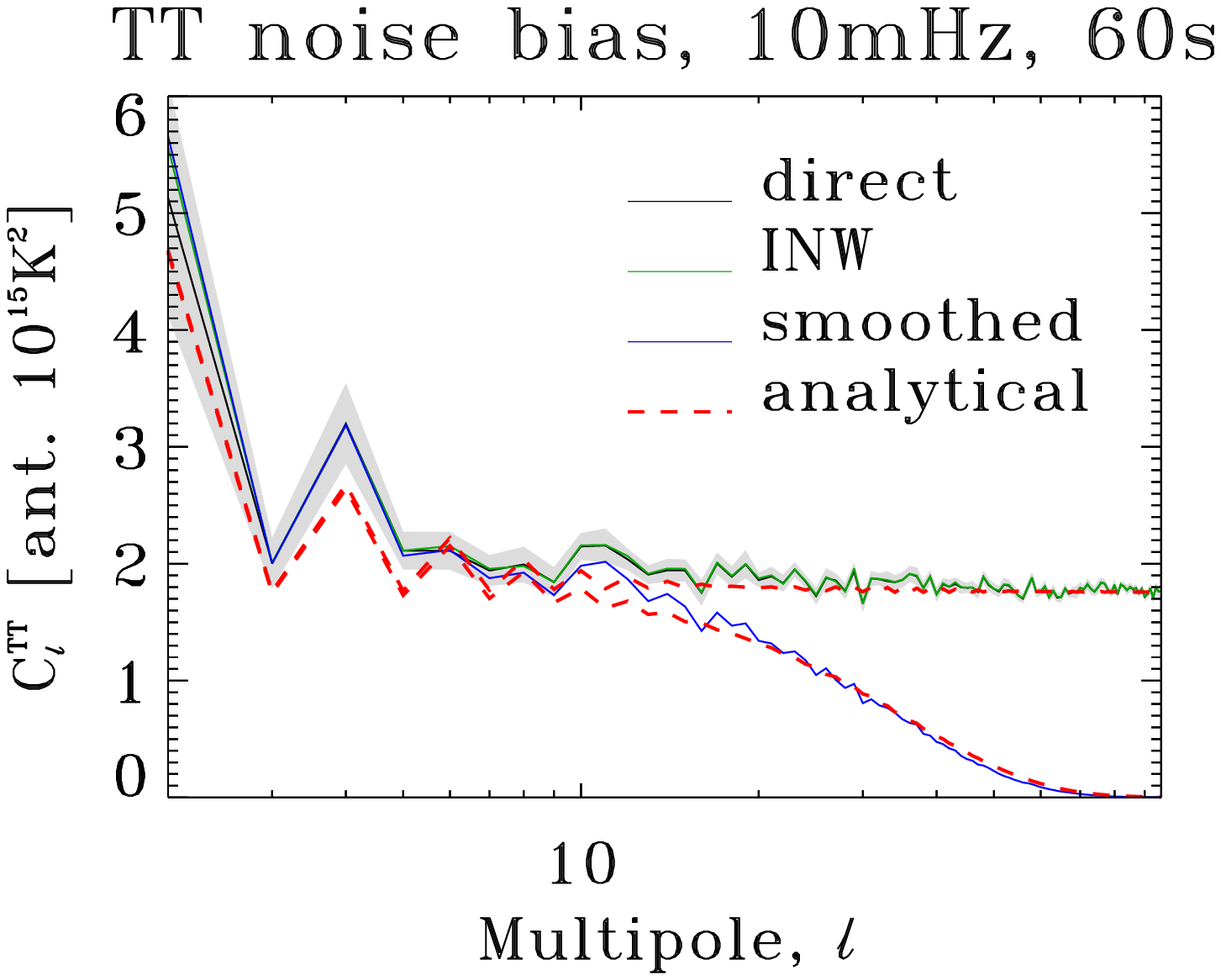}
    \includegraphics[trim=50 20 5 0,clip]
    {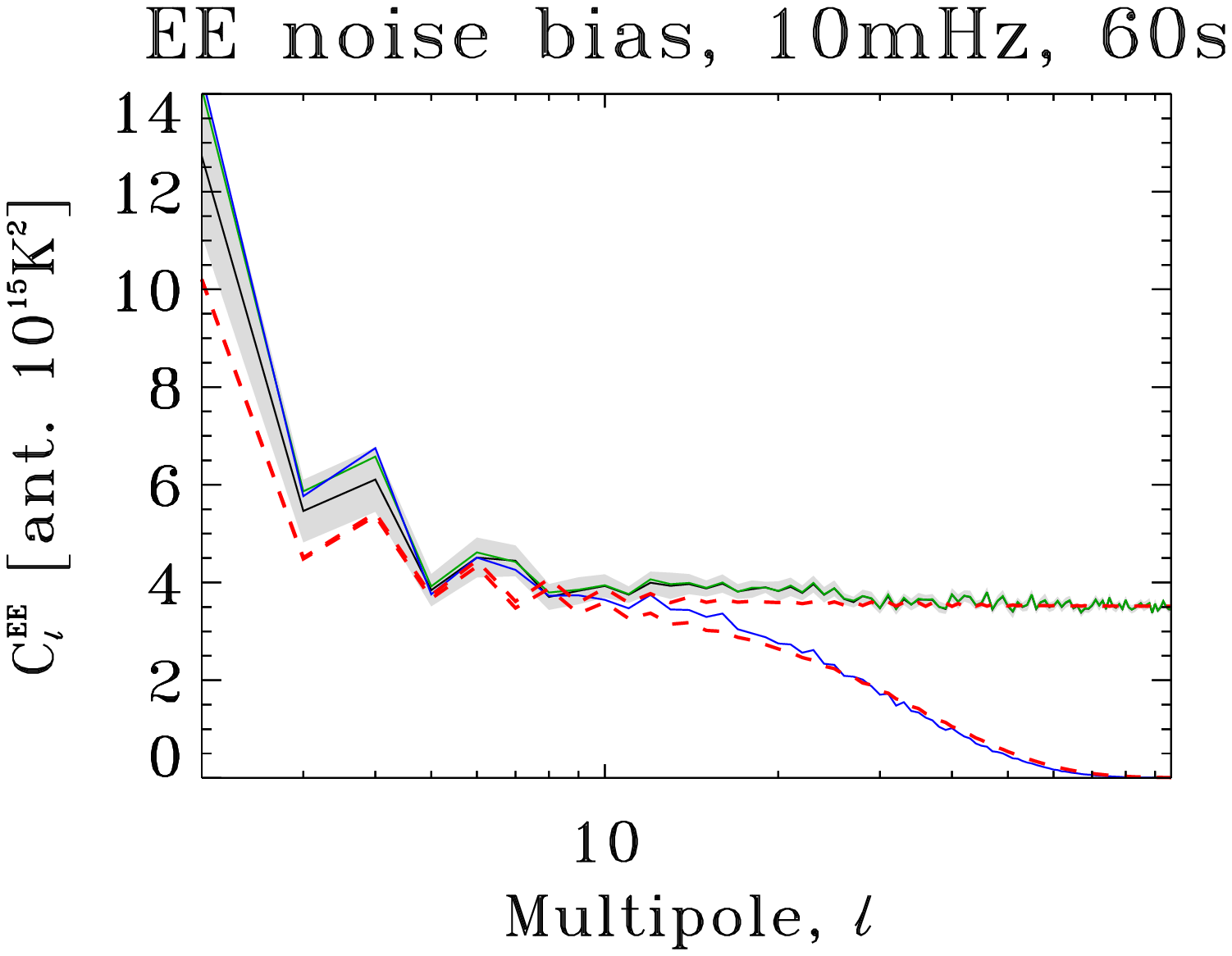}
    \includegraphics[trim=50 20 5 0,clip]
    {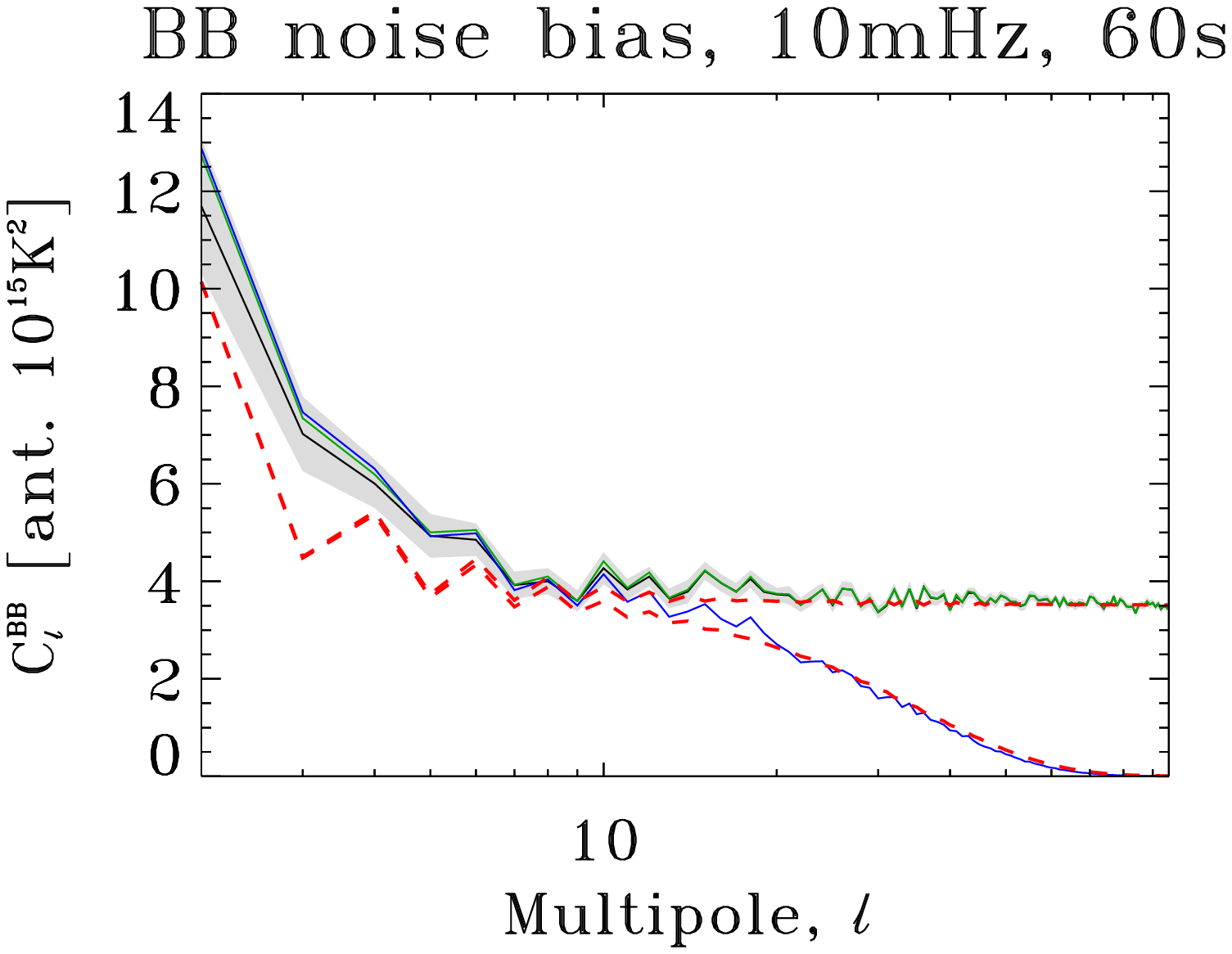}
  }
  \caption{
    Analytical and mean Monte Carlo noise biases from Madam runs at 
    $f_\mathrm{knee}=10\,$mHz.
  }
  \label{fig:Madam_mean_bias_3}
\end{figure}

For completeness, we present in Fig.~\ref{fig:madmap_mean_bias} the noise
bias computed from the MADmap NCM and the $25$ corresponding noise maps
and in Fig.~\ref{fig:Madam_mean_bias_3} the Madam bias for
$f_\mathrm{knee}=10\,$mHz and $60\,$s baseline.

\begin{figure}[!tbh]
  \centering
  \resizebox{\hsize}{!}{
    \includegraphics[trim=50 20 5 0,clip]
    {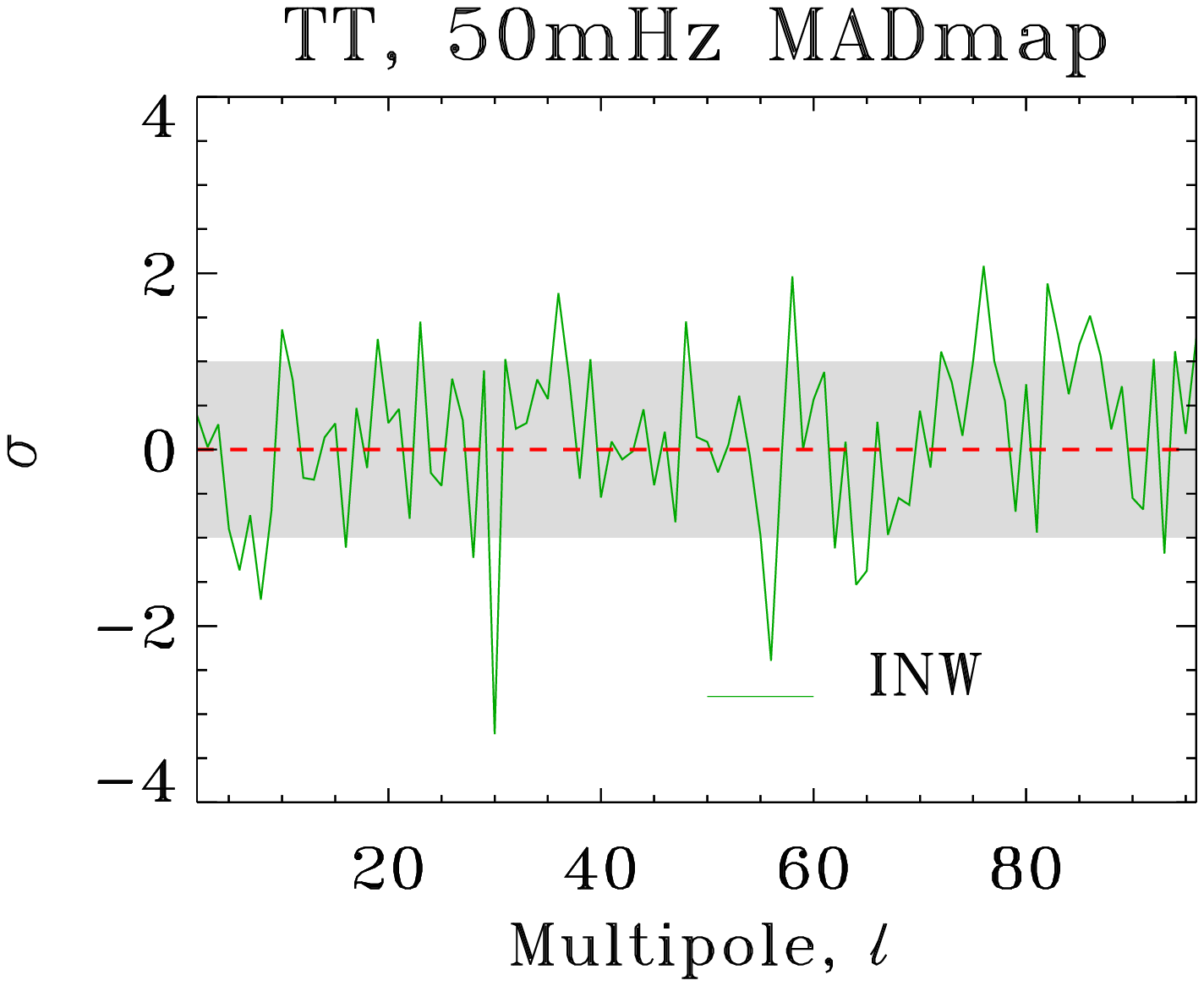}
    \includegraphics[trim=50 20 5 0,clip]
    {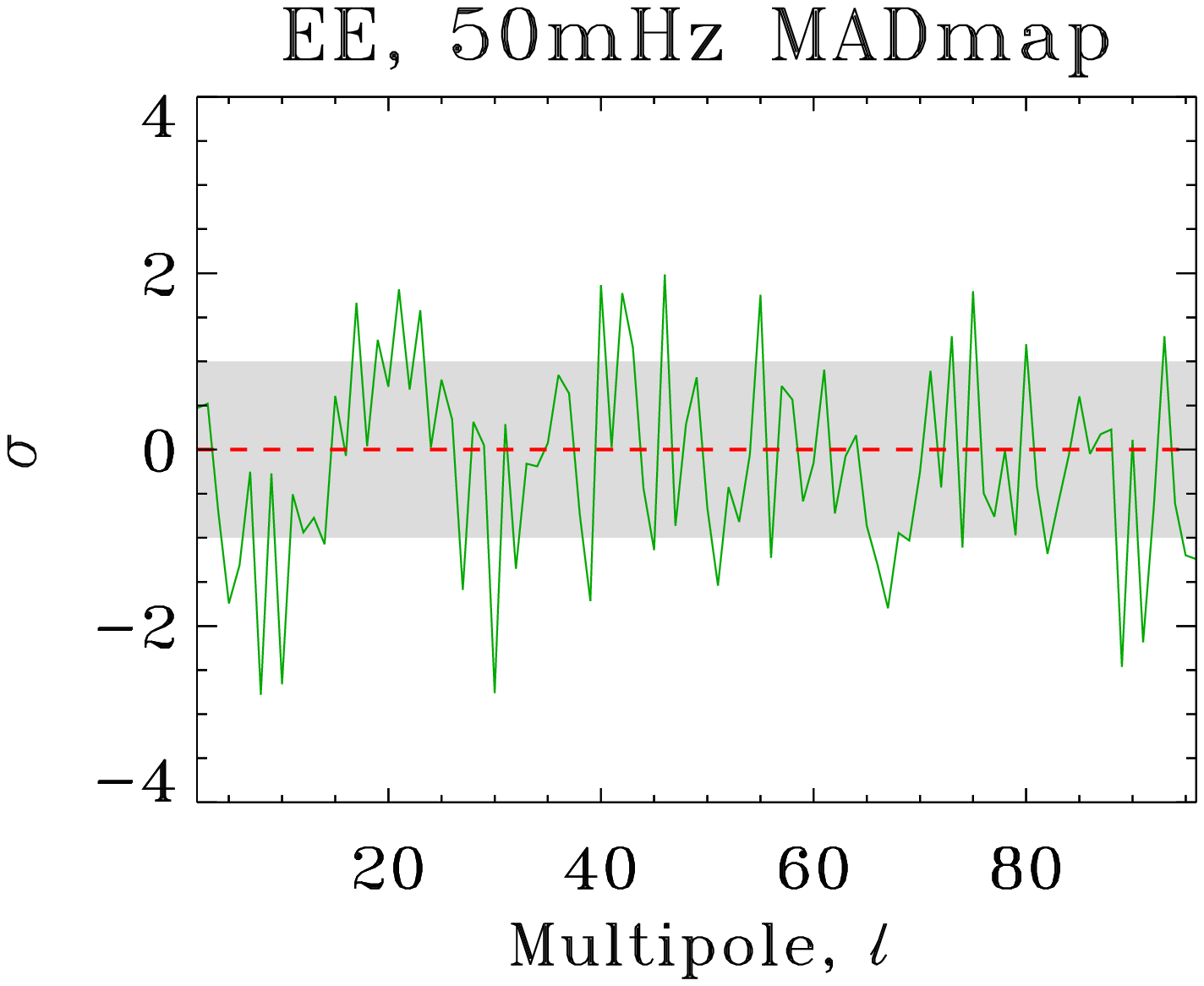}
    \includegraphics[trim=50 20 5 0,clip]
    {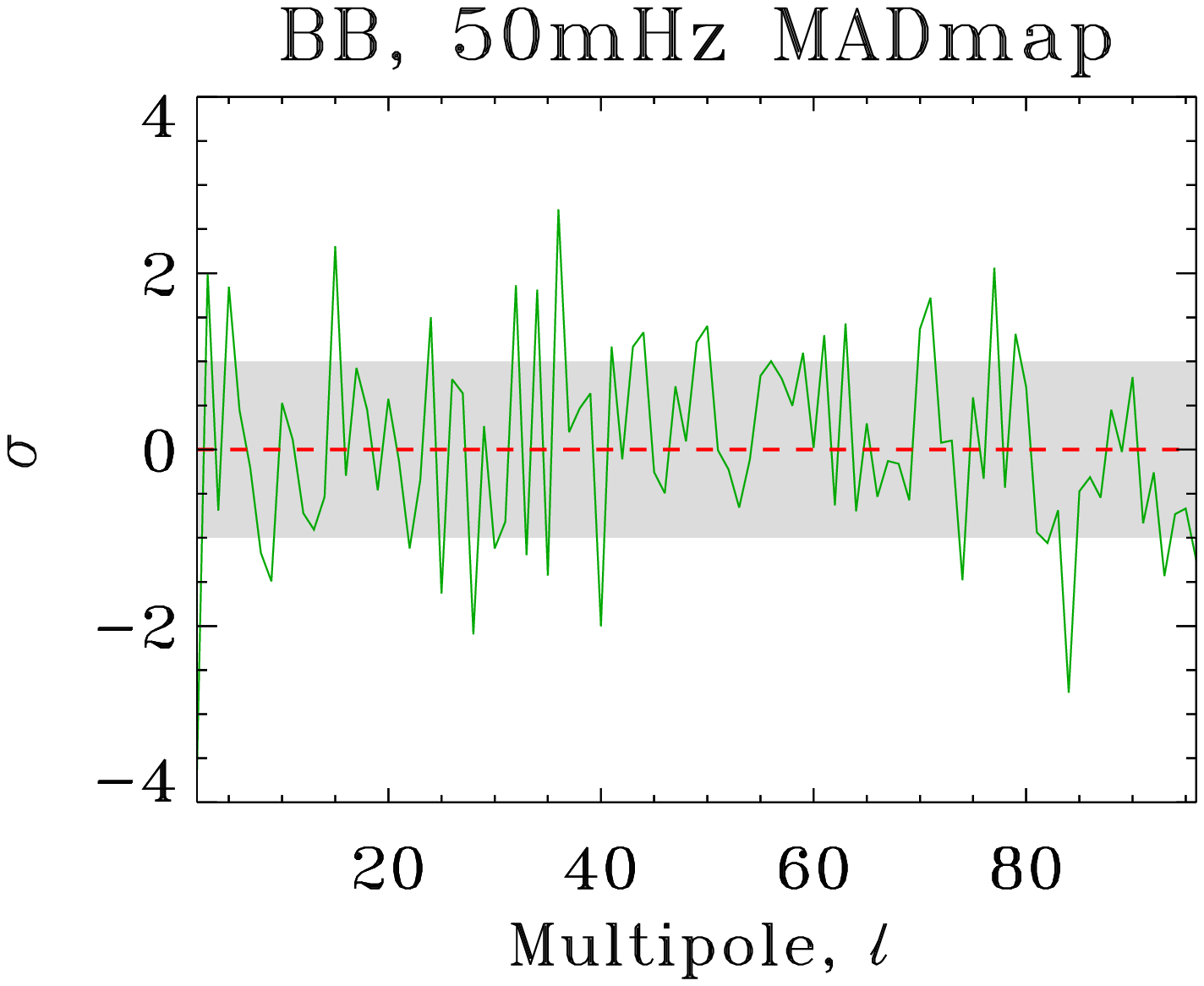}
  }
  \caption{
    Averaged noise biases after subtracting the analytical estimate
    and normalizing with the standard deviation. This plot contains
    the same curves as Fig.~\ref{fig:madmap_mean_bias}.
  }
  \label{fig:madmap_mean_bias_norm}
\end{figure}

\begin{figure}[!tbh]
  \centering
  \resizebox{\hsize}{!}{
    \includegraphics[trim=50 20 5 0,clip]
    {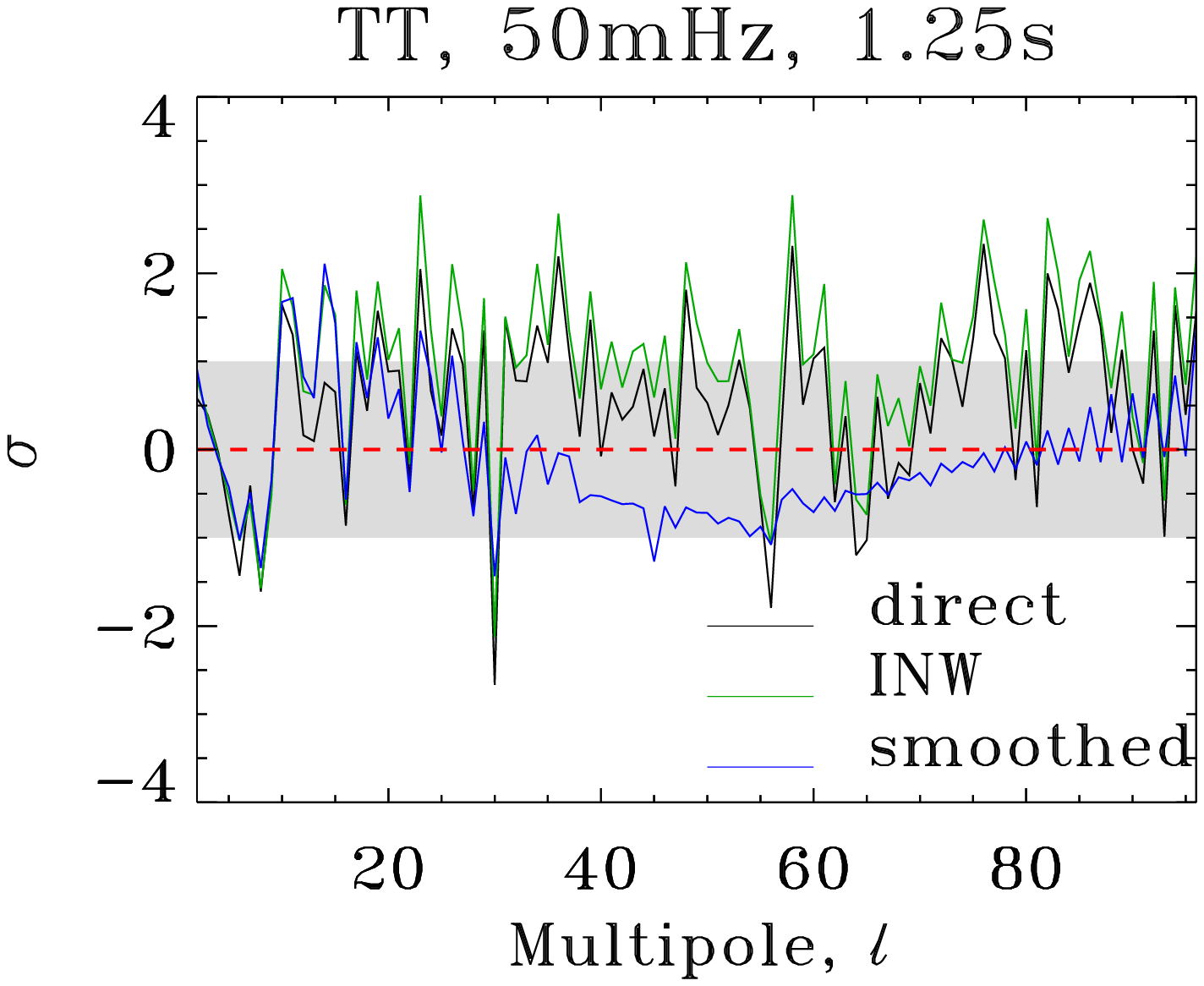}
    \includegraphics[trim=50 20 5 0,clip]
    {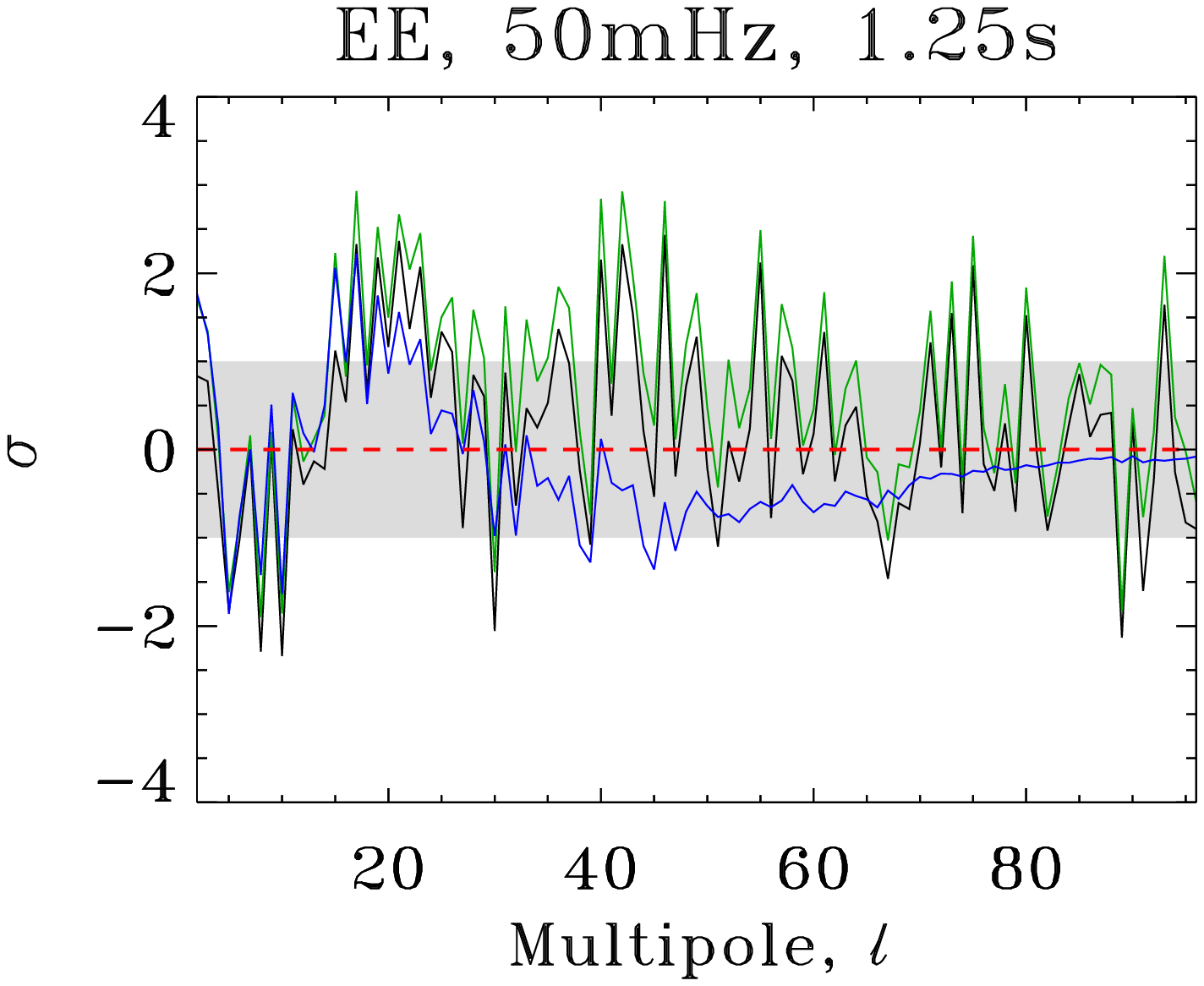}
    \includegraphics[trim=50 20 5 0,clip]
    {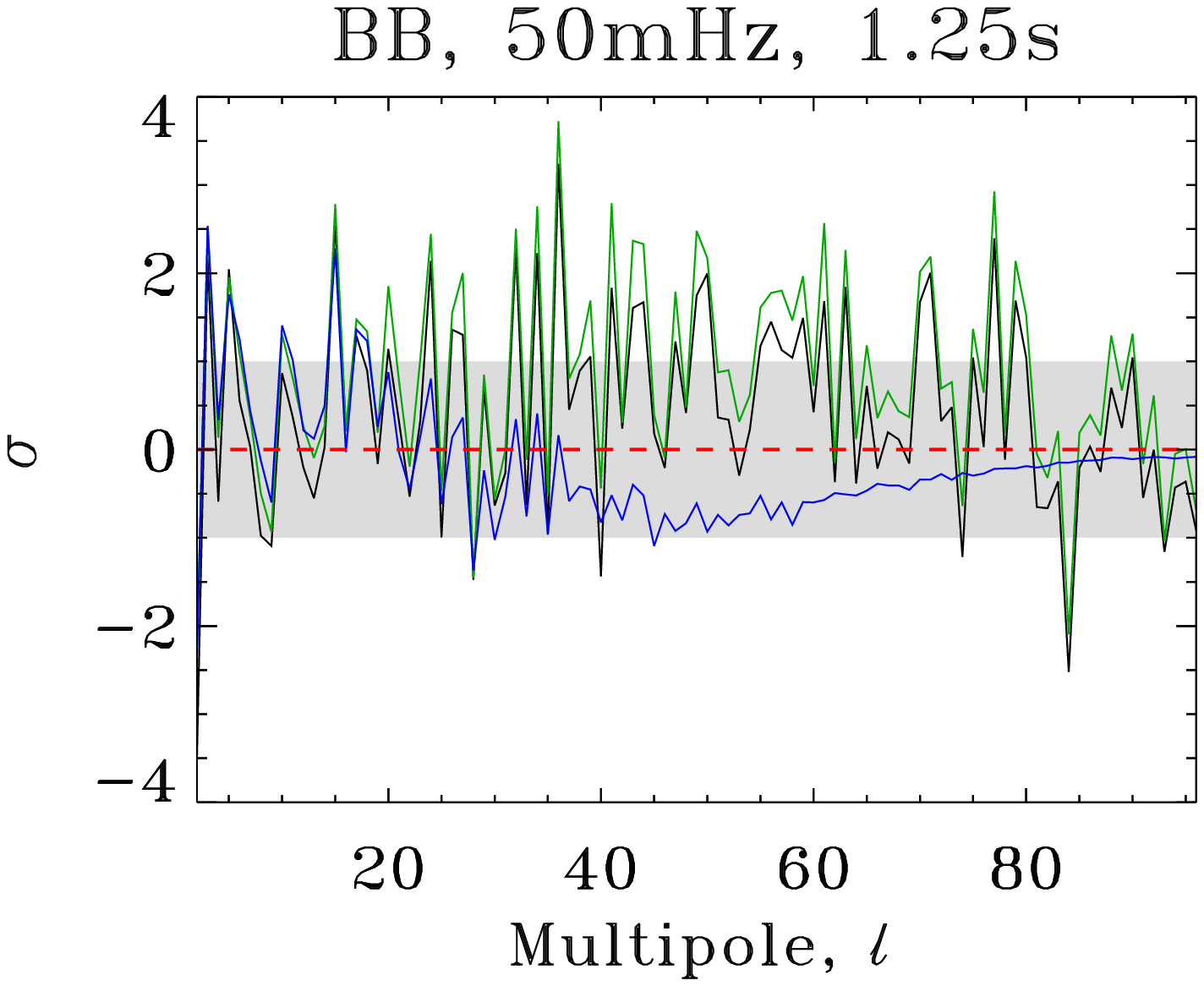}
  }
  \caption{
    Averaged noise biases after subtracting the analytical estimate
    and normalizing with the standard deviation. This plot contains
    the same curves as Fig.~\ref{fig:Madam_mean_bias}.
  }
  \label{fig:Madam_mean_bias_norm}
\end{figure}

\begin{figure}[!tbh]
  \centering
  \resizebox{\hsize}{!}{
    \includegraphics[trim=50 20 5 0,clip]
    {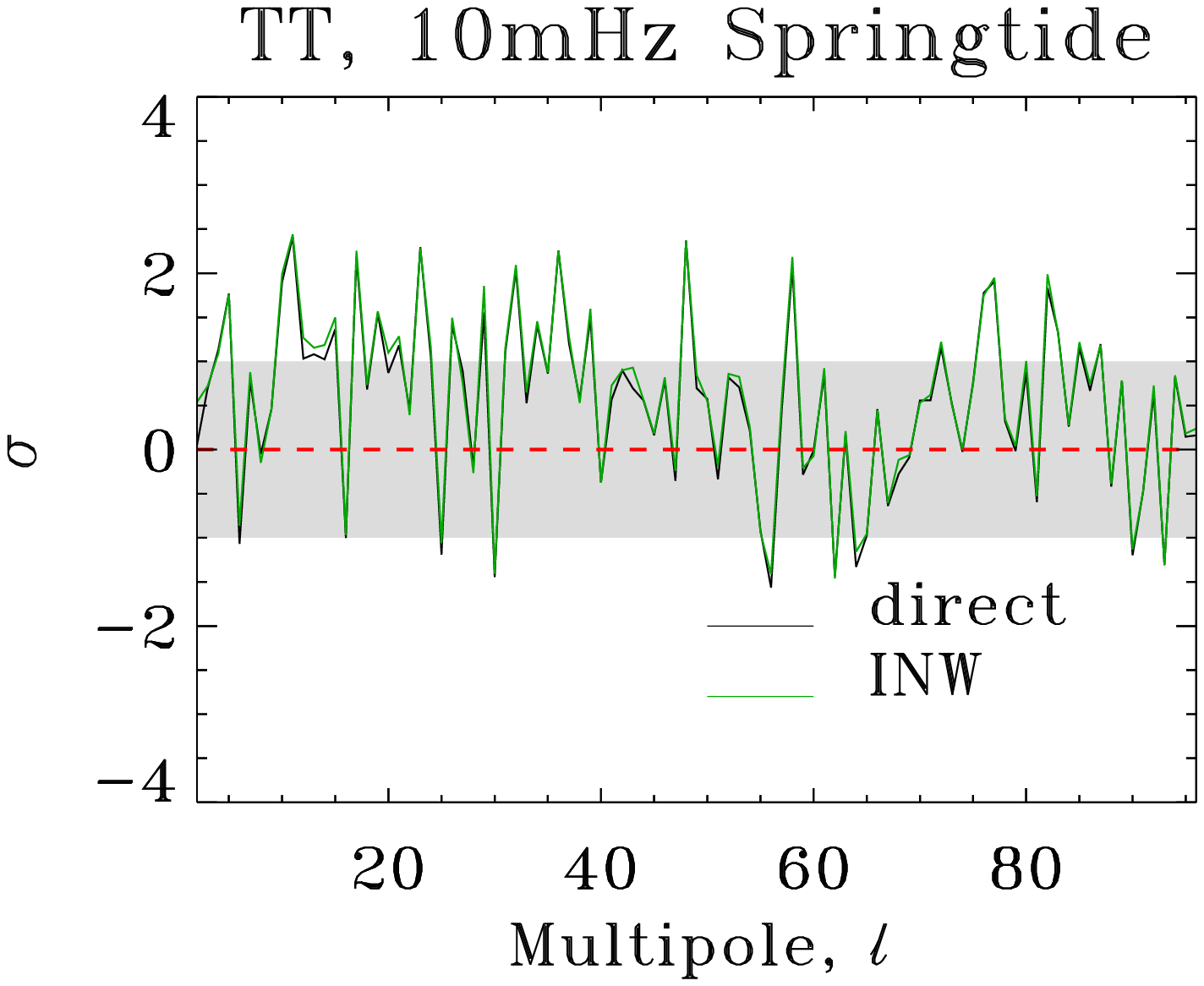}
    \includegraphics[trim=50 20 5 0,clip]
    {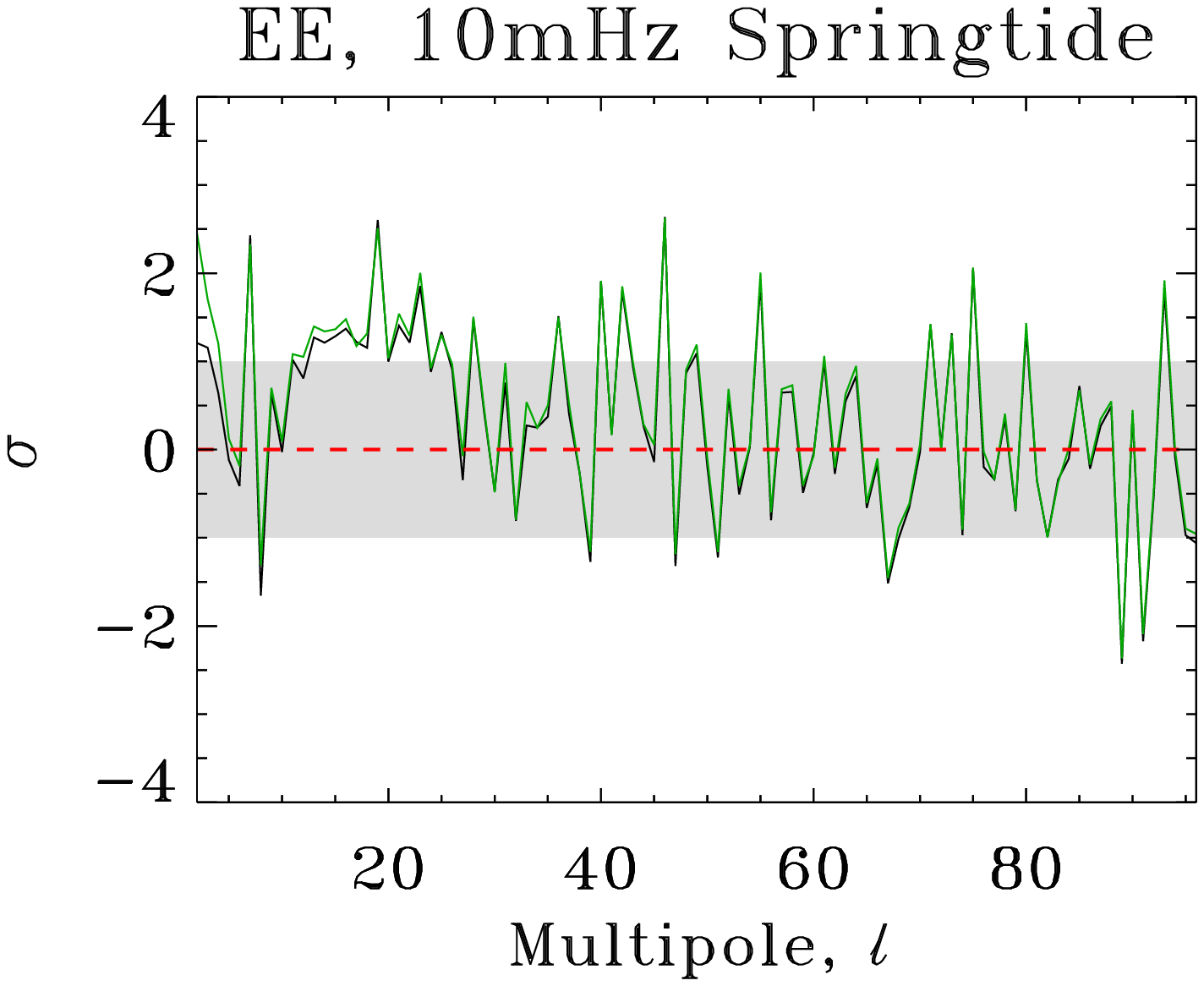}
    \includegraphics[trim=50 20 5 0,clip]
    {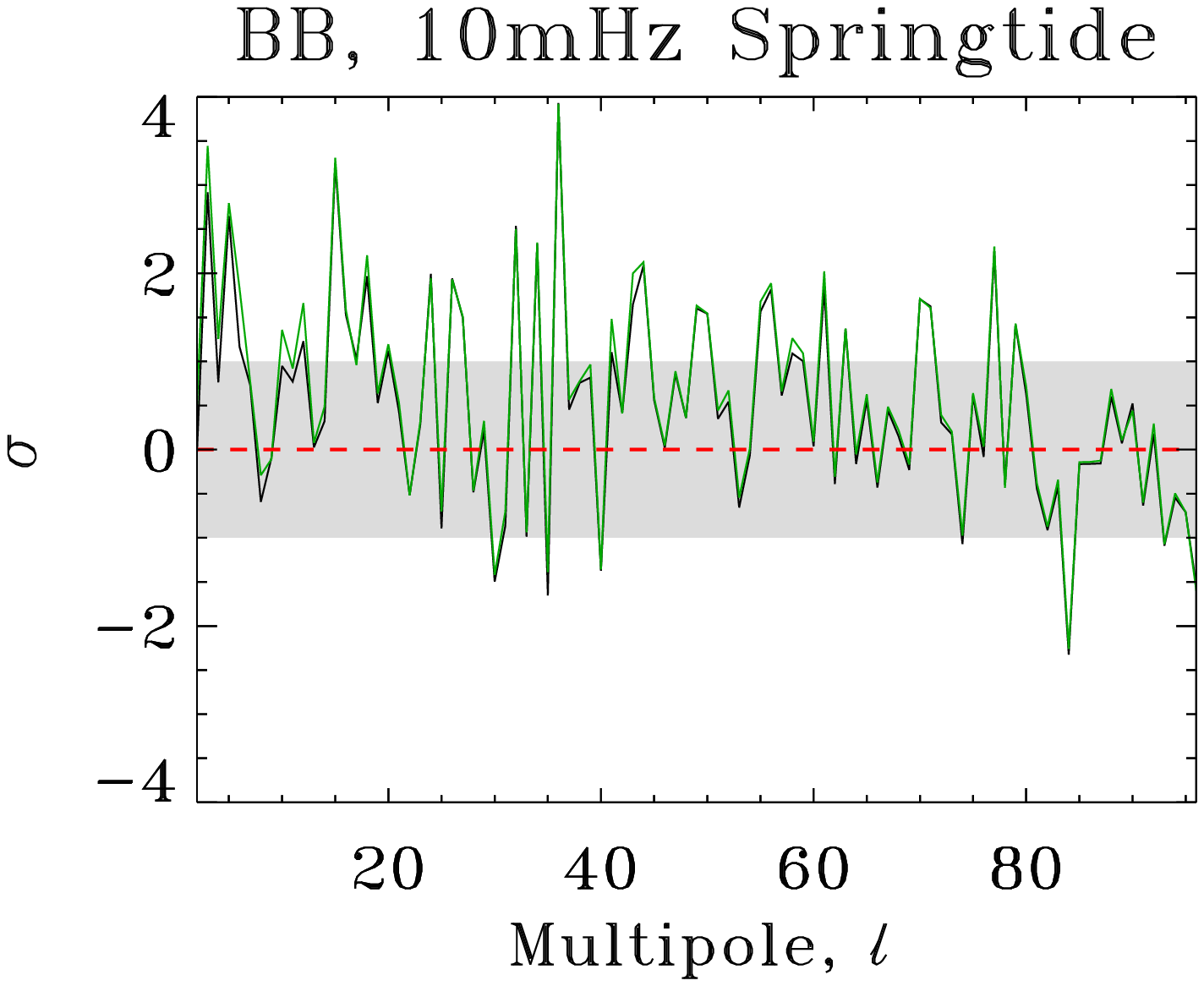}
  }
  \caption{
    Averaged noise biases after subtracting the analytical estimate
    and normalizing with the standard deviation. This plot contains
    the same curves as Fig.~\ref{fig:Springtide_mean_bias}.
  }
  \label{fig:Springtide_mean_bias_norm}
\end{figure}

To highlight differences between the estimates and simulated noise maps
we also show the fractional differences in
Figs.~\ref{fig:madmap_mean_bias_norm}--\ref{fig:Springtide_mean_bias_norm}.
These plots complete the ones presented in Sect.~\ref{subsec:noise_bias}.

\subsection{Power spectra} \label{app:ps}

\begin{figure}[!tbhp]
  \centering
  \resizebox{11cm}{!}{\includegraphics[trim=5 10 15 0,clip]
    {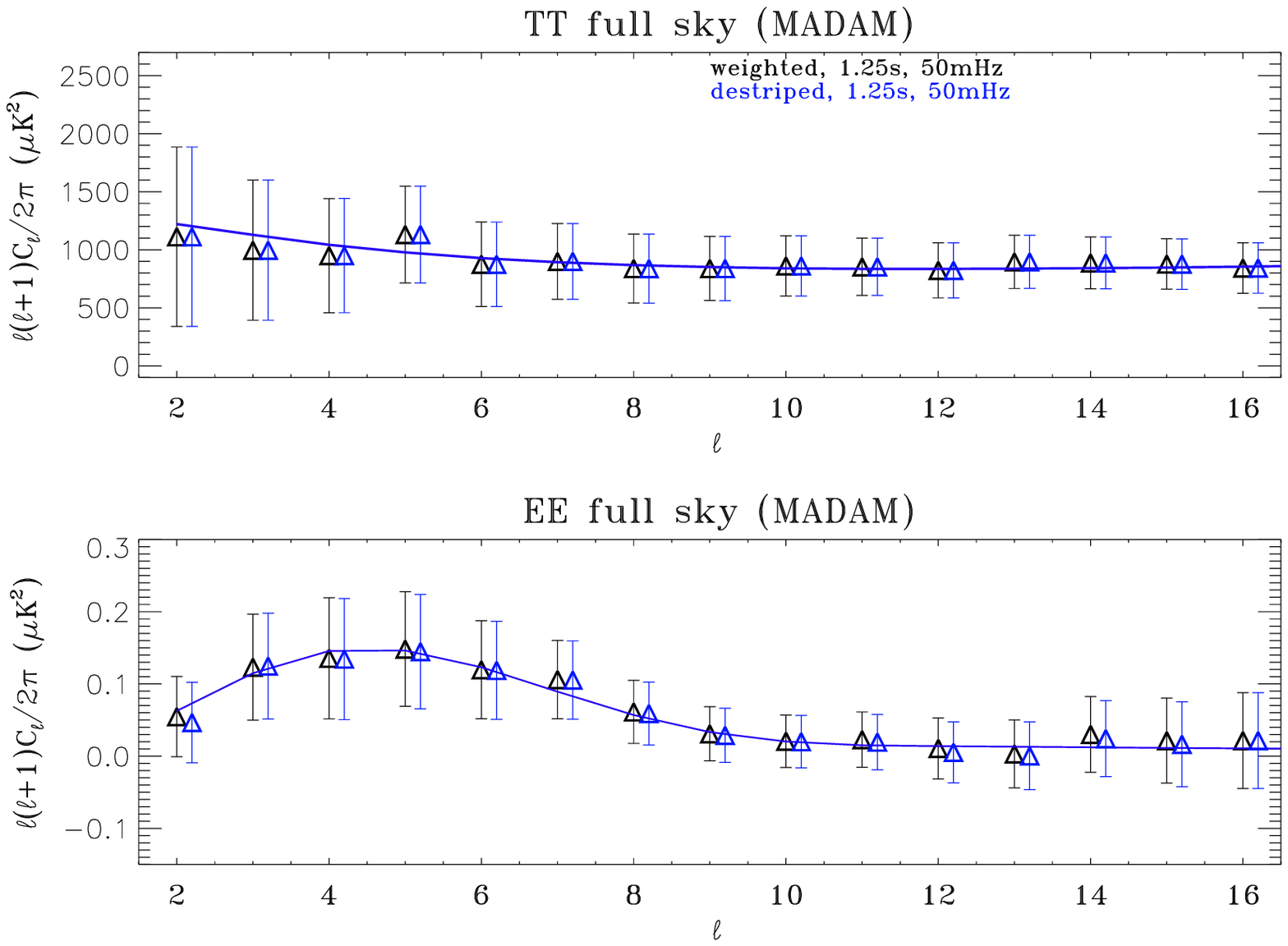}
  }
  \resizebox{11cm}{!}{\includegraphics[trim=5 10 15 0,clip]
    {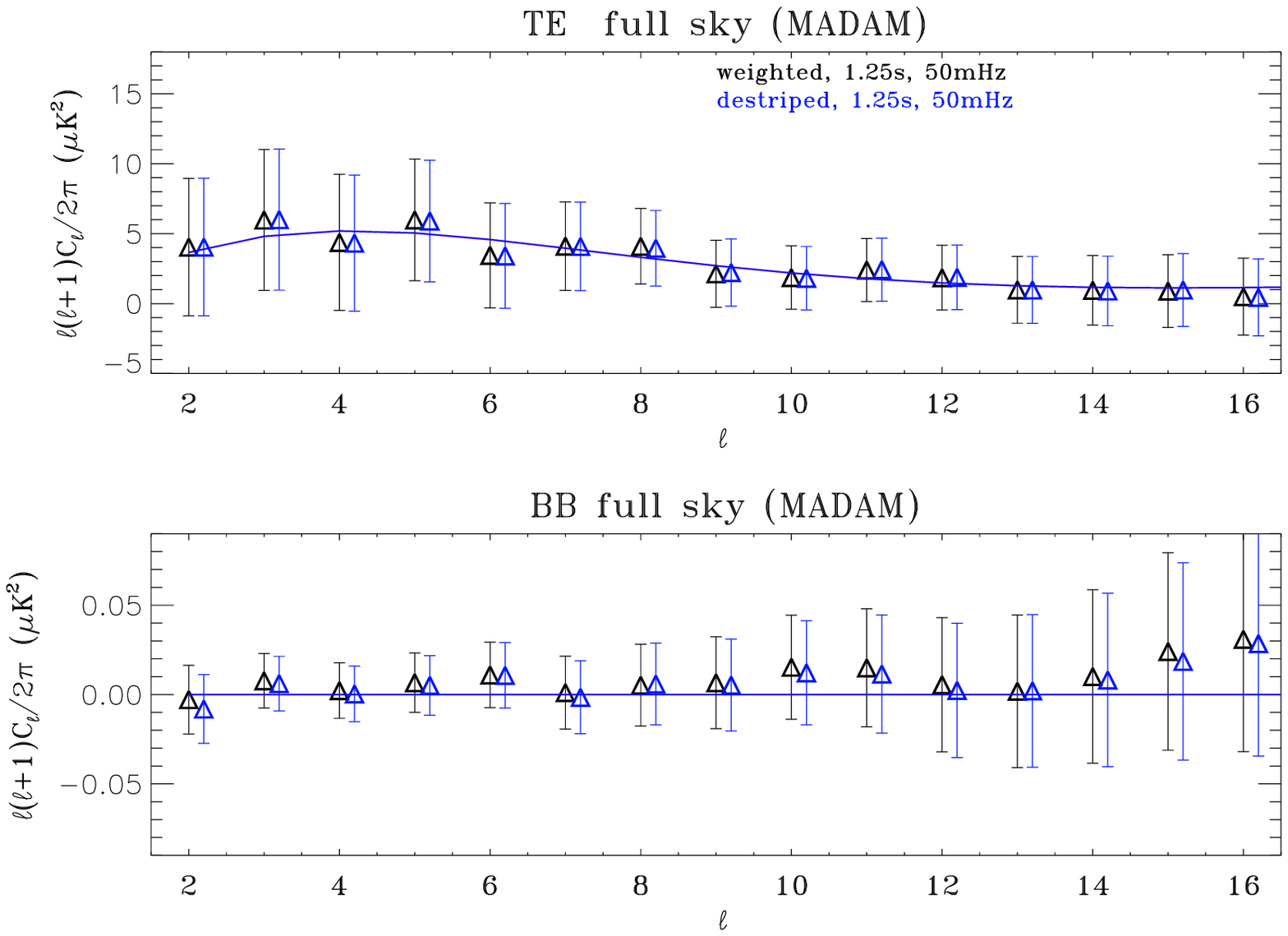}
  }
  \resizebox{11cm}{!}{\includegraphics[trim=5 10 15 0,clip]
    {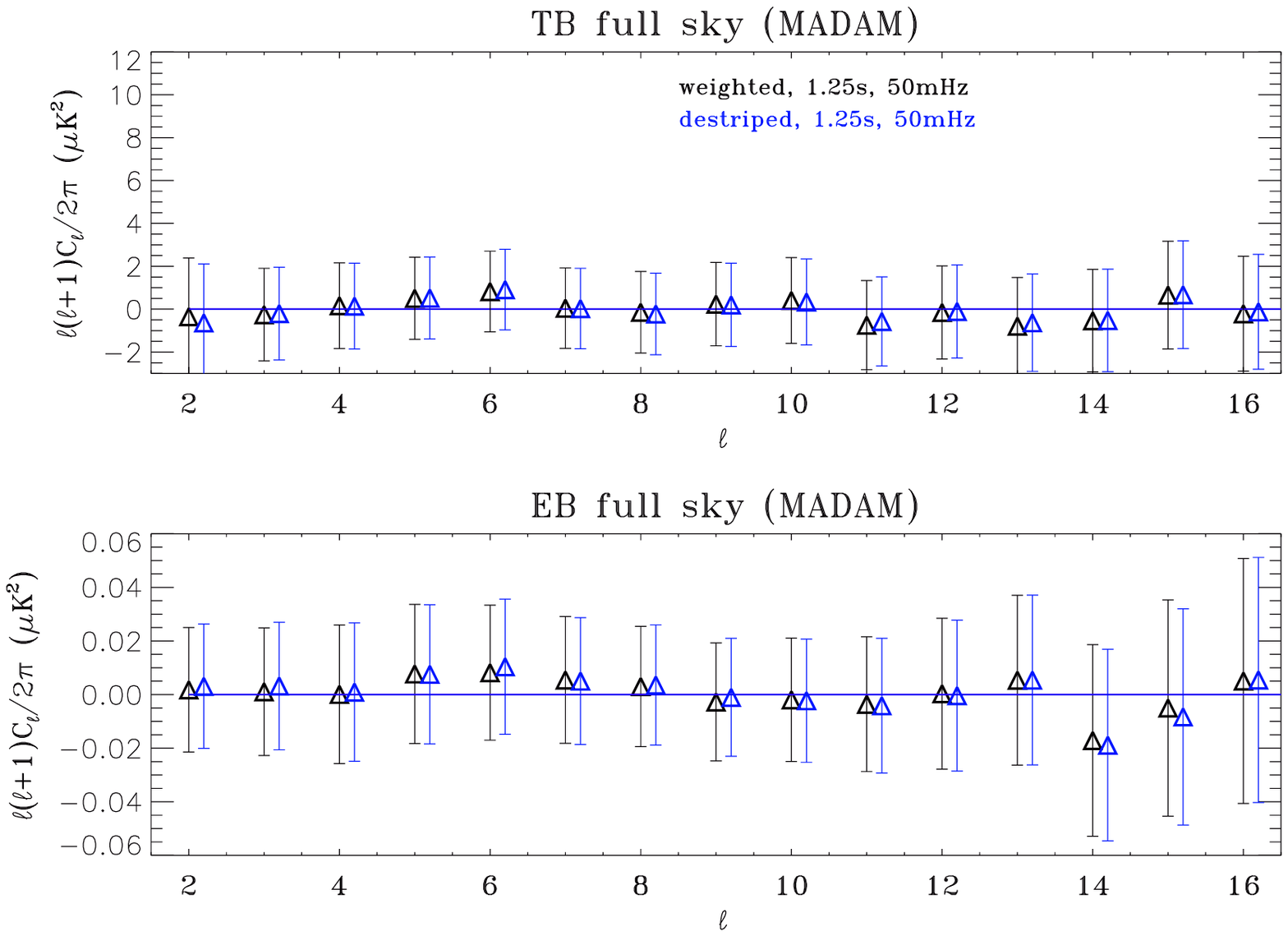}
  }
  \caption{
    Averaged power spectrum estimates over 25 noise and CMB realizations.
    The noise has a $50\,$mHz knee frequency.
  }
  \label{pic:PSE1}
\end{figure}

\begin{figure}[!tbhp]
  \centering
  \resizebox{11cm}{!}{\includegraphics[trim=5 10 15 0,clip]
    {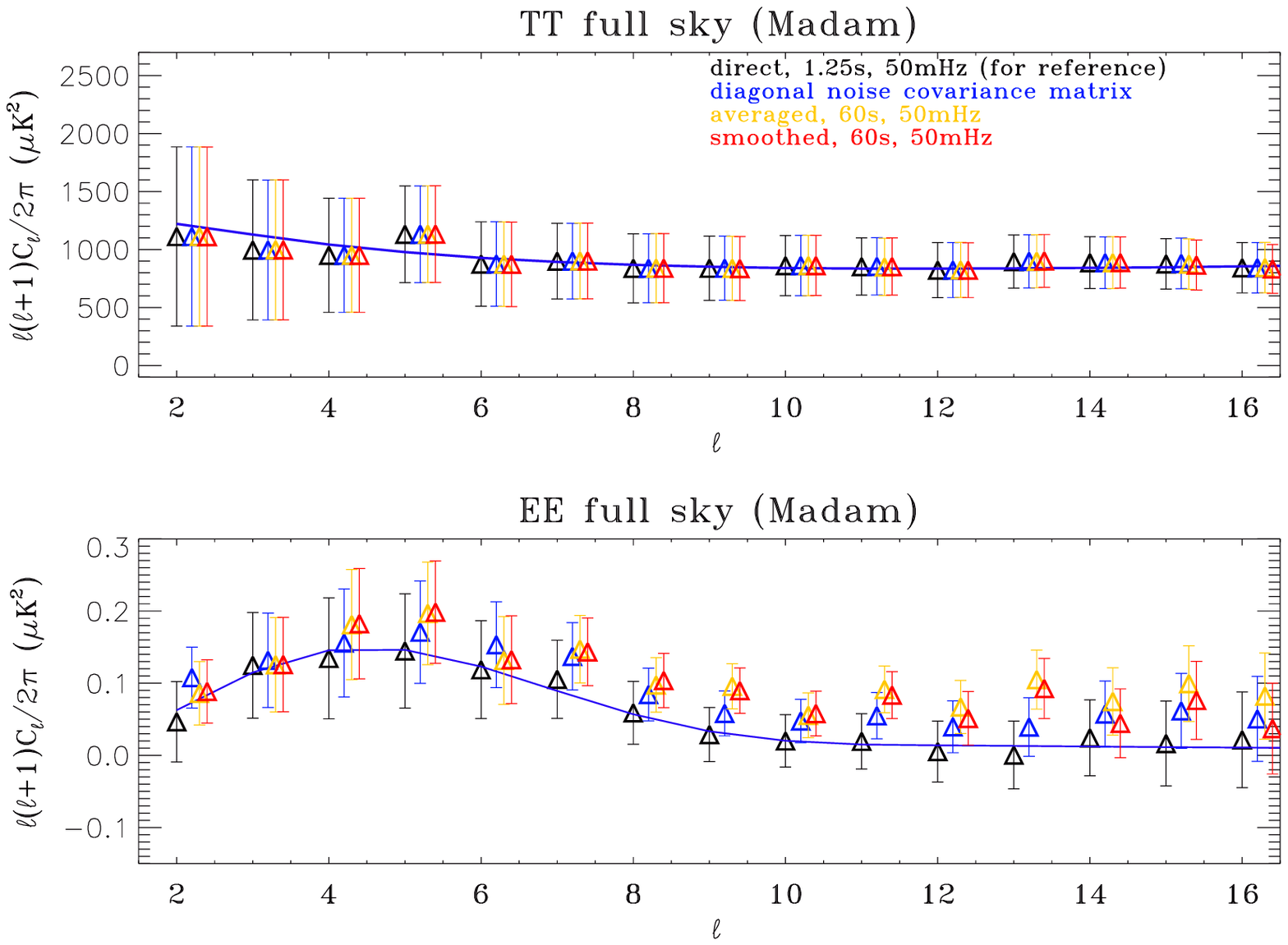}
  }
  \resizebox{11cm}{!}{\includegraphics[trim=5 10 15 0,clip]
    {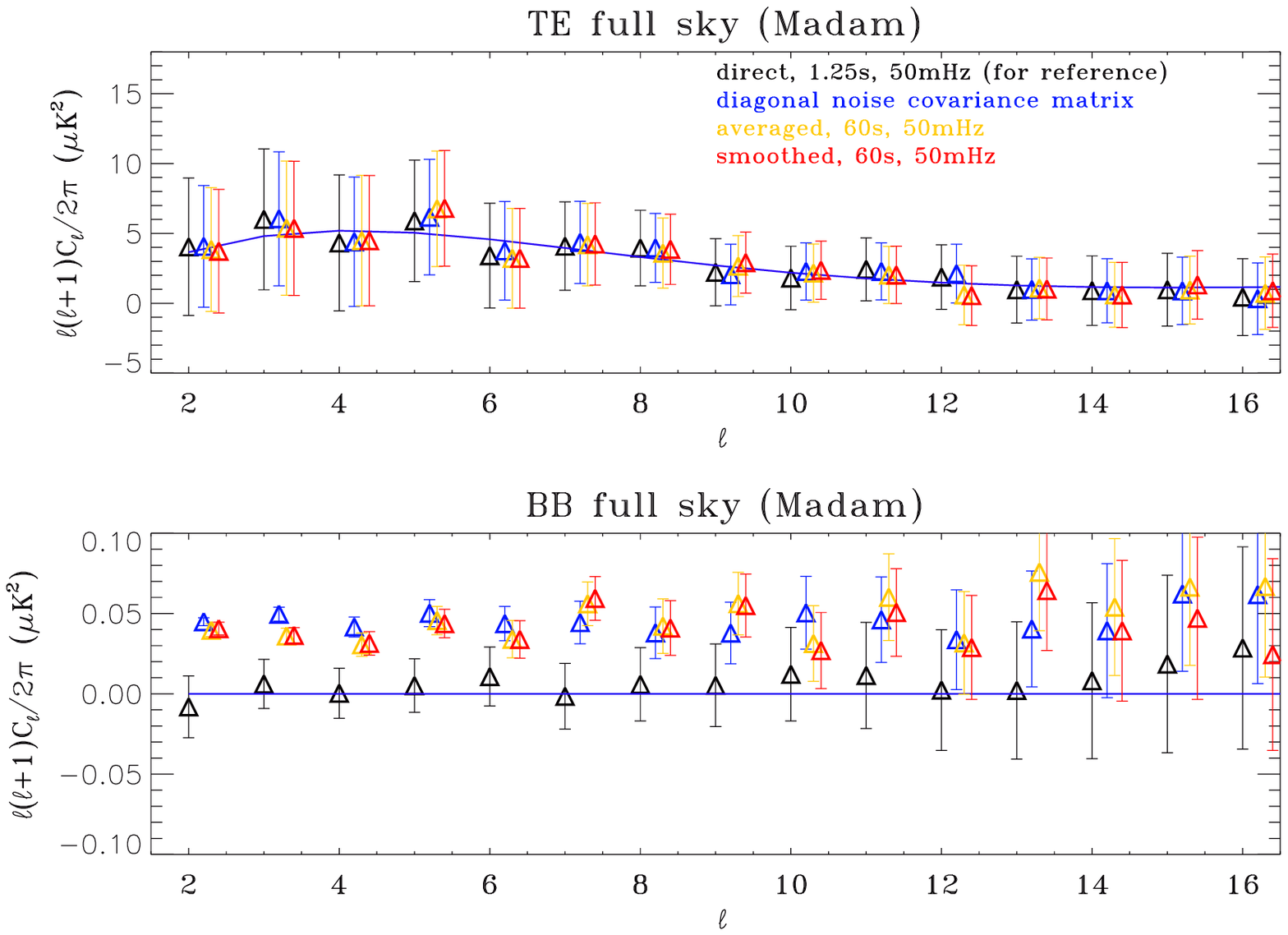}
  }
  \resizebox{11cm}{!}{\includegraphics[trim=5 10 15 0,clip]
    {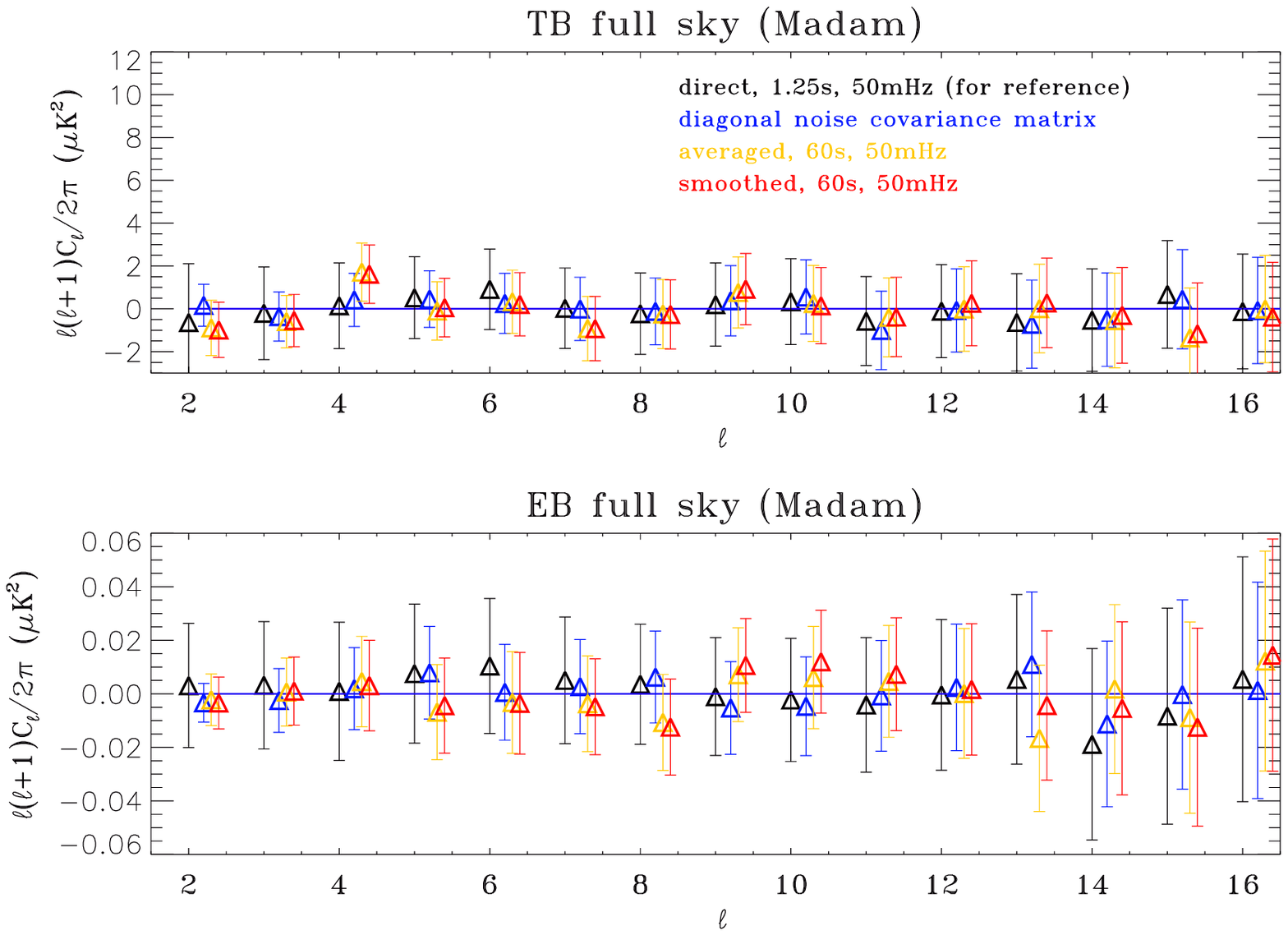}
  }
  \caption{
    Averaged power spectrum estimates over 25 noise and CMB realizations.
    The noise has a $50\,$mHz knee frequency.
  }
  \label{pic:PSE3}
\end{figure}

Here we present another successful test of the covariance matrix
used in power spectrum estimation, Fig.~\ref{pic:PSE1}. We also
show how the power spectrum estimates can be used to pick out
inaccurate residual noise covariances in Fig.~\ref{pic:PSE3}.

\end{document}